\documentclass[aps,rmp,reprint,amsmath,amssymb,graphicx,longbibliography]{revtex4-2}
\usepackage[dvipsnames]{xcolor}
\usepackage[utf8]{inputenc}
\usepackage{ragged2e}
\usepackage{url}
\usepackage{tcolorbox}
\usepackage{amsmath}
\usepackage{tabularx}
\usepackage{caption}
\usepackage{subcaption}
\usepackage{adjustbox}
\usepackage{mathtools}
\usepackage{amssymb}
\usepackage{braket}
\usepackage{bbold}
\usepackage{soul}
\usepackage{pgfplots}
\usepackage{tikz}
\pgfplotsset{compat=newest}
\usetikzlibrary{3d}
\usepackage[T1]{fontenc}

\newtcolorbox{roundedbox}[1][]{
    colback=gray!10,
    colframe=gray!50,
    rounded corners,
    boxrule=0.5pt,
    arc=3mm,
    title=#1,
    fonttitle=\bfseries,
    coltitle=black,
    toptitle=1mm,
    bottomtitle=1mm,
    left=3mm,
    right=3mm,
    parbox=false
}

\newtcolorbox{roundedboxw}[1][]{
    colback=white,
    colframe=gray!50,
    rounded corners,
    boxrule=0.5pt,
    arc=3mm,
    title=#1,
    fonttitle=\bfseries,
    coltitle=black,
    toptitle=1mm,
    bottomtitle=1mm,
    left=3mm,
    right=3mm,
    parbox=false
}

\newtcolorbox{roundedboxwide}[1][]{
    colback=gray!10,
    colframe=gray!50,
    rounded corners,
    boxrule=0.5pt,
    arc=3mm,
    title=#1,
    fonttitle=\bfseries,
    coltitle=black,
    toptitle=1mm,
    bottomtitle=1mm,
    left=3mm,
    right=3mm,
    parbox=false
}

\newtcolorbox{roundedboxfloat}[1][]{
    colback=gray!10,
    colframe=gray!50,
    rounded corners,
    boxrule=0.5pt,
    arc=3mm,
    overlay={\node[fill=gray!20,draw=gray!50,rounded corners,font=\bfseries] 
             at ([xshift=8mm,yshift=-2mm]frame.north west) {#1};}
}

\newtcolorbox{roundedboxbreak}[1][]{
    colback=gray!10,
    colframe=gray!50,
    rounded corners,
    boxrule=0.5pt,
    arc=3mm,
    title=#1,
    fonttitle=\bfseries,
    coltitle=black,
    toptitle=1mm,
    bottomtitle=1mm,
    left=3mm,
    right=3mm,
    parbox=false,
    breakable,
    break at=\textheight/3
}

\usepackage{mdframed}
\newmdenv[
    backgroundcolor=white,
    linecolor=gray!50,
    linewidth=0.5pt,
    roundcorner=5pt,
    innertopmargin=8pt,
    innerbottommargin=8pt,
    innerleftmargin=3mm,
    innerrightmargin=3mm,
    skipabove=\baselineskip,
    skipbelow=\baselineskip,
    tikzsetting={draw=gray!50,fill=white,rounded corners=3pt}
]{mdbox}

\newcommand{\mdboxtitle}[1]{%
    \noindent\textbf{#1}\par\vspace{-0.5\baselineskip}%
}

\usepackage{hyperref}
\hypersetup{
    colorlinks,
    allcolors = blue,
    linktoc=page 
    }

\begin{document}

\preprint{CERN-TH-2025-128}
\title{Krylov Complexity}
\author{Eliezer Rabinovici}
\email{eliezer@mail.huji.ac.il}
\affiliation{Racah Institute of Physics, The Hebrew University, Jerusalem 9190401, Israel}
\affiliation{CERN, Theoretical Physics Department, CH-1211 Geneva 23, Switzerland}
\author{Adrián Sánchez-Garrido}
\email{A.Sanchez-Garrido@soton.ac.uk}
\affiliation{School of Mathematical Sciences, University of Southampton, SO17 1BJ, United Kingdom}
\author{Ruth Shir}
\email{ruth.shir@uni.lu}
\affiliation{Department of Physics and Materials Science, University of Luxembourg, L-1511 Luxembourg}
\author{Julian Sonner}
\email{julian.sonner@unige.ch}
\affiliation{Jefferson Physical Laboratory, Harvard University, Cambridge, MA 02138, USA}
\affiliation{CERN, Theoretical Physics Department, CH-1211 Geneva 23}
\affiliation{Department of Theoretical Physics, University of Geneva, 24 quai Ernest-Ansermet, 1211 Genève 4, Suisse}
\date{\today{}}

\begin{abstract}
   We introduce and review a new complexity measure, called `Krylov complexity', which takes its origins in the field of quantum-chaotic dynamics, serving as a canonical measure of operator growth and spreading. Krylov complexity, underpinned by the Lanczos algorithm, has since evolved into a highly diverse field of its own right, both because of its attractive features as a complexity, whose definition does not depend on arbitrary control parameters, and whose phenomenology serves as a rich and sensitive probe of chaotic dynamics up to exponentially late times, but also because of its relevance to seemingly far-afield subjects such as holographic dualities and the quantum physics of black holes. In this review we give a unified perspective on these topics, emphasizing the robust and most general features of K-complexity, both in chaotic and integrable systems, state and prove theorems on its generic features and describe how it is geometrised in the context of (dual) gravitational dynamics. We hope that this review will serve both as a source of intuition about K-complexity in and of itself, as well as a resource for researchers trying to gain an overview over what is by now a rather large and multi-faceted literature. We also mention and discuss a number of open problems related to K-complexity, underlining its currently very active status as a field of research.\vskip2em
   CERN-TH-2025-128
\end{abstract}

\maketitle

\pagebreak

\tableofcontents

\section{Introduction: what is Krylov complexity?}\label{sect:Intro}
The term {\it complexity} has many facets and as many different meanings. Consequently, the  history of the study of complexity is almost as complex as the phenomena it attempts to understand. The idea that complex phenomena tend to emerge from simple ingredients and basic principles goes as far back as Aristotle\footnote{Other authors have even invoked the pre-Socratics in this regard \cite{aaronson2013quantum}.}, who expressed it in his `Metaphysics' in the statement that "The whole is more than the sum of its parts" (Metaphysics, Book VIII). In comparison the modern theory of complexity is a rather young subject, and it is only with the advent of the computing age that the number of studies referring to one or another notion of complexity has waxed significantly\footnote{See for example the non-exhaustive list of complexity measures  \href{https://web.mit.edu/esd.83/www/notebook/Complexity.PDF}{here}.}.

In this review we will mainly focus on the properties of a special form of measuring complexity, Krylov complexity or ``K-Complexity''. It is named after the Imperial Russian (and later Soviet) naval engineer and General of the fleet, Aleksey N. Krylov, who produced deep and long-lasting contributions in applied mathematics, including pioneering computational methods in linear algebra that today bear his name, and which inspired the notion of complexity explored herein.
We will highlight the reasons we think that it is a measure of complexity that is both fascinating and fruitful in its own right and will demonstrate that it has important applications to gravity, in the context of the AdS/CFT correspondence. 
We will start from a bird’s eye view of some topics in quantum information and high-energy physics and continue on to provide a detailed exposition of the tools needed to study aspect of chaos and integrability with the help of K-complexity. We will discuss how the AdS/CFT correspondence allows to geometrize complexity, and in particular K-complexity, in a natural way. At various points, when we supply material that is useful background for the main thread of the discussion we display it in a titled box with light grey background. The grey boxes can be skipped by a more experienced Reader, and serve as additional information for the Curious. We will also display some key results and methods in titled boxes, but this time with white background. The white boxes form part of the main text, and are meant as a visual guide to key results and methods.

This review is structured as follows. We begin with two introductory sections, the first one, Section \ref{sect:Intro} gives an overview of the notion of complexity in general, touching on computational and algorithmic complexity, gate complexity, Nielsen complexity, and Krylov/Lanczos methods in particular. We give intuition of how Krylov complexity significantly extends the idea of size complexity and use this to relate Krylov complexity to physical notions of operator growth in the Heisenberg picture, and state evolution in the Schrödinger picture. We introduce basic concepts in quantum chaos, such as information scrambling, and couch the Krylov complexity in these notions. Section \ref{sec.HoloComplexity} develops the necessary background to formulate the expectations of `holographic complexity' which embeds the notion of complexity in the theory of gravity. We review the basic notions of the so-called holographic duality between gravity in anti-de Sitter space and its boundary dual description, and explain the intuition behind the holographic complexity conjectures that will feature prominently in Section \ref{sect:Holography}. Section \ref{sec.CloserKrylov} is fully dedicated to the details of Krylov complexity, as well as the Lanczos algorithm underpinning it. We describe the universal operator growth hypothesis, and review foundational optimality theorems distinguishing Krylov complexity as an ideal complexity in a certain sense. We prove a number of standard results bounding Krylov complexity and review efficient and powerful numerical and analytical methods to study it in general quantum systems. We finish Section \ref{sec.CloserKrylov} by giving a number of extensions of the basic `plain' implementation of Krylov methods and the Lanczos algorithm. Section \ref{Sec:Krylov_Pheno} reviews important results on Krylov complexity as a probe of chaos, including a thorough analysis of the behavior of this quantity in integrable, that is non-chaotic, systems. We describe in particular short-time chaos as it relates to scrambling dynamics as well as long-time dynamics which distinguishes spectral manifestations of chaos (or its absence). The late-time saturation behavior is argued to be related to the phenomenon of Krylov localization of the wave function on the Krylov chain of Lanczos coefficients. Section \ref{sect:Holography} reviews results on the bulk manifestation of Krylov complexity, mostly in the context of Jackiw-Teitelboim (JT) gravity, and includes a discussion of extensions, as well as non-perturbative corrections at late times from a bulk perspective. The final Section \ref{Sect:Discussion} gives a discussion of the state of Krylov complexity and Lanczos methods in chaotic quantum systems and holographic duality, before concluding with a brief outlook on future directions of interest in the field.

\subsection*{What is complex? And what is simple?}
Measures of complexity have the goal of assigning numerical values to the commonplace idea of how complicated something is. Naturally, in order to proceed, we need to be more precise about what we assign complexity to -- what objects or processes are being evaluated -- and what characteristics of these objects or processes are to be considered when we assign them a certain value of complexity. Part of the wide appeal of complexity indeed stems from the fact that many systems under study both in the hard sciences and the social sciences (and beyond) can rightfully be evaluated as complex systems. Examples abound, ranging from the patterns of living organisms, to the organisation of cells to the social structures of colonies of animals and societies, and even financial markets, which all show some of the characteristics that are associated with the term complexity. Those characteristics include concepts such as {\it emergence}, where the properties of a {\it large number} of constituents are not reducible to the behavior of each individual element, {\it non-linearity}, meaning that small changes in initial conditions can produce unexpectedly large changes in outcomes, as well as {\it self-organisation} and {\it feedback}.

Similarly in physics, many phenomena qualify to be termed as complex, such as turbulent fluid flow, self-organized criticality, disordered many-body systems (e.g. spin glasses), and not least chaotic dynamical systems with their sensitive non-linear dependence on initial conditions. In this review we focus in on a specific context of complexity, namely that of complexity in chaotic quantum many body systems and moreover with a focus on Krylov complexity. Among the characteristic features of chaotic many-body quantum systems, we find several tell-tale elements of complexity, such as the multitude of microscopic ingredients, the property of emergence, the self-organization (phase structure), as well as the chaotic non-linearity we already mentioned as one of our complexity criteria. In fact, one of our main motivators in the study of complexity is also tied to the idea of emergence in complex systems, namely the fact the certain many-body systems have such non-trivial collective behavior that it ends up being best described in terms of emergent gravitational degrees of freedom, not present in their microscopic description (see Section \ref{sect:Holography})

This answers, to some extent, the question of what is complex? What then, is simple? Just like complexity, simplicity depends on the context. For example, algorithmic simplicity signifies situations where a typical output can be generated by a few basic instructions, rather than needing an elaborate program. Or more generally if a given system requires only few resources, such as energy or computation time or similar, to prepare. Finally one may associate simplicity with systems that require fewer components to specify, with fewer interactions and more regular patterns of structure or behavior. When characterizing complexity in the preceding paragraph, a core principle that emerged from the discussion was that of chaos. In defining simplicity, the analogous core principle would be integrability; indeed in (quantum) physics integrable systems are those which show regular non-chaotic behavior. It is thus an important litmus test for any measure of complexity, and in particular in our chosen context of many-body quantum physics and gravity, to be able to distinguish between chaotic and integrable behavior.

\subsection*{Quantum complexity}
Given its broad appeal, it is no surprise that the notion of complexity is receiving  attention from a number of different communities in theoretical physics, whose common denominator is the need to understand the dynamics of {\it complex} systems and states wherever they may arise, as reviewed for example in \cite{Chapman:2021jbh,Baiguera:2025dkc}. With the advent of the quantum-technology revolution paired with impressive progress in quantum computer science, complexity has emerged as a major tool in quantum physics, where one is interested in the dynamics of states or operators, with respect to a given Hamiltonian. Quantum complexity has emerged as a key tool for several communities, among them high-energy physicists studying black holes and quantum gravity more generally, as well as many-body physicists who study the evolution of strongly correlated many-body states, often in the context of controlling and maintaining quantum coherence for the purpose of quantum computation. The ability to efficiently create such complex states by time evolution is typically associated to chaotic Hamiltonians, opening up the possibility of using a suitably defined notion of complexity in order to characterize quantum chaos, both qualitatively and quantitatively.

Indeed, one of the main motivations of formulating a well-defined notion of quantum complexity is the desire to quantify the intuitive notion of how complicated a quantum state or a Hamiltonian system is, or equivalently to be able to define the computational complexity of certain tasks for a quantum quantum computer. In this review we will mainly address recent developments on complexity from the point of view of dynamics of quantum chaotic systems with applications to holographic duality, but we will make comments about their relation to information-theoretic considerations in quantum complexity  whenever they may be elucidating for the Reader. 

There are by now a number of different notions and definitions of quantum complexity. For the sake of giving a broader context, we will briefly review a selection of these below, all the while putting our main focus of attention on Krylov complexity. Let us begin by giving a bird's eye perspective on this notion, before delving into more details later on in this review.

\subsubsection*{Operator size as complexity measure}

In quantum mechanics, we characterize physical systems by assigning them a state, essentially a complex-valued vector in Hilbert space. Asking how complex a state is, therefore means that we need to assign a complexity value to states in Hilbert space. Observables are represented by operators acting on states in Hilbert space, so if we want to assign a notion of complexity to observables, we need to assign a complexity value to operators acting on Hilbert space. Krylov complexity, also denoted K-complexity, can be applied both as a measure of operator complexity, and as a measure of state complexity. Let us begin with the former, returning to the latter further below in this introduction (both notions will be developed in more depth in Section \ref{sec.CloserKrylov}).

We begin by noticing that Heisenberg evolution of a quantum operator $W(t)$ in an interacting quantum system, will drive the operator to be complex in a rather elementary sense as follows. Time evolution simply adds more and more commutators with the Hamiltonian, which generates more complicated expressions, that is more an more complicated strings composed of the basic operators appearing in the Hamiltonian. Consider, for illustration, a simple operator $W$, propagated forward in time through Heisenberg evolution, $W(t) = e^{iHt} W(0) e^{-i H t}$,
\onecolumngrid
\begin{equation}\label{eq.HeisenbergEvolution}
    W(t) = W(0) +it [H,W] -\frac{t^2}{2} \left[H,\left[H,W \right] \right] - \frac{it^3}{3!} \left[H,\left[H,\left[H,W \right] \right]\right]+ \cdots
\end{equation}
\begin{figure}
    \centering
    \includegraphics[width=0.75\linewidth]{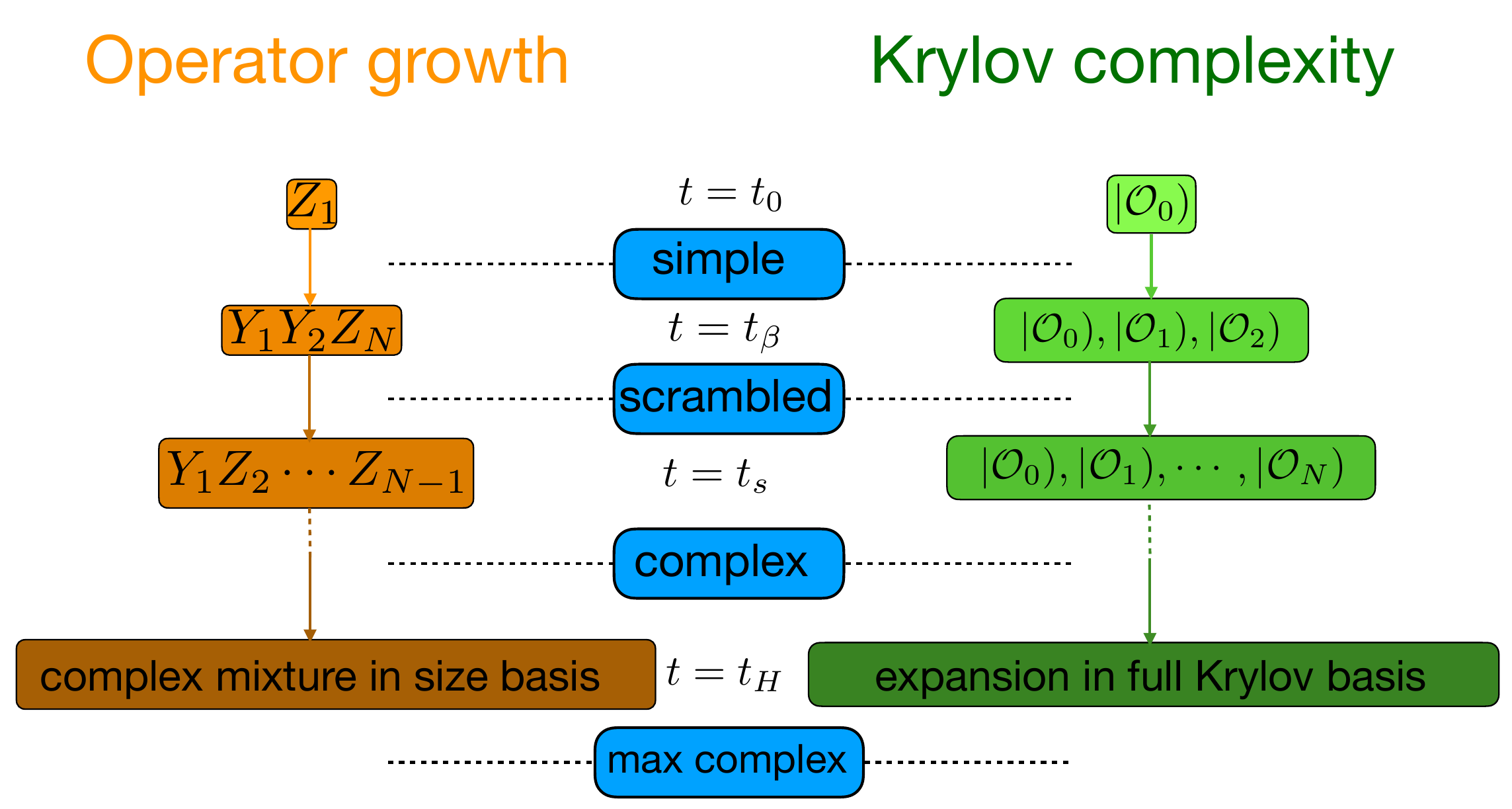}
    \caption{ Krylov complexity quantifies operator growth, measured in an optimal basis, here denoted $\{|{\cal O}_n) \}$ and defined in Section \ref{sec:Lanczos_operator}. Scrambling dynamics of quantum information proceeds via operator spreading, as shown in the Figure above. This finds its expression in the universal operator growth hypothesis of \cite{Parker:2018yvk}, relating the scrambling exponent $\lambda_L$ to the slope of the asymptotic growth rate of the Lanczos coefficients, $\alpha$. In situations where the Hilbert space is (effectively) finite, this linear growth cannot last forever, instead the different epochs of behavior that follow, reveal further characteristic properties of the underlying system. Measuring the complexity and quantifying the different stages of this process constitutes the subject of this review. K-complexity shows non-trivial evolution over the full time space covered in the above diagram, while `naive' size complexity already saturates at its maximal value by the scrambling time, $t_s$. }
    \label{fig:KrylovScrambling}
\end{figure}
\twocolumngrid
Suppose, for concreteness, that we work with a simple spin system, for example \cite{Roberts:2014isa}
\begin{equation}\label{eq.SimpleSpinHam}
    H = \sum_i Z_i Z_{i+1} + g X_i + h Z_i\,,
\end{equation}
where $X_i,Y_i,Z_i$ are the Pauli spin matrices at site $i$ in the respective spatial directions of their labels. Say we start with a simple\footnote{Note that, as anticipated above, we have now already thrice invoked the adjective `simple', in our quest to define complexity. The operator is simple because it is composed of very few local ingredients, for example, while the Hamiltonian is deemed simple because it is again made up of a small number of local interactions of elementary ingredients.} operator, for example $W(0) = Z_1$ the third Pauli matrix on the first site. An elementary calculation then demonstrates that up to third order we generate the following monomials of Pauli operators
\begin{equation}\label{eq.OperatorGrowthElementary}
    Z_1 \longrightarrow Y_1\,, Y_1 Z_N\,, Y_1 Z_2\,, Y_1Z_2Z_N  \,.
\end{equation}
The initial seed operator has both spread over the real-space chain, and has spread in Hilbert space, in the sense that it has evolved into a more complicated operator that is a composite of simple Pauli matrices. One can convince oneself that this spreading will continue as we consider higher and higher order contributions to the CBKH expansion of the Heisenberg evolution of the simple operator $Z_1$. In order to formalize this idea, it is necessary to define more carefully the notion of `size', which means defining a basis of increasingly `complex' operators, and then to expand and measure the position of the operator $W(t)$ in this basis. In the present case, it would be natural to define
\begin{equation}\label{eq.OpInSizeBasis}
    W(t) = \sum_{i_1, i_2\,\ldots,i_n} c_{i_1 i_2\cdots i_s} (t) \sigma_{i_1, \cdots i_s} \,,
\end{equation}
where the notation $\sigma_{i_1, \cdots i_n}$ is meant to indicate all possible monomials made from the Pauli matrices at each site and the time-dependent coefficient $c_{i_1 i_2 \ldots, i_n}(t)$ encodes the evolution of the operator in this basis and its absolute value squared measures the probability weight of being supported on a particular Pauli string at time $t$. The probability distribution for the `size' of the operator is then
\begin{equation}
    P_s(t) = \sum_{i_1, i_2\,\ldots,i_s} \left| c_{i_1 i_2\cdots i_s} (t)\right|^2
\end{equation}
and a natural definition of `size' would be the average position of the operator $W(t)$ with respect to this probability distribution \cite{Roberts:2014isa,Roberts:2018mnp},
\begin{equation}\label{eq.sizeComplexity}
 {\rm size}= \sum_s P_s(t)\,.
\end{equation}
Note, however, that the definition of this distribution relies on a number of somewhat arbitrary definitions, for example what constitutes a simple or elementary operator, or what is a good basis, which would seem to depend on the particular microscopics of the system at hand. Krylov complexity generalizes this intuitive notion of operator size and its spreading with respect to a given basis, and at the same time fixes a number of natural definitions regarding the basis and probability distributions involved, as we will describe in detail in this review. Furthermore, by definition in a quantum system with $S$ degrees of freedom, whose Hilbert space therefore is ${\cal O}\left( e^S \right)$ dimensional, size complexity reaches a maximal value of ${\cal O}(S)$, after which it cannot grow any further. Supposing that size grows exponentially -- as in fact it will turn out to do at early times -- it will only take a time $t_s\sim \log S$ to reach saturation. For reasons we will make clear from both a complexity-theoretic and a gravitational perspective below, it is in fact necessary to develop complexity notions that can keep growing way past $t_s$ and that can attain saturation values of ${\cal O}\left(e^S \right)$. As the Reader may guess, Krylov complexity is satisfies all of the criteria above.

To give a first flavour of Krylov complexity, we return to the simple Heisenberg evolution, \eqref{eq.HeisenbergEvolution}. Let us say, we want to use the successive commutators, as a basis of the space of operators. We would then have a basis of the form
\begin{eqnarray}
   {\cal B}' :=\left\{ |Z_1) \,, \left| \left[H,Z_1\right] \right)\,,\left| \left[H, \left[ H, Z_1\right] \right] \right)\,,\left| \left[H, \left[ H,  \left[ H, Z_1\right] \right] \right] \right)\,,\cdots \right\}\,\nonumber
\end{eqnarray}
where the rounded bra-ket notation is used for elements of the Hilbert space of linear maps on the physical Hilbert space ${\cal H}$, which can be given an inner product (see below) and rigorously defined as a bona-fide Hilbert space in its own right (see for example \cite{reed1980functional}). An immediate issue arises due to the fact that the basis ${\cal B}$ is not orthonormal and therefore not the most efficient construction possible. This can be easily remedied, by using Gram-Schmidt orthogonalization
\begin{eqnarray}
    {\cal B}' \quad  \xrightarrow[\text{Schmidt}]{\text{Gram}} \quad {\cal B}
\end{eqnarray}
which produces the canonical Krylov basis ${\cal B}$. In practice, the Krylov basis is constructed iteratively layer by layer in $n$, a technique referred to as the Lanczos algorithm (see below), which is more efficient than a simple application of the Gram-Schmidt algorithm to the full set ${\cal B}'$. This basis has a number of desirable properties, but for now we simply note that we produce at each step, $k$, in the order of commutators a linear combination of elements of ${\cal B}'$ of nested commutators of size up to $k$ and a string of normalization coefficients 
\begin{equation}
  L =  \left\{b_1, \, b_2 \,, \cdots b_k \right\}\,,
\end{equation}
called the Lanczos sequence. Loosely speaking, the pair $\left( {\cal B}\,, L  \right)$ encodes the information on operator complexity we wish to explore in this review; it is, for the time being, an efficient way of characterizing the subspace of Hilbert space the operator explores under time evolution.

To give a first example of an application to complexity, it is often very useful to investigate the asymptotic rate of growth with $n$ of the Lanczos sequence. There is by now good evidence that this asymptotic growth rate is linear
\begin{equation}
    b_n \sim \alpha n + {\cal O}(1) \qquad n\gg 1 \,
\end{equation}
for chaotic quantum systems, with the asymptotic linear growth rate $\alpha$ a sensitive measure of the degree of chaos in the system, valid in the thermodynamic limit. 

The pair $\left( {\cal B}\,, L  \right)$ also allows us to give our first definition of Krylov complexity, as the average position of the seed operator at a given time $t$ when measured with respect to the ordered Krylov basis. To do this, one expands a given operator 
\begin{equation}
    | W(t) ) = \sum_{n=1}^{K-1} i^{n}\varphi_n(t) | W_n )\,,
\end{equation}
where ${\cal B} = \left\{ | W_1 ),| W_2 ) ,\ldots  | W_{K-1} )  \right\}$ is the Gram-Schmidt orthogonalized Krylov basis above, and $K$ denotes the dimension of Krylov space, while the factor $i^n$ is included for later convenience. Note that this is the K-generalization of the expansion of the operator in the size basis we gave in Equation \eqref{eq.OpInSizeBasis} above.

Equipped with the wave-function coefficients $\varphi_n(t)$ we are now able to define K-complexity for operators as the average position of the system with respect to the ordered Krylov basis ${\cal B}$:
\begin{equation}
    C_K (t) = \sum_n n |\varphi_n (t)|^2\,.
\end{equation}
 Note that this definition is very similar in spirit to the size complexity defined in \eqref{eq.sizeComplexity} above\footnote{We should compare it to the quantity $C_{\rm size}(t) = \sum_s s P_s(t)$. We will perform a more rigorous comparison around Equation \eqref{eq.DefSizeComplexity}.}, but measures a generalized notion of `size' with respect to an optimal basis. For now we have not given any arguments for the optimality of the Krylov basis, in order to keep this introduction technically lighter, but the interested Reader may find the relevant details  in Section \ref{subsect:theorems} below. 

Given how naturally Krylov complexity fits with the physics of operator spreading (see Fig. \ref{fig:KrylovScrambling}), it is no surprise that it is intimately connected to the physics of scrambling, as encoded in out-of-time-order correlation functions (OTOCs). The simplest avatar of such OTOCs is the 4-pt function
\begin{equation}\label{eq.OTOC4pt}
    {\cal F}_{\rm OTOC}(t,x) := \left\langle W^\dagger(t,x) V^\dagger W (t,x) V \right\rangle_\beta \,,
\end{equation}
and the name ``OTOC" expresses the ordering of the operators within the expectation value, which effectively corresponds to evolving backwards in time at intermediate stages. This correlation function captures the sensitivity to small perturbations of `initial conditions' for a quantum system, which allows to define a quantum version of a Lyapunov exponent, \cite{larkin1969quasiclassical}, $\lambda$
\begin{equation}\label{eq:OTOCscrambling}
    {\cal F}_{\rm OTOC}(t,x) \sim 1 - \hbar e^{2\lambda \left(t- \frac{|x|}{v_B}\right)}\,,
\end{equation}
Here $v_B$ is the so-called butterfly velocity, and we have assumed that the system has a small$-\hbar$ type semi-classical limit, which allows a parametric separation between the local thermalization scale $t_\beta \sim \beta$. The scrambling time $t_*$ is defined as the time at which the OTOC has decayed to zero for all spatial locations $x$ in the system. This time is conjectured to scale $t_* \sim {\rm log}N$ for `fast scramblers'.
 Coming back to Krylov space, it has been advocated in \cite{Parker:2018yvk} that the asymptotic rate of linear growth of Lanczos coefficients $b_n$ in a chaotic system upper bounds the Lyapunov exponent as
\begin{eqnarray}
\label{eq:bound_Lyapunov_KrylovExp_intro}
    \lambda \le 2\alpha\,,
\end{eqnarray}
a statement that they were able to prove at infinite temperature, and which remains a (plausible) conjecture at generic finite temperature. We discuss this so-called `universal operator growth hypothesis' in a later Section.

\subsection*{Schrödinger picture Krylov complexity}
Having given an overview of Krylov complexity as applied to operators, we will end this part of the introduction by briefly presenting the closely related concept of Krylov complexity for states. In this case, we start with an arbitrary reference state $|\phi(t=0)\rangle = |0\rangle \in {\cal H}$, whose time evolution in the Schrödinger picture is 
\begin{equation}
    | \phi(t) \rangle = e^{-i H t} | \phi(0) \rangle = \sum_k \frac{(-it)^k}{k!}H^k | 0 \rangle\,.
\end{equation}
The state at any time $t$ is a linear combination over the basis ${\cal B'}=\{|0\rangle\,,\, H |0\rangle\,,\, \ldots \,,H^k |0\rangle  \}$. As in the operator case, the Krylov basis ${\cal B}$ is again given by the Gram-Schmidt orthogonalized version of the naive basis ${\cal B}$. We note again, that in practice the basis ${\cal B}$ is constructed iteratively layer-by-layer, using an appropriate version of the Lanczos algorithm, which at the same time also produces the series of Lanzcos coefficients $\{b_k \}$, in analogy with the operator case. The Lanczos algorithm is much more efficient than first constructing all of ${\cal B}'$ and subsequently applying the Gram-Schmidt procedure. We then expand the time-evolving state in the Krylov basis
\begin{equation}
    |\phi(t) \rangle = \sum_n \phi_n(t) | n\rangle \,,
\end{equation}
where the basis ${\cal B}$ is spanned by the vectors $\{ |n\rangle  \}$. Krylov state complexity is defined as the average position of the time-evolved quantum state along the Krylov chain
\begin{equation}
    C_K(t) = \sum n |\phi_n(t)|^2\,.
\end{equation}

Having outlined the flavour of the main ideas and concepts behind Krylov space techniques and associated complexities, we now wish to take a step back, and discuss briefly the wider picture of chaotic quantum dynamics and quantum complexity, before returning in more detail to the field of Krylov complexity, the main subject of this review.
\begin{figure}
    \centering
    \includegraphics[width=0.9\linewidth]{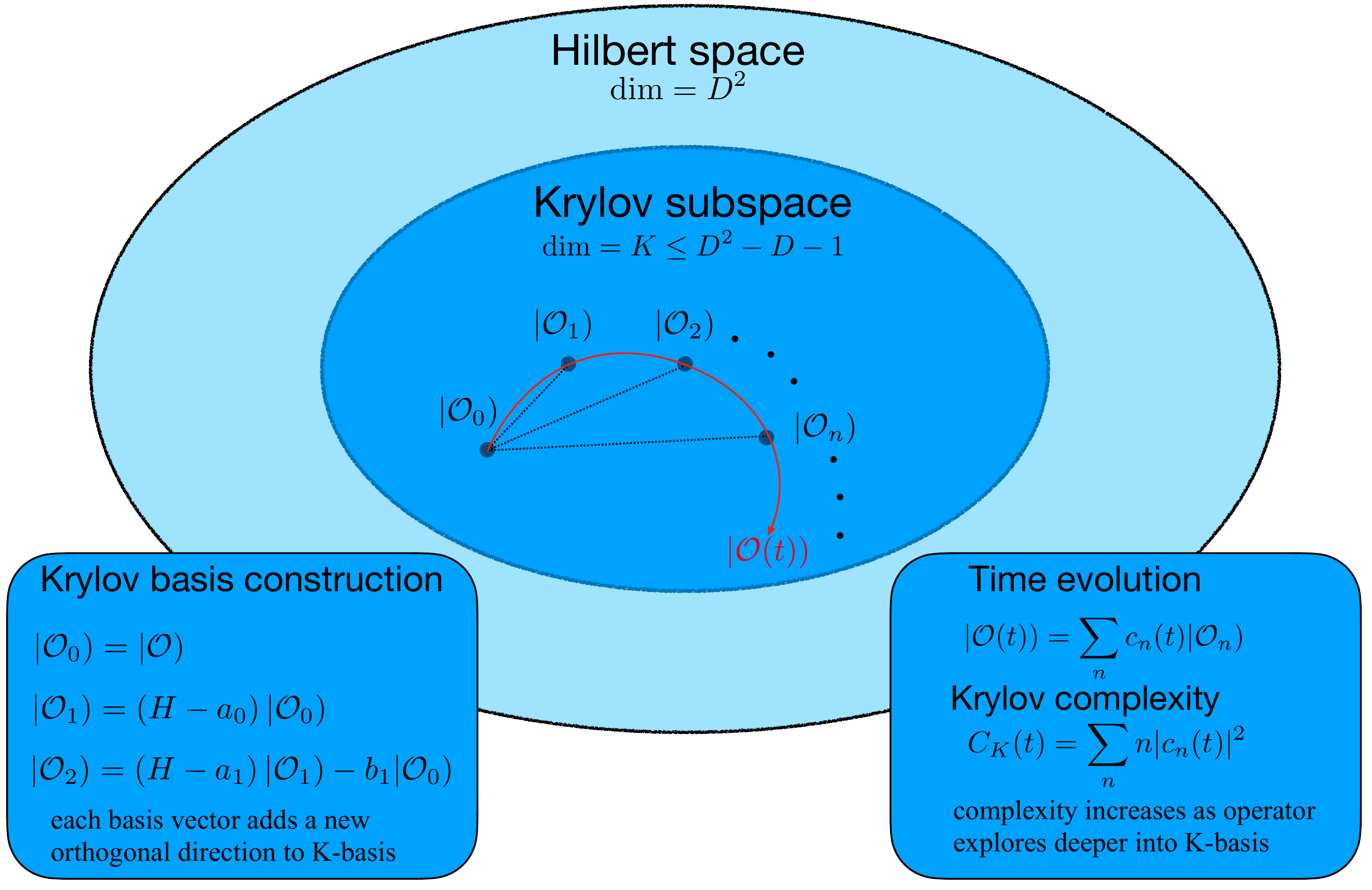}
    \caption{Krylov space is a subspace of the full Hilbert space, optimally chosen to accommodate time evolution of states and operators. Without loss of generality we show the operator case. Here the relevant Hilbert space has dimension $D^2$ and one can upper-bound the dimensions of the Krylov subspace by $D^2 - D -1$. Krylov complexity is defined by the spread of an operator or state along the Krylov chain.}
    \label{fig:KrylovOverview}
\end{figure}

\subsection{Computational origins of complexity}
Despite motivating complexity from what may be referred to as the ``chaos perspective", we have already seen that there exist close links to the physics of quantum information and its spreading across physical systems, exemplified by the connection between scrambling and operator growth. Hence one often sees motivations of complexity from an information-theoretic point of view, as well. We shall now turn to such an approach, mainly to help the Reader to orient the subsequent discussion in the wider context.

For the purposes of this review, the notion of complexity traces its origins to the modern field of computational complexity, a subfield of theoretical computer science. For example, in this context it is natural to ask about how complex a given bit string is. Intuitively, one would want to say that a bit string of the form $\left\{ 10101010101010\ldots \right\}$ is less complex than a bit string of the form $\left\{11001001010011\ldots \right\}$ because the former consists of a simple short string, $01$, and the simple instruction to repeat it, while the latter is not produced from a similar `low-entropy' set of input instructions. An early notion of complexity which addresses this question was proposed by Kolmogorov \cite{kolmogorov1963tables}, who assigns a complexity to such strings\footnote{and any other mathematical object, for that matter.} as the length of the minimal program that produces the string as an output, for example when run using a universal instruction set on a Turing machine. Even more concretely one may want to express computational complexity in terms of the resources, i.e. the number of operations needed to produce a given output state from the input state. Moreover, what is important in this context is more often than not the notion of asymptotic scaling with input size $n$, e.g. ${\cal O} \left( n^\alpha\right)$ or ${\cal O}\left( e^n \right)$, according to which certain complexity classes may be defined (see for example \cite{mertens2002computational}).

Loosely speaking such an algorithmic notion of complexity can be extended to the quantum realm, where information is now encoded in fundamental qubits\footnote{A qubit is an abstraction of any two-level quantum system with states $| \textrm{``up"}\rangle = |1\rangle $ and $| \textrm{``down"}\rangle = |0\rangle $, which may be realized physically as spins, photon polarizations, levels of atoms, flux states etc.}, and hence qubit strings, and quantum algorithms can be expressed as sequences of unitary operations, so-called gates, acting on subsets of qubits. One would now like to assign values of complexities to quantum states such as $|\psi\rangle = \left| 10101010101010\ldots \right\rangle$ or $|\psi'\rangle = \left| 11001001010011\ldots \right\rangle$ by asking the question about the minimal set of operations needed to be performed using the universal gate set in order to producec a given output state, say $\left| \psi \right\rangle$  or $\left| \psi' \right\rangle$ , from a fixed input state $\left| \psi_{\rm IN} \right\rangle$. The successive and parallel application of these basic quantum gates, and thus how the associated resources scale asymptotically with the input size, leads naturally to the idea of gate complexity (see, for example \cite{nielsen2010quantum}) which we will now  describe.

\subsection{Gate and Nielsen complexity}\label{subsec:gateAndNielsen}
\begin{figure}
    \centering
    \includegraphics[width=0.9\linewidth]{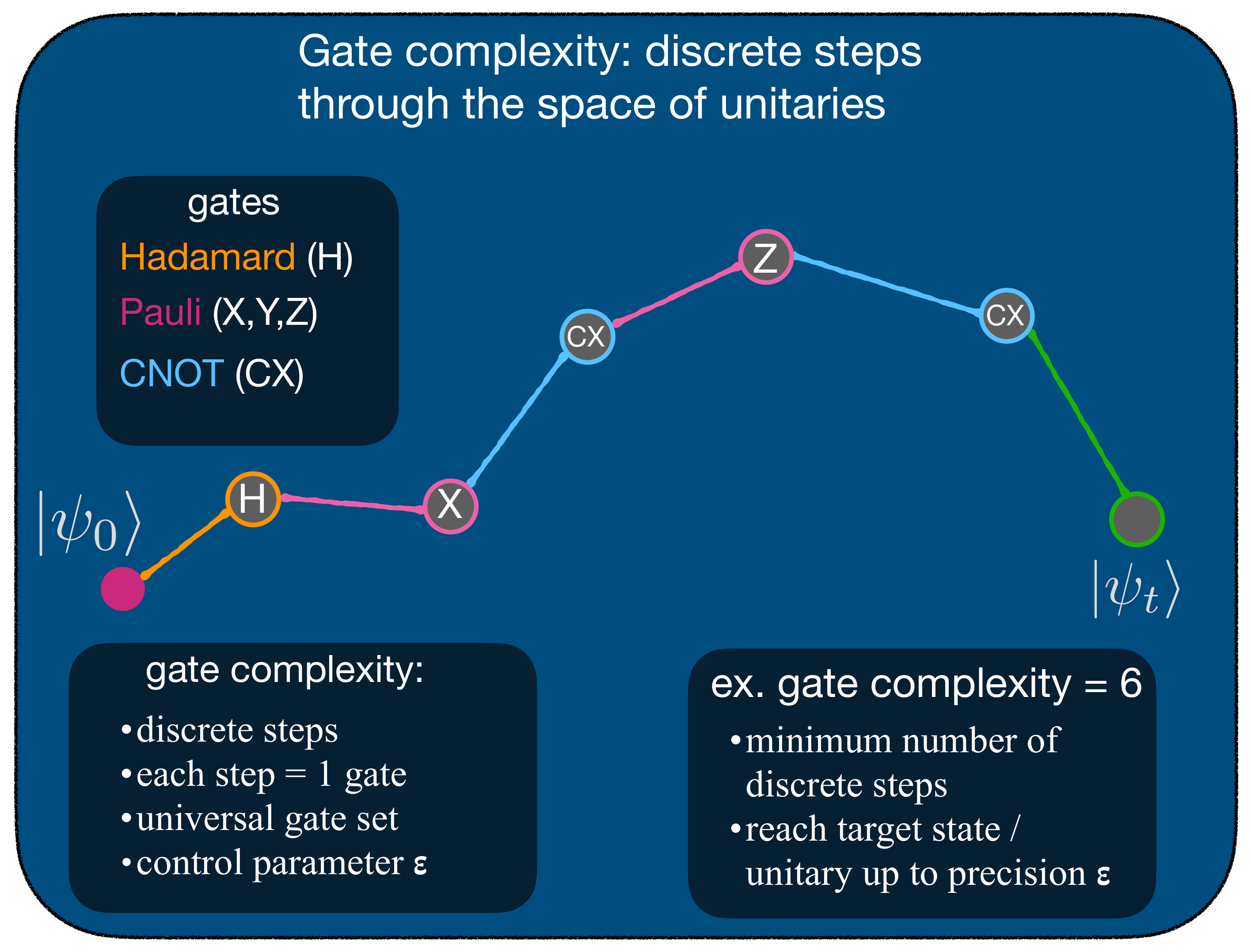}
    \caption{Gate complexity as discrete motion through the space of unitaries (shown in dark blue). At each step we apply a single gate among a small set of universal quantum gates, for example those given in the inset. The total number of gates needed to reach the final state from a given initial state, within a tolerance $\epsilon$ is the sought-after gate complexity. In this case there are six steps, and we assign a complexity of six.}
    \label{fig:GateComplexity}
\end{figure}
Let us now more formally define gate complexity, the perhaps most familiar notion of complexity as applied to quantum systems. Let us say that we want to quantify the complexity of the time evolution, represented by the operator $U(t) = e^{-i H t}$ of some quantum system, characterized by the Hamiltonian operator $H$ and some Hilbert space ${\cal H}$ of dimension $D$. For simplicity we may think about a collection of spins, i.e. about a collection of qubits $\psi_i$ residing at sites $i=1,\ldots N$, so that $D=2^N$. The idea of gate complexity then relies on the elementary observation that we can approximate an arbitrarily complicated Hamiltonian $H$ and its associated evolution operator $U(t)$ by successive applications of simple unitary operations \cite{nielsen2010quantum}, the so-called gates (see Fig. \ref{fig:GateComplexity}),
\begin{equation}
    U(t) \approx U_{\rm approx}(t)=U_n U_{n-1} \cdots U_1\,,
\end{equation}
where each $U_i$ of course still depends on the elapsed time $t$ as a parameter. Furthermore, we should be more precise about the sense of our approximation, by imposing that $|| U(t) - U_{\rm approx}(t)|| < \epsilon$ for some small positive $\epsilon$ with respect to a suitable matrix norm, $\| 
\, \cdot \,\|$. Moreover, one can show that each $U_i$ can be approximated to arbitrary precision by repeated application of a simple and finite set of unitary operators, say ${\cal U} = \left\{ U_a^{\rm el} \right\}$ acting only on single qubits and pairs of qubits \cite{lloyd1996universal}. This means that we can define the quantum computational complexity of an operator\footnote{We choose to work here exclusively with the time evolution operator $U(t)$, but it should be clear from the definitions that any other unitary operator can just as well be associated a quantum computational complexity in the same spirit.} $U(t)$, denoted ${\cal C}[U(t)]$ as the minimal $n$, such that
\begin{equation}
    \| U(t) - \prod_{a=1}^n U_i^{\rm el} \| < \epsilon\,,
\end{equation}
where each $U_i^{\rm el}$ is chosen from the small finite set of allowed elementary unitaries.
Intuitively this resonates well with the idea of computational complexity being represented by the length of the most efficient logical circuit that can produce a given operation, or a given output from a reference input, adapted to the quantum realm. It is amusing to point out that this coincides with the Kolmogorov complexity introduced above, if we take the programming language to be defined by the finite set of elementary bit-wise (qubit-wise) operations on a collecting of $N$ bits (qubits). Note that in most circumstances the asymptotic scaling of $n$, the number of gates needed, with system size, $N$, is the most interesting aspect of gate complexity. One perhaps less desirable feature of gate complexity, as defined here, is its intrinsic dependence on the tolerance parameter $\epsilon$, and furthermore the fact that it is really a natively discrete notion. The discreteness combined with the arbitrary control parameter $\epsilon$ also means that there is no absolute upper bound on gate complexity. This issue can be addressed by instead discussing a slightly more abstract, but highly useful notion of complexity, namely Nielsen complexity.
\subsection*{Nielsen complexity}
\begin{figure}
    \centering
    \includegraphics[width=0.9\linewidth]{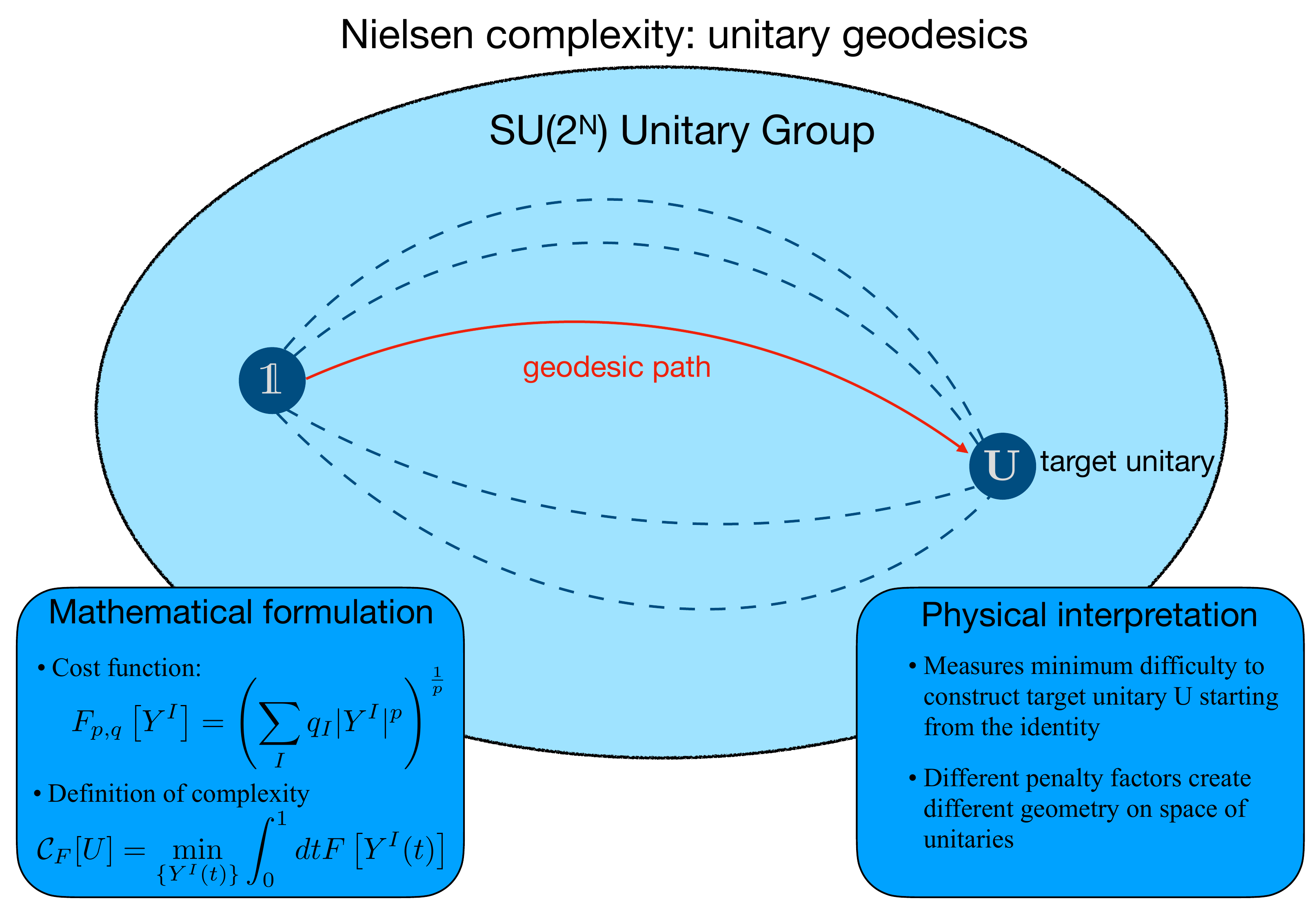}
    \caption{Nielsen complexity is defined in terms of the optimal path connecting two states in the space of unitaries with respect to some cost function, $F[Y^I]$, depending on a set of parameters $Y^I$. Different cost functions may be appropriate in different contexts (see Section \ref{subsect:KC_geometry}).}
    \label{fig:NielsenComplexity}
\end{figure}
Nielsen's approach to complexity has two notable features as compared to circuit complexity. Firstly, it is intrinsically a continuous notion: in some sense circuit complexity with its gate-by-gate approach approaches the space of unitaries along discrete trajectories, each size being roughly of ${\cal O}(\epsilon)$. Here, ``space of unitaries" should be taken to mean the continuous Lie group manifold of unitary operators equipped with some norm, a simple and elegant example being the bi-invariant metric $ds^2 =  {\rm tr} \left((dU U^\dagger)(dU^\dagger U)\right)$. Crucially the actual choice of norm is arbitrary and needs to be discussed more carefully. Since the space of unitaries, that is the group manifold SU$(D)$, is compact, Nielsen complexity is naturally upper bounded, and can furthermore be described without recourse to the arbitrary control parameter $\epsilon$. We will see later a connection between Krylov and Nielsen complexities, which however relies on a more complicated choice of metric on the space of unitaries. Secondly, Nielsen complexity offers an appealing differential-geometric interpretation of quantum complexity, and in fact associates the notion of optimal paths in the space of unitaries with geodesics in a suitable metric.

Let us now describe this idea in some more detail. Given the manifold structure of the space of unitaries, it is natural to define the complexity of an operator $U$ as the length of the path connecting it to some reference operator $V_R$, usually taken to be the unit operator id. One then defines the length of a path connecting the target operator $U$ to the reference operator as the length of a certain program that prepares the target unitary (cf Kolmogorov). One associates the Nielsen complexity of the target operator as the length of the minimal path connecting id to $U$, i.e. as the the length of a geodesic connecting the two in the manifold of unitaries. We have already seen that Krylov complexity is a natural generalization of size complexity. It is less obvious what the relation is to gate or Nielsen complexity (if any), but interesting proposals in the literature have already related the latter to Krylov (e.g. \cite{Craps:2023ivc}). We will return to this below.

\subsection{Chaos produces complexity and pseudorandomness}
We have already introduced the idea (see Fig. \ref{fig:KrylovScrambling}) that Krylov complexity captures, among other things, the notion that operators (and states) grow to be more complex as time progresses, leading to the relation between aspects of Krylov complexity and the Lyapunov exponent of OTOC 4-point functions, which is recognized to be a characteristic of chaotic quantum systems \cite{Maldacena:2015waa}. At the same time, such chaotic dynamics also cause operators and states to look more and more random, where a truly random operator, would be one that is drawn from the Haar ensemble\footnote{For a definition of Haar random operators and states see the intuition box around Equation \eqref{eq.HaarMoments} below.}. On the other hand, deterministic time evolution, such as unitary quantum dynamics, can never produce a truly random operator, but must instead results in a pesudorandom state or operator that behaves like a truly random one in for certain quantities. A convenient way of measuring such pseudorandom behavior is via unitary $k-$designs. A unitary $k-$design is an ensemble, ${\cal E}$, of unitaries, $\{p_j, U_j  \}$, with the property that its moments match those of the Haar ensemble up to order $k$. The connection to chaos on complexity is seen as follows. A chaotic system will grow an initial seed operator to larger and larger size (or Krylov length), leading eventually to the decay of the OTOC , \eqref{eq:OTOCscrambling}, with characteristic Lyapunov exponent $\lambda$, as in \eqref{eq:OTOCscrambling}. One can show \cite{Roberts:2016hpo} that 
\begin{equation}
    \left\langle {\rm OTOC}\right\rangle_{2-\rm design} \sim \left\langle{\rm OTOC} \right\rangle_{\rm Haar} \sim \frac{1}{D^2}\,,
\end{equation}
which is the floor value of the OTO-correlator after it has scrambled for $t>t_*$, $D$ being the dimension of the Hilbert space. Averaging the OTOC over a unitary $1-$design produces simply the factorized expression $\langle W^\dagger W \rangle \langle V^\dagger V\rangle \gg 1/D^2$. We thus get the (rough) picture under time evolution of a chaotic quantum system that
\begin{eqnarray}
    \left({\rm K}-\right){\rm complexity}: \quad\nearrow\nonumber\\
    {\rm pseudorandomness}:\quad  \nearrow\nonumber\\
    {\rm OTOC}:\quad \searrow\,,
\end{eqnarray}
meaning that (pseudo-)randomness and complexity generally increase\footnote{If one had started with an $4m-$fold OTOC correlator, a generalization of the simple $4-$fold OTOC \eqref{eq:OTOCscrambling}, one would see that averaging over $k-$designs for $k<m$ would produce values larger than the OTOC$^{(m)}$ floor value after saturation. This gives a more fine-grained notion of scrambling and pseudorandomness.}, while OTOCs decrease, with relations between them given via $k-$designs, Lyapunov exponents and Lanczos growth. Throughout the evolution, K-complexity gives a continuous and fine-grained notion of the increase at all time scales.
\vskip1em

Having given an introductory exposition of Krylov complexity, as well as the wider framework of algorithmic, information-theoretic and quantum complexity, we move on to a pedagogical introduction of the role complexity plays in studies of gravity, a subject which we will return to in Section \ref{sect:Holography}.

\section{Holographic origins of complexity}\label{sec.HoloComplexity}
Having given an introductory overview of the subject from a complexity-theoretical point of view, we will now connect complexity to a seemingly far-away subject, namely (quantum) gravity.  In the following Section, we will tie some knots between high energy particle physics,  the toolkit used to understand it, and the ideas and concepts of complexity. 
We start with motivations, the relevant setup, and some general observations.

\subsection{Motivation and lightning review of holography}

One of the main motivations leading to the study of complexity is the desire to understand the very extreme properties of gravity that can result from the incorporation of quantum mechanics in the theory of gravity, potentially leading us far astray from the familiar semi-classical limit, where Einstein gravity gives a good approximate description of the physics.

The quantities we have in mind here are as large as $\exp(S)$ and as small as $\exp(-S)$, where $S$ is the entropy of a system whose energy is in the neighborhood of some value $E$. Often $S(E)$ will be the entropy of a black hole, when thinking about gravity.  
Where do such extreme quantities appear? We will explain in this section  that the large quantity $\exp(S)$ is related to upper bounds of various complexities. The extremely small value $\exp(-S)$, on the other hand, is tied to the extremely small but non-zero values of correlation functions at late times, in some cases capturing the tiny deviations from zero of the overlap of what were supposed to be orthogonal states. 
 
A major hindrance in the attempt to understand such extreme effects is the fact that in four dimensions (Einstein) gravity is perturbatively non-renormalisable\footnote{From this perspective it is thus perhaps surprising that recent years have seen important progress on the microphysics of black holes, relying only on semi-classical that is EFT arguments.}. In an effort to address and overcome this issue, string theory was formulated, a theory based on one-dimensional fundamental objects, rather than the zero-dimensional point particles of standard QFT.

This theory which originally resulted from attempts to explain the 
structure of the copious number of hadrons came to suggest a way to obtain a perturbative consistent theory of quantum gravity in ten space-time dimensions. 

Despite going far beyond point-particle physics, to some extent, QFTs are part and parcel of the more general framework, for example, they are recovered as effective low-energy descriptions of String Theory\footnote{The role effective field theory can play in a String Theory is subtle, see
for example \cite{Banks:2010tj}.}. This is an example of the rather familiar situation where a theory turns out to be a certain limit of a new, qualitatively different theory.

\begin{figure}
    \centering
    \includegraphics[width=0.7\linewidth]{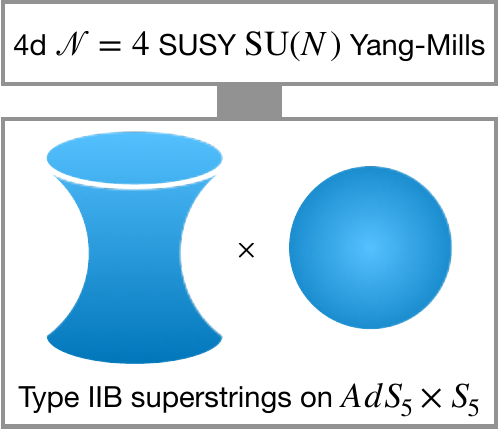}
    \caption{$\mathcal{N}=4$ super Yang-Mills in 4d is conjectured to be dual to type IIB string theory on AdS$_5$ $\times$ $S_5$ containing $N$ units of Ramond-Ramond flux. The salient feature of this holographic duality is that five-dimensional gravity in the ``AdS-bulk" is dually described by four-dimensional Yang-Mills theory on the boundary. The compact $S^5$ factor is dually related to the internal symmetries of the SYM theory.}
    \label{fig:AdS/CFT}
\end{figure}

In a breakthrough development is has been realized that there are circumstances when a QFT is not just a limit of String Theory but actually dual to it, giving in effect an equivalence between certain QFTs and quantum gravity \cite{Maldacena:1997re}. This has come to be known as the  AdS/CFT duality \cite{Maldacena:1997re,Gubser:1998bc,Witten:1998qj}, where AdS stands for `Anti-de Sitter space', a negative constant-curvature solution of Einstein's equations with a negative cosmological constant. This duality was tested under a very large number of circumstances that are described, for example, in \cite{Aharony:1999ti}, \cite{AmmonErdmenger2015}, \cite{Nastase2015}. A main achievement of AdS/CFT duality is the demonstration that the idea of holography \cite{thooft1993dimensional,susskind1995world}, motivated by the the behavior of black holes, which carry an entropy that is proportional to their surface (not their volume), can be concretely realized \cite{Maldacena:1997re}.  The original conjecture states that a string theory on a particular ten-dimensional  background, namely $AdS_5\times S^5$ is dual to a $\mathcal{N}=4$, SU($N$), supersymmetric gauge theory, living on the four-dimensional boundary of AdS$_5$, with parameters on both sides matched by an appropriate `dictionary'. In Figure \ref{fig:AdS/CFT} we summarize the correspondence between the parameters of the 10d theory and the CFT living on its boundary. In this context the AdS spacetime is usually referred to as the ``bulk".
This correspondence has been further extended to a wide array of other setups, relating gravity in AdS spaces of different dimensions to a vast array of gauge theories  (see \cite{Itzhaki:1998dd,Aharony:1999ti,Gauntlett:2025vnh}.

\begin{figure*}[ht]
\justifying
\begin{roundedbox}[Anti-de Sitter background]
Anti-de Sitter space in dimension $d+1$ is the unique maximally symmetric solution to Einstein's equations with a negative cosmological constant, $\Lambda<0$:
\begin{equation}
    R_{\mu\nu} - \tfrac{1}{2}R g_{\mu\nu} + \Lambda g_{\mu\nu} = 0\,.
\end{equation}
A geometric way to construct this solution is by considering the embedded hyperboloid of constant curvature in flat space of dimension $d+2$, obeying the quadratic
\begin{equation}\label{eq.AdSHyperboloid}
    -X_{0}^2 - X_{d+1}^2 + X_1^2  +\cdots + X_d^2 = -\ell^2\,.
\end{equation}
\noindent For the case of AdS$_2$ this hyperboloidal embedding can be visualized as follows:\vskip1em
\begin{centering}
\begin{tikzpicture}
    \begin{axis}[
      view={30}{20},
      axis lines=center,
      xlabel={$X_1$},
      ylabel={$X_0$},
      zlabel={$X_3$},
      ticks=none,
      domain=-1.8:1.8,       
      y domain=0:2*pi,
      samples=15,            
      samples y=20,
      z buffer=sort,
    ]
      \addplot3[
        surf,
        shader=flat,
        fill=gray!40,         
        draw=black,           
        mesh,
        opacity=1.0,
      ]
      ({sinh(x)},
       {cosh(x)*cos(deg(y))},
       {cosh(x)*sin(deg(y))});
    \end{axis}
  \end{tikzpicture}
  
    \end{centering}
    
Note that we can see this quadratic equation as a bilinear with the somewhat unusual `two-time' signature quadratic form $(-,+,\cdots, + -)$. If this were instead the quadric form with all pluses, we would recognize this as the inner product of two standard vectors in $\mathbb{R}^{d+2}$ and we would conclude that the quadratic form is invariant under the orthogonal group in $d+2$ dimensions. In fact, the same happens here, except that the group in question is SO$(2,d)$, which is inherited by AdS$_{d+1}$ itself as its group of geometric symmetries (`isometries'). Remarkably this group is exactly the conformal group appearing in the study of $d-$dimensional conformal field theories (CFT). 

In order to write down a metric on AdS$_{d+1}$ it suffices to solve the embedding constraint \eqref{eq.AdSHyperboloid} in terms of any set of $d+1$ independent coordinates. Each inequivalent set of coordinates will give rise to a different coordinate system on AdS, with arguably the most popular one being given by `global coordinates', in terms of which the metric reads
\begin{equation}
ds^2 = \ell^2 \left( -\cosh^2 \rho\, d\tau^2 + d\rho^2 + \sinh^2 \rho\, d\Omega_{d-1}^2 \right)\,.
\end{equation}
The last factor $d\Omega_{d-1}^2$ is the metric on a unit-radius round sphere in dimension $d-1$. An interesting geometric fact about anti-de Sitter spacetime is the following: if we consider a sphere centered at the origin of coordinate radius $R$, then its proper volume $V(R) \sim R^{d-1} \ell$, while its surface area scales like $A(R)\sim R^{d-1}$. Thus for large $R$, the ratio of surface to volume in AdS actually becomes a constant. In flat space, where there is no additional length scale that could take on the role of $
\ell$, we of course simply have $V^{\rm flat}(R)\sim R^d$, while $A^{\rm flat}(R)\sim R^{d-1}$ and the scale of volume over surface diverges linearly with $R$. In this sense the claim that a bulk theory is holographically represented by a theory living on its boundary appears slightly less astonishing in AdS than it does using flat-space intuition.\footnote{Of course this geometrical fact should not deter people from searching for a holographic dual of flat space itself.}
 Anti-de Sitter space is a particular kind of spacetime which plays a central role in the AdS/CFT correspondence. This space has many special properties, starting with the fact that it is the unique maximally symmetric solution of Einstein equations with a negative cosmological constant, whose geometric isometry group realizes the conformal group of the CFT on the boundary. Furthermore, AdS space can serve as a Faraday cage for energy and although its volume is infinite, massless particles sailing to infinity return from their voyage in a finite amount of time.
\end{roundedbox}
\end{figure*}

Classical and semi-classical properties of gravity in the bulk can give insights about properties of the boundary QFT at strong coupling, while weakly coupled boundary computations correspond to quantum effects in bulk gravity, reflecting the strong-weak coupling nature of the duality. By this one means that whenever one side is weakly coupled, the other side is strongly coupled. The natural coupling on the gravity side is Newton's constant, $G_N$, while the coupling constant of the boundary CFT depends on the specific setup. In ${\cal N}=4$  SYM with gauge group SU$(N)$ in four spacetime dimensions there is the gauge coupling $g$, as well as the rank of the gauge group $N$, which can be combined into the 't Hooft coupling parameter $\lambda = g^2 N$. Weakly coupled Einstein-like gravity is a good approximation in the double-scaling limit of $N\rightarrow \infty$ and $\lambda\rightarrow \infty$.

During the development of the Standard Model in the 1970s, a subset of researchers turned their attention to the theoretical properties of black holes, despite the lack of direct experimental evidence for their existence at the time. It was proposed that, under conditions of sufficiently high and localized energy density capable of forming a black hole, the entropy of the system would not behave extensively—as is typical for thermodynamic systems—but would instead scale with the surface area of the black hole horizon \cite{Bekenstein:1972tm, Bekenstein:1973ur}. This idea introduced a significant revision to the conventional understanding of entropy in the presence of gravity.

Assigning thermodynamic properties, such as entropy or temperature, to black holes naturally leads to several foundational questions. For example: do black holes exist in astrophysical contexts? Are they fundamental constituents or do they represent in some sense average quantities? What are the relevant time scales for thermalization following black hole formation? Do black holes exhibit chaotic dynamics? What is the fate of global symmetries in the presence of black holes?

Information and its absence have long been recognized as foundational principles of statistical mechanics. In the context of black holes, this amounts to asking the question about the maximum amount of information that can be stored within a given region of spacetime \cite{BekensteinBound}. A related question inquires whether event horizons destroy information, leading to potential violations of unitarity in quantum mechanics \cite{Hawking1976}. Researchers explored the mechanisms through which information might be stored or lost in black holes, and whether spacetime geometry and information could be fundamentally linked.

\begin{figure}
     \centering
     \includegraphics[width=0.7\linewidth]{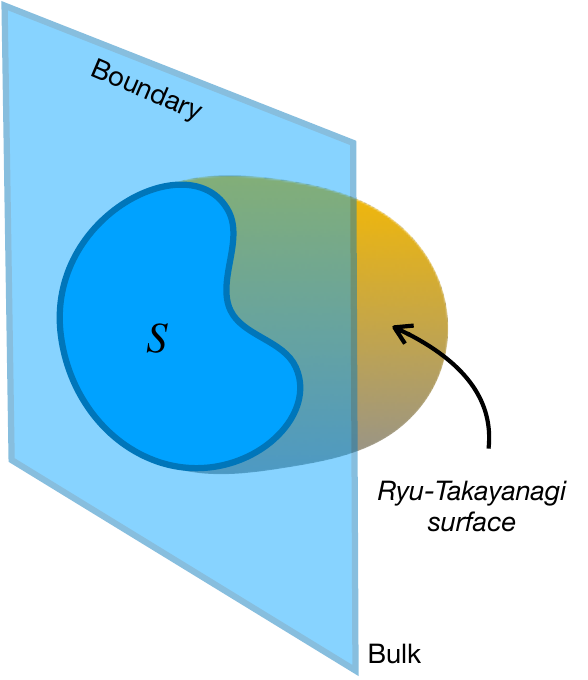}
     \caption{Bulk geometry (area in this case) = Boundary entanglement entropy, $S$.}
     \label{fig:RT}
 \end{figure}

This line of investigation gained traction with the development of the AdS/CFT correspondence. A concrete formulation of the relationship between geometry and information was provided by the Ryu-Takayanagi proposal, which connects the entanglement entropy in a quantum field theory (QFT) to the area of minimal surfaces in the corresponding gravitational theory \cite{Ryu:2006bv}. These authors have shown that if one wants to find the geometric entanglement between a region and its complement, its value can be obtained by calculating a geometrical quantity in the bulk.  The quantity in question is the area of the minimal surface that emanates from the boundary, anchored at the surface separating one region from the other, see Figure \ref{fig:RT}.  
 This result stimulated a large amount of research on relating information to geometry, see e.g. \cite{Swingle:2009bg}, \cite{VanRaamsdonk:2010pw}, \cite{Harlow:2014yka}, \cite{RangamaniTakayanagi2017}. 

\subsection{Time and space scales: the long, the small, the large and the complex } \label{subsect:Time_And_Space_Scales}
 
 \subsubsection*{Long time scales and very small but not zero time correlation functions: all you need is patience} 
 We will now describe a case in which both the extremely large and the extremely small values make an appearance. This happens when considering the long time behavior of correlation functions in the framework of QFT. 
 Consider a system in thermal equilibrium, and let it be represented by a canonical or a micro-canonical ensemble. Perturb it by the action of one or a few operators (as long as their number is much smaller than the entropy of the system) and inquire about the behavior in time of this perturbation. One might think that waiting long enough, the memory of  this perturbation will be well forgotten and forgiven, which is essentially true, but not quite.  
  
 We can see this, for example, by considering Poincar\'{e} recurrences. Consider a time-dependent correlation function $C(t_0)=\braket{\mathcal{O}(t_0,x_1)\mathcal{O}(0,x_2)}$.
 Classically the statement established by the Poincaré recurrence theorem is that if a dynamical system has compact phase space, and the flow preserves phase space volume, then for any tolerance $\epsilon >0$, there exists a time $t_P(\epsilon)$
 such that  
 \begin{eqnarray}
     \| C(t_P(\epsilon)) - C(t_0)\|<\epsilon
 \end{eqnarray}
 Quantum mechanically an analogous formula holds under the assumptions that the energy spectrum of the system is discrete and the system evolves unitarily \cite{QuantumRecurrenceTh} 
 This feature alone forbids a correlation function that exactly vanishes in the infinite-time limit. The recurrence theorem therefore tells us that the correlation function itself cannot approach zero in the $t\rightarrow \infty$ limit. However, often it is more convenient to talk about time-averaged correlation functions, in which case a concrete estimate of the lower bound can be obtained for chaotic systems.

In order to establish this estimate, we now discuss what a generic local operator, $\mathcal{O}$, looks like in a theory which is considered `chaotic'. If the unitary matrix, $U$, needed to diagonalize the operator looks like a random unitary matrix, we say that the spectrum of the theory is `chaotic' with respect to the operator $\mathcal{O}$. The Eigenstate Thermalization Hypothesis (ETH), \cite{PhysRevA.30.504,PhysRevA.43.2046,Srednicki_1999} \cite{DAlessio:2015qtq} provides the general structure of the matrix elements of such an operator in the energy basis:
\begin{eqnarray} \label{eq:ETH_Op}
    \braket{E_a|\mathcal{O}|E_b}= O(\bar{E})\delta_{ab} + e^{-S(\bar{E})/2} f(\bar{E},\omega)r_{ab}
\end{eqnarray}
where $\bar{E}=(E_a+E_b)/2$, $\omega=E_a-E_b$; the functions $O(\bar{E})$ and $f(\bar{E},\omega)$ are the one-point function and spectral function respectively; $S(\bar{E})$ is the microcanonical entropy; and $r_{ab}$ are random numbers with zero mean and unit variance.

\begin{roundedbox}[Haar random unitaries]
Haar random unitaries often arise in discussions of quantum physics of chaotic systems, and especially in quantum statistical physics. Physically a Haar random unitary can stand in for a unitary transformation of a sufficiently chaotic quantum system, so as to give an idea of the typical, or generic behavior of the system with respect to the given unitary. Often one may have in mind that a chaotic system scrambles states and operators strongly enough, so that the time-evolution operator $U(t)$ becomes sufficiently (pseudo-) random, that we can in effect replace it by a Haar-random unitary.

Mathematically, one constructs Haar random unitaries by sampling operators $V$ from the group U$(D)$, the compact Lie group of unitary $D\times D$ matrices, with a specific measure $\mu(S)$ that is uniquely defined (up to scaling) by requiring left invariance\footnote{One could also define a right-invariant Haar measure, but for a unimodular group, such as U$(D)$, these coincide.}, $\mu(S) = \mu(gS)$ under the unitary group, that is for $g\in $ U$(D)$, and $S$ a measurable subset of U$(D)$. Additonally, one requires the measure to be normalized $\mu(U(D))=1$. In practical applications, one often makes use of the so-called Weingarten calculus, which allows one to evaluate averages of moments of uniatries, such as 
\begin{equation}\label{eq.HaarMoments}
    \int_{U(D)} \mu(U) U \, \cdot \, U^\dagger  \ldots U \, \cdot \, U^\dagger
\end{equation}
in terms of the so-called Weingarten functions Wg($\sigma, D$). For example
\begin{equation}
    \int_{U(D)} \mu(U) U_{ij} \,\cdot \, U^*_{kl} = {\rm Wg}(1,D) \delta_{ik}\delta_{jl} = \frac{\delta_{ik}\delta_{jl}}{D}\,,
\end{equation}
identifying the first Weingarten function Wg$(1,D) = 1/D$. A much more detailed account is given in \cite{collins2006integration}.
Finally, with the aid of Haar random unitaries, we can also define Haar random states. For this we take a reference state, $|\psi\rangle$,  and then consider the orbit
\begin{equation}
    |\psi_U\rangle := U\left| \psi \right\rangle
\end{equation}
over the Haar measure, $\mu(U)$. A physical context in which Haar random states and the Weingarten calculus are extremely useful is that of Page's theorem, about the dynamics of reduced density matrices in uniatry quantum systems, see \cite{Page:1993wv}.
\end{roundedbox}
  The time dependence of correlation functions of such operators, as described below, is characteristic of time correlations of a large class of operators. In the presence of black holes there is evidence that such operators are generic.  Consider first the time evolution of a correlation function, 
\begin{eqnarray} \label{eq:CorrFunc}
    C(t)\!=\!\mathrm{Tr}[\rho\,\mathcal{O}(t)\mathcal{O}(0)] \!=\! \sum_{mn}\rho_m \mathcal{O}_{mn}\mathcal{O}_{nm}e^{-it(E_m-E_n)}\nonumber\\
\end{eqnarray}
of operators of the form \eqref{eq:ETH_Op}. Define $L(t)=|C(t)/C(0)|^2$ and its long-time average
\begin{eqnarray}
    \bar{L} = \lim_{T\to \infty}\frac{1}{T}\int_0^T dt\, L(t).
\end{eqnarray}
It can be shown \cite{Barbon:2003aq} that when the theory is unitary, the energy spectrum is discrete, and the operators satisfy ETH, that the time average, $\bar{L}$, must always be positive, that is non-zero. In essence correlations are bounded from below and an estimate of the  bound, under the hypotheses listed above, can be give as
\begin{eqnarray}
    \bar{L} \sim e^{-S}.
\end{eqnarray}
The existence of Poincar\'{e} recurrences are an indication that correlations in a quantum system cannot decay asymptotically to zero, which is made precise by the ETH-type argument above.  We have thus encountered a physical situation where the extremely small number $e^{-S}$ has made an appearance.
This estimate was shown \cite{Maldacena:2001kr}, \cite{Barbon:2003aq}, not to be respected by cases where one of the assumptions for the fluctuation bound did not hold. For example the spectrum 
of a particle propagating  in the presence of a classical black hole background is continuous and in that case the average of the correlation function indeed vanishes. The consequences of this behavior will be elaborated upon below. A non-unitary theory may violate the bound as well. 
It is important to recall that this bound as well as the other entropy-related bounds and estimates we discussed are of significance only as long as the entropy is finite.

\begin{figure}
    \centering
    \includegraphics[width=1\linewidth]{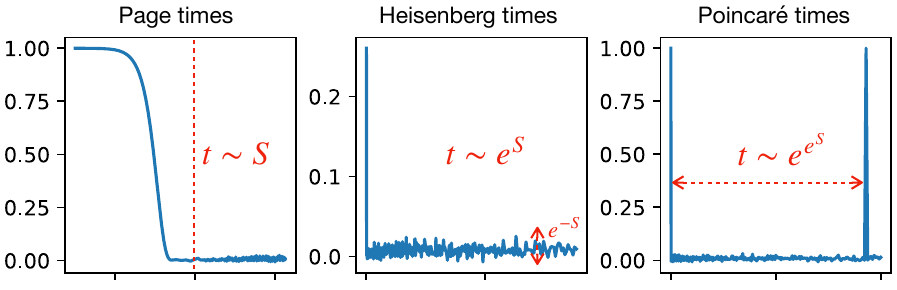}
    \caption{The time scales associated with a correlation function of the form \eqref{eq:CorrFunc} for a system with $S$ degrees of freedom, e.g. $S$ fundamental fermions in the Majorana SYK model.}
    \label{fig:ACor_times}
\end{figure}

We now turn to describe the more detailed time dependence of the correlation function.
The time dependence of the correlation function of the operator is loosely described by considering different time scales related logarithmically to each other. A pedagogical 
descriptions to extracting the different time scales appears in \cite{Barbon:2003aq, Barbon:2014rma}. See Figure \ref{fig:ACor_times} for a summary of what follows \footnote{See also \cite{Altland:2020ccq} for a definition of relevant timescales in quantum chaotic systems and furthermore \cite{Liu:2013iza,Hartman:2013qma} for entanglement time scales.}.

\begin{enumerate}
    \item $t \leq S$: The correlation function decays exponentially in time, for most systems, as $e^{-a\,t}$; the coefficient $a$ is not universal and may depend on the operator and the system. For integrable systems, the decay may be only power-like, indicating that in such systems it takes longer to propagate information between degrees-of-freedom.
Because of the estimated value of the time-averaged correlation function, this decay must end at times of order $S$ (times a characteristic  scale of the problem) . 
The time scale of order $S$ is also the time scale after which one can start to extract meaningful information from a black body or a black hole with entropy $S$. 
In the case of black bodies this time scale is called the Page time \cite{Page:1993wv, Page:2013dx}.
We remark that there is also the time scale of order log$(S)$ which is the scale it takes to distribute information about the perturbation among the degrees of freedom of the system, called the scrambling time \cite{Sekino:2008he}, mentioned in Section \ref{sect:Intro} (see discussion around Equation \eqref{eq:OTOCscrambling}).

\item $S< t< e^S $:
In this time  window the correlation function becomes increasingly sensitive to and aware of the discrete nature of the energy spectrum. 
The continuous approximation, applicable for shorter time-scales, starts failing to capture the correct behavior of the correlation function. For some correlation functions it is known \cite{Alhassid1992, Leviandier1986, Wilkie1991, Cotler:2016fpe} that
the correlation function plunges below its time average and then climbs up to the time average at times of order $e^S$; this time scale is called the Heisenberg time. 
A rough estimate of the average distance between energy levels, $\Delta E\sim e^{-S}$, coupled to the energy width uncertainty principle leads to this time scale, 
\begin{equation}\label{eq.HeisenbergTime}
t_H=\frac{2\pi \hbar}{\Delta E}
\end{equation}

In Figure \ref{fig:ACor_times} we show the autocorrelation function of a local operator in complex SYK. Note that the clear time-dependent profile is only visible after averaging over many realizations (or time averaging) since such quantities are not evidently self-averaging. 
Notice that the Heisenberg time introduces us to the extremely large number $e^S$ . It will reappear when we discuss complexity values.

\item $e^S< t < e^{e^S}$:
This is a period of absence of major changes while rich with tiny, order $e^{-b\,S}$ fluctuations around the time average value.  These fluctuations are sometimes characterized as 
noise, see \cite{Barbon:2014rma}.
This is the epoch where the system spends by far most of its time. It is the period that determines the time average value of the correlation function. 
In this time window one encounters a continuity of double exponential time scales.

\end{enumerate}

\begin{roundedbox}[The Poincaré scale simplified]
One way to appreciate the significance of time scales of the order $\exp(\exp(bS))$ is to imagine each of the phases in Eq.~\eqref{eq:CorrFunc} as a point on a circle defined by its angular position $\phi$. Let us assume that there are $e^S$ such phases, representing the energy levels of a quantum system with $S$ degrees of freedom. Consider the collections of such points on the circle at for example time $t=0$. Then we ask: at what time will all the phases return to within a distance at most $\delta \phi$ from their original position on the circle? This is equivalent to the requirements that at a time $t$, all phases $\theta_it$ return to a distance at most $\delta\phi$ within their original position mod $2\pi$, which mathematically is a Diophantine approximation problem. A simple estimate, using Dirichlet's theorem, reveals this scale to be 
\begin{equation}\label{eq.PoincareTime}
    t_{\rm P}\sim\exp(\exp(bS)) \,, 
\end{equation}
where $b= \delta\phi/(2\pi)$.
\end{roundedbox}

While the results discussed above were obtained in the framework of quantum mechanics, they do carry over to answering fundamental questions in a quantum theory containing gravity.
The vehicle for transporting this knowledge from QFT to a system including gravity is the AdS/CFT correspondence.

 \subsubsection*{ Holography requires that gravity respect topological diversity}

 In holographic duality a  state on the boundary that is dual to a particular geometrical background, in the semi-classical limit\footnote{Note, however, that not every state can be dual to a smooth geometry.}. A poster-boy case is the Thermofield Double (TFD) state.

 The boundary is taken to consist of two disjoint surfaces. It could be a four sphere.  The kets $|n_L\rangle, |n_R\rangle$ form an energy  basis for the theories on the boundaries  Left (L) and Right (R) with Hamiltonians $H_L$ and $H_R$  correspondingly. $E_n$ corresponds to the respective energy eigenvalues of the states $|n_L\rangle, |n_R\rangle$.  The TFD state is given by 
\[\ket{{\rm TFD}} \equiv \frac{1}{\sqrt{Z(\beta)}}\sum_n e^{-\beta E_n/2}\ket{n_L}\ket{n_R}\]
 where $Z(\beta)=\sum_n e^{-\beta E_n}$.
 If one traces out the information available on one of the boundaries the reduced density matrix of the other boundary is the thermal ensemble with  temperature $T=1/\beta$. 
 The TFD state is invariant under the boost operator $H_R-H_L$, and we can make it evolve in time by applying the total Hamiltonian, $H=H_R+H_L$. 
 What geometry corresponds to this state in the AdS/CFT correspondence? This is not straightforward.
 
\begin{figure}
    \centering
    \includegraphics[width=1\linewidth]{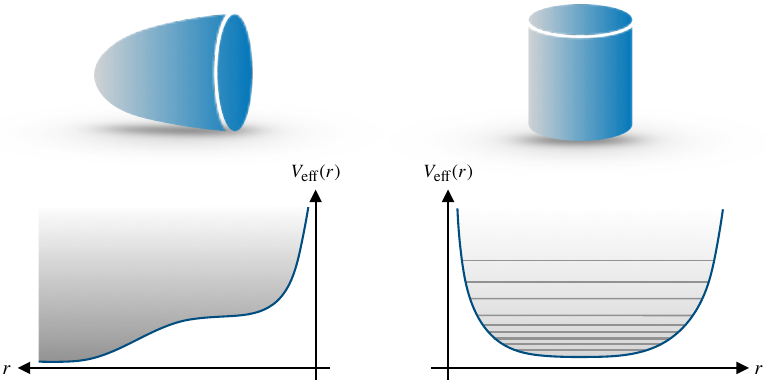}
    \caption{A black hole in AdS can be described by a `cigar' topology and a massless particle propagating in this background will have a continuous spectrum (left). Thermal AdS is described by a cylindrical topology; a massless particle moving in this background will have a discrete spectrum (right).}
    \label{fig:AdS_potential}
\end{figure}

 In Euclidean AdS, \citet{Hawking:1982dh} discovered two solutions to the classical equations of motion above a certain temperature.  One is a thermal AdS background, that has a non-trivial topology in Euclidean space-time, the other is a  black hole in AdS that classically has the trivial  topology of a cigar. Moreover they discovered that there is a first order phase transition, in the canonical ensemble,  between the two classical backgrounds. The transition temperature is called the Hawking-Page (HP) temperature, $T_{\rm HP}$, it occurs slightly above the temperature above which black holes become possible Euclidean solutions. The corresponding temperature scale is determined by the value of the negative cosmological constant, $\Lambda$. When the temperature is below the HP temperature the dual background is best described by a cylinder topology, see Figure \ref{fig:AdS_potential} (right) known as ``thermal AdS". This spacetime has non-trivial topology. Above the $T_{HP}$ the background is the black hole in AdS, see Figure \ref{fig:AdS_potential} (left). The Thermofield Double (TFD) state on the boundary therefore corresponds to the thermal Anti-de Sitter (AdS) geometry below the Hawking-Page transition temperature ($T_{\text{HP}}$).
Above $T_{\text{HP}}$, a common initial approach is to consider only the leading classical solution---the black hole in AdS---and to disregard thermal AdS, even though thermal AdS also remains a classical solution with the same conformal boundary.

Maldacena pointed out that this would lead to a breakdown of unitary in the holographic correspondence. The reason is that calculating the averaged time dependent correlation function on the boundary reveals that it does not decay exactly to zero at any temperature. Of course, this is just a concrete manifestation of the requirements imposed by unitarity of late-time correlations in any finite quantum system.

On the other hand, in the bulk, a calculation of the correlation function considering only the dominant black hole saddle yields a contradictory result.
Calculating the correlator in the black hole background leads to a continuous energy spectrum\footnote{This is strictly correct for a massless particle propagating in the black hole background. In the case that the particle has a non-zero mass, the continuous spectrum arises above a certain energy.} (see Fig.~\ref{fig:AdS_potential}, left).
This continuous spectral density, which falls outside the assumptions going into the unitarity bound, implies an exponentially decaying correlation function, and thus a vanishing time average -- an apparent contradiction.

To resolve this discrepancy, \citet{Maldacena:2001kr} suggested moving beyond considering exclusively the dominant “master field” (or saddle-point) and instead investigating the consequences of including both. Before detailing these consequences, it is pertinent to consider the broader significance of this inclusive approach. In the context of gravity, a long-standing question was whether quantum gravity could be consistently formulated within a single topological sector, or if it necessitates the inclusion of all topologies (in this instance, those sharing the same conformal boundary). Incorporating the highly suppressed thermal AdS component into the geometry implies that all such topologies should be considered, particularly if the principle of holography is to be upheld.

Indeed, with this inclusion, the time average calculated in the bulk geometry approaches more closely the unitary boundary result, while not agreeing (yet) at a fully quantitative level.
The contribution from thermal AdS introduces a discrete energy spectrum for a particle propagating within the potential walls imposed by AdS (see Fig.~\ref{fig:AdS_potential}, right), and gives a contribution to the signal that remains constant at late times. Further investigations have updated this picture somewhat, but still rely on topological diversity to resolve the tension with unitarity \cite{Saad:2018bqo}, \cite{Saad:2019pqd}, \cite{Cotler:2020ugk}, \cite{Altland:2020ccq,Altland:2022xqx}.
In these instances, semiclassical effects successfully capture these extremely small values (e.g., of order $e^{-S}$) characteristic of averaged, inclusive, coarse-grained quantities\footnote{In a different, but closely related development, topological diversity was also a key ingredient in the successful semiclassical computation of the unitary Page curve from gravity \cite{Almheiri:2019psf,Penington:2019kki,Penington:2019npb,Almheiri:2019qdq}}.

\begin{figure}
    \centering
    \includegraphics[width=0.8\linewidth]{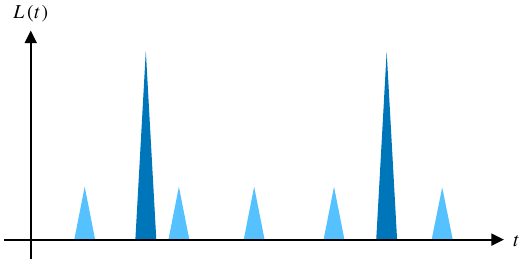}
    \caption{While the thermal AdS contribution to the time-averaged correlation function given the correct value it alone does not describe the exclusive correlation, that is the behavior at every time $t$ of the correlation function. According to the QFT, the average results from many very small contributions and not from a few sparsely spaced ones.}
    \label{fig:Lt}
\end{figure}

However, this success does not immediately extend to the corresponding fine-grained quantity.
Specifically, the detailed temporal evolution of the correlation function at late times (of order $O(S)$) cannot be reproduced merely by summing these two geometric contributions, see Figure~\ref{fig:Lt}.
(Up to times of order $S$, the dominant black hole contribution adequately describes this exclusive time behavior \cite{Barbon:2003aq}).

While the ability of semiclassical methods to yield such extremely small numbers for averaged quantities is impressive, a concern arises that capturing the corresponding exclusive, fine-grained properties may prove significantly more challenging for these approaches, see however \cite{Cotler:2016fpe} for some simpler cases.

The reason we entered in some detail into this issue in the case at hand is that we wanted to set the scene for the closely related but not identical case of calculating complexities.

\begin{figure*}[ht]
\justifying
\begin{roundedbox}[Chaotic timescales: from local thermalization to Heisenberg]
\begin{center}
\includegraphics[width=0.8\textwidth]{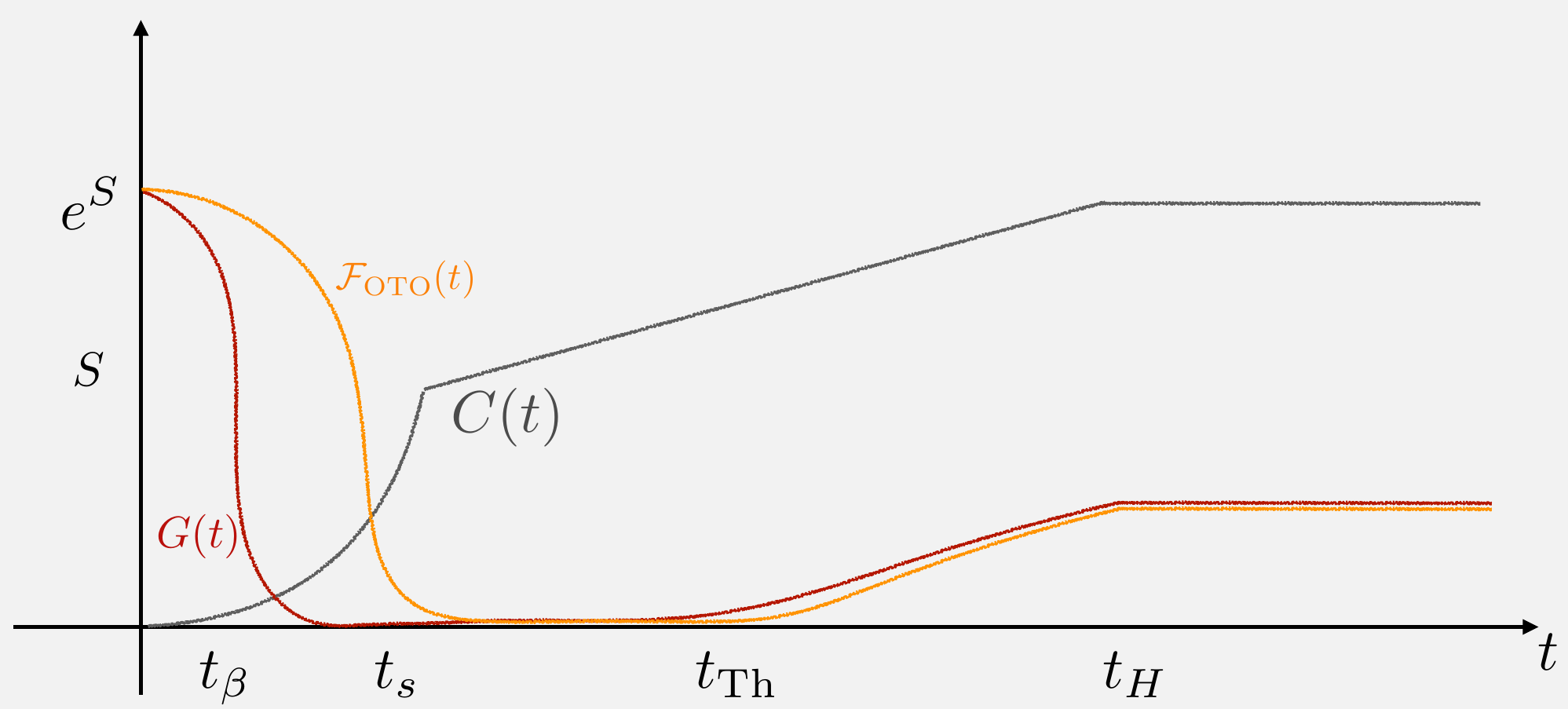}\end{center}
    \vskip1em
    We depict the behavior of an (un-normalized) correlation function $G(t) = {\rm Tr} \left[e^{-\beta H} {\cal O}(t) {\cal O}(0) \right]$, the out-of-time order correlator  ${\cal F}_{\rm OTO}(t)$, defined in \eqref{eq.OTOC4pt}, as well as complexity, $C(t)$, highlighting universal time scales present in typical chaotic systems. The first time scale shown is $t_\beta \sim \beta$, the local thermalization timescale. At this point time-ordered correlations will have decayed to very small values. Next, we see $t_s \sim \log S$, the scrambling time, the typical timescale on which a quantum system `forgets' its initial conditions. This is diagnosed by the exponential decay of certain OTO correlations, as shown \cite{Maldacena:2015waa}. Complexity (e.g. Krylov) shows a crossover from exponential to linear growth at this scale, while a two-point correlation function has no special feature. The past-scrambling complexity rises linear in time. The next timescale shown is $t_{\rm Th}$ the so-called Thouless time, which corresponds to the beginning of the linear ramp seen in correlation functions and spectral probes, characteristic of level repulsion of the underlying chaotic system . A generally accepted definition of the Thouless time, defines $t_{\rm Th}$ as the earliest timescale on which a system can be well approximated by a random-matrix ensemble in the Wignerian sense. Moving on in time post Thouless, both correlation functions and complexity plateau at the Heisenberg time, $t_{H} \sim e^S$, as shown. The curves shown are schematic, and correspond to averaged versions; the average could be over a sliding time window, for example, or an artificially (or not) introduced source of disorder. The microscopic versions would have strong fluctuations around the mean values, starting around the Thouless time. More detailed descriptions of the behavior of correlation functions in chaotic systems can be found in \cite{Delacretaz:2020nit}, with an emphasis on hydrodynamics, and in \cite{Altland:2020ccq, Altland:2021rqn} with an emphasis on quantum chaotic behavior of holographic systems.
\end{roundedbox}
\end{figure*}

 \subsubsection*{Very large Volumes, Time scales and Actions: identifying the characteristics }

\begin{figure*}
    \centering
    \includegraphics[width=1\linewidth]{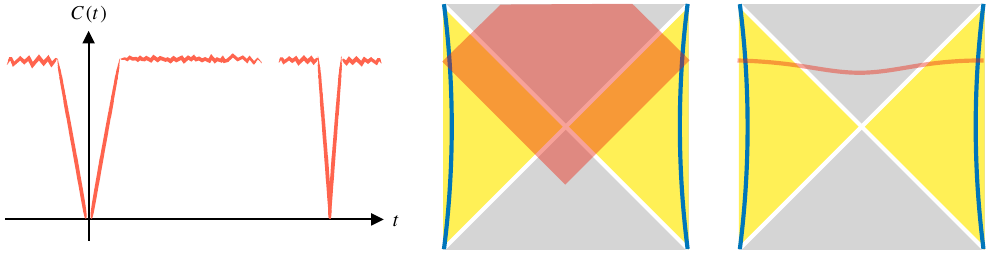}
    \caption{The Einstein-Rosen bridge (right) and the Wheeler-de Witt patch (middle) are two possible descriptions for holographic complexity. They both grow linearly in time initially, and are expected to exhibit a similar time-dependent profile (left).}
    \label{fig:hologrphic_complex}
\end{figure*}

We now move to study large spatial distance features in which extremely large and extremely small quantities emerge.
We consider again the Penrose diagram and define what is called the Einstein-Rosen bridge (ERB), which is also referred to as the (Lorentzian) wormhole, and focus on its volume. 
The definition starts by picking  one point on the AdS boundary in CFT$_1$ with coordinates $(x_1,t_1)$ and another in CFT$_2$ with coordinates $(x_2,t_2)$. To be concrete, the choice can be made that $x_1=x_2$ and $t_1=t_2$. One defines the volume of the ERB as the maximal volume that extends between these two boundary points. Due to geometric divergence near the AdS boundary, this volume needs to be renormalized\footnote{This is an example of a standard procedure in AdS/CFT, called `holographic renormalization'. See e.g. \cite{Skenderis:2002wp} for a review of this subject.}. Intuitively one would like to define as the volume of the ER bridge the value of that part of this maximal volume that extends between the parts where it crosses the respective horizons, see Fig.~\ref{fig:hologrphic_complex} (right). In practice one applies a regularization scheme that attempts to implement the intuition that in particular one would like to assign zero value to the volume at the time the horizons intersect that is at the bifurcation surface\footnote{Note that by injecting certain types of matter in the past, one can resolve the bifurcation point increasing the shortest distance between the two branches as much as one wants \cite{PhysRevX.14.011024, Barbon:2025bbh}.}. That time is usually chosen to be zero.
Let us now follow the evolution of the value of the ERB's volume as a function of time. We first consider the evolution for positive time beginning at $t=0$. 
The calculations can be performed for a variety of black hole backgrounds in AdS, with and without charge and in various space-time dimensions \cite{Brown:2015lvg}. They are done in the region of validity of semi-classical gravity, in particular away from any singularities.\footnote{In effect, requiring a maximal volume pushes them away from singularities, so they do not effectively probe them (see e.g. \cite{Barbon:2015ria})}  The results have some common features:
In these cases the behavior of the time derivative of the volume, once the volume is slightly different from zero, is constant, proportional to $TS$, were $T$ and $S$ are the temperature and  the entropy, respectively, that are associated with the background. It is negative for negative values of $t$ and positive for positive values of $t$. The behavior for small values of $t$ 
is less universal. 
A particularly notable feature of this behavior is that apparently the volume of the ERB corresponds to a physical quantity which keeps growing (linearly) up to timescales exponentially large in entropy, far after the thermalisation and scrambling processes have reached their stationary equilibrium.

A salient and motivating question for this review is what physical quantity this should correspond to precisely.

Moreover, in addition to the ERB, there are other objects that were found to exhibit a similar behavior. They include the classical WdW action, which is the action calculated over the causality diamond of the 
initial data, see Fig.~\ref{fig:hologrphic_complex} (middle). In \cite{Bolognesi:2018ion} one sees that both prescriptions share some universal behavior but are far from being identical.  In fact, \cite{Belin:2021bga} have enlarged the class of prescriptions that give rise to similar complexity time-evolution profiles for a given background. In Figure \ref{fig:hologrphic_complex}
we remind the Reader of some different bulk prescriptions. These give rise to the same asymptotic behavior of complexity but may have non-leading prescription-dependent behaviors. 
At this stage there are no clear criteria differentiating them and in principle each of them should have a translation into some CFT boundary object. 

In any case, all of them contain objects of interest whose order of magnitude is proportional to $\exp(S)$ and whose value continues to evolve up to exponentially large times.

Recall that we have already encountered a time scale of the order  $\exp(S)$. This scale, the Heisenberg time, was introduced earlier in Equation \eqref{eq.HeisenbergTime}. In the present context, this time scale is far beyond the time scale it takes a BH to thermalize, and even larger than the time scale after which the black hole starts radiating the stored information -- that time scale is of order $S$ -- the Page time.

Talking about the ERB volume makes sense at temperatures above the respective $T_{\rm HP}$ temperatures of the different 
backgrounds. What happens when the temperature is below $T_{\rm HP}$? The appropriate saddle-point solution is then thermal  AdS, dual to the boundary TFD at temperatures below the Hawking-Page transition. In this background the ERB volume is not  really well defined due to the absence of horizons. The time derivative would change discontinuously as one crosses the HP transition. In addition when, as is the case for the BH in AdS, the boundary theory has a discrete spectrum and is unitary, one should observe, for $N$ large but not infinite,  the Sisyphic cycle of Poincare recurrences.

One may ask whether we would ever be able to observe Poincaré recurrences in a theory of gravity, as a question of principle, not simply because the time scale it implies is so cosmologically large. As we have already emphasized, a necessary condition for a unitary quantum theory to show recurrences is a discrete spectrum and an effectively finite Hilbert space of states contributing to our observables\footnote{This means that if the Hilbert space is not finite, we should consider individual microcanonical windows or a similar physical truncation.}. In this case an argument along the lines we gave in the Poincaré info box above, will ensure recurrences on a doubly exponential time scale.
On the other hand, if as has been discussed extensively in recent years, the bulk theory of gravity is in fact fundamentally an ensemble average, then even if we have an average over individually unitary theories with an effectively finite Hilbert space and a discrete spectrum, the ensemble average will introduce an exponential damping in real-time correlation functions that suppresses recurrences. Note that extensive work has shown that Thouless-time and even Heisenberg-time signatures in correlation functions and spectral form factors are still discernible even if the bulk is fundamentally an average \cite{Saad:2019lba},\cite{Saad:2018bqo}, \cite{Altland:2020ccq}. This suggests that a strategy to demonstrate that a given bulk theory is {\it not} an ensemble average would be to find such Poincaré recurrences from a bulk calculation. A similar argument was put forth in \cite{Barbon:2003aq}, which pointed out that dropping unitarity in the usual proof of Poincaré recurrences also leads to an effective damping that suppresses these recurrences, making these recurrences an effective diagnostic of unitarity itself.

The recurrences also suggest that the volume of the bridge should reach a maximum at some place. Another indication of the saturation of the volume of the ERB is obtained by considering the following spatial correlation function
extending across the horizon $\langle {\cal O}(r_1, t){\cal O}(r_2, t)\rangle =\exp (-\text{length} (ER))$. The average value of this correlation function can be argued, using ETH-type logic to saturate at late times to a small value, of ${\cal O}\left(e^{-S}  \right)$. This corresponds to the saturation behvior of the volume. (note in the time-averaged correlator there appears $S$ and not $\exp(S)$)
This leads to the following expected time dependence of the bulk dual of complexity. 

\begin{enumerate}
    \item The region $0<t< \log S$: in the period until the scrambling time the volume grows very fast, in fact exponentially.
    \item The region $\log (S)< t < \exp (S)$: in the period until the Heisenberg time the volume grows linearly with a slope proportional to $TS$.
    \item The region $\exp(S)< t <  \exp(b \exp(S))$: 
    if one were to trust semi-classics, similarly to time correlation functions, the volume continues to grow linearly. But although there is no immediate indication that this approximation is invalid, it is expected that the volume saturates at a value proportional to $\exp(S)$ times the length scale in the problem. The coefficient $b$ was already  described earlier in Eq.~\eqref{eq.PoincareTime} in a simple example. For every $b$ there are regions of measure $\exp(S)/\exp(b\exp(S))$ for which the volume drops back to near zero and then rebounds in time of order $\exp(S)$, as described in the 
time windows above, to it plateau value of $\exp(S)$. This scenario eventually repeats itself. Looking at this figure one eventually realizes that it is only for those ``tiny'' $\exp(S)$ times that one is fortunate to have (at large $N$) a semiclassical description to hang onto. These are measure zero time intervals compared to the time the system fluctuates with noise around the plateau. 

The background chosen has a $t\to -t$  symmetry which we now summarize.
\item The value starts being very large, of  ${\cal O}(\exp(S))$, at least for negative times of order $-{\cal O}(\exp(S))$, it decreases until a time $t=0$ where it vanishes whereupon it climbs up at least to
times of order $+\exp(S)$ where it reaches values of order $\exp(S)$. The extremely large number $\exp(S)$ we were seeking, appears both in the value of the volume reliably calculated semi-classically and the time it takes to reach this value, forward or backwards from $t=0$.
\end{enumerate}

In summary then, we have exhibited several geometric features of semiclassical black holes that has a number of striking properties in common, including the fact that they continue to evolve in time until exponentially large times, far beyond local thermalization and scrambling time scales\footnote{It is interesting to note that entanglement entropy also shows non-linear growth followed by linear growth and saturation, but its bulk mechanism is different from that of complexity, and consequently the time scales do not work out correctly. In particular, entanglement entropy saturates already a time scale of ${\cal O}(S)$, which is reflected in it being used as a standard probe of thermalization, rather than the comparatively much later-time phenomenology of complexity.}. Laying our trust into the holographic correspondence of bulk gravity and boundary CFT, we should thus look for a corresponding physical property of the dual quantum system that exhibits the same qualitative features, and ultimately attempt to formulate a precise match between a particular geometric quantity in the bulk and a particular physical quantity on the boundary.

It was suggested \cite{Susskind:2014rva} that this concept is akin to the concept of complexity in quantum information theory.
With the guidance of hindsight, we may thus describe the period beyond scrambling,  the complexification process, analogously to the thermalization process, which goes on until the theory reaches complexity equilibrium.
  It may be surprising that the precise details involved in a system thermalizing is still an active field of research, although in some sense
the field of complexification has just been born.

\subsubsection*{The complexity plateau} 
We have motivated that the linear in time growth of the complexity needs to stop and move into a very long phase when the complexity saturated, up to small noise, at the value of $\exp(S)$. When we discuss the various concepts of Complexity in a boundary QFT with finite entropy, we find that the estimate of $\exp(S)$ appears in one way or another in all of them. Thus the responsible adult  QFT directs the whatever corresponds in the bulk to the boundary complexity should indeed saturate for a long time at that extremely large value.  The question and the burden of an answer bounces back to the bulk. Is there a semi-classical description of the saturation process. When we discussed the time average correlation function we mentioned that such an effective description appears once one always topological diversity of the master fields. This notion of topological diversity was employed from several points of view to attempt to explain the saturation. Allowing for baby universes and wormholes in the bulk suggested mechanisms leading to the length saturation. In the framework of low dimensional bulk theories of gravity also effects rising from the summation of all genera of two dimensional world sheet is claimed to lead to the saturation. We will return to the issue of bulk saturation in Section \ref{sec.BulkSaturation} below. We note that the assumed time evolution of the complexity behaves like one over the time correlation graph. In particular, long-lived small plateaus transform into long lived large saturation values with the corresponding estimates of $\exp(-S)$ and $\exp(S)$. An additional object of interest of magnitude $\exp(-S)$ was identified in a similar context in the bulk. Some authors suggest to view the
magnitude of the ERB to represent the number of states or operators the respective state or operators mix with a time goes on. We emphasized that our interest lies in large but
finite entropies, this means that in the picture of mixing, for example with other states, there comes a time when one has saturated the possibility to mix with new states and one is just repeating on and on mixing with states that have already been explored and are thus mot independent. An explicit calculation \cite{Barbon:2025bbh} shows in some case that indeed the states one thought as independent of each other and emerge as time crosses the Heisenberg time, actually are not orthogonal but have as an overlap the extremely small magnitude of $\exp(-S)$.

The complexity evolution profile presented in Figure~\ref{fig:hologrphic_complex}, 
can be generic under certain circumstances, but by any stretch it is not the only profile corresponding to all bulk configurations. In the presence of certain types of singularities, the volume actually decreases and one can identify complexity that decreases in time, $dC/dt<0$. This can be shown with the volume prescription for complexity \cite{Barbon:2015ria} and the WdW prescription \cite{Bolognesi:2018ion}. 
We discussed thermal AdS where the complexity is essentially constant in time $dC/dt\approx 0$. There is a background \cite{Hartnoll:2020rwq, Hartnoll:2020fhc, Auzzi:2022bfd, An:2022lvo} for which is seems that the ERB collapses at a certain time. It turns out to be rather complicated and not necessarily illuminating to try and detect the very specific features of this background using the bulk prescription for complexity.

In general, when dealing with a specific semi-classical gravity issue, the task is to identify the appropriate background that would reveal the relevant time-evolution of complexity in that case. Notice that the basic properties of the bulk descriptions discussed, do not depend on any tolerance parameter, although various regulators need to be introduced to obtain finite numbers. This is, as we explained before, a feature shared with the notion of Krylov complexity. We now proceed to an in-depth discussion of this quantity.

\section{Krylov complexity}\label{sec.CloserKrylov}

We now embark on a more detailed and in-depth exploration of the notion of Krylov complexity, which will also revisit some of the ideas already encountered in the Introduction.
Krylov complexity, at first sight, approaches the notion of quantum complexity from a very unconventional angle, not shared by most other notions of complexity. From the point of view of quantum mechanics, on the other hand, the idea behind Krylov complexity is both very natural and simple. Like circuit complexity, Krylov complexity can be defined both for operators and for states.
In the former case, we will see that Krylov complexity formalizes, and in fact generalizes, the idea of quantifying the growth of an operator with respect to a basis of reference or `simple' operators, which can be chosen canonically. In the case of states, it quantifies the spread in time of an initial state over a naturally preferred basis of the Hilbert space which can be unambiguously identified.

This Section will provide the specific definition of Krylov complexity and a comparative analysis that will illustrate the extent to which it can not only be seen as an instance of quantum complexity, but in fact as a generalization. We shall also provide a thorough review of the abundant numerical and analytical tools that have been developed in the recent years to study Krylov complexity. This exposition may be used as a toolkit for the Reader's own research, and to understand the physical results that will constitute the subject of section \ref{Sec:Krylov_Pheno}.

This Section is structured as follows: subsection \ref{subsect:KC_and_size} will give a brief discussion of size complexity, its main features and drawbacks, as a motivation for Krylov complexity, which will be presented as a generalization of this concept that avoids the aforementioned weak points; subsections \ref{subsect:KrylovSpaceOps} and \ref{sec:Lanczos_operator} provide the elementary ingredients used to define Krylov complexity, namely the concept of Krylov space, the Lanczos algorithm and its associated Lanczos coefficients; with this, subsection \ref{Sec:Krylov_Dynamics} presents the definition of Krylov complexity and introduces several tools that are useful for the study of the dynamics of time evolution in Krylov space; most of the text is written using a notation adapted to the operator formalism, but subsection \ref{sect:State_formalism_framework} will review the specific formulation of the Lanczos algorithm and the definition of Krylov complexity in the states formalism; subsection \ref{subsect:theorems} will review two important theorems that formalize the fact that Krylov complexity is an \textit{optimal} representative of a suitably defined class of generalized quantum complexities; in subsection \ref{subsect:KC_geometry} we shall review some results that compare Krylov complexity to Nielsen's geometric approach to quantum complexity; subsection \ref{subsect:toolbox} provides a number of tools that have been extensively used in the literature in order to analyze properties of the Lanczos coefficients and Krylov complexity; to complement the latter, subsection \ref{subsect:numerical_implementations} provides a detailed discussion of the state of the art of the numerical algorithms with which the Lanczos algorithm may be efficiently implemented; finally, extensions of this algorithm to other contexts of physical interest (such as multiple-seed analyses, time-dependent Hamiltonians, open systems and discrete dynamics) are gathered in subsection \ref{subsect:Extensions}, which shall compactly review the corresponding frameworks and point the Reader to the relevant references.

\subsection{Krylov complexity and size complexity} \label{subsect:KC_and_size}

One of the most basic elements of the theory of quantum mechanics is the Heisenberg evolution of an operator ${\cal O}$ with time,
\begin{equation}\label{eq:Heisenberg_evol_op_introCloserLook}
    {\cal O}(t) = e^{i H t} {\cal O} e^{-iHt}\,.
\end{equation}
Consider for example an operator ${\cal O}$ that is constructed from elementary building blocks, $\{\Phi_i \}$. Cases of such fundamental building block are elementary fields\footnote{The Reader may use the Majorana fermions of the SYK model as an instructive example, or alternatively a basis of on-site Pauli operators in the context of spin chains.} appearing in the definition of $H$, conformal blocks, and symmetry-preserving operators. 
If the theory is interacting, as in the elementary example of the introduction \eqref{eq.OperatorGrowthElementary}, successive commutators will increase the number of elementary building blocks of the operator as time progresses\,,
\begin{equation}
\overbrace{\Big[H,\big[H,\cdots , [H,}^{(n)}{\cal O}]\Big] \sim {\cal O} \Phi \Phi \cdots {\cal OO}\Phi\,,
\end{equation}
where the right hand side schematically denotes any set of words of length up to $n$ that can be made from the operator itself and the set of elementary operators that appear in the Hamiltonian. 

It may be more illuminating to specialise this idea to a quantum system made up of a number of sites hosting local qubits and a set of simple Pauli operators acting on them, such as the Ising Hamiltonian \eqref{eq.SimpleSpinHam} considered in the introduction. In this case we can be more concrete and define a basis of Pauli strings \cite{Roberts:2014isa}, i.e. monomials of the form $\prod_i \sigma^{a_i}_i$ where $\sigma^{a_i}_i$ is a Pauli matrix ($a_i\equiv x,y,z$) or the $2\times 2$ unit matrix ($a_i\equiv 0$) at site $i$. Then, for a Hamiltonian $H$ made up of local (or more generally $k-$local) terms in the Pauli operators, we find that, for a specific on-site Pauli matrix $\sigma^{a_j}_j$: 
\begin{equation}\label{eq:commutators_Pauli_Strings}
    \left[H, [ H, \ldots    [H,\sigma^{a_j}_j] \right] = \sum_{s,I} a_I^{s} \pi_I^{s}\,,
\end{equation}
where $\pi_{I}^{s}$ denotes the set of all Pauli strings of length $s$ and ``$I$'' should be viewed as a multi-index, as will become clear below. We see that the seed operator $\sigma^{a_j}_j$ has grown into the linear superposition of larger Pauli strings by successive commutation with the Hamiltonian. Time evolution \eqref{eq:Heisenberg_evol_op_introCloserLook} is, by the Baker-Kramer-Campbell-Hausdorff formula, an expansion in ever increasing numbers of such commutators and, for interacting $k$-local systems, these cause the operator to evolve into an ever expanding superposition of Pauli strings of growing size\footnote{It should be noted that in free (quadratic) systems, where $k=2$, a commutator of $H$ with an on-site Pauli matrix $\sigma_j^{a_j}$ will yield some linear superposition of still on-site Pauli matrices, in which case the seed operator will not grow in size during its time evolution.}.
In order to make this qualitative discussion more precise, let us now review the formal definition of operator `size' applicable in this context.
\subsubsection*{Size complexity}
The definition of operator complexity that we shall consider follows from the desire to quantify operator size by its projection over the elements of the basis

\begin{eqnarray}\label{eq:size_basis}
  \Bigl\{\pi_I^{s}=\sigma_{i_1}^{a_1}\dots\sigma_{i_s}^{a_s}~,\quad 1\leq i_1<i_2<\dots<i_s\leq N~,\Bigr.\nonumber\\
  \Bigl. a_s\in\{x,y,z\}~\Bigr\},
\end{eqnarray}
where $N$ is the length of the spin chain. By convention, we denote $\pi^{0}=\mathbb{1}$, and note that we exclude $a_i=0$ in the range of the spin index at sizes $s>0$ in order to avoid over-counting the identity. As promised, we see that the multi-index $I$ at fixed $s$ would then be the collection $I=\{i_1,a_1,\dots,i_s,a_s\}$. For \textit{qudits} (instead of qubits) the range of the index $a_s$ at $s>0$ in the basis above is a discrete set of cardinality $d^2-1$.

Let us now expand the time-evolving operator in the Pauli string basis,
\begin{equation}
{\cal O}(t) = \sum_{s,I}c_I^{s}(t)\pi_{I}^{s}
\end{equation}
which then gives rise to the probability distribution for size $s$
\begin{equation}\label{eq:size_probability}
P_{s}(t) = \sum_{I}|c_I^{s}(t)|^2\,.
\end{equation}

\begin{figure}
    \centering
    \includegraphics[width=0.9\linewidth]{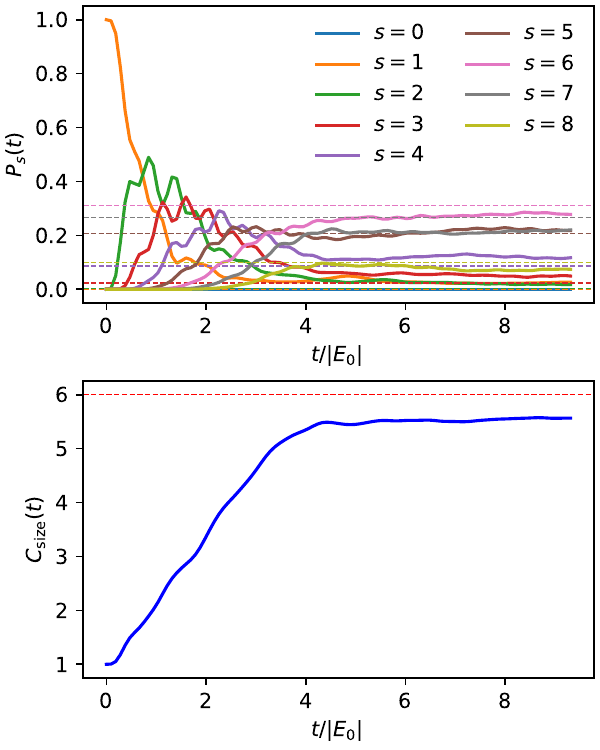}
    \caption{Probability distributions (top) defined in \eqref{eq:size_probability} and size complexity (bottom) defined in \eqref{eq.DefSizeComplexity} for the operator $\sigma_1^z$ in the Ising model, $H=-\sum_{i}\left[\sigma_i^z\sigma_{i+1}^z+g\sigma_i^x+h\sigma_i^z\right]$ with parameters $g=-1.05,h=0.5$ and $N=8$ spins; in this parameter regime the Hamiltonian is chaotic \cite{Bauls2010StrongAW}. The dashed lines show the estimates \eqref{eq:size_prob_late_times} (top) and \eqref{eq:size_comp_late_times} (bottom). At all times, the probabilities add up to 1. The time axis is normalized by the absolute value of the lowest energy eigenvalue. Note that the size definition used here is different from that in \cite{Roberts:2014isa} and agrees with that in \cite{Roberts:2018mnp}. We note that size complexity in this spin chain is not seen to grow exponentially, in resonance with the behavior of out-of-time-order correlation functions in the same model, which is known not to be chaotic in the sense of scrambling, cf. \citet{Roberts:2014isa}; \citet{Chen:2018hjf}. In contrast, \citet{Roberts:2018mnp} do find an exponentially growing size complexity in the SYK model.}
    \label{fig:sizeComplexity}
\end{figure}

With this probability distribution in hand one can define the average size of the operator by 
\begin{equation}\label{eq.DefSizeComplexity}
C_{\rm size}(t) = \sum_{s=0}^N s P_s(t)\,,
\end{equation}
where we have chosen the notation $C_{\rm size}(t)$ to indicate the notion of `size complexity'. In a typical quantum system, such as the simple spin Hamiltonian \eqref{eq.SimpleSpinHam} to which we have been alluding so far, when initialized on a simple on-site operator,  the probability $P_s(t)$ starts at size $s=1$, reflecting the initial choice of simple operator, and then subsequently moves its maximum to larger values of $s$ as time evolves. The aggregate effect is to cause $C_{\rm size}(t)$ to increase until eventually saturating for a finite system. An estimate for the saturation value can be given following similar arguments to those in \citet{Roberts:2018mnp}: at late times, in a sufficiently chaotic system, the operator should be well described by a random choice among all possibilities at every size $s$, in which case the probabilities $P_s$ can be computed by dividing the number of independent Pauli strings of size $s$ by the total number of inequivalent strings. For a chain of $N$ qudits, where each chain site is described by a $d$-dimensional Hilbert space, we can use eq. (2.66) in \citet{Sanchez-Garrido:2024pcy} and we find:
\begin{equation}
    \label{eq:size_prob_late_times}
    \lim_{t\to\infty}P_s(t)\approx  \frac{(d^2-1)^s}{d^{2N}}\binom{N}{s}~,
\end{equation}
which then yields the following size complexity saturation value at late times:
\begin{equation}
\label{eq:size_comp_late_times}
\lim_{t\to\infty}C_{\text{size}}(t)\approx \sum_{s=0}^{N}s\frac{(d^2-1)^s}{d^{2N}}\binom{N}{s}=\frac{d^2-1}{d^2}N~.
\end{equation}
Note that for the case of a qubit system we have $d=2$ and therefore the size saturation value is equal to $\frac{3N}{4}$, while in the limit $d\to\infty$ it becomes equal to $N$, reflecting the fact that in this limit the number of strings of size $s$ becomes infinitely larger than the number of strings of size $s-1$, for any $s=1,\dots,N$, and in particular $\lim_{d\to\infty}\frac{(d^2-1)^s}{d^{2N}}\binom{N}{s}=\delta_{s,N}$. In any case, for general $d$ size saturates at a value of order $\mathit{O}(N)$, and even if at late times the operator was not a random superposition of chains of all possible sizes but just some Pauli string of large size, the size complexity saturation value would never surpass the system size $N$, which bounds it from above\footnote{To give another example, we can consider the size saturation value for a Majorana fermion in a system of $N$ Majoranas with SYK$_4$ interactions \cite{kitaev2015kitp}: in this case we have $P_s \propto \binom{N}{s}$ and, since only odd strings of Majoranas contribute to successive nested commutators of the Hamiltonian with the initial fermion,  we find $\lim_{t\rightarrow \infty}C_{\rm size}(t) = \sum_{s\,\,\textrm{odd}}^N  \frac{s}{2^{N-1}}\binom{N}{s}=\frac{N}{2}$.}. We show a numerical example of size complexity, its projections on fixed size and its overall time evolution, for a simple spin system in Figure \ref{fig:sizeComplexity}.

A comment is necessary at this point: In order to be allowed to call the $P_s(t)$ in \eqref{eq:size_probability} a \textit{probability}, we have made the tacit assumptions that the size basis \eqref{eq:size_basis} is \textit{orthonormal} (so that the probability of a fixed size is given by the sum of the modulus-squared of the coefficients $c_I^s$ corresponding to independent strings of the same size) and that the operator $\mathcal{O}(t)$ is normalized at all times (so that the probabilities sum up to one). This illustrates the fact that we \textit{need} to introduce the notion of an operator inner product in order to formalize the definition of size complexity. Therefore, we re-express the above developments in way that manifestly makes use of a Hilbert space structure. It is well known (see e.g. \cite{reed1980functional}) that the space of linear operators acting on Hilbert space ${\cal H}$ can itself be seen as a Hilbert space, denoted $\hat{\cal H}$. Elements of this Hilbert space are in one-to-one correspondence with the linear operators acting on ${\cal H}$
\begin{equation}\label{eq:operator_Hilbert_space}
 {\cal O} \,\, \textrm{acts on}\,\, {\cal H} \quad \longleftrightarrow \quad |{\cal O}) \in \hat{\cal H}\,.
\end{equation}
This Hilbert space is endowed with an inner product, which may be taken to be
\begin{equation}\label{eq:op_inner_product_introSectIII}
( {\cal O} | {\cal W}) = \frac{{\rm tr} \left[ \rho {\cal O}^\dagger {\cal W} \right]}{{\rm tr}\left[\rho\right]}\,,
\end{equation}
for any $|\mathcal{W}),|\mathcal{O})\in\widehat{\mathcal{H}}$.
Often the operator $\rho$ is taken to be the canonical density matrix $\rho = e^{-\beta H}$, or even more conveniently its infinite temperature limit, whence $( {\cal O} | {\cal W}) = \frac{1}{D}{\rm tr} \left[ {\cal O}^\dagger {\cal W} \right]$ (i.e. the conventional Frobenius inner product\footnote{Note that the Pauli strings of the basis \eqref{eq:size_basis} are indeed orthonormal with respect to the Frobenius inner product thanks to the identity $\sigma^{a}\sigma^b=\delta^{ab}\mathbb{1}+i\varepsilon^{abc}\sigma^c$ and to the fact that $\text{Tr}(\sigma^{a})=0$ for all $a=x,y,z$.}), where $D=\dim(\mathcal{H})$.
The inner product with which operator Hilbert space is defined remains a choice, but it admits the physical interpretation of a tool that can be used to focus on different sectors of the spectrum of the theory by controlling through $\rho$ the weight with which they contribute to \eqref{eq:op_inner_product_introSectIII}.
The inner product in operator space allows to define an operator norm which one can then decompose in terms of projections onto the subspaces spanned by strings of fixed size, yielding the following more formal (yet equivalent) version of the probabilities $P_s(t)$ in \eqref{eq:size_probability}:
\begin{equation}\label{eq.sizeProbabilityAbstract}
P_s(t) = \sum_{\{{\pi} \textrm{ has size }s \}} \frac{\left|\left(\pi | {\cal O}(t)\right)\right|^2}{(\pi|\pi) ({\cal O} | {\cal O})}
\end{equation}

Size complexity is, to summarize, a useful observable that measures operator growth and hence constitutes a probe of chaos, which has been directly related to infite-temperature out-of-time-order correlators \cite{Roberts:2018mnp} and has found application in holographic contexts (e.g. \cite{Roberts:2014isa}). The fact that it does not necessitate a tolerance parameter to be well-defined and finite is another of its attractive features. However, it has some drawbacks that pose obstacles for both its generalization and its applications. For example, for the description of the growth of the Einstein-Rosen bridge in holography. Even though we have already addressed them in the previous discussion, let us compactly summarize such drawbacks:
\begin{itemize}
    \item The ``size eigenbasis'' is useful to define a notion of complexity in situations in which the building blocks of the Hamiltonian that generate time evolution are themselves elements of the size basis with small size, as otherwise the behavior of complexity as a function of time would be extremely erratic and saturate at very early times. The size basis \eqref{eq:size_basis} is well adapted to describe evolution in spin systems such as \eqref{eq.SimpleSpinHam}, but this is not guaranteed to be the case in general.
    \item As we have noted in our previous discussion, size complexity is bounded from above by system size, $N$ in the example that we addressed, and it indeed saturates at a value of order $\mathit{O}(N)$ once chaotic dynamics have efficiently randomized the operator $\mathcal{O}(t)$. It has been widely studied \cite{Roberts:2018mnp, Shenker:2013pqa, Roberts:2014isa,Maldacena:2015waa}  that the size of typical operators in sufficiently chaotic systems grows exponentially fast in time, $C_{\rm size}\sim e^{\lambda t}$ (where $\lambda$ is the Lyapunov exponent), implying that saturation is attained at the \textit{scrambling time}, $t_s\sim\lambda^{-1}\log N $. From the holographic point of view (See Section \ref{sec.HoloComplexity}), in order to describe the Einstein-Rosen bridge growth we would need a boundary observable that behaves non-trivially up to exponentially late times $t\sim e^{\mathit{O}(N)}$, \cite{Barbon:2003aq}, after which the full Hilbert space should have been probed and the classical, geometrical description of gravity would cease to apply \cite{Maldacena:2001kr, Brown:2015lvg,Susskind:2018pmk}. Therefore, size complexity fails to fulfill this crucial requirement for being a holographic complexity in this sense. We can in fact pinpoint the reason why size saturates \textit{too early} from what would be desirable in this context: Its eigenvalues are highly degenerate, in the sense that many independent Pauli strings in the basis \eqref{eq:size_basis} have the same size: Because of this, even though after the scrambling time, the operator $\mathcal{O}(t)$ may continue to probe independent basis elements not previously explored, all the different size eigenvalues will have already been probed by $t_s$.
\end{itemize}

\subsubsection*{Krylov complexity is better}

Given the above-mentioned drawbacks of size complexity, an immediate question would be if there is a different canonical basis in $\hat{\cal H}$ in which the probability \eqref{eq.sizeProbabilityAbstract} and associated complexity is richer and natural in the sense of being straightforward to generalize to arbitrary Hamiltonians. Indeed, we will argue that the so-called Krylov basis is the sought-after basis, and the resulting notion of complexity has rich and revelatory features \cite{Barbon:2019wsy}. 

Instead of relying on a choice of basis that, while natural from certain points of view, is a pure choice -- for example the set of on-site Pauli monomials, or Majorana monomials we introduced above, or any other such choice of ``fundamental constituent'', a more canonical approach would be to choose the basis by appealing to the notion of complexity of time evolution itself. As we will see throughout this section, Krylov complexity quantifies complexity as the position of the time evolving operator, or state, with respect to a very generic basis that can be defined \textit{a priori}, without knowing the specific details of the Hamiltonian of the theory. Krylov complexity offers a universal definition for the time-evolution of any operator or state under any given Hamiltonian. Such a very generic basis is nothing but a suitable orthonormal rearrangement of the consecutive powers of the time evolution generator acting on the initial condition. The algorithm by which this orthonormal basis is produced is the Lanczos algorithm \cite{Lanczos:1950zz}, on which we will shortly elaborate. This algorithm, often also referred to as the recursion method, has traditionally been used in the context of many-body physics due to its very intimate relation to spectral properties and to quantities relevant to linear response theory. We will say some words about this; \citet{ViswanathMuller} is an excellent review for these applications. As reviewed in \citet{ViswanathMuller}, the so-called Lanczos algorithm may be applied to states, for which the time evolution generator is the Hamiltonian, in the so-called \textit{Hamiltonian formalism}, or to operators in their corresponding Hilbert space, for which time evolution is generated by the adjoint action of the Hamiltonian, i.e. the Liouvillian, in the so-called \textit{Liouvillian formalism}. The idea of using the Krylov basis output by this algorithm to define a notion of quantum complexity was put forward by \cite{Parker:2018yvk}, where they focused on the case of operator complexity due to its richer structure and relation to properties of two- and four- point functions. In this spirit, we will begin by introducing the Krylov complexity formalism directly in operator space, and we will use Section \ref{sect:State_formalism_framework} to give details on the state formalism, which has been studied to a lesser extent but which also is of great interest for holography, given its relation to the thermofield double state and the spectral form factor.

\subsection{Krylov space and Krylov space dimension}\label{subsect:KrylovSpaceOps}

Formulating the precise algorithm to calculate K-complexity starts by taking 
 a closer look at the structure of the subspace of the operator Hilbert space that is relevant to time evolution, which we shall refer to as Krylov space. As mentioned earlier, for now we will focus on the operator formalism, while the state formalism will be discussed in section \ref{sect:State_formalism_framework}\footnote{Historically, the seminal papers by A. N. Krylov \cite{Krylov:1931} and C. Lanczos \cite{Lanczos:1950zz} introduced Krylov methods and the Lanczos algorithm (respectively) in what we would nowadays refer to as the Hamiltonian, or states, formalism. See \cite{Sanchez-Garrido:2024pcy} for an historical contextualization. For a systematic comparison between the Lanczos algorithm in the Hamiltonian and Liouvillian formalisms, see \cite{ViswanathMuller}.}.

Let us consider a hermitian operator, $\mathcal{O}=\mathcal{O}^\dagger$, evolving in the Heisenberg picture under a time-independent hermitian Hamiltonian, $H$:
\begin{equation}\label{Opt}
    \mathcal{O}(t) = e^{iHt}\,\mathcal{O}\,e^{-iHt} = \mathcal{O} + it\, [H,\mathcal{O}] +\frac{(it)^2}{2}[H,[H,\mathcal{O}]]+\dots
\end{equation}
where the Baker-Kramer-Campbell-Hausdorff formula was used in the second equality.  Before proceeding further we can already notice a few things. First, if $[H,\mathcal{O}]\neq 0$, we identify a buildup of linear combinations -- as time increases, terms with more and more commutators with the Hamiltonian become important in this linear combination.  Second, as already analyzed in the earlier discussion on size complexity,
if the system has a notion of $k$-locality and the operator is initially simple, commuting with the Hamiltonian will cause it to grow.  
As can be seen in \eqref{Opt}, the operator's time evolution happens in a subspace of the operator Hilbert space. This subspace is known as the \textit{Krylov space}:
\begin{equation}
    \label{Krylov_sp}
    \mathcal{K} := \mathrm{span}\Bigl\{ \mathcal{O},\, [H,\mathcal{O}],\, [H,[H,\mathcal{O}]], \dots \Bigr\} ~.
\end{equation}
To simplify the notation, it is often useful to introduce the \textit{Liouvillian super-operator}, $\mathcal{L} \equiv [H,\cdot]$. In the spirit of the definition of operator Hilbert space in \eqref{eq:operator_Hilbert_space}, we shall use the term \textit{super-states} to refer to elements of operator space, and \textit{super-operators} to refer to linear operators acting on $\widehat{\mathcal{H}}$. Formally, the Liouvillian is defined through its action on an arbitrary super-state:
\begin{equation}
    \label{eq:Loouvillian_action_def}
    \mathcal{L}|\mathcal{O}) = \Big|[H,\mathcal{O}]\Big)~,
\end{equation}
for any $|\mathcal{O})\in\widehat{\mathcal{H}}$. We can now rewrite the definition of Krylov space as:
\begin{equation}
    \label{Krylov_sp_L}
    \mathcal{K} = \mathrm{span}\{ |\mathcal{O}),\, \mathcal{L}|\mathcal{O}),\, \mathcal{L}^2|\mathcal{O}), \dots \} ~.
\end{equation}
By construction, the Krylov space is a subspace of operator space, which we may denote as $\mathcal{K}\leq\widehat{\mathcal{H}}$, and we may define it formally as the minimal subspace of operator space (in the sense of minimal dimension) that contains $\big|\mathcal{O}(t)\big)$ for any $t\in\mathbb{C}$ (i.e. it is also suitable for Euclidean time evolution). An immediate question that arises at this stage, and whose answer will indeed provide a good deal of physical insight, is the specific value of the dimension of $\mathcal{K}$. Let us denote the \textit{Krylov dimension} as 
\begin{equation}
    \label{eq:Krylov_dimension}
    K:=\dim\left(\mathcal{K}\right)~.
\end{equation}
The naive set $\left\{\mathcal{L}^n|\mathcal{O)}\right\}_{n\geq 0}$ whose span defines the Krylov space as in \eqref{Krylov_sp_L} has an infinite cardinality but, as we have just discussed, by definition Krylov space is always a subspace of operator space. In particular, in systems with a finite number of degrees of freedom, the dimension of operator space $\widehat{\mathcal{H}}$ will be finite, while the cardinality of the naive set $\left\{\mathcal{L}^n|\mathcal{O)}\right\}_{n\geq 0}$ will remain infinite, implying that some (in fact, infinitely many) elements of the latter will not be linearly independent. In this spirit, we can provide a first straightforward bound for the Krylov dimension $K$. Using $D$ to denote the dimension of the Hilbert space of states, $D:=\dim(\mathcal{H})$, which typically scales exponentially with the number of degrees of freedom $D\sim e^S$ and may therefore be finite or infinite, it immediately follows that $\dim(\widehat{\mathcal{H}})=D^2$ and, as a consequence of the fact that $\mathcal{K}\leq \widehat{\mathcal{H}}$, we have that
\begin{equation}
    \label{eq:Op_Krylov_dim_naive_bound}
    K\leq \dim(\widehat{\mathcal{H}})=D^2~.
\end{equation}
In \cite{I}, a tighter (and yet completely generic) upper bound on the Krylov space dimension was found. Its derivation is based on expressing the operator in the energy basis $\{|E_a\rangle\}_{a=1}^D$, whose elements are Hamiltonian eigenstates verifying $H|E_a\rangle = E_a |E_a\rangle$. With this, we can write
\begin{equation}\label{eq:Operator_energy_basis}
    \mathcal{O} = \sum_{a,b=1}^D O_{ab} \, |E_a\rangle \langle E_b|,
\end{equation}
where $O_{ab}$ are the matrix elements of the operator in coordinates over this basis. Defining the energy differences or \textit{phases} as
\begin{equation}
    \label{eq:phases_def}
    \omega_{ab}:=E_a-E_b~,
\end{equation}
we can see that commuting $H$ with the independent elements $|E_a\rangle\langle E_b|$ in \eqref{eq:Operator_energy_basis} yields the corresponding phase as an overall prefactor, namely
\begin{eqnarray}
    \label{eq:Commutator_ket_bra}
    \Big[ H,|E_a\rangle\langle E_b| \Big]=\omega_{ab}|E_a\rangle\langle E_b|~.
\end{eqnarray}
Now, recalling the definition of operator space \eqref{eq:operator_Hilbert_space} and the notation we use for it, we shall represent the ket-bras $|E_a\rangle\langle E_b|$ as elements $|\omega_{ab})\in\widehat{\mathcal{H}}$ labeled by the relevant energy difference. In this notation, the identity \eqref{eq:Commutator_ket_bra} is manifestly equivalent to the statement that the super-states $|\omega_{ab})$ are eigenstates of the Liouvillian with eigenvalue $\omega_{ab}$. Putting everything together, we have that \eqref{eq:Operator_energy_basis} gives a decomposition of $|\mathcal{O})$ in terms of Liouvillian eigenstates: 
\begin{align}
    &|\mathcal{O}) = \sum_{a,b=1}^D O_{ab}|\omega_{ab})~,\label{eq:O_L_estates_decomp} \\
    &\mathcal{L}|\omega_{ab})=\omega_{ab}|\omega_{ab}) \label{eq:L_estates}
\end{align}
Equipped with this\footnote{Note that in passing from \eqref{eq:Operator_energy_basis} to \eqref{eq:O_L_estates_decomp} there is an implicit redefinition of $O_{ab}$, since we take the super-states $|\omega_{ab})$ to have unit norm, while the ket-bra $|E_a\rangle\langle E_b|$ is in general not normalized according to the inner product \eqref{eq:op_inner_product_introSectIII}. Since it will not affect our discussion, let us ignore this subtlety and accept that the coordinates $O_{ab}$ are defined through \eqref{eq:O_L_estates_decomp}.}, we may now express the elements $\mathcal{L}^n|\mathcal{O})$ that span Krylov space in coordinates over the Liouvillian eigenbasis:
\begin{equation}
    \label{eq:LnO_L_ebasis}
    \mathcal{L}^n|\mathcal{O}) = \sum_{a,b=1}^D \omega_{ab}^n O_{ab}|\omega_{ab})~,\qquad\text{for any  }n\geq 0~.
\end{equation}
The above expression tells us that successive powers of the Liouvillian acting on $|\mathcal{O})$ have the net effect of amplifying the projection of the latter on the corresponding Liouvillian eigenspaces, in particular not rotating such projections within each eigenspace. Using $|\sigma(\mathcal{L})|$ to denote the number of distinct eigenvalues in the spectrum of the Liouvillian, gathered in the set $\sigma(\mathcal{L})\equiv\left\{\omega_j\right\}$, we can rewrite \eqref{eq:LnO_L_ebasis} more suggestively as:
\begin{eqnarray}
    \label{eq:Ln_O_projections}
    \mathcal{L}^n|\mathcal{O}) = \sum_{j=1}^{|\sigma(\mathcal{L})|}\omega_j^n \sum_{\substack{ (a,b) ~{\rm s.t} \\\omega_{ab}=\omega_j}}O_{ab}|\omega_{ab})~.
\end{eqnarray}
To make the discussion clearer, let us define the \textit{eigenstate representatives} as the (non-normalized) projection of $|\mathcal{O})$ over each eigenspace:
\begin{equation}\label{eq:eigenspace_representatives}
    |K_\omega) = \sum_{\substack{ (a,b) ~{\rm s.t} \\\omega_{ab}=\omega}} O_{ab}|\omega_{ab})~.
\end{equation}
Note that the spectral theorem\footnote{In order to apply the spectral theorem, we need $\mathcal{L}$ to be hermitian with respect to the operator space inner product. It can be shown that, for any canonical density matrix $\rho=e^{-\beta H}$, this property is indeed fulfilled given the inner product \eqref{eq:op_inner_product_introSectIII}.} ensures that these projections are orthogonal, i.e.
\begin{equation}
    \label{eq:eigenspace_reps_orthog}
    \left(K_{\omega}|K_{\omega^\prime}\right)=0\qquad \text{if  }\omega\neq\omega^\prime
\end{equation}
because by definition they belong to different eigenspaces of the Liouvillian.
Using these super-states, equation \eqref{eq:Ln_O_projections} becomes:
\begin{eqnarray}
    \label{eq:Ln_O_Kw}
    \mathcal{L}^n|\mathcal{O})=\sum_{j=1}^{|\sigma(\mathcal{L})|}\omega_j^n|K_{\omega_j})~.
\end{eqnarray}
Finally, we see that the Krylov space is nothing but the span of all these orthogonal eigenspace projections:
\begin{eqnarray}\label{eq:Kspace_span_projections}
    \mathcal{K} = \text{span}\left\{\mathcal{L}^n|\mathcal{O})\right\}_{n\geq 0} =\mathrm{span}\{ |K_\omega) \}_{\omega\in\sigma(\mathcal{L})}~.
\end{eqnarray}
If there is an eigenspace over which the starting $|\mathcal{O})$ has no projection, then it will not contribute with any independent directions to the Krylov space \eqref{eq:Kspace_span_projections}. An immediate corollary is the following: The Krylov dimension $K$ is equal to the number of Liouvillian eigenspaces over which the initial operator $\mathcal{O}$ has a non-zero projection \cite{I}.

A further corollary of the above result, also given in \cite{I}, is the promised tighter upper bound on the Krylov space dimension. Given that the spectrum of the Liouvillian is given by the phases $\omega_{ab}=E_a-E_b$, for $a,b=1,\dots,D$, it follows that the zero eigenvalue is always at least $D$ times degenerate, since $\omega_{aa}=0$ for any $a=1,\dots D$ regardless of the details of the Hamiltonian. Since the corresponding eigenspace can only contribute with one direction to the Krylov space \eqref{eq:Kspace_span_projections}, this shrinks the upper bound on the Krylov dimension $K$ from $D^2$ in \eqref{eq:Op_Krylov_dim_naive_bound} down to:
\begin{equation}
    \label{Krylov_dimension_bound}
    K \leq D^2 -D -1 ~.
\end{equation}

For completeness, let us provide a more algebraic and visual proof for the value of the Krylov dimension.
Arranging in a matrix the coordinates over each linearly independent direction in expression \eqref{eq:LnO_L_ebasis} we obtain: 
\begin{equation}
    \label{Vandermonde}
    \begin{pmatrix}
        O_{11} & \dots & O_{DD} & O_{12} & O_{13} & \dots & O_{D-1,D} \\
        0 & \dots & 0 & O_{12}\omega_{12} & O_{13}\,\omega_{13} & \dots & O_{D-1,D}\,\omega_{D,D-1}\\
        0 & \dots & 0 & O_{12}\omega_{12}^2 & O_{13}\,\omega_{13}^2 & \dots & O_{D-1,D}\,\omega_{D,D-1}^2\\
        0 & \dots & 0 & O_{12}\omega_{12}^3 & O_{13}\,\omega_{13}^3 & \dots & O_{D-1,D}\,\omega_{D,D-1}^3\\
        \vdots & & \vdots & \vdots & \vdots & & \vdots
    \end{pmatrix}~.
\end{equation}
The above is a Vandermonde matrix, whose rank by construction gives the number of independent elements in the set $\left\{\mathcal{L}^n|\mathcal{O})\right\}_{n\geq 0}$, and hence the Krylov dimension $K$. Using the properties of Vandermonde matrices, we conclude that $K$ is equal to the number of \textit{distinct} phases $\omega_{ab}$ corresponding to non-zero matrix elements $O_{ab}$ of the operator in the energy basis, which is equivalent to the statement below equation \eqref{eq:Kspace_span_projections}.

Noticeably, the value of the Krylov dimension $K$ and its proximity to the bound \eqref{Krylov_dimension_bound} is very sensitive to the spectral structure of the Liouvillian (and hence the Hamiltonian) and to the form of the operator in the energy basis. In section \ref{Sec:Krylov_Pheno} we will revisit this aspect in relation to the discussion about chaotic and integrable systems.

\subsection{Lanczos algorithm, Krylov basis and Lanczos coefficients} \label{sec:Lanczos_operator}
The previous section, where the definitions of Krylov space and its dimension were provided, may be seen as the description of the kinematic setup. In order to describe dynamics, and to eventually define a notion of complexity that describes operator growth in an optimal manner, we need to give ourselves an orthonormal basis for Krylov space, which we will eventually dub the \textit{Krylov basis} \cite{ViswanathMuller,Parker:2018yvk} and which, as we shall discuss, will turn out to be an optimal generalization of the size eigenbasis discussed in \eqref{eq:size_basis}.

The Krylov space spanning set, $\{ \mathcal{O},\, \mathcal{L}\mathcal{O},\, \mathcal{L}^2\mathcal{O}, \dots \}$, is generally not guaranteed to be orthogonal with respect to the inner product in operator space, as the construction of these states is independent from the latter Section \ref{Sec:Inner_prod} will provide a discussion of a family of inner product choices that generalize the example given in \eqref{eq:op_inner_product_introSectIII}, but for now let us just denote the inner product by $(\mathcal{A}|\mathcal{B})$ and assume that, besides fulfilling the usual inner product axioms, it has the following properties \cite{Hornedal:2022pkc} (which are indeed satisfied by the family of inner products discussed in \ref{Sec:Inner_prod}):
\begin{enumerate}
    \item $(\mathcal{A} |\mathcal{L}|\mathcal{B})= (\mathcal{L}\mathcal{A} |\mathcal{B})$ for any operators $\mathcal{A}$ and $\mathcal{B}$, i.e. the Liouvillian is a \textit{hermitian super-operator}.
    \item $(\mathcal{A} |\mathcal{L}|\mathcal{A})= 0$ for hermitian operators $\mathcal{A}=\mathcal{A}^\dagger$. It can be shown that this property eventually implies that $\left(\mathcal{A}|\mathcal{L}^{2n+1}|\mathcal{A}\right)=0$ for any non-negative integer $n$. 
\end{enumerate}
The norm of an operator will be denoted by $\|\mathcal{A}\|:=\sqrt{(\mathcal{A}|\mathcal{A})}$.  With these properties of the inner product, an orthonormal basis for Krylov space. can be achieved by performing a Gram-Schmidt-like process known as the Lanczos algorithm \cite{Lanczos:1950zz}. In order to build some intuition, let us perform explicitly a couple of \textit{Lanczos steps}. We will assume that the initial operator is an observable and is therefore Hermitian, $\mathcal{O}=\mathcal{O}^\dagger$. The \textit{zeroth} step is to normalize the initial operator $\mathcal{O}$, defining $\mathcal{O}_0=\mathcal{O}/\|\mathcal{O}\|$.  To define the next operator -- orthogonal to it -- we act with the Liouvillian and subtract the projection of the new direction over $\mathcal{O}_0$:
\begin{equation}
   |\mathcal{A}_1 ) = \mathcal{L} |\mathcal{O}_0) - (\mathcal{O}_0|\mathcal{L}|\mathcal{O}_0) \mathcal{O}_0 = \mathcal{L} | \mathcal{O}_0)
\end{equation}
where in the last equality we used the hermiticity of $\mathcal{O}_0$ and the second property of he inner product, listed above, which in this case ensures that $\mathcal{L}|\mathcal{O}_0)$ is directly orthogonal to $|\mathcal{O}_0)$. The next task is to normalize this new orthogonal direction (if it is not null) thereby defining the next normalized direction in the Krylov space. Thus, if $\|\mathcal{A}_1\|\neq 0$ we define the first \textit{Lanczos coefficient}, $b_1=\|\mathcal{A}_1\|$, and the next \textit{Krylov element}, $\mathcal{O}_1 = \mathcal{A}_1/\|\mathcal{A}_1\|$. For the next step, we find the direction orthonormal to both $\mathcal{O}_0$ and $\mathcal{O}_1$ by acting with $\mathcal{L}$ on $\mathcal{O}_1$ and performing the adequate subtractions:
\begin{align}
    |\mathcal{A}_2) &= \mathcal{L} |\mathcal{O}_1) - (\mathcal{O}_1|\mathcal{L}| \mathcal{O}_1) |\mathcal{O}_1) - (\mathcal{O}_0|\mathcal{L}| \mathcal{O}_1) |\mathcal{O}_0) \nonumber\\&= \mathcal{L} |\mathcal{O}_1) - b_1| \mathcal{O}_0)
\end{align}
where both of the assumed properties of the inner product have been used. Again, if the new direction is not zero, we define, the second Lanczos coefficient $b_2 =\|\mathcal{A}_2\|$ and the next Krylov element $\mathcal{O}_2 = \mathcal{A}_2/\|\mathcal{A}_2\|$. Interestingly \cite{Lanczos:1950zz}, subsequent iterations of this procedure will remain structurally simple in the sense that only one subtraction will need to be performed at each Lanczos step in order to obtain a new Krylov basis element orthogonal to all the previously constructed ones, as one can show inductively.

\begin{roundedboxw}[The Lanczos algorithm for operators]
Let us now give the generic form of the \textit{Lanczos algorithm} in the operator formalism \cite{ViswanathMuller}: \\

\noindent{\textbf{Input:} $\mathcal{O}, \mathcal{L}$}

\begin{enumerate}
    \item Define $|\mathcal{O}_0) = \frac{1}{\|\mathcal{O}\|}|\mathcal{O})$
    \item Compute $|\mathcal{A}_1)=\mathcal{L}|\mathcal{O}_0)$\\
    If $\|\mathcal{A}_1\|\neq 0$ 
    define $b_1 = \|\mathcal{A}_1\|$ and $|\mathcal{O}_1)=\frac{1}{b_1}|\mathcal{A}_1)$ \\
    Otherwise, end the process.
    \item For $n>1$:\\
    Compute $|\mathcal{A}_n) = \mathcal{L}|\mathcal{O}_{n-1})-b_{n-1}|\mathcal{O}_{n-2})$\\
    If $\|\mathcal{A}_n\|\neq 0$ define $b_n = \|\mathcal{A}_n\|$ and $|\mathcal{O}_n)=\frac{1}{b_n}|\mathcal{A}_n)$  \\
    Otherwise, end the process.
\end{enumerate}
\textbf{Output:} \textit{Lanczos coefficients} $\{b_n\}_{n=1}^{K-1} $ and orthonormal \textit{Krylov basis} $\{|\mathcal{O}_n)\}_{n=0}^{K-1} $.\\

The process ends at $n=K$, when $\|\mathcal{A}_K\|=0$ and all independent orthonormal directions are exhausted. As a remark, let us note that it can be proved inductively that performing a Gram-Schmidt process on the set $\{ |\mathcal{O}),\, \mathcal{L}|\mathcal{O}),\, \mathcal{L}^2|\mathcal{O}), \dots \}$ results in the same orthonormal basis $\{|\mathcal{O}_0), |\mathcal{O}_1), \dots, |\mathcal{O}_{K-1})\} $, with the advantage that the Lanczos algorithm only requires a sequence of $K-1$ orthogonalization coefficients (the Lanczos coefficients $b_n$) instead of a $K\times K$ upper-triangular coefficient matrix as is generically the case for the Gram-Schmidt process.
\end{roundedboxw}

It is worth to point out that the output Krylov basis $\{|\mathcal{O}_n)\}_{n=0}^{K-1} $ preserves a notion of ordering in the sense that each state $\mathcal{L}^n|\mathcal{O})$ is a linear combination of Krylov elements $|\mathcal{O}_m)$ with $m=0,\dots,n$, i.e. the Krylov basis is an orthonormal basis of Krylov space that the time-evolving operator progressively explores during its time evolution. In this sense, it generalizes the size eigenbasis \eqref{eq:size_basis}, which is only useful for theories in which the building blocks of the Hamiltonian are $k$-local operator strings. In section \ref{subsect:theorems} we will review more formal arguments establishing that the Krylov basis is, in fact, an \textit{optimal} generalization of the size, or similar, bases.

So far, the Lanczos coefficients $b_n$ have appeared as some rather obscure sequence of orthogonalization coefficients, but they turn out to carry a great deal of physical information.
As we will see below and in section \ref{Sec:Krylov_Dynamics}, they characterize the dynamics of the spreading of the operator $\mathcal{O}(t)$ over the ordered Krylov basis in a very direct manner. 
They also encode important information about the 2-point autocorrelation function, $C(t) \equiv (\mathcal{O}|\mathcal{O}(t))$, and its moments \cite{ViswanathMuller} as we shall also review in section \ref{Sec:LancCoef_Moments}.

Let us now discuss some features and implications of the Lanczos algorithm.

\subsubsection{A hopping model on the Krylov chain} \label{subsect:Krylov_Chain_Hopping}
The recurrence step in the Lanczos algorithm reads $\mathcal{L} |\mathcal{O}_{n-1})= b_n |\mathcal{O}_n) + b_{n-1}|\mathcal{O}_{n-2})$, and the orthogonality of the Krylov basis $(\mathcal{O}_n|\mathcal{O}_m)=\delta_{nm}$ means that in the Krylov basis, the Liouvillian has the tridiagonal form of size $K\times K$:
\begin{equation} 
\label{eq:Liouvillian_tridiag}\mathcal{L}\overset{*}{=}
        \begin{pmatrix}
            0 & b_1 &  &  &  & \\
            b_1 & 0 & b_2 &  &  & \\
             & b_2 & 0 & b_3 &  & \\
             &  & b_3 & 0 & \ddots & \\
             &  &  & \ddots & \ddots & b_{K-1} \\
             &  & & & b_{K-1} & 0
        \end{pmatrix} ~,
\end{equation}
where the star above the equal sign is used to denote that the right-hand side is the matrix representation of the (super-)operator in the left-hand side in coordinates over a specific basis, in this case the Krylov basis, i.e. the right-hand side is a the matrix $\Big((\mathcal{O}_m|\mathcal{L}|\mathcal{O}_n)\Big)_{m,n=0}^{K-1}$.

More compactly, we may write the following basis-independent expression which follows directly from the Lanczos algorithm:
\begin{equation}
    \label{eq:L_Krylov_hopping_model}
    \mathcal{L}=\sum_{n=0}^{K-2}b_{n+1}\Big(|\mathcal{O}_n)(\mathcal{O}_{n+1}|+|\mathcal{O}_{n+1})(\mathcal{O}_n|\Big)~.
\end{equation}
Recalling that the operator $\mathcal{O}$, seen as an element of the operator Hilbert space $|\mathcal{O})\in\widehat{\mathcal{H}}$, evolves in the Heisenberg picture as $|\mathcal{O}(t))=e^{it\mathcal{L}}|\mathcal{O})$, and inspecting expression \eqref{eq:L_Krylov_hopping_model}, we see that we have mapped operator evolution to a one-dimensional quantum mechanical hopping problem on the so-called \textit{Krylov chain}, where the hopping amplitudes are nothing but the Lanczos coefficients. This fact, known since \cite{ViswanathMuller,Parker:2018yvk} led to recognizing a relation between Krylov complexity dynamics and the Anderson localization problem by \cite{II,III}, as we shall review in section \ref{subsect:Krylov_localization}.

\subsubsection{Lanczos coefficients and moments}
\label{Sec:LancCoef_Moments}
Besides having a physical interpretation as hopping amplitudes in the Krylov chain, the Lanczos coefficients are in one-to-one correspondence with the Taylor series coefficients of the operator two-point function, which in turn implies that they are deeply related to the analytical structure of the latter, as well as its various cousins (Laplace and Fourier transforms). In order to review this, let us start by considering the 2-point autocorrelation function:
\begin{eqnarray}
    C(t) &:=& (\mathcal{O}|\mathcal{O}(t)) = \sum_{m=0}^\infty \frac{(it)^{2m}}{(2m)!}\mu_{2m}, \label{eq:Op_two_pt} \\
    &&\mu_{2m} := (\mathcal{O}|\mathcal{L}^{2m}  |\mathcal{O})~,
\end{eqnarray}
which can be directly obtained from \eqref{Opt}. Provided that $\mathcal{O}=\mathcal{O}^\dagger$, by the second property of the inner product, i.e. $(\mathcal{O}|\mathcal{L}|\mathcal{O})=0$, it can be shown that all odd \textit{moments} $\mu_{2n+1}$ are zero (i.e. the autocorrelation function is an even function of time), which has been directly assumed in the first line of the equation above. Note that, for an initially normalized operator (which we assume), $\mu_0=1$.

The subset of moments, $\left\{\mu_{2k}\right\}_{k=0}^m$ is in one-to-one correspondence to the subset of Lanczos coefficients, $\{b_n\}_{n=1}^m$ through a combinatorial relationship involving a sum over \textit{Dyck paths} of length $2m$ \cite{ViswanathMuller,Parker:2018yvk}:
\begin{eqnarray}\label{eq:moments_Lanczos_Dyck}
    \mu_{2m} = \sum_{\{n_1,n_2,\dots,n_m\}} (b_1^2)^{n_1}(b_2^2)^{n_{2}} \dots (b_m^2)^{n_{m}}
\end{eqnarray}
where the set $\{n_1,n_2,\dots,n_m\}$ defines a Dyck path of length $2m$ consisting of $2n_1$ steps \textit{at} (i.e. up to or down from)  level 1, $2n_2$ steps at level 2, etc.
Specifically, a Dyck path of length $2m$ is a discrete trajectory along a two-dimensional grid that starts at the origin $(0,0)$ and ends at $(2m,0)$ where each step increases the first (horizontal) coordinate by one unit and can either increase or decrease the second (vertical) coordinate by one unit (respectively, an \textit{up step} or a \textit{down step}), with the condition that the vertical coordinate is never negative along the path.
Since all paths start and end at level $0$, they must all have even length $2m$ and we have that $n_1+n_2+\dots+n_m=m$, the maximum possible height being level $m$.
In equation \eqref{eq:moments_Lanczos_Dyck}, each Dyck path is assigned a certain weight according to the following evaluation rule: each step up to, or down from, level $n$ corresponds to a multiplication by the Lanczos coefficient $b_n$.
In particular, \eqref{eq:moments_Lanczos_Dyck} contains always a contribution of the form $(b_1^2)^m$ and a term of the form $b_1^2b_2^2\dots b_m^2$. As an example we note that the first few moments are given by $\mu_2=b_1^2$, $\mu_4 = b_1^4+b_1^2b_2^2$ and $\mu_6=b_1^6+2b_1^4b_2^2+b_1^2 b_2^4+b_1^2b_2^2b_3^2$. The number of Dyck paths of length $2m$ is equal to the Catalan number $C_m=\frac{(2m)!}{(m+1)!m!}$.
 
As announced, the previous combinatorial relation can be inverted in order to express the $n$-th Lanczos coefficient $b_n$ in terms of the subset of moments $\left\{\mu_{2k}\right\}_{k=0}^n$.
This is done via the following iterative procedure, historically known as the \textit{recursion method} \cite{ViswanathMuller}:
\begin{eqnarray}
    \mu_{2k}^{(m)} &=& \frac{\mu_{2k}^{(m-1)}}{b_{m-1}^2} - \frac{\mu_{2k-2}^{(m-2)}}{b_{m-2}^2} , \label{eq:recursion_method_line1} \\
    &&m=1,2,\dots,n \quad k=m,m+1,\dots,n \nonumber\\
    b_n^2 &=& \mu_{2n}^{(n)}~, \label{eq:recursion_method_line2}
\end{eqnarray}
where the initial conditions are $\mu_{2k}^{(0)}=\mu_{2k}$ for $k=0,\dots,n$, $\mu_{2k}^{(-1)}=0$ and $b_{-1}\equiv1\equiv b_0$. As an example, we have that the first few Lanczos coefficients are given in terms of the moments via: $b_1^2=\mu_2$, $b_2^2=\mu_4/\mu_2-\mu_2$ and $b_3^2=(\mu_6/\mu_2-\mu_4)/(\mu_4/\mu_2-\mu_2)-\mu_4/\mu_2$. It is possible to give a closed-form, not recursive, solution of the recursion \eqref{eq:recursion_method_line1}-\eqref{eq:recursion_method_line2} in terms of ratios of Hankel determinants, as reviewed in \cite{ViswanathMuller,Parker:2018yvk}. We may conclude this discussion by noting that the relation between Lanczos coefficients and two-point function moments implies in turn that the Lanczos coefficients encode information about the analytical structure of the two-point function and, as announced earlier, related quantities such as the spectral function $\Phi(\omega)$, defined as the Fourier transform of $C(t)$,
\begin{equation}
    \label{eq:Spectral_function_def}
    \Phi(\omega):=\int_{\mathbb{R}}dt e^{-i\omega t} C(t)~,
\end{equation}
or the operator resolvent (or Green function) $G(z)$,
\begin{eqnarray}
    \label{eq:Spectral_resolvent}
    G(z):=\Big(\mathcal{O}\Big|\frac{1}{z-\mathcal{L}}\Big|\mathcal{O}\Big)~,
\end{eqnarray}
which is a multi-valued function in the complex plane of its argument $z$, having a branch cut along the support of spectral function (which plays the role of an \textit{operator-weighted} density of states of the Liouvillian), the discontinuity across which is precisely equal to $\Phi(\omega)$:
\begin{equation}
     \Phi(\omega)=\lim_{\varepsilon\to 0^+} \Big(G(z+i\varepsilon)-G(z-i\varepsilon)\Big)~,
\end{equation}
as one can show noting that $G(z)$ is the Laplace transform of $C(t)$. The latter fact can further be used to prove that the Lanczos coefficients control the continued fraction expansion of the operator resolvent, which we mention without derivation:
\begin{equation}
    \label{eq:Gz_continued_fraction}
    G(z)=\frac{1}{z-\frac{b_1^2}{z-\frac{b_2^2}{z-\dots}}}~.
\end{equation}
For an extensive review on these and related matters, the Reader is referred to \cite{ViswanathMuller,Magnus}.

\subsubsection{Choice of inner-product}
\label{Sec:Inner_prod}

In order to have a well-defined operator Hilbert space, whose structure is required for making formal sense of notions such as operator size, the autocorrelation function, and Krylov complexity itself, one needs to be specific about the inner product with which such a Hilbert space is equipped. 
The axioms that an inner product $(\mathcal{A}|\mathcal{B})$ needs to satisfy are: sesqui-linearity (i.e. $(\mathcal{A}|\mathcal{B})=(\mathcal{B}|\mathcal{A})^*$ and linearity in the second argument), positive semidefiniteness $(\mathcal{A}|\mathcal{A})\geq 0$ and absence of null states. A commonly used instance of operator inner product is the Frobenius inner product,
\begin{eqnarray} \label{eq:InProd_inf_temp}
    (\mathcal{A}|\mathcal{B}) = \frac{1}{D} \mathrm{Tr}(\mathcal{A}^\dagger \mathcal{B})
\end{eqnarray}
where the normalization makes sure that the identity operator has norm 1. This inner product can be thought of as the \textit{infinite temperature} inner product, while at finite temperature, $T=1/\beta$, one can define the general inner product \cite{ViswanathMuller}:
\begin{eqnarray} \label{eq:InProd_beta}
    (\mathcal{A}|\mathcal{B})_{\beta} := \frac{1}{Z(\beta)} \int_0^\beta d\lambda g(\lambda)\mathrm{Tr}[e^{(-\beta+\lambda)H}\mathcal{A}^\dagger e^{-\lambda H}\mathcal{B}]~, \nonumber \\
\end{eqnarray}
where $Z(\beta):= e^{-\beta H}$ and the weight function $g(\lambda)$ satisfies $g(\lambda)\geq0$, $g(\beta-\lambda)=g(\lambda)$ and $\int_0^\beta d\lambda g(\lambda)=1$.
Some commonly used choices for $g(\lambda)$ are $g(\lambda)=1$, $g(\lambda)=(1/2)[\delta(\lambda)+\delta(\beta-\lambda)]$ \cite{ViswanathMuller} and $g(\lambda)= \delta(\lambda-\beta/2)$ \cite{Parker:2018yvk}. The infinite temperature inner-product \eqref{eq:InProd_inf_temp} can be obtained in the limit $\beta\to0$ for any of these choices, i.e. regardless of the function $g$. 

All inner products in the family defined in \eqref{eq:InProd_beta} (where each representative amounts to a choice of the $g$ function) satisfy the inner product axioms, and additionally also satisfy conditions stated at the beginning of section \ref{sec:Lanczos_operator}, i.e. the Liouvillian is a hermitian super-operator with respect to the inner product, and hermitian operators are orthogonal to anti-hermitian ones. There exist other other choices of operator inner-products in the literature. Some examples relevant to Krylov complexity may be found in works by \citet{Dymarsky:2019elm,Kundu:2023hbk}. We additionally mention that for physical reasons that will be emphasized in section \ref{Sec:Krylov_Pheno}, such as the seek for universal dynamics and probing spectral correlations, it will often be desirable in the context of Krylov complexity to work with operators with a zero one-point function, the latter being defined as 
\begin{equation}
    \label{eq:One_Point_Function}
    \left(\mathcal{A}\right)_\beta:=\frac{1}{Z(\beta)}Tr[e^{-\beta H}\mathcal{A}]=\left(\mathbb{1}|\mathcal{A}\right)_\beta~.
\end{equation}
Practically, this can be achieved by restricting the operator inner product \eqref{eq:InProd_beta} to operators with a vanishing one-point function (which is a well-defined subspace of operator space) and shifting the operators of interest by $\mathcal{A}\mapsto\mathcal{A}-(\mathcal{A})_\beta \mathbb{1}$, or equivalently by not restricting the domain of the inner product and instead complementing \eqref{eq:InProd_beta} with the explicit subtraction of the product $(\mathcal{A}^\dagger)_\beta (\mathcal{B})_\beta$, which does not conflate with the well-definedness of the inner product. Either way, in both cases the two-point function \eqref{eq:Op_two_pt} to which the Lanczos coefficients are in correspondence will now be a \textit{connected} two-point function. 

The thermal operator inner product can be understood as a physical tool used to control the weight with which different sectors of the theory's spectrum contribute to the computation of observables. At finite temperatures $\beta>0$, the existence of various (inequivalent) choices of $g$ function poses a potential ambiguity that needs to be physically understood on a case-to-case basis. In holographic contexts relevant to size or Krylov complexity, for instance, the thermal operator inner-product may be selected from the overlap of operator-perturbed thermofield double states, cf. \citet{Sanchez-Garrido:2024pcy}, and in the more generic framework of quantum field theory it may be chosen in the context of a certain regularization scheme of thermal correlators, as we will review in section \ref{Sec:Krylov_Pheno}. For completeness, let us note that there exists yet another possible definition of the thermal inner product, which is natural in certain quantum-mechanical contexts and which does not belong to the family \eqref{eq:InProd_beta} because it cannot be defined through an even function $g$: It is the \textit{canonical} thermal inner product,
\begin{eqnarray}
    \label{eq:Canonical_thermal_inner}
    \left(\mathcal{A}|\mathcal{B}\right) = \frac{1}{Z(\beta)}\text{Tr}\left[e^{-\beta H}\mathcal{A}^\dagger \mathcal{B}\right]~.
\end{eqnarray}
With respect to this inner product, the Liouvillian $\mathcal{L}$ is still a hermitian super-operator, but in general $\left(\mathcal{O}|\mathcal{L}|\mathcal{O}\right)\neq 0$ for hermitian operators $\mathcal{O}$. I.e. property 1 at the beginning of section \ref{sec:Lanczos_operator} is satisfied, but not property 2. This implies, in turn, that the Lanczos algorithm presented in that section needs to be extended if implemented when the notion of orthogonality is dictated by the canonical inner product \eqref{eq:Canonical_thermal_inner}: The Lanczos orthogonalization step requires an additional subtraction controlled by a new sequence of $a_n$ coefficients; this is in fact the most generic form of the Lanczos algorithm (for self adjoint time-evolution generators) \cite{Lanczos:1950zz}, which we shall review in section \ref{sect:State_formalism_framework}.

Finally, we mention that, analogously to the definition of thermal inner products describing a  canonical ensemble, it is also possible to consider a \textit{microcanonical} inner product \cite{Kar:2021nbm}, which may be conveniently used to focus on specific sectors of the spectrum of the theory. Such an inner product can be defined through a microcanonical trace defined on some energy window centered at some energy $E$ and with some width\footnote{In practical implementations, the width $\delta E$ may be chosen to be some quantity that goes to zero in the large system size limit, but which still allows for an extensive number of energy eigenvalues to be contained in the window centered around $E$.} $\delta E$:
\begin{equation}
    \label{eq:Micro_canonical_Inner_Prod_Trace}
    \left(\mathcal{A}|\mathcal{B}\right)_E := \frac{1}{Z(E)}\text{Tr}_E\left[ \mathcal{A}_E^\dagger \mathcal{B}_E\right]~,
\end{equation}
where $Z(E)$ denotes the microcanonical partition function, $\text{Tr}_E$ is the trace within the energy window defined above, and the operators $\mathcal{A}_E^\dagger$ and $\mathcal{B}_E$ have explicitly been restricted to such a window\footnote{The action of the operators over a given microcanonical window need not, strictly speaking, be closed, so by their \textit{restriction} we just mean considering the operators that one can define through the matrix elements of $\mathcal{A}^\dagger$ and $\mathcal{B}$ that only involve energy eigenstates contained in the microcanonical window.}. It is worth to note that \citet{Kar:2021nbm} formalized further the notion of microcanonical inner product by noting that the super-operator $\mathcal{E}$, defined such that $|E_a\rangle\langle E_b|$ are its eigenstates with eigenvalue $E_a+E_b$, commutes with the Liouvillian, i.e. $[\mathcal{E},\mathcal{L}] = 0$. The eigenstates of $\mathcal{L}$ can therefore be labeled with two numbers, $|E,\omega)$, where the energy difference $\omega$ is their eigenvalue under $\mathcal{L}$, and the total energy $E$ is their eigenvalue under $\mathcal{E}$. Operator matrix elements may be defined, in this discussion, as $\mathcal{A}(E,\omega):=(E,\omega|\mathcal{A})$ using the infinite-temperature inner product \eqref{eq:InProd_inf_temp}, with which the microcanonical inner product can be formally defined as an inner product inside an eigenspace of $\mathcal{E}$:
\begin{equation}
\label{eq:MicroCanonical_Inner_Kar}
    (\mathcal{A}|\mathcal{B})_E := \int_{-2E}^{2E} d\omega \, \rho(E,\omega) \mathcal{A}^*(E,\omega) \mathcal{B}(E,\omega)~,
\end{equation}
where no $Z(E)$ normalization factor is needed because it is incorporated in the definition of the matrix elements $\mathcal{A}(E,\omega)$ through the inner product, and $\rho(E,\omega)$ stands for the density of states, which in the case of a discrete Liouvillian spectrum becomes $\rho(E,\omega):=\sum_{i,j}\delta\left[E-(E_i+E_j)/2\right]\delta(\omega-\omega_{ij})$, where $\omega_{ij}$ are the Liouvillian eigenvalues introduced in \eqref{eq:phases_def} and the sum ranges over the eigenvalues of the Hamiltonian.

\subsection{Wave function dynamics and Krylov complexity}
\label{Sec:Krylov_Dynamics}
We are now in position to study the dynamics of $\mathcal{O}(t)$ in Krylov space by analyzing its progressive spreading over the Krylov basis. This may be done by explicitly expanding $|\mathcal{O}(t))$ over the Krylov basis produced by the Lanczos algorithm:
\begin{equation}
    \label{Op_Krylov_basis}
    \big|\mathcal{O}(t)\big) = e^{it\mathcal{L}}|\mathcal{O}) = \sum_{n=0}^{K-1} \phi_n(t) |\mathcal{O}_n)\, , \quad \phi_n(t) := (\mathcal{O}_n|\mathcal{O}(t))~,
\end{equation}
where we have defined the Krylov space wave function $\phi_n(t)$, which measures the projection of the time-evolving operator $\mathcal{O}(t)$ on each element of the Krylov basis. Since $\mathcal{L}$ is a hermitian (super-)operator, time evolution is unitary and we have that
\begin{eqnarray}
    \label{eq:unitarity_krylov_wave_fns}
    \sum_{n=0}^{K-1}|\phi_n(t)|^2=1
\end{eqnarray} 
at all times. Additionally,
by construction of the Krylov basis we have that $\phi_n(t=0) = \delta_{n0}$, since $\big|\mathcal{O}(0)\big)=|\mathcal{O}_0)$. 

From the Heisenberg equation generating the evolution of $\big|\mathcal{O}(t)\big)$,
\begin{equation}
    \frac{d}{dt}\big|\mathcal{O}(t)\big) = i\, \mathcal{L} \, \big|\mathcal{O}(t)\big)~,
\end{equation}
we can derive a discrete Schrödinger equation for the Krylov space wave function,
\begin{equation}
    \label{DifRecEq}
    -i \dot{\phi}_n(t) = b_n \phi_{n-1}(t) + b_{n+1}\phi_{n+1}(t)
\end{equation}
which may equivalently be interpreted as a recursive-differential equation for a discrete set of functions of time, $\left\{\phi_n(t)\right\}_{n=0}^{K-1}$. The boundary conditions of this recursion are $\phi_{-1}(t)=0=\phi_K(t)$ at all times and the initial condition is, as we have already mentioned, $\phi_n(t=0)=\delta_{n,0}$.
It is also useful to note that the autocorrelation function is the wave function evaluated at the position $n=0$:
\begin{eqnarray}\label{eq:AC_phi0}
    C(t) = (\mathcal{O}_0|\mathcal{O}(t)) = \phi_0(t) ~.
\end{eqnarray}
Moments of the autocorrelation function \eqref{eq:AC_phi0} are related to the Lanczos coefficients through intricate combinatorial objects, as reviewed above.

Eq.~\eqref{DifRecEq} shows that the dynamics of the operator on the Krylov basis is fully determined by the Lanczos coefficients. These are in fact the dynamics of wave function spreading through the one-dimensional Krylov chain introduced in the previous section, starting from a wave function initially localized on the \textit{leftmost} site $n=0$.
Having more support on Krylov elements $|\mathcal{O}_n)$ of larger and larger $n$ means that the operator is further withing its time evolution, requiring more actions of the Liouvillian to reproduce efficiently $|\mathcal{O}(t))$. With this hindsight, the authors of \citet{Parker:2018yvk} proposed to use position over the Krylov basis as a generalized notion of quantum complexity optimally capturing operator growth.
\begin{roundedboxw}[Defintion: Krylov complexity]
\textit{Krylov complexity} (or \textit{K-complexity}) is defined as a quantum observable, i.e. it is the expectation value of some operator $\widehat{C_K}$, which is taken to be the position operator in Krylov space, $\hat{n}$:  
\begin{eqnarray} \label{eq:KC_nhat}
    \widehat{C_K}=\widehat{n} = \sum_{n=0}^{K-1} n \, |\mathcal{O}_n)(\mathcal{O}_n|~,
\end{eqnarray}
whose expectation value in the time-evolving state $\big|\mathcal{O}(t)\big)$ defines Krylov complexity:
\begin{eqnarray}
\label{KC_def}
    C_K(t) := (\mathcal{O}(t)|\hat{n}|\mathcal{O}(t)) = \sum_{n=0}^{K-1} n\, |\phi_n(t)|^2 ~.
\end{eqnarray}
Note that the value of $C_K(t)$ is always bounded from above by the Krylov space dimension
\begin{eqnarray}
    0\leq C_K(t) \leq K, \quad \text{for all $t$}.
\end{eqnarray}
Given that, as we discussed, $\big|\mathcal{O}(0)\big)=|\mathcal{O}_0)$, we have that K-complexity always starts from zero:
\begin{eqnarray}
\label{eq:KC_starts_from_zerp}
    C_K(t=0) = 0.
\end{eqnarray}

\end{roundedboxw}
K-complexity provides a precise quantitative value for the time evolution of any operator or state, given the Hamiltonian of the system. It does not assign an invariant quantity to the initial complexity value of the operator or state. 
The phenomenology of Krylov complexity in various types of systems and its relation to chaotic or integrable dynamics will be reviewed in section \ref{Sec:Krylov_Pheno}. Let us now define some additional tools that are widely used in the literature to study quantitatively the properties of the spreading of the Krylov wave function through Krylov space.

\subsubsection{Krylov entropy}
While K-complexity measures the expectation value of position of the wave function spreading over the Krylov chain, another important measure is the K-entropy of the wave function, defined by \cite{Barbon:2019wsy} to be the Shannon entropy of the probability $|\phi_n(t)|^2$:
\begin{eqnarray}\label{eq:K_entropy_def}
    S_K(t) = -\sum_{n=0}^{K-1} |\phi_n(t)|^2 \ln |\phi_n(t)|^2 ~.
\end{eqnarray}
K-entropy is a measure of the amount of randomness in the wave function. At $t=0$ the wave function is a delta-function localized at $n=0$ and $S_K(t)=0$. The maximum value of $S_K(t)$ corresponds to the maximally uniform (normalized) wave function, $\phi_n=1/K$, whose associated entropy is equal to $\log(K)$, implying the upper bound:
\begin{equation}
    \label{eq:KS_bound}
    S_K(t)\leq \log(K)~,\qquad\text{for all  }t~.
\end{equation}

\subsubsection{Long-time average}\label{subsec.timeAverageKrylov}
Let us now give some details on a tool that can be used to probe the late-time behavior of Krylov complexity which often reveals itself useful: The long-time average.

As explained above, the eigenbasis of the Liouvillian in Krylov space is composed of the non-zero eigenspace projections of the initial $|\mathcal{O})$, given in \eqref{eq:eigenspace_representatives}. Upon adequate normalization, the basis $\left\{|K_{\omega_j}\right\}_{i=0}^{K-1}$ provides an orthonormal basis of operator space, whose elements are by construction eigenstates of the Liouvillian,
\begin{eqnarray}
    \label{eq:K_eigenstate_representatives_evectors_L}\mathcal{L}|K_{\omega_i})=\omega_i |K_{\omega_i}),
\end{eqnarray}
whose associated eigenvalues are non-degenerate, because each of these projections corresponds to a different Liouvillian eigenspace and captures the specific direction within such an eigenspace contributes to the Krylov space, as argued in \ref{subsect:KrylovSpaceOps}.

The time-evolution of the operator in this basis is given by
\begin{eqnarray}
    |\mathcal{O}(t)) =\sum_{i=0}^{K-1}e^{i\omega_i t}|K_{\omega_i})(K_{\omega_i}|\mathcal{O})~,
\end{eqnarray}
and we can now see that the long-time average of the transition amplitude $|\phi_n(t)|^2=|\big(\mathcal{O}_n\big|\mathcal{O}(t)\big)|^2$ takes the form:
\begin{eqnarray}
    \label{eq:Wavefn_long_time}\overline{|\phi_n|^2} :=\lim_{T\to \infty} \frac{1}{T}\!\! \int_0^T \!\! \! \!dt |\phi_n(t)|^2 \!=\! \!\sum_{i=0}^{K-1}|(\mathcal{O}_0|K_{\omega_i})|^2|(K_{\omega_i}|\mathcal{O}_n)|^2,\nonumber\\
\end{eqnarray}
where non-degeneracy of the Liouvillian eigenvalues corresponding to the projections $|K_\omega)$ has been used. With this, we can in turn define the long-time average of K-complexity
\begin{eqnarray} \label{KC_late-time_average}
    \overline{C_K} = \sum_{n=0}^{K-1} n \overline{|\phi_n|^2}~.
\end{eqnarray}
As mentioned earlier, this quantity can be used to study aspects of the late-time dynamics of Krylov complexity, and as we shall review in section \ref{subsect:Krylov_localization}, it has been used to quantitatively analyze the effects of localization on the Krylov basis.

For completeness, let us note that, defining the K-complexity of each individual eigenspace projection as 
\begin{equation}
    \label{eq:KC_of_K_representative}
    C_K^{(i)}:=\sum_{n=0}^{K-1}n |(\mathcal{O}_n|K_{\omega_i})|^2~,
\end{equation}
then equations \eqref{eq:Wavefn_long_time} and \eqref{KC_late-time_average} can be alternatively manipulated in order to obtain \cite{II}:
\begin{eqnarray}
    \label{eq:KC_long_time_avg_spectral_KC}
    \overline{C_K}=\sum_{i=0}^{K-1} |(\mathcal{O}|K_{\omega_{i}})|^2 C_K^{(i)}~,
\end{eqnarray}
which admits the interpretation of a \textit{spectral average} over the \textit{spectral K-complexities} of each eigenspace projection, weighted by the overlap of the initial condition $|\mathcal{O})$ with each spectral projection.

\subsection{State formalism}\label{sect:State_formalism_framework}
Krylov methods are completely applicable to any Hilbert space, as the only ingredients that they require are a \textit{state} (i.e. an element of) the Hilbert space of interest, a \textit{linear operator} acting on such a Hilbert space (which in our contexts of interest plays the role of the time evolution generator), and a notion of orthogonality (which by definition exists in a Hilbert space). As such, Krylov space can be constructed whenever this structure is available, and Krylov complexity can in turn be defined as the position expectation value over the corresponding orthonormal Krylov basis. The seminal paper by Lanczos \cite{Lanczos:1950zz} in fact put forward an algorithm for building an orthonormal basis for the space spanned by the successive powers of a matrix (not necessarily self-adjoint) acting on a vector, and indeed even the original paper by Krylov \cite{Krylov:1931} addressed the problem of the construction of the secular equation of the restriction of some matrix to what later came to be dubbed as the Krylov subspace of a given vector to which this matrix is iteratively applied in his algorithm (although he did not directly address the construction of an orthonormal basis for such a subspace). A manifestation of this generality is the fact that, in order to define the Krylov space of an operator in section \ref{subsect:KrylovSpaceOps}, we needed to explicitly write operators as \textit{states} in the Hilbert space $\widehat{\mathcal H}$, and the adjoint action of the Hamiltonian $[H,\cdot]$ as a \textit{linear (super-)operator} acting on $\widehat{\mathcal{H}}$ (i.e., the Liouvillian $\mathcal{L}$). With this hindsight, using the generic form of the Lanczos algorithm written in \citet{Lanczos:1950zz} one can indeed apply these methods to the Hilbert space of states of a physical system, which also meets the afore-mentioned requirements. The mathematical structure is completely analogous to what we have been discussing in the case of operator Krylov space, with the difference that the time evolution generator is the theory's Hamiltonian, rather than the Liouvillian. For this reason, in \citet{ViswanathMuller} they distinguish the application of the recursion method to operator space from its application to state space by referring to them as \textit{Liouvillian} and \textit{Hamiltonian} formalisms, respectively. The motivation for using either formalism is the relation of the corresponding sequences of Lanczos coefficients (and, in turn, K-complexity) to other observables of potential interest. In the Liouvillian formalism, as we have discussed, there is a very intimate connection between the Lanczos coefficients and properties of the operator two-point, or spectral, function, cf. subsection \ref{subsubsect:Lanczos_Spectral_Probes}. On the other hand, implementing the Lanczos algorithm in the Hamiltonian formalism in order to compute the Krylov complexity of the thermofield double state unveils a connection between the latter quantity and the theory's spectral form factor, as noted in \cite{Balasubramanian:2022tpr}, as we will review in section \ref{Sec:TFD_state}.

In this subsection, we will review the Lanczos algorithm and Krylov complexity for states evolving in the Schrödinger picture.
The formalism is very similar to the operator case, with the main difference being that the Liouvillian is replaced by the Hamiltonian, $\mathcal{L}\to H$, as the operator responsible for the time-evolution, and the initial operator is replaced by an initial state, $\big|\mathcal{O}(0) \big) \to \ket{\psi(0)}$. As announced, the replacement is straightforward given that we had already presented the operator formalism by treating operators as states in their corresponding Hilbert space. The relevant Lanczos algorithm in this construction is the most generic form of this recursive method, which involves an additional sequence of orthogonalization coefficients $a_n$. As already mentioned in Sec.~\ref{Sec:Inner_prod}, these coefficients may also be needed in the operator (Liouvillian) formalism for some choices of inner product, and in these cases the Krylov basis should be constructed with the Lanczos algorithm that we review in this section (modulo the relevant replacements of the objects that participate in it).

Other, more technical differences between the Hamiltonian and Liouvillian formalisms include the bound on the Krylov space dimension, which is $K\leq D$, on which we shall elaborate below, and the presence of diagonal elements in the Lanczos algorithm because $\braket{\psi|H|\psi}$ is not necessarily zero (which is related to the sequence of $a_n$ coefficients). This setup was detailed in \citet{ViswanathMuller} and, in fact, already by Lanczos himself \cite{Lanczos:1950zz}. Using this well-established framework, Krylov complexity was first computed for states (i.e. in the Hamiltonian formalism) in \citet{Balasubramanian:2022tpr}, where it was dubbed \textit{spread complexity}. We prefer to refer to both the operator and state cases as Krylov complexity.

\begin{roundedboxw}[The Lanczos algorithm for states]
Let us now review the most generic form of the Lanczos algorithm. Given an initial state, $\ket{\psi}$, evolving unitarily under a hermitian Hamiltonian, $H$, the Krylov basis $\{\ket{\psi_0}, \ket{\psi_1}, \dots, \ket{\psi_{K-1}} \} $ can be constructed via the Lanczos algorithm:
\begin{enumerate}
    \item Define $\ket{\psi_0}= \ket{\psi}/\|\psi\|^2$
    \item Compute $a_0\!=\!\braket{\psi_0|H|\psi_0}$ and $\ket{A_1}\! =\! (H-a_0) \ket{\psi_0}$ \\
    If $\|A_1\|\neq 0$ define $b_1 =\|A_1\|$ and $\ket{\psi_1}=\ket{A_1}/b_1$\\
    Otherwise, end the process.
    \item For $n>1$:\\
    Compute $a_{n-1}\!=\!\braket{\psi_{n-1}|H|\psi_{n-1}}$ and $\ket{A_n}\! =\! (H-a_{n-1}) \ket{\psi_{n-1}}-b_{n-1}\ket{\psi_{n-2}}$ \\
    If $\|A_{n}\|\neq 0$ define $b_n = \|A_n\|$ and $\ket{\psi_{n}}=\ket{A_n}/b_n$\\
    Otherwise, end the process.    
\end{enumerate}
\end{roundedboxw}

The output of this process are \textit{two} sets of Lanczos coefficients: $\{a_0, a_1,\dots a_{K-1}\}$ and $\{b_1, b_2,\dots, b_{K-1}\}$.
The Hamiltonian has a tridiagonal form in the Krylov basis, $\{\ket{\psi_n}\}_{n=0}^{K-1}$:
\begin{equation}
\label{eq:H_tridiag}
H\overset{*}{=}
        \begin{pmatrix}
            a_0 & b_1 &  &  &  & \\
            b_1 & a_1 & b_2 &  &  & \\
             & b_2 & a_2 & b_3 &  & \\
             &  & b_3 & a_3 & \ddots & \\
             &  &  & \ddots & \ddots & b_{K-1} \\
             &  & & & b_{K-1} & a_{K-1}
        \end{pmatrix} ~,
\end{equation}
where the star above the equality sign denotes, similarly to equation \eqref{eq:Liouvillian_tridiag}, that the right-hand side is a representation of the operator $H$ in coordinates over the Krylov basis. 
Equation \eqref{eq:H_tridiag} can be equivalently written as:
\begin{equation}
    \label{eq:H_Krylov_hopping_model}
    H=\sum_{n=0}^{K-1}a_n|\psi_n\rangle\langle\psi_n| + \sum_{n=0}^{K-2}b_{n+1}\Big(|\psi_n\rangle\langle\psi_{n+1}|+|\psi_{n+1}\rangle\langle\psi_n|\Big)~,
\end{equation}
which generalizes the Krylov cain hopping model \eqref{eq:L_Krylov_hopping_model} to one that also includes a position-dependent potential controlled by the $a_n$ coefficients. 

Note that if the initial state $\ket{\psi}$ is an eigenstate of the Hamiltonian, the Lanczos algorithm terminates at $n=1$, and the Krylov space dimension is $K=1$. On the other hand,
in this case, the Krylov space dimension upper bound coincides with the total Hilbert space dimension, i.e. $K\leq D$, because, unlike in the case of the Liouvillian formalism, complete absence of degeneracies is not forbidden generically in the spectrum of the Hamiltonian.
Saturation of this bound will be achieved by a Hamiltonian without any degeneracies in its spectrum and an initial state with non-zero projections over every eigenstate of the Hamiltonian. 

The time evolution of the initial state on the Krylov basis is given by
\begin{eqnarray}\label{eq:states_Krylov_wavefn}
    \ket{\psi(t)} = \sum_{n=0}^{K-1} \psi_n(t) \ket{\psi_n}, \quad \psi_n(t) := \braket{\psi_n|\psi(t)}
\end{eqnarray}
where by construction we have $\psi_0(t=0)=1$, since $|\psi_0\rangle = |\psi(0)\rangle$, and unitarity enforces $\sum_{n=0}^{K-1}|\psi_n(t)|^2=1$. The Schr\"{o}dinger equation $\frac{d}{dt}\ket{\psi(t)}=-iH\ket{\psi(t)}$ becomes a recursive-differential equation for $\psi_n(t)$:
\begin{eqnarray}
\label{eq:Diff_rec_states}
    i \dot{\psi}_n(t) =  b_n\psi_{n-1}(t) +a_n \psi_n(t)+b_{n+1}\psi_{n+1}(t)~,
\end{eqnarray}
with boundary conditions $\psi_{-1}(t)=\psi_K(t)=0$ for any time $t$, and initial condition $\psi_n(t=0)=\delta_{n0}$. Note that $\psi_0(t)$ can be understood as the \textit{survival amplitude} of the initial state $\ket{\psi}$:
\begin{eqnarray}
\label{eq:Survival_amplitude}
    \psi_0(t) := \braket{\psi_0|\psi(t)}=\braket{\psi|\psi(t)}~.
\end{eqnarray}
In a similar manner to the operator case, a Krylov basis position operator can be introduced,
\begin{eqnarray}
    \hat{n} := \sum_{n=1}^{K-1} \ket{\psi_n}\bra{\psi_n},
\end{eqnarray}
through which \textit{K-complexity} is defined as usual:
\begin{eqnarray}
    C_K(t):= \braket{\psi(t)|\hat{n}|\psi(t)}= \sum_{n=0}^{K-1} n |\psi_n(t)|^2 .
\end{eqnarray}

\subsubsection{Lanczos coefficients and moments of the survival amplitude} \label{subsect:state_formalism_moments}

Finally, just like in the Liouvillian formalism, moments of the survival amplitude \eqref{eq:Survival_amplitude} are related to the two sets of Lanczos coefficients $\{a_n\}_{n=0}^{K-1}$ and $\{b_n\}_{n=1}^{K-1}$. Let us review this relationship.

The survival amplitude for states plays the same role as the autocorrelation function for operators. The moments, i.e. Taylor series coefficients, of \eqref{eq:Survival_amplitude}, are given by
\begin{eqnarray}
    M_n = \braket{\psi|H^n|\psi}~.
\end{eqnarray}
These are in turn related to the two sets of Lanczos coefficients through sums over \textit{Motzkin paths}\footnote{See \citet{MotzkinRef} for a review of these combinatorial objects and their properties.} of length $n$. A Motzkin path is a generalization of a Dyck path, reviewed in section \ref{sec:Lanczos_operator}, which also allows for horizontal steps. We have:
\begin{eqnarray}
\label{eq:Moment_survival_Motzkin}
    M_n =\!\!\!\! \sum_{\{k_i\}_{i=0}^R,\{l_j\}_{j=1}^L} \!\!\!\! (a_0)^{k_0}\!\dots\! (a_R)^{k_R} (b_1^2)^{l_1}\!\dots\! (b_L^2)^{l_L}
\end{eqnarray}
where $R\equiv\lfloor \frac{n-1}{2}\rfloor$ and $L\equiv\lfloor \frac{n}{2}\rfloor$, and the above sum is constrained by the condition $\sum_{i=0}^R k_i+\sum_{j=1}^L l_j=n$.
In equation \eqref{eq:Moment_survival_Motzkin}, we see that each Motzkin path contributing to $M_n$ is evaluated by a monomial where $a_n$ represents a horizontal step at level $n$, and $b_n$ corresponds to a vertical step up to or down from level $n$.
We note that, just like Dyck paths, each Motzkin path starts and ends at level $0$.
The transformation from Lanczos coefficients to moments can be inverted.
When some subset the moments, $M_1, \dots, M_{2{n_{*}}+1}$, of the survival amplitude are known, for some $n_{*}$, it is possible to extract from them the Lanczos coefficients, $a_0,\dots, a_{n_*}$ and $b_1, \dots b_{n_*}$ iteratively \cite{ViswanathMuller}. First an  initialization step is performed: $M_k^{(0)} = (-1)^kM_k$, $ L_k^{(0)} = (-1)^{k+1} M_{k+1}$ for $k=0,\dots, 2{n_*}$, after which the iteration proceeds as follows:
\begin{eqnarray}
    M_k^{(n)} &=& L_k^{(n-1)}-L_{n-1}^{(n-1)}\frac{M_k^{(n-1)}}{M_{n-1}^{(n-1)}} \label{eq:recursion_method_states_line1}\\
    L_k^{(n)} &=& \frac{M_{k+1}^{(n)}}{M_n^{(n)}}-\frac{M_k^{(n-1)}}{M_{n-1}^{(n-1)}}~, \label{eq:recursion_method_states_line2}
\end{eqnarray}
which runs following an outer loop $n=1,\dots,n_{*}$
and an inner loop\footnote{The source \citet{ViswanathMuller} has some typos in the ranges of both the outer and inner loops which were corrected by \citet{Sanchez-Garrido:2024pcy}, from where the ranges shown in this text have been taken.} at each $n$ where $k=n,n+1,\dots,2n_{*}-n+1$.
With this, the Lanczos coefficients can be evaluated from:
\begin{eqnarray}
     b_n^2 &=& M_n^{(n)}, \quad a_n=-L_n^{(n)}, \quad n=0,\dots,n_{*}~. \label{eq:recursion_method_states_finalstep}
\end{eqnarray}

Note that when the survival probability is a symmetric function in $t\to -t$, the odd moments $M_{2n+1}=0$. In this case, the diagonal Lanczos coefficients vanish, $a_n=0$ for all $n=0,\dots,K-1$, and the relationships between the moments and Lanczos coefficients reduce to ones shown in section \ref{Sec:LancCoef_Moments}.

\subsubsection{States and density matrices}\label{subsect:States_vs_density}

	For completeness, it is worth to note that pure states of a quantum system can be described both in terms of elements of the Hilbert space of the theory, $\mathcal{H}$, or density matrix operators belonging to the operator Hilbert $\widehat{\mathcal{H}}$. In particular, there is a well-known one-to-one correspondence between kets $|\psi\rangle$ and density matrices whose square has unit trace, namely:
	\begin{equation}
		\label{eq:State_vs_density_matrix}
		|\psi\rangle \in \mathcal{H}\qquad\longleftrightarrow\qquad\rho=|\psi\rangle\langle \psi|\in\widehat{\mathcal{H}}~.
	\end{equation}
	Now, in the Shcrödinger picture, evolution of the pure state may be described either in the Hamiltonian formalism, where time evolution is governed by the Schrödinger equation, or in the Liouvillian formalism, where it is generated by the von Neumann equation\footnote{Not to be confused with the Heisenberg equation, which generates time evolution of \textit{observables in the Heisenberg picture.}}:
	\begin{eqnarray}
    &\text{Schrödinger:}\qquad i\partial_t |\psi(t)\rangle = H|\psi(t)\rangle \label{eq:Schr_eqn} \\
        &\Longrightarrow \qquad|\psi(t)\rangle = e^{-itH}|\psi\rangle~,  \label{eq:Schr_eqn_sol}\\
		&\text{von Neaumann:}\qquad i\partial_t\big|\rho(t)\big) = \mathcal{L}\big|\rho(t)\big)\label{eq:vN_eqn} \\
        &\Longrightarrow\qquad \big|\rho(t)\big) = e^{-it\mathcal{L}}|\rho)~. \label{eq:vN_eqn_sol} 
	\end{eqnarray}
	It is worth to reiterate that the solutions to both equations \eqref{eq:Schr_eqn} and \eqref{eq:vN_eqn} describe evolution \textit{in the Schrödinger picture}, hence the sign difference between the exponent in the right side of \eqref{eq:vN_eqn_sol} and that in \eqref{Op_Krylov_basis}, and both lines \eqref{eq:Schr_eqn_sol} and \eqref{eq:vN_eqn_sol} describe the time evolution of the same (pure) quantum state of the system \eqref{eq:State_vs_density_matrix}. This is why often the terminology \textit{Hamiltonian vs Liouvillian formalism} \cite{ViswanathMuller} may turn out to be more accurate than \textit{states vs operators formalism}. Given an observable $\mathcal{A}=\mathcal{A}^\dagger\in\widehat{\mathcal{H}}$, its expectation value in the time-evolving state can be computed as:
	\begin{equation}
		\label{eq:ExoVal_state_rho}
		{A}(t) = \langle \psi(t)|\mathcal{A}|\psi(t)\rangle = \text{Tr}\left[\rho(t)\mathcal{A}\right]~,
    \end{equation} 
	In any case, the point of the present discussion is to highlight the fact that at this point one is now free to choose whether to quantify the complexity of the time-evolving quantum state of the system by the (state) K-complexity of $|\psi(t)\rangle$, or, as first suggested by \citet{Alishahiha:2022anw}, via the (operator) K-complexity of $\big|\rho(t)\big)$, which we may denote as $C_K^{(\psi)}(t)$ and $C_K^{(\rho)}(t)$, respectively. Both quantities are conceptually similar, and computable via completely analogous procedures, but in fact turn out to be quantitatively distinct. Let us provide some intuition for this:
	\begin{itemize}
		\item As we know from section \ref{Sec:LancCoef_Moments}, the Lanczos coefficients associated to the tuple $\left\{ |\rho),\mathcal{L} \right\}$ are in one-to-one correspondence to the autocorrelation $C(t)=\Big(\rho\Big|\rho(t)\Big)$, which therefore also specifies  $C_K^{(\rho)}(t)$. On the other hand, we saw in section \ref{subsect:state_formalism_moments} that $C_K^{(\psi)}(t)$ is entirely specified by the Lanczos coefficients associated to $\{|\psi\rangle, H\}$, which one can derive applying the recursion method to the survival amplitude $S(t)=\langle \psi|\psi(t)\rangle$. For pure states, it is possible to show that the self-fidelity of the state is related to the autocorrelation function of its associated density matrix through:
		\begin{equation}
			\label{eq:Ct_St_squared}
			C(t) = |S(t)|^2~.
		\end{equation}
		This implies that the moments of $C(t)$ are different from those of $S(t)$ (although they are related to them), implying a difference in the corresponding sequences of Lanczos coefficients. As an illustration, a quick computation presented by \citet{Caputa:2024vrn} shows that the first Lanczos coefficient of the density matrix is related to that of the ket by:
		\begin{equation}
			\label{eq:states_densityMatrices_b1}
			b_1^{(\rho)}=\sqrt{2}b_1^{(\psi)}~,
		\end{equation}
		from which the authors claim that $C_K^{(\rho)}(t)$ grows faster than $C_K^{(\psi)}(t)$ locally near $t\sim 0$.
		\item While $|\psi(t)\rangle$ is an element of the states Hilbert space $\mathcal{H}$, whose dimension we denote by $D$, $\big|\rho(t)\big)$ belongs to operator space $\widehat{\mathcal{H}}$, of dimension $D^2$. The upper bounds for their respective Krylov dimensions are, therefore:
		\begin{eqnarray}
			\label{eq:KDIMS_state_vs_densityMatrix}
			& K_{\psi}^{(\text{bound})} = D~, \\
			&K_{\rho}^{(\text{bound})}=D^2-D+1~.
		\end{eqnarray}
        This implies that $C_K^{(\rho)}(t)$ has a larger scope of growth as compared to $C_K^{(\psi)}(t)$. In particular, the upper bounds on the Krylov dimension in both cases are related at large system size by:
        \begin{equation}
            \label{eq:Kbound_states_vs_densityMatrix}
            K_{\rho}^{(\text{bound})}\sim \left(K_\psi^{(\text{bound})}\right)^2\qquad (D\gg 1)~.
        \end{equation}
        More specifically, given a decomposition of the state $|\psi\rangle$ in terms of the Hamiltonian eigenbasis,
        \begin{equation}
            \label{eq:state_Ebasis}
            |\psi\rangle=\sum_{n=1}^D \psi_n|E_n\rangle~,
        \end{equation}
        we can decompose its associated density matrix as:
        \begin{equation}
            \label{eq:DensityMatrix_ebasis}
            \rho = \sum_{m,n=1}^D \psi_m\psi_n^{*}|E_m\rangle\langle E_n|~,
        \end{equation}
        from where we see that $\rho$ does not necessarily belong to a diagonal subspace of operator space and, in particular, if the Krylov dimension of \eqref{eq:state_Ebasis} equals the upper bound $K_\psi^{\text{bound}}$, then the Krylov dimension of \eqref{eq:DensityMatrix_ebasis} will in turn also saturate its bound and will be equal to $K_{\rho}^{(\text{bound})}$, implying that, in this case, $C_K^{(\rho)}$ is indeed allowed to take (much) bigger values than $C_K^{(\psi)}$.
	\end{itemize}

So far the discussion concerned exclusively pure states, which can be described both using kets in the Hilbert space $\mathcal{H}$ or density matrix operators belonging to $\widehat{\mathcal{H}}$ whose square has unit trace. Besides this, there also exist some works in the literature that have applied Krylov complexity methods to density matrices representing \textit{mixed} states. In particular, in contexts that are often of interest for gravity and holography, a mixed-state density matrix $\rho$ can be interpreted as the exponential of a so-called \textit{modular Hamiltonian} $H_{\text{mod}}$, defined through the relation $\rho = e^{-H_{\text{mod}}}$ \cite{takesaki1970}. For some applications of Krylov complexity methods to modular evolution, see e.g. \citet{Caputa:2023vyr}.
Furthermore, using a suitable purification scheme allows to apply the states formalism to the computation of Krylov complexity of mixed states. 
In situations of interest for holography, one would mainly be interested in using the thermofield-double state as the purified version of a thermal density matrix \cite{Maldacena:2001kr}: This was the state considered by the various works \cite{IV,Ambrosini:2024sre,Heller:2024ldz} in which the bulk dual of Krylov complexity was constructed in the low-dimensional instance of holography between the double-scaled SYK model and two-dimensional gravity, as we shall review in section \ref{sect:Holography}.
A more generic study of the dependence of the K-complexity of the purified state on the purification scheme, proposed by \citet{Das:2024zuu}, has not been extensively explored yet. This is, in a sense, the reciprocal of the discussion around equation \eqref{eq:State_vs_density_matrix}: In that case we considered the possibility of quantifying the Krylov complexity of a pure state by considering the density matrix that can be built out of it, while now we are discussing the possibility of quantifying the complexity of a mixed state represented by some arbitrary density matrix, via the state K-complexity of the pure state constructed by purifying the latter. In particular, we stress that the purified state associated to a density matrix is in general not unique, as it strongly depends on the chosen purification scheme. 

\subsection{K-complexity is best: two optimality theorems}\label{subsect:theorems}

So far we have introduced Krylov methods and various related tools that constitute the framework in which Krylov complexity is defined. At first glance, the Reader might have the impression that quantifying complexity as average position over the Krylov basis seems like a rather heterodox notion of quantum complexity. In our presentation, however, we have tried to make clear that it is in fact a natural generalization of other complexity notions which, like size complexity, quantify operator (or state) growth by measuring its position with respect to a basis built out of elements to which a fixed complexity value is associated (i.e. complexity eigenstates). In this section, we will collect results that argue that Krylov complexity is, furthermore, an \textit{optimal} generalization of these complexities.

These results came in the form of two optimization theorems presented by \citet{Parker:2018yvk} and \citet{Balasubramanian:2022tpr} which, despite having technically different formulations, share the same spirit: In both cases a class of \textit{generalized complexities} is defined, by requiring a generalized complexity to, roughly speaking, consist on the expectation value of some \textit{complexity observable} that has some eigenbasis and some associated set of complexity eigenvalues. In both works Krylov complexity is a representative within the class of generalized complexities and is shown to stand out as an optimal representative in the sense that it saturates specific inequalities that bound the class of complexities.

The work of \citet{Parker:2018yvk} was presented in the operator formalism, while that of \citet{Balasubramanian:2022tpr} built upon the Hamiltonian formalism; nevertheless, both theorems apply to the definition of Krylov complexity in \textit{any} Hilbert space, and in particular both results apply to both operator and state Krylov complexity. Let us now present, without proof, the statements of both theorems, using notation ot the operator formalism, although the general applicability will be manifest and one can reformulate both results in the states formalism by making the notational change of going from ``smooth kets'' $|\cdot)$ to usual Dirac notation $|\cdot\rangle$, and replacing $H$ by $\mathcal{L}$ when needed.

\subsubsection*{A Q-complexity theorem}

The authors of \citet{Parker:2018yvk} proposed to consider a class of generalized complexities, dubbed \textit{q-complexities}\footnote{The \textit{q} stands for the French word \textit{quelconque} \cite{Parker:2018yvk}.}, where each representative consists on a complexity defined as the expectation value of a hermitian (super-)operator $\mathcal{Q}=\mathcal{Q}^{\dagger}$ in the time-evolving (super-)state $\big|\mathcal{O}(t)\big)$. Denoting schematically the sets of eigenvalues of $Q$ as $\{q_\alpha\}$, where $\alpha$ is some abstract index labeling the spectrum, and their corresponding eigenstates as $\{|q_{\alpha})\}$, we have the following spectral decomposition of the q-complexity operator:
\begin{equation}
    \label{eq:Q_compl_spectral_decomp}
    \mathcal{Q}=\sum_{\alpha}q_{\alpha} |q_{\alpha})(q_\alpha|~.
\end{equation}
Note that the eigenstates $|q_\alpha)$ can be taken to be orthonormal thanks to the spectral theorem, since we have assumed that $\mathcal{Q}$ is hermitian.
Q-complexity is then defined as the expectation value of $\mathcal{Q}$ measured in $\big|\mathcal{O}(t)\big)$, that is:
\begin{equation}
    \label{eq:q_compl_def}
    \mathcal{Q}(t):= \big(\mathcal{O}(t)\big|\mathcal{Q}\big|\mathcal{O}(t)\big)=\sum_{\alpha}q_\alpha \big|\big(q_\alpha | \mathcal{O}(t)\big)\big|^2
\end{equation}

The theorem assumes the following axioms:
\begin{itemize}
    \item $\mathcal{Q}$ is positive-semidefinite, i.e. $q_\alpha\geq 0$ for all $\alpha$.
    \item There exists a positive constant $M$ such that:
    \begin{itemize}
        \item $(q_\alpha|\mathcal{L}|q_\beta)=0$ if $|q_\alpha-q_\beta|>M$, i.e. the action of the Liouvillian on a complexity eigenstate does not increase arbitrarily its complexity.
        \item $(q_\alpha|\mathcal{O})=0$ if $|q_\alpha|>M$, i.e. the initial seed operator is not arbitrarily complex.
    \end{itemize}
\end{itemize}
In particular, K-complexity is a representative of this class, characterized by complexity eigenvalues $q_n=n\in\mathbb{N}_0$ and eigenstates $|q_n)=|\mathcal{O}_n)$, and for which $M=1$.
Roughly speaking, the assumptions above can be understood as defining a basis that is some sort of a ``dilated'' version of the Krylov basis, and in fact the authors of \citet{Parker:2018yvk} proved that all q-complexities are \textit{at most} a multiple of K-complexity. In particular, their theorem states that any q-complexity is bounded by:
\begin{equation}
    \label{eq:q_theorem}
    \mathcal{Q}(t)\leq \overline{M} C_K(t)\qquad \text{at all times}~,
\end{equation}
where $\overline{M}$ is some constant\footnote{The proof of this theorem by \citet{Parker:2018yvk} used $\overline{M}=2M$, which results in the bound $\mathcal{Q}(t)\leq 2M C_K(t)$, which may be understood as a conservative bound by noting that, when the q-complexity is K-complexity itself, in which case $\mathcal{Q}(t)=C_K(t)$ and $M=1$, it takes the form $C_K(t)\leq 2 C_K(t)$. \citet{Sanchez-Garrido:2024pcy} argued that the same theorem can still be proved with $\overline{M}=M$, yielding an optimal bound in the sense that when the q-complexity is K-complexity it takes the form $C_K(t)\leq C_K(t)$.} proportional to $M$.

This q-complexity bound is very powerful because, as \citet{Parker:2018yvk} argued, both size complexity and out-of-time-order correlation functions constitute representatives of q-complexity, being therefore bounded from above by \eqref{eq:q_theorem}. In section \ref{Sec:Krylov_Pheno} we will elaborate on further implications of this bound. In particular, we shall see that it provides a bound on the Lyapunov exponent of OTOCs that can improve on the universal chaos bound by \citet{Maldacena:2015waa}.

\subsubsection*{An optimal basis theorem}

The authors of \citet{Balasubramanian:2022tpr} presented another theorem whose nature is qualitatively similar to the one discussed above, but whose slightly different formulation puts explicitly K-complexity as a lower bound for the class of generalized complexities, rather than giving an upper bound related to a multiple of it. The article by \citet{Balasubramanian:2022tpr} was the same in which Krylov methods were used for the first time in order to define the K-complexity of the thermofield double state (referred to as \textit{spread complexity} by the authors), and in this spirit the theorem was presented in the states formalism. As an exercise to illustrate the generality of this theorem, and to put it on the same footing as the result by \citet{Parker:2018yvk} summarized earlier, let us review here this theorem in the operator formalism. 

The starting point is again to define a class of generalized complexities $C_{\mathcal{B}}(t)$ as the position expectation value of $\big|\mathcal{O}(t)\big)$ with respect to an orthonormal complexity eigenbasis $\left\{|\mathcal{B}_n)\right\}$ with some associated set of complexity eigenvalues $\left\{c_n\right\}$, that is:
\begin{equation}
    \label{eq:Compl_basis}
    C_{\mathcal{B}}(t)=\sum_n c_n\left|\big(\mathcal{B}_n\big|\mathcal{O}(t)\big)\right|^2~,
\end{equation}
where we have schematically omitted the summation range for $n$, but it suffices to say that the basis $\mathcal{B}:=\left\{|\mathcal{B}_n)\right\}$ should be complete. Now, the approach by \citet{Balasubramanian:2022tpr} differs slightly from that by \citet{Parker:2018yvk}: In this case we shall consider the set of complexity eigenvalues $\left\{c_n\right\}$ to be \textit{fixed}, and we seek the \textit{optimal} choice of orthonormal basis $\mathcal{B}$ that minimizes, in a suitable sense, the complexity \eqref{eq:Compl_basis} given the fixed set of complexity eigenvalues. Besides the requirement that the acceptable bases $\mathcal{B}$, over which optimization will be performed, should be orthonormal, the other assumption that \citet{Balasubramanian:2022tpr} made is the fact that the fixed set of complexity eigenvalues $\left\{c_n\right\}$ should be non-negative and sorted increasingly. With this, they showed that the optimal basis is in fact the Krylov basis $\mathcal{B}=\mathcal{K}$, with $\mathcal{K}:=\left\{|\mathcal{O}_n)\right\}_{n=0}^{K-1}$, for which the particularization of \eqref{eq:Compl_basis} we shall denote\footnote{We denote the particularization of \eqref{eq:Compl_basis} to the Krylov basis $\mathcal{B}=\mathcal{K}$ as $C_\mathcal{K}$ instead of the usual notation for K-complexity, $C_K(t)$, because in \eqref{eq:Compl_basis} the complexity eigenvalues are $c_n$. $C_\mathcal{K}(t)$ coincides with the usual K-complexity $C_K(t)$ in the specific case $c_n=n$.} as $C_{\mathcal{K}}$. The precise optimization statement is the following: Given a fixed set of complexity eigenvalues $\left\{c_n\right\}$, for any complete orthonormal basis $\mathcal{B}$ there always exists a $\mathcal{B}$-dependent time $t_{\mathcal{B}}$ such that
\begin{equation}
    \label{eq:Optimal_basis_bound}
    C_{\mathcal{K}}(t)\leq C_{\mathcal{B}}(t)\qquad\text{for all  }t\in[0,t_{\mathcal{B}}]~.
\end{equation}

The above is therefore a bound in the functional sense, since the Krylov complexity will bound from below a given generalized complexity within a time-interval that depends on the latter. The strategy to prove this consists on directly studying the Taylor series coefficients of the function \eqref{eq:Compl_basis} and using the fact that the Krylov basis is uniquely defined by requiring that (i) it is orthonormal and (ii) $\mathcal{L}^n|\mathcal{O})$ is a linear combination of the first $n$ elements of the basis. We may also understand this result as a statement of \textit{local optimization} around $t=0$. Note, as an illustration, that if we choose a basis $\mathcal{B}$ such that $|\mathcal{B}_0)=\big|\mathcal{O}(t_*)\big)$ for some $t_{*}>0$, then at $t=t_*$ we have that $C_{\mathcal{B}}$ reaches its minimum allowed value, $C_{\mathcal{B}}(t_{*})=c_0$, and therefore $t_\mathcal{B}$ in this case is smaller than or equal to $t_{*}$. In particular, $t_{*}$ can be chosen as close to zero as one wishes, as this amounts to defining the basis $\mathcal{B}$.

\subsubsection*{Comparative summary}

For the sake of clarity, let us gather in two bullet points the assumptions leading to each of the two bounds \eqref{eq:q_theorem} and \eqref{eq:Optimal_basis_bound} and their corresponding regime of applicability:
\begin{itemize}
    \item The bound \eqref{eq:q_theorem} holds at all times $t$ for any generalized complexity as long as its set of complexity eigenvalues are non-negative (not necessarily sorted) and its eigenvectors satisfy the second axiom of the theorem by \citet{Parker:2018yvk}, which ensures a progressive growth of complexity as a function of time and morally has a similar role to the requirement of eigenvalue ordering of the second theorem, not assumed here. Note that these complexity eigenvalues need not be the same as the eigenvalues of the K-complexity on the right-hand side. The $\overline{M}$ constant in \eqref{eq:q_theorem} is nevertheless dependent on the q-complexity choice.
    \item The bound \eqref{eq:Optimal_basis_bound} takes a \textit{fixed} set of sorted, non-negative complexity eigenvalues used to define both the $C_{\mathcal{K}}(t)$ on the right-hand side of \eqref{eq:Optimal_basis_bound} and the $C_{\mathcal{B}}$ on the left-hand side. With this, the bound \eqref{eq:Optimal_basis_bound} does not hold at all times, but for a $\mathcal{B}$-dependent time interval around $t=0$.
\end{itemize}
For a more detailed comparative analysis of both theorems, see \citet{Sanchez-Garrido:2024pcy}, where the special case of a generalized complexity with $c_n=n\in\mathbb{N}_0$ is considered, allowing to combine the bounds \eqref{eq:q_theorem} and \eqref{eq:Optimal_basis_bound}. 

In summary, these works put Krylov complexity in a privileged position within the notion of generalized complexities roughly understood as the propagation, or spreading, of an initial state over a complexity eigenbasis. It minimises the rate of growth locally near $t=0$, but at the same time it also bounds from above the late time growth of any such complexity definition. The fact that K-complexity is picked by an optimization argument also makes it more similar in nature to other historical notions of quantum complexity, having the additional features of not requiring any extrinsic choices such as tolerance parameters or arbitrary unitary gates, as we have extensively discussed already in this text. In section \ref{Sec:Krylov_Pheno} we will also review that it has a non-trivial behavior up to exponentially late times in sufficiently chaotic (including holographic) systems.

\subsection{Is Krylov complexity geometrical?} \label{subsect:KC_geometry}

We have just reviewed in which sense Krylov complexity is an optimal generalized notion of complexity that can measure both operator growth and the spreading of a state through the system's Hilbert space. Another natural question that one may ask is whether Krylov complexity admits a geometrical interpretation in terms of properties of paths along the complex projective manifold associated to the Hilbert space, similarly to other familiar notions of quantum complexity. Such a geometric interpretation is not generally available yet, mainly due to the fact that in K-complexity one does not fix the reference state ``once and for all'', since it is meant to quantify the complexity of the time evolution of an initial seed, which plays the role of the reference state every time K-complexity is computed. Nevertheless, in this section we shall see that there are some structural connections to geometric Nielsen complexity and, furthermore, there even exist situations in which Krylov complexity can be computed as a geometric property of a geodesic trajectory over the complex projective space. 

\subsubsection{Relating Krylov and Nielsen complexity}
At face value the geometric definition of Nielsen complexity (see Section \ref{sect:Intro}), has no obvious relation to the notion of Krylov complexity, as described above in Section \ref{subsec:gateAndNielsen}.  Nielsen complexity is always given as a geodesic in a suitable metric on the space of unitaries. While it may be hard to find the specific metric which would correspond to Krylov complexity, one may try to argue using general principles of metric spaces to rule in or rule out a Nielsen complexity metric whose geodesics would correspond to Krylov complexity. For example, any complexity given by geodesic distance in a metric space, would have to satisfy a triangle equality of the form
\begin{equation}
    {\cal C}(t_A,t_B) + {\cal C}(t_B, t_C) \ge {\cal C}(t_A, t_C)\,,
\end{equation}
where ${\cal C}(t_A, t_B)$ is the complexity between two states related by time evolution by times $t_A, t_B$. Along these lines, \cite{Aguilar-Gutierrez:2023nyk} show that in the case of Krylov complexity this triangle equality would reduce to requiring sub-additivity of $C_K(t_A - t_C)$, a property to which \cite{Aguilar-Gutierrez:2023nyk} present counterexamples. While compatible with the bulk dual of Krylov complexity being bulk length, as shown in \cite{IV}, this seems to forbid a direct connection between Krylov and Nielsen complexities. However, in \cite{Craps:2023ivc}, a relation between time-averaged Krylov complexity (see Section \ref{subsec.timeAverageKrylov}) is established. The authors couch their argument in the language of state complexity, but an analogous argument equally applies to operator complexity. Taking a time-average of Krylov complexity gives the expression
\begin{equation}
    \overline{C}_K = \sum_{n=1}^D w_n \overline{|\phi_n|^2}\,,
\end{equation}
where we have generalized the usual expression which has $w_n=n$ to allow for a more general non-decreasing sequence of coefficients, $w_n$. In order to make the promised connection between Krylov and Nielsen complexities, we first need to give a reformulation of the latter in terms of a trace over a suitably defined `Q-matrix'. 

Following \cite{Craps:2022ese}, we introduce a set of group generators $T_\alpha$, which we use to expand the velocity $V(\tau) = i \frac{dU(\tau)}{d\tau} U^\dagger(\tau)$ of a given path $U(\tau)$ through the space of unitaries. We then write
\begin{equation}
    V = V^\alpha T_\alpha \,,\qquad \textrm{with}\qquad V^\alpha = {\rm Tr}\, \left( T_\alpha^\dagger V \right)\,.
\end{equation}
Instead of using the bi-invariant metric\footnote{The Reader may want convince themselves that this in fact corresponds to the `naive' first example of a Nielsen metric we gave in Section \ref{sect:Intro}.} $ds^2 = {\rm Tr}\, \left(V^\dagger V \right)d\tau^2$, one considers a more general family, parametrized by the penalty weights $\mu_\alpha$:
\begin{equation}\label{eq.mualphaNielsenMetric}
    ds^2 = \sum_\alpha \mu_\alpha {\rm Tr}\, |T^\dagger_\alpha V|^2\,.
\end{equation}
Intuitively one may think of the different $\mu_\alpha$ as defining some generators (think of quantum gates) as hard, and others as easy to implement. The quantity we want to link with Krylov complexity is the upper bound on Nielsen quantity, obtained by minimizing the distance \eqref{eq.mualphaNielsenMetric} over only a subset of geodesics, namely those of constant velocity. This upper bound is given by the the minimization problem
\begin{equation}
    C_b = 2\pi {\rm min}_{k \in \mathbb{Z}^D} \sqrt{y^T Q y}\,,\quad y = \frac{E t}{2\pi} - k\,,
\end{equation}
where the matrix $Q$ is given by
\begin{equation}\label{eq.NielsenQmatrix}
    Q_{nm} = \sum_\alpha \mu_\alpha \langle n | T_\alpha | n\rangle \langle m | T_\alpha^\dagger | m \rangle\,,
\end{equation}
$E (E_0, E_1,\ldots , E_{D-1})$ is the vector of eigenenergies and $k$ is the $D-$dimensional vector of parameters which is being minimized over. Now we show how to connect these results to the time-averaged Krylov complexity above. 
To this end, let us define a second `q-matrix',
\begin{equation}
    q_{nm} =\sum_{j=0}^{D-1} \frac{w_j}{2} \left( \langle n | v_0 \rangle \langle v_j | n\rangle \langle m|v_j \rangle \langle v_0 | m\rangle + c.c.  \right)\,.
\end{equation}
This second $q-$matrix is a special case of the first one \eqref{eq.NielsenQmatrix} with the identification of parameters and generators, \cite{Craps:2023ivc}, given in table \ref{tab.NielsenWeights}.
    \begin{table}[h!]
\centering
\begin{tabular}{|c|c|}
\hline
$T_a$ & assignment of $\mu_\alpha$ \\
\hline
$|v_0 \rangle \langle v_0 |$ & $w_0$ \\
\hline
$|v_0 \rangle \langle v_j|\,, j\ge 1$ & $w_j/2$ \\
\hline
$|v_i \rangle \langle v_j |\,,i,j \ge 1$ & $0$ \\
\hline
\end{tabular}
\caption{Weight assignments to connect Nielsen and Krylov complexities.}
\label{tab.NielsenWeights}
\end{table}
With the aid of this quantity, one has the simple relation
\begin{equation}
    \overline {C_K} = {\rm Tr} q\,,
\end{equation}
establishing the searched-after connection between the two notions of complexity. The particular penalty schedule chosen in table \ref{tab.NielsenWeights} is a bit peculiar, and we refer the Reader to \cite{Craps:2023ivc} for further discussion adding some intuition of the values needed. We conclude this section by mentioning that is very natural to seek a similar relation for operator K-complexity, which we leave as an interesting open problem.

\subsubsection{Generalized coherent states on the Krylov chain}\label{subsubsect:coherent_states}

There exists a body of work related to the previous discussion, initiated by \citet{Caputa:2021sib}, that investigates the possibility of giving a geometric description to the evolution of Krylov complexity as a function of time. This work considers the special cases of what the authors refer to as \textit{symmetry-dominated} systems for which the propagation of the wave packet along the Krylov chain happens to be conveniently described by a generalized coherent state, which admits a semiclassical description in terms of a point particle propagating along a geodesic trajectory on a certain manifold equipped with the Fubini-Study metric \cite{Zhang_RMP_coherent}. In such special cases, Krylov complexity is found to be related to a geometrical quantity computed in the Hilbert space's complex projective manifold, namely the volume of the region explored by the time-evolving operator (or state).

The starting point for these considerations is the observation that the Liouvillian is tridiagonal in coordinates over the Krylov basis. In particular, we can take equation \eqref{eq:L_Krylov_hopping_model} and rewrite it as:
\begin{equation}
    \label{eq:L_Lplus_Lminus}
    \mathcal{L}=\mathcal{L}_{+}+\mathcal{L}_{-}~,
\end{equation}
where $\mathcal{L}_{+}$ may be understood as some sort of generalized ladder operators:
\begin{equation}
    \label{eq:Generalized_ladder_Lplus_Lminus}
    \mathcal{L}_{+}=\sum_{n=0}^{K-2}b_{n+1}|\mathcal{O}_{n+1})(\mathcal{O}_n|~,\qquad\mathcal{L}_{-}=\mathcal{L}_{+}^\dagger~.
\end{equation}
The above expression is very revealing because it suggests that, \textit{if} the operators $\mathcal{L}_{+}$ and $\mathcal{L}_{-}$, together with potentially a small finite set of others (such as $J = [\mathcal{L}_{+}, \mathcal{L}_{-}]$), form a closed algebra, then the time evolution of the operator $\big|\mathcal{O}(t)\big)$ is manifestly expressible as the propagation of a generalized coherent state on the Krylov chain, namely:
\begin{equation}
    \label{eq:Krylov_generalized_coherent}
    \big|\mathcal{O}(t)\big)=e^{it(\mathcal{L}_{+}+\mathcal{L}_{-})}|\mathcal{O})=e^{\xi\mathcal{L}_{+}-\overline{\xi}\mathcal{L}_{-}}|\mathcal{O})~,
\end{equation}
where in the last equality we have introduced $\xi=it$ in order to rewrite $\big|\mathcal{O}(t)\big)$ in a more familiar form from the perspective of generalized coherent states. This connection can be made even more accurately by noting that $|\mathcal{O})=|\mathcal{O}_0)$ is a lowest-weight state of the ``Krylov algebra'', since $\mathcal{L}_{-}|\mathcal{O})=0$.

Whether the operators $\mathcal{L}_{+}$ and $\mathcal{L}_{-}$ do actually form part of a closed algebra is, of course, strongly dependent of the specific values of the Lanczos coefficients $b_n$ of the system under consideration, and in fact such a situation is not generic. The approach by \citet{Caputa:2021sib} is, in a sense, the reciprocal to this discussion: They propose to consider toy models in which the Liouvillian (or Hamiltonian) can be written as a linear combination of raising and lowering operators of a given algebra, taking precisely the form of \eqref{eq:L_Lplus_Lminus}. If the initial condition for time evolution is precisely a lowest-weight state of the algebra, then there is no need to solve the Lanczos algorithm in the sense that the Krylov basis elements are given by the natural basis generated by successive application of the raising operator, where the corresponding Lanczos coefficients can be read off from the action of the raising operator, c.f. \eqref{eq:Generalized_ladder_Lplus_Lminus}. The authors of \citet{Caputa:2021sib} proposed to consider these models as benchmarks for the behavior of systems in time regimes for which the dynamics are driven by a specific symmetry group. They further provided explicit results for various symmetry groups, whose phenomenology we shall gather in section \ref{subsect:SpeedLimits_ComplexityAlgebra}.

For the cases in which the Lanczos algorithm requires diagonal $a_n$ coefficients, expression \eqref{eq:L_Lplus_Lminus} gets uplifted to 
\begin{equation}
    \label{eq:L_Lplus_Lminus_Lnot}
    \mathcal{L} = \mathcal{L}_{+}+\mathcal{L}_{-}+\mathcal{L}_0~,
\end{equation}
where:
\begin{equation}
    \label{eq:Lnot}
    \mathcal{L}_0 = \sum_{n=0}^{K-1} a_n|\mathcal{O}_n)(\mathcal{O}_n|~,
\end{equation}
and the discussion on generalized coherent states carries through analogously provided that the Lanczos coefficients $b_n$ and $a_n$ are such that $\mathcal{L}_{\pm,0}$ close an algebra. This framework can be applied to the states formalism as one can see by performing the notational replacement $\mathcal{L}\mapsto H$, as studied e.g. by \citet{Balasubramanian:2022tpr}.

Works that have made similar manipulations of symmetry-dominated dynamics and/or generalized coherent states in order to study Krylov complexity evolution, as well as related quantum-information-theoretic quantities, include those by \citet{Patramanis:2021lkx,Caputa:2021ori,Lv:2023jbv,Patramanis:2023cwz,Caputa:2023vyr,Caputa:2022zsr,Caputa:2024xkp,Caputa:2022eye,Caputa:2022yju}.

As a final comment, it is worth discussing the interplay between the work by \citet{Caputa:2021sib} and the one by \citet{Aguilar-Gutierrez:2023nyk}, already discussed in the preceding subsection. In particular, the latter presents an objection to the possibility of interpreting Krylov complexity as a measure of distance between states, because it need not satisfy the triangle inequality that all distance measures must satisfy by definition. At first sight, it might seem that this is in tension with the fact that propagation of generalized coherent states in the Krylov chain is equivalently described by a geodesic trajectory in the complex projective manifold of the Hilbert space, as established by \citet{Caputa:2021sib}. The tension is, however, only apparent, because assessing the triangle inequality for Krylov complexity involves computing different instances of Krylov complexity, each of them computed out of a different seed state, which from the point of view of the complex projective manifold means changing the metric each time the reference state is changed. From this perspective, it does not make sense to even assess a triangle inequality because the three quantities concerned by it are never computed in the same geometry. Besides this, a further technical subtlety that circumvents the apparent tension is the fact that \citet{Caputa:2021sib} relate the Krylov complexity of the generalized coherent state to the \textit{volume} explored by the point particle in the manifold\footnote{We refer the Reader to the original article by \citet{Caputa:2021sib} for the specific definition of the volume explored by the (one-dimensional) trajectory. Roughly speaking, for the symmetry groups considered, the trajectory travels in a radial direction through manifolds with positive, negative or zero curvature, and the volume ``explored'' by the trajectory is computed as the volume of the ball centered at the coordinate origin (which coincides with the reference state) and whose radius coincides with the position of the point particle traveling along the geodesic trajectory at a given time.}, rather than the geodesic length.

\subsection{Toolbox}\label{subsect:toolbox}

This section reviews a collection of results that either serve to better understand features of Krylov complexity, or which provide analytical tools to compute objects like the Lanczos sequence, which have been widely used in the Krylov complexity literature.

\subsubsection{Integrable Toda hierarchy} \label{subsubsect:Toda}

As we reviewed in sections \ref{Sec:LancCoef_Moments} and \ref{subsect:state_formalism_moments}, the Lanczos coefficients that determine entirely the Krylov complexity can be computed out of the moments of the operator two-point function (or the self-fidelity in the states formalism), via the recursion method. Often in practical cases one has analytical access to such a two-point function and its moments, and it might turn out to be more direct to calculate the Lanczos coefficients out of the latter without explicitly constructing the Krylov basis elements that characterize the underlying kinematic structure of Krylov space. This approach is often preferable in quantum field theory contexts \cite{Dymarsky:2021bjq,Avdoshkin:2022xuw,Camargo:2022rnt,Kundu:2023hbk,Aguilar-Gutierrez:2025kmw}, where the aforementioned kinematic structure is somewhat harder to understand (at least at the moment of the writing of this review) than in the case of many-body quantum mechanical systems, due to the different nature of the algebra of observables, cf. for instance \citet{Witten:2021jzq}. 

A numerical construction of the Lanczos coefficients out of the recursion method, cf. equations \eqref{eq:recursion_method_line1}-\eqref{eq:recursion_method_line2} and \eqref{eq:recursion_method_states_line1}-\eqref{eq:recursion_method_states_finalstep} is complicated due to the important numerical instability of the recursion: The Lanczos coefficients are bounded by the spectral width \cite{ViswanathMuller,Barbon:2019wsy}, while the two-point function moments $\mu_{2n}$ typically grow exponentially in time, meaning that very precise cancellations need to be resolved by the implementation of the recursion method, which therefore suffers from important instabilities when implemented at finite machine precision. Similar considerations apply to the states formalism. As of today there are no efficient algorithms controlling this numerical instability, and the re-orthogonalization algorithms used to improve the Lanczos algorithm that constructs the Krylov basis, which we will review in section \ref{subsect:numerical_implementations}, do not apply to this case.

Fortunately, \citet{Dymarsky:2019elm} proposed a tool that allows in many cases to obtain a closed, analytical expression for the sequence of Lanczos coefficients provided that the two-point function moments are known analytically. These authors pointed out the fact that the recursion method between moments $\mu_{2n}$ and Lanczos coefficients $b_n$ can be mapped to the recursive relation between the so-called tau functions of the integrable Toda hierarchy. There exist known families of analytic functions that solve this hierarchy, out of which the Lanczos coefficients can be directly obtained via simple evaluation. The method is also applicable to the states formalism, or in other words, it is applicable to the most generic formulation of the recursion method, which also features diagonal $a_n$ Lanczos coefficients. This has given rise to numerous analytical results on Krylov complexity in field theories in two or more spacetime dimensions, mainly by the authors cited further above, which we shall review in section \ref{Sec:Krylov_Pheno}.

Let us now give, without proof, some additional details on how this connection to the Toda hierarchy works, for the sake of orientation. A classical Toda chain is described by a Hamiltonian of some number of degrees of freedom $q_n$ (let us be inespecific about the number of such degrees of freedom, which gives the range of $n=0,1,\dots$) with canonical quadratic kinetic terms and an interaction potential with the structure $\sum_n e^{2(q_n-q_{n+1})}$, see \citet{Olshanetskii_Perelomov_TodaChain} and references therein. This classical system is known to be integrable, and the so-called tau functions, defined as
\begin{equation}
    \label{eq:tau_functions_Toda}
    \tau_n = e^{\sum_{k=0}^n q_k}~,
\end{equation}
can be shown to satisfy the Hirota bilinear relation,
\begin{equation}
    \label{eq:Hirota_bilinear_relation}
    \tau_n\ddot{\tau}_n-\dot{\tau}_n^2 = \tau_{n+1}\tau_{n-1}~,
\end{equation}
with the boundary condition ${\tau}_{-1}=1$. This follows directly from the classical equations of motion of the coordinates $q_n$. The connection between the recursion method and this structure, put forward by \citet{Dymarsky:2019elm}, boils down to constructing a ``time-dependent Krylov basis'', achieved by absorbing the (Euclidean) time evolution into the inner product, but fixing the initial condition of the Lanczos algorithm. This results in a time-dependent sequence of Lanczos coefficients $b_n(\tau)$ that can be related to the $q_n(\tau)$ functions of a Toda chain, and their corresponding $\tau_n(\tau)$ functions, via the redefinition
\begin{equation}
    \label{eq:Lanczos_vs_tau_functions}
    b_n(\tau)^2= e^{q_{n+1}(\tau) - q_n(\tau)} = \frac{\tau_{n+1}(\tau)~\tau_{n-1}(\tau)}{\tau_n^2(\tau)}~.
\end{equation}
With this, the \textit{actual} Lanczos coefficients $b_n$ are given by the evaluation $b_n=b_n(0)$. All that is left is an initial condition for the tau functions associated to the Lanczos coefficients, which satisfy \eqref{eq:Hirota_bilinear_relation}. This is provided by the Euclidean correlation function $C_E(\tau):=C(-i\tau)$, where $C(t)$ is defined in \eqref{eq:Op_two_pt}. Denoting its $k$-th derivative as $C_E^{(k)}(\tau)$, we define the following family of Hankel matrices $\left\{\mathcal{M}^{(n)}\right\}$:
\begin{equation}
    \label{eq:Hankel_matrices_tau}
    \mathcal{M}^{(n)}_{ij}(\tau):=C_E^{(i+j)}(\tau)~,
\end{equation}
where for each $n$, $\mathcal{M}^{(n)}$ is an $(n+1)\times(n+1)$ matrix whose indices in \eqref{eq:Hankel_matrices_tau} range through $i,j=0,\dots n$. The tau functions are now given by the corresponding Hankel determinants:
\begin{equation}
    \label{eq:tau_function_Hankel}
    \tau_n(\tau)=\det\mathcal{M}^{(n)}(\tau)~. 
\end{equation}
This can be used either to provide initial conditions for the Hirota bilinear relation \eqref{eq:Hirota_bilinear_relation} or, more operationally, to build the first few tau functions and eventually try to guess their form and prove the Ansatz via an inductive argument. As already mentioned, the Lanczos coefficients will be obtained out of them by plugging them in \eqref{eq:Lanczos_vs_tau_functions} and evaluating the resulting $b_n(\tau)$ at $\tau=0$. It is interesting to note that, in fact, combining \eqref{eq:tau_function_Hankel} with \eqref{eq:Lanczos_vs_tau_functions} and setting $\tau=0$ retrieves the usual closed-form solution of the recursion method in terms of Hankel determinants built out of moments of the two-point function, which can be consulted in \citet{ViswanathMuller}. For simplicity, here we have reviewed the case in which there are no diagonal Lanczos coefficients, $a_n=0$, but the method applies in full generality when these coefficients do not vanish, cf. \citet{Dymarsky:2019elm}. For a concise summary on how to use this method in practical applications, the Reader is also encouraged to consult the first appendix of \citet{Dymarsky:2021bjq}. Finally, we stress that this method is general and, in particular, also applicable in the states formalism, upon the replacement of the two-point function $C(t)$ by the survival amplitude $S(t)$, defined in \eqref{eq:Survival_amplitude}, in the manipulations above.

\subsubsection{Continuous approximation of Krylov space}\label{subsect:Cont_Approx}

The previous subsection described an exact method to compute the sequence of Lanczos coefficients. If one has access to the moments of the operator two-point function or of the state's self-fidelity, the corresponding Lanczos sequences can be computed either numerically via direct application of the recursion method, cf. \eqref{eq:recursion_method_line1}-\eqref{eq:recursion_method_line2} or \eqref{eq:recursion_method_states_line1}-\eqref{eq:recursion_method_states_finalstep}, or analytically if the Toda method is easy to implement in practice. There also exist abundant results relating properties (in particular, the asymptotics) of the Lanczos coefficients to the analytical structure of the spectral function \eqref{eq:Spectral_function_def} that build up on the recursion algorithm and of the expression of the Lanczos coefficients as sums over Dyck or Motzkin paths \textit{à la} \eqref{eq:moments_Lanczos_Dyck} or \eqref{eq:Moment_survival_Motzkin}, cf. \citet{Magnus} or the book by \citet{ViswanathMuller}, which even includes a chapter referring to the Lanczos sequence as a \textit{genetic code} of the spectral function.

Recently, some works have used the recursion method to estimate the asymptotics of the Lanczos coefficients in terms of the asymptotics of the moments with which they are in correspondence, see e.g. \citet{Parker:2018yvk,Barbon:2019wsy}. We shall elaborate a bit more on this in section \ref{Sec:Krylov_Pheno}, but for now it is important to note that such results only apply if the Lanczos sequence $b_n$ is sufficiently \textit{smooth} in $n$. For instance, as pointed out by \citet{Avdoshkin:2019trj,Dymarsky:2021bjq}, some exact solutions that can obtained by the Toda method feature ``staggering'' of the Lanczos sequences, namely the even and odd Lanczos coefficients subsequences, $\{b_{2n}\}$ and $\{b_{2n+1}\}$ respectively, follow distinct sequences whose large-$n$ asymptotics can even be different\footnote{Such a possibility was in fact known in the mathematical literature, and it is associated to the existence of a gap in the spectral density, cf. \citet{ViswanathMuller}.}.
Nevertheless, in Lanczos sequences where such a staggering phenomenon is absent, there may be some scope for treating the smooth sequence $b_n$ as a continuous function, potentially simplifying some analytical manipulations. As an insightful illustration, \citet{Avdoshkin:2019trj} showed that the two-point function moments $\mu_{2k}$, for sufficiently large $k$, can be expressed as a \textit{path integral} over continuous Dyck paths, i.e. giving a continuous version of \eqref{eq:moments_Lanczos_Dyck} that builds up on random walk techniques summarized by \citet{Okounkov:2015rir}. Using a saddle-point approximation (controlled by the large parameter $k$) the authors were able to translate efficiently asymptotics of the Lanczos coefficients into asymptotics of the moments\footnote{\citet{Avdoshkin:2019trj} performed this study for the case in which diagonal Lanczos coefficients vanish, i.e. $a_n=0$, but the path integral formalism should carry through to the more general case \eqref{eq:Moment_survival_Motzkin} whenever the $a_n$ coefficients have a sufficiently smooth behavior.}. 

While it is intuitively easy to accept that the staggering example described above, in which different subsequences of the Lanczos sequence feature different asymptotic behavior at large $n$, is an instance of a \textit{non-smooth} Lanczos sequence, the notion of \textit{smoothness} of a discrete sequence is, however, somewhat ill-defined, since analytic continuation cannot be invoked. In this subsection, we will address a specific class of situations where there is a sense in which a continuous limit of Krylov space can be formally defined. When such a continuous limit applies, there is a variety of analytical tools that simplify very significantly the computation of Krylov complexity. For simplicity, let us focus for now on cases for which there are no diagonal Lanczos coefficients, $a_n=0$, and we will make some comments on the more general case at the end. The starting point for this discussion is the discrete Schrödinger equation that describes the propagation of the Krylov space wave packet, cf. equation \eqref{DifRecEq}, which we restate here for clarity:
\begin{equation}
    \label{eq:Rec_eqn_phi_n_no_a}
    -i \dot{\phi}_n(t) = b_n \phi_{n-1}(t) + b_{n+1}\phi_{n+1}(t)~.
\end{equation}
This equation describes the propagation of the packet $\phi_n(t)$ along the Krylov chain, where position is labeled by the \textit{discrete} variable $n$. As discussed in section \ref{subsect:Krylov_Chain_Hopping}, the Lanczos coefficients play the role of hopping amplitudes. We shall now focus on systems in which there exists a small parameter, say $\varepsilon$, and whose Lanczos coefficients satisfy the property that $b_n$ only depends on $n$ through the product $\varepsilon n$. In particular, let us assume the following form of the Lanczos coefficients:
\begin{equation}
    \label{eq:Lanczos_continuous_Ansatz}
    b_n = \frac{1}{2\varepsilon} v(\varepsilon n)~,
\end{equation}
where the factor of $2$ has been introduced for later convenience, and $v$ is a continuous function of one variable\footnote{For a physical example of a Lanczos sequence satisfying the continuity Ansatz \eqref{eq:Lanczos_continuous_Ansatz}, see \citet{IV,Ambrosini:2024sre}, where the small parameter $\varepsilon$ is the 't Hooft coupling of the double-scaled SYK model, as we will review in section \ref{sect:Holography}.}. Note that a Lanczos sequence featuring staggered subsequences will fail to fulfill this Ansatz. Given this assumption, the continuous limit of Krylov space can be defined introducing the continuous variable $x$ defined as:
\begin{equation}
    \label{eq:Krylov_space_cont_limit}
    n\to\infty~,\qquad \varepsilon\to 0~,\qquad\varepsilon n=x~~~~~\text{fixed .}
\end{equation}
With this, $v(x)$ will be a continuous function defined on $\mathbb{R}^+_0$. We may now manipulate the differential recursion equation \eqref{eq:Rec_eqn_phi_n_no_a} and show that it becomes an ordinary differential equation in the limit \eqref{eq:Krylov_space_cont_limit}, but before this, it will turn out to be insightful to redefine the Krylov space wave function as follows:
\begin{equation}
    \label{eq:wavefn_redef_plus}
    \phi_n(t) = i^n\varphi_n(t)~,
\end{equation}
which has the advantage of turning the recursion \eqref{eq:Rec_eqn_phi_n_no_a} into a real equation, i.e.:
\begin{equation}
    \label{eq:Rec_eqn_varphiPlus_n_no_a}
    \dot{\varphi}_n(t) = b_n\varphi_{n-1}(t)-b_{n+1}\varphi_{n+1}(t)~,
\end{equation}
with initial condition $\varphi_n(t=0)=\delta_{n,0}$ and boundary conditions $\varphi_{-1}(t) = \varphi_K(t)=0$ at all times, just like for \eqref{DifRecEq}. Now, given a Lanczos sequence \eqref{eq:Lanczos_continuous_Ansatz} that takes a smooth limiting form in the continuous limit \eqref{eq:Krylov_space_cont_limit}, we see that as $\varepsilon$ goes to zero $b_n$ only effectively changes over a scale $\sim \frac{1}{\varepsilon}$, and consequently so will the wave function $\varphi_n(t)$ that solves \eqref{eq:Rec_eqn_varphiPlus_n_no_a}: This is a heuristic justification for accepting that in the limit \eqref{eq:Rec_eqn_varphiPlus_n_no_a} the Krylov space wave functions also become continuous. In particular, we pose:
\begin{equation}
    \label{eq:varphiPlus_cont_Ansatz}
    \varphi_n(t) =  f(t,,\varepsilon n)~,
\end{equation}
where $f$ is a continuous function of its two variables, which therefore becomes continuous and supported over $\mathbb{R^+_0}$ with respect to its second argument in the continuum limit \eqref{eq:Krylov_space_cont_limit}. Plugging \eqref{eq:Lanczos_continuous_Ansatz} and \eqref{eq:varphiPlus_cont_Ansatz} into the recursion \eqref{eq:Rec_eqn_varphiPlus_n_no_a}, taking the continuum limit \eqref{eq:Krylov_space_cont_limit} and expanding in $\varepsilon$ yields, after the dust settles:
\begin{equation}
    \label{eq:Continuous_Limit_Chiral_eq}
    \left( \partial_t + v(x)\partial_x + \frac{v^\prime(x)}{2} \right)f(t,x) = \mathit{O}(\varepsilon)~,
\end{equation}
which can be seen to explicitly take the form of a chiral wave equation by re-defining the continuous position coordinate as $x\mapsto y(x)$, where $y$ is uniquely specified by the requirements that $y(0) = 0$ and $dy = \frac{dx}{v(x)}$, as well as re-defining the wave function $f\longmapsto g$ via:
\begin{equation}
    \label{eq:Cont_Limit_Wavefn_Redef}
    \sqrt{v(x)} f(t,x) = g\big(t,y(x)\big)~.
\end{equation}
This turns \eqref{eq:Continuous_Limit_Chiral_eq} into:
\begin{equation}
    \label{eq:Chiral_eq_Right_movers}
    \big(\partial_t+\partial_y\big)g(t,x)=\mathit{O}(\varepsilon)~.
\end{equation}
In the strict continuous limit \eqref{eq:Krylov_space_cont_limit}, which in particular implies $\varepsilon\to 0$, the equation above yields ballistic propagation of the initial condition, i.e.
\begin{equation}
    \label{eq:chiral_wave_eq_rightMover_sol}
    g(t=0,y)=g_0(y)\quad\Longrightarrow\quad g(t,y) = g_0(y-t)~.
\end{equation}
In particular, the initial condition for the Krylov problem $\varphi_n(t=0)=\delta_{n,0}$ gives $g_0=\delta(x)$ which, when applied to the general solution \eqref{eq:chiral_wave_eq_rightMover_sol}, gives $g(t,y)=\delta(y-t)$: That is, a peak propagating ballistically through Krylov space with unit velocity in the $y$ coordinate, following the trajectory $y(t)=t$. Transforming back to the $x$ coordinate, which was explicitly the continuous limit of $\varepsilon n $, we have that the trajectory $x(t)$ on the continuous Krylov space is given by the solution of the following integral equation:
\begin{equation}
    \label{eq:Integral_eq_cont_lim}
    t=\int_0^{x(t)}\frac{dx^\prime}{v(x^\prime)}~.
\end{equation}
In the continuous limit \eqref{eq:Krylov_space_cont_limit} and for systems with a smooth sequence of Lanczos coefficients in the sense of \eqref{eq:Lanczos_continuous_Ansatz}, wave packet propagation on the Krylov chain \textit{classicalizes} and is well described by the propagation of a point particle following a trajectory whose velocity profile $v(x)$ is dictated by the (continuous limit of the) Lanczos coefficients, $v(x) = \lim_{\substack{\varepsilon\to 0 \\ \varepsilon n = x}}2\varepsilon b_n$.  
Krylov complexity is then given by the position of this point particle on the continuous Krylov space, $x(t)$, but note that the original definition of K-complexity \eqref{KC_def} is an expectation value of the originally discrete Krylov position label $n$, so there is actually a relative factor $\varepsilon$ between $C_K(t)$ and $x(t)$, and so the formally correct statement in the continuous limit is in fact:
\begin{equation}
    \label{eq:KC_cont}
    \lim_{\substack{\varepsilon\to 0}}\Big(\varepsilon~ C_K(t)\Big) = x(t)\quad\text{(continuum limit).}
\end{equation}

This kind of continuous approximation was first developed by \citet{Barbon:2019wsy}, where the authors proposed to directly promote $n$ to a continuous variable driven by a velocity field $v(n)\equiv 2b_n$. This approximation need not be generally valid, especially at early times where the initial condition for the Krylov wave packed $\varphi_n(t=0)=\delta_{n,0}$ makes it probe the discrete nature of the Krylov chain. The authors \citet{Barbon:2019wsy} investigated the validity of this approximation by analyzing numerically the growth of tails in the wave function $\varphi_n(t)$ that make it depart from the ballistic approximation. The more formal analysis of the role and system-dependence of the small parameter $\varepsilon$, which we followed in this presentation, was given by \citet{IV}. The latter work also noted that in systems where there is no obvious small system parameter that might be amenable to this continuum approximation, one such small parameter may be universally identified if the systems have finite size since, as we shall see in section \ref{Sec:Krylov_Pheno}, often the Lanczos sequences of finite systems have a very long regime, the so-called \textit{descent} \cite{I}, where the Lanczos coefficients $b_n$ are a function of $\frac{n}{K}$, where $K$ is the Krylov space dimension, and hence there is a scope for treating $\frac{1}{K}$ as a small parameter $\varepsilon$. The ballistic propagation \eqref{eq:Integral_eq_cont_lim} is, however, only acceptable as an approximation because, by the time the Krylov wave packet enters the regime where $b_n$ is a function of $n/K$, it need no longer be localized, and therefore one does not have a delta-function initial condition.

At this point, it is worth to make a comment that connects with the finite-size considerations made at the end of the previous paragraph. In numerical studies such as \citet{I,Balasubramanian:2022tpr} it has been reported that, in finite Krylov chains, which in particular have a \textit{right edge} at $n=K$, the Krylov wave packet may bounce back when reaching the edge of Krylov space, causing Krylov complexity to decrease for some interval of time\footnote{This phenomenon is generally responsible for late-time complexity oscillations \cite{I}, and in the particular case of the K-complexity of the thermofield double state, reviewed in Section \ref{Sec:TFD_state}, it was identified by \citet{Balasubramanian:2022tpr} as the explanation for the K-complexity's absolute maximum, or \textit{peak}.}. In order to describe this effect out of a wave equation, one would need \textit{left-moving} solutions, but as can be seen in \eqref{eq:chiral_wave_eq_rightMover_sol}, equation \eqref{eq:Chiral_eq_Right_movers} only admits right-moving solutions. This might give the erroneous impression that Krylov space propagation is only described by right movers but, as pointed out by \citet{Erdmenger:2023wjg}, this is an artifact of the wave function re-definition \eqref{eq:wavefn_redef_plus}: It can be shown that if, instead, we chose the re-definition
\begin{equation}
    \label{eq:wavefn_redef_minus}
    \phi_n(t) = i^{-n}\varphi_n(t)~,
\end{equation}
then the differential recurrence equation for $\varphi_n(t)$ would differ from \eqref{eq:Rec_eqn_varphiPlus_n_no_a} by an extra minus sign on the left-hand side which, upon the previously described continuum limit manipulations, would yield a different chiral wave equation, differing from \eqref{eq:Chiral_eq_Right_movers} by $\partial_t\leftrightarrow-\partial_t$, whose solutions turn out to be \textit{left-movers} instead of the previously found right-movers. In summary, right- and left- movers are found as solutions of the wave equations that describe the continuum limit of the wave packets $\varphi_n$ defined out of the original $\phi_n$ by striping out a pre-factor of $i^n$ and $i^{-n}$, respectively. Note that in the continuum limit we have that $i^{\pm n}\sim e^{\pm i\pi x / \varepsilon}$: This indicates that right-movers and left-movers are \textit{non-perturbatively} different sectors of the solutions of the continuum limit, which explains why one sector cannot be accessed from the chiral wave equation describing the other sector in the $\varepsilon\to 0$ limit. The authors \citet{Erdmenger:2023wjg} propose a unified treatment of both sectors by posing a second order differential equation for the Krylov wave function $\phi_n(t)$, obtained by taking its second derivative with respect to time and iterating the recursion \eqref{eq:Rec_eqn_phi_n_no_a}, which in the continuum limit yields a second-order differential equation that admits both right- and left-moving solutions. Note that the continuum analysis by \citet{Erdmenger:2023wjg}, which builds up on works by \citet{Muck:2022xfc} and \citet{Alishahiha:2022anw}, is slightly different from the derivation presented here, consisting on a WKB-like approximation scheme where the integral equation \eqref{eq:Integral_eq_cont_lim} arise as the equation defining the characteristic curve of the wave function that solves the corresponding continuous wave equation.

Let us close this subsection by mentioning, as promised, that the continuum-limit manipulations carry through in a qualitatively similar manner to the more generic case where there are also non-vanishing diagonal Lanczos coefficients $a_n$. Assuming that $a_n=\alpha(\varepsilon n)$ for some continuous function $\alpha$, the continuous limit of the recursion equation for the right-movers takes the following form:
\begin{equation}
    \label{eq:Continuous_Limit_Chiral_eq_with_a}
    \Big( \partial_t +v(x)\partial_x +\frac{v^\prime(x)}{2} +i\alpha(x)\Big)f(t,x)=\mathit{O}(\varepsilon)~.
\end{equation}
The solution of the above equation can be found using similar function and variable redefinitions as the ones employed earlier, together with the method of separation of variables, yielding \cite{Alishahiha:2022anw,Erdmenger:2023wjg}:
\begin{eqnarray}
    \label{eq:Cont_WKB_sol}
    f(t,x)=\frac{1}{\sqrt{b(x)}} F(t_{-}(t,x))\exp\left\{i\int_0^x\frac{\alpha(x^\prime)dx^\prime}{v(x^\prime)}\right\}~, \nonumber \\
\end{eqnarray}
where $F$ is some function related to the initial condition and $t_{-}(t,x)$ is the function that labels the (right-moving) characteristic curves, namely:
\begin{equation}
    \label{eq:Charac_curves_label}
    t_{-}(t,x) = t-\int_0^x\frac{dx^\prime}{v(x^\prime)}~.
\end{equation}
Similar considerations apply to the left-movers. Note that the potential term $\alpha(x)$ does not affect the characteristic curves \eqref{eq:Charac_curves_label}, as it only enters through the oscillating phase in \eqref{eq:Cont_WKB_sol}. We can additionally see that, if $\alpha(x)$ were itself of order $\sim1/\varepsilon$, the phase of \eqref{eq:Cont_WKB_sol} would become wildly oscillatory and localize on its stationary point (if there is one). This is the reason why further above we referred to the methods developed by \citet{Muck:2022xfc}, \citet{Alishahiha:2022anw} and \citet{Erdmenger:2023wjg} as a WKB-like approximation scheme. 

\subsubsection{Lanczos coefficients as spectral probes}\label{subsubsect:Lanczos_Spectral_Probes}

Krylov subspace methods and the Lanczos algorithm were historically developed as numerical techniques for the resolution of the eigenvalue problem, suitable for diagonalizing large matrices and, as such, applicable to the computation of the spectrum of discretized differential operators defining differential equations of physical interest. This was the motivation of the seminal papers by \citet{Krylov:1931} and \citet{Lanczos:1950zz}. We will not discuss here the details of the application of the Lanczos algorithm to the numerical eigenvalue problem, a topic on which the interested Reader may find useful information in references such as \citet{Parlett,KrylovBook_Liesen,KrylovBook_Vorst}. Nevertheless, the fact that the Lanczos coefficients are very intimately related to the spectrum of the time evolution generator is of crucial relevance for the study of quantum complexity dynamics. For this reason, in this section we shall review some recent studies in the context of Krylov complexity and holography where the interplay between the Lanczos coefficients and the spectral density plays a protagonist role.

An important tool in this discussion is the family of orthogonal polynomials $\left\{P_n(\omega)\right\}_{n=0}^{K-1}$ generated by the Lanczos algorithm. These polynomials are defined such that:
\begin{equation}
    \label{eq:Orthog_polyns_def}
    |\mathcal{O}_n) = P_n(\mathcal{L})|\mathcal{O})~,
\end{equation}
where for simplicity we are working in the operator formalism, while the discussion is completely generalizable. These polynomials where introduced by \citet{Lanczos:1950zz}, but we may refer to them as \textit{Krylov polynomials} because they are in one-to-one correspondence to the Krylov basis elements $|\mathcal{O}_n)$. Seen as functions of a real variable $\omega$, i.e. $P_n(\omega)$, the support of these polynomials is the same as that of the spectral function $\Phi(\omega)$ defined in \eqref{eq:Spectral_function_def}, and it can be shown that the Lanczos algorithm written in section \ref{sec:Lanczos_operator} implies for them the following recursion:
\begin{equation}
    \label{eq:Recursion_polynomials}
    b_{n+1}P_{n+1}(\omega) = \omega P_n(\omega)-b_{n}P_{n-1}(\omega)~,
\end{equation}
with initial conditions $P_0(\omega)=1$ and $P_1(\omega)=\omega/b_1$. Orthonormality and completeness\footnote{Note that the Krylov basis spans the Krylov space, which is a subset of the total operator space. In this sense, completeness of the Krylov basis refers to the fact that it is a complete basis \textit{within Krylov space}.} of the Krylov basis in turn implies the following relations for the Krylov polynomials:
\begin{eqnarray}
    &\int_{\sigma(\mathcal L)}d\omega \rho(\omega)|O(\omega)|^2P_m(\omega)P_n(\omega)=\delta_{mn}~,\label{eq:Polynimials_orthog} \\
    &\sum_{n\geq 0} P_n(\omega)P_n(\omega^\prime) = \frac{\delta(\omega-\omega^\prime)}{\rho(\omega)|O(\omega)|^2}~.\label{eq:Polynomials_completeness}
\end{eqnarray}
In the expressions above, $\sigma(\mathcal{L})$ should be understood as the Liouvillian spectrum contained in Krylov space, $\rho(\omega)$ is the corresponding density of states, and $O(\omega)=(\omega|\mathcal{O})$ is the ``operator matrix element'' or projection along the Liouvillian eigenstate $|\omega)$, cf. \eqref{eq:L_estates}. In the completeness relation \eqref{eq:Polynomials_completeness}, the eigen-phases $\omega$ and $\omega^\prime$ belong to $\sigma(\mathcal{L})$. 
Note that for simplicity we have considered an infinite Krylov dimension probing a continuous spectrum, although the discrete and finite cases can be deduced by setting a finite Krylov dimension in the upper limit of the sum in \eqref{eq:Polynomials_completeness}, and replacing the density of states $\rho(\omega)$ by a sum of delta functions\footnote{In the discrete case, however, there will not be a density-of-states factor in the denominator of \eqref{eq:Polynomials_completeness} due to a subtlety related to the normalization of states in the continuum case, cf. \citet{Sanchez-Garrido:2024pcy}, and the numerator will just be a Kronecker delta $\delta_{ij}$, where $\omega_i$ and $\omega_j$ are frequencies contained in the discrete spectrum. This is consistent with the statement that the continuous version of a Kronecker delta is a Dirac delta divided by the density of states: This prescription is compatible with the fact that $\sum_i \delta_{ij}=1$ for any $j$ within the summation range, which in the continuum becomes $\int_{\sigma(\mathcal{L})} d\omega \rho(\omega)\frac{\delta(\omega-\omega^\prime)}{\rho(\omega)}=1$, for any $\omega^\prime\in\sigma(\mathcal L)$.}. See \citet{Muck:2022xfc} for a summary of the properties of these orthogonal polynomials, as well as some analytical examples. Crucially, it can be seen \cite{Lanczos:1950zz} that each polynomial $P_n(\omega)$ is proportional to the characteristic polynomial of the $n$-dimensional truncation of the truncation of the tridiagonal Liouvillian \eqref{eq:Liouvillian_tridiag} built out of the Lanczos subsequence $\left\{b_n\right\}_{m=1}^{n-1}$. See \citet{Sanchez-Garrido:2024pcy} for a wider analytical and numerical study on how these successive approximations reconstruct the spectrum: The results, compatible with previous observations on the eigenvalue problem \cite{Parlett}, indicate that isolated eigenvalues are the first to be reconstructed in a scheme of successive truncations, while clustered, or quasi-degenerate eigenvalues need more Lanczos coefficients to be resolved. \citet{Lanczos:1950zz} originally proposed that the largest eigenvalues are resolved first but, as noted in \citet{Sanchez-Garrido:2024pcy}, the relevant quantity that discriminates whether an eigenvalue is easier or harder to resolve, which is dictated by the symmetries of the Lanczos algorithm, is not its magnitude, but its distance to its nearest neighbor, normalized by the spectral width. In other words, in systems with a finite Krylov dimension, the Lanczos coefficients $b_n$ with larger $n$ probe successively finer energy differences. Expressed in terms that might resonate more to an audience interested in late-time dynamics of quantum chaos, this implies that Krylov complexity at late times is sensitive to energy differences of order of the mean-level spacing, to the extent that the Krylov wave packet will be probing the sector of the Krylov chain corresponding to the above-mentioned ``last'' Lanczos coefficients. In section \ref{Sec:Krylov_Pheno} we shall be more specific about the relevant time scales when discussing in detail chaotic and integrable dynamics.

In relation to the intuition described above, the authors \citet{Kar:2021nbm} proposed a \textit{renormalization-group-like} formalism. Namely, given a known sequence of Lanczos coefficients, they propose a limiting protocol that effectively zooms onto specific regions of the Lanczos coefficients $b_n$, out of which they use properties of orthogonal polynomials to derive the compatible density of states that is effectively described by the zoomed-in region. We refer the Reader to their article \cite{Kar:2021nbm} for details. Additionally, the authors also propose to apply these techniques in combination with the use of the microcanonical inner product defined in \eqref{eq:MicroCanonical_Inner_Kar}, which allows to focus on energy windows and, in particular, to make the Hilbert space effectively finite (if the corresponding microcanonical entropy is also finite), a prescription that is judged of interest for systems with infinite-dimensional Hilbert spaces and continuous spectral densities, which is the case in higher-dimensional instances of quantum field theory and gravity.
As pointed out in \citet{ViswanathMuller}, all systems whose spectral density has a compact support have a Lanczos sequence $b_n$ that asymptotes to a constant $b$ at large $n$. The density of states associated to a constant Lanczos sequence by \citet{Kar:2021nbm} is an inverse Wigner semicircle, namely $\rho(\omega)\propto (4b^2-\omega^2)^{-1/2}$, which is in fact the density of states of a Toeplitz matrix\footnote{A Toeplitz matrix is a matrix whose entries are constant along each diagonal. In this case, we are referring to a tridiagonal, real and symmetric Toeplitz matrix with zero diagonal.}.
In section \ref{Sec:Krylov_Pheno} we shall discuss to what extent this is related to chaotic systems, but at this point it is worth to mention that there exist previous studies of the relation between the Lanczos sequences and the spectrum of the tri-diagonal matrices built ouf of them, cf. \eqref{eq:Liouvillian_tridiag} or \eqref{eq:H_tridiag} for the more general case including diagonal $a_n$ coefficients: For instance, \citet{SANCHEZDEHESA1978275} worked out several analytical examples that, among others, include the relation between constant Lanczos coefficients and the inverse Wigner semicircle.

Finally, recent works by \citet{Balasubramanian:2022dnj,Balasubramanian:2023kwd} have used the continuum approximation, described in subsection \ref{subsect:Cont_Approx}, in combination with the ideas discussed in the current subsection in order to obtain an explicit relation between the density of states and the continuous limit of the Lanczos coefficients, when such a limit exists. The underlying qualitative idea is to note that every Lanczos sequence is ``locally constant'', yielding a Toeplitz matrix, whose density of states is an inverse Wigner semicircle, that describes some sector of the spectrum. Building up on this, \citet{Balasubramanian:2022dnj} propose that the density of states $\rho(\omega)$ that corresponds to a certain (continuous) Lanczos sequence is given by a suitable convolution of inverse Wigner semicircles, which we reproduce here without proof:
\begin{equation}
    \label{eq:rho_convolution}
    \rho(\omega) = \int_{0}^1 dx \frac{\Theta\Big(4b(x)^2-\big(\omega-a(x)\big)^2\Big)}{\pi\sqrt{4b(x)^2-\big(\omega-a(x)\big)^2}}~,
\end{equation}
where $x=\frac{n}{K}$, $K$ being the Krylov space dimension (which has been taken to infinity, together with $n$, in order to define the continuous variable $x$), and $\Theta$ is the Heaviside theta function.
Note that \eqref{eq:rho_convolution} includes potential dependence on diagonal $a_n$ coefficients, whose continuous limit is denoted as $a(x)$. The authors \citet{Balasubramanian:2022dnj} developed this expression in the states formalism of the Lanczos algorithm, but for notational consistency we have adapted it to the operator formalism without loss of generality.
Additionally, the authors \citet{Balasubramanian:2022dnj} proposed an algorithm to invert \eqref{eq:rho_convolution} in order to extract Lanczos coefficients out of a known density of states, which they applied to instances of random matrix theory with polynomial potentials, finding good agreement between their approximations and numerical simulations, as well as previously known results on the Lanczos sequences of random matrix theories with Gaussian potentials, cf. \citet{Dumitriu_Matrix_Models_beta_ensembles} and \citet{KrishnapurNotesRMT}. In \citet{Balasubramanian:2023kwd}, similar techniques are also used to analyze the statistics of the Lanczos coefficients for various random matrix theory ensembles, in relation to the Krylov complexity of the thermofield double state, on which we shall elaborate further in section  \ref{Sec:TFD_state}. Let us close this section by noting that, in the context of many-body physics, there exist works that use similar approaches where a continuous approximation of the Lanczos coefficients in the thermodynamic limit is used to extract properties of the spectrum, cf. \citet{Witte_1999,Hollenberg:1993bp,Witte_Hollenberg,Witte_1997}.

\subsection{Numerical implementations}\label{subsect:numerical_implementations}

In practical situations in which an analytical expression for the Lanczos coefficients is not directly available, which may often be the case due to the highly non-linear relation between such coefficients and the initial operator or state, one is confronted with the need to implement numerically either the Lanczos algorithm as described in sections \ref{sec:Lanczos_operator} and \ref{sect:State_formalism_framework} or, alternatively, the recursion method given in equations \eqref{eq:recursion_method_line1}-\eqref{eq:recursion_method_line2} and \eqref{eq:recursion_method_states_line1}-\eqref{eq:recursion_method_states_finalstep}. Both options suffer from the same problem: They are recursive algorithms and, as such, numerically unstable in the sense that numerical error propagates and builds up across the iterations of the algorithm. The recursion method has the advantage that it only computes Lanczos coefficients directly out of moments of the operator two-point function (or state survival amplitude in the Hamiltonian formalism), which are at the end of the day the only information required to compute Krylov complexity through the differential recursions \eqref{DifRecEq} and \eqref{eq:Diff_rec_states}, but on the other hand countering its numerical instability requires the use of high precision arithmetics, cf. \citet{ViswanathMuller}, which come at the cost of high memory requirements and a loss in computational efficiency. On the other hand, in the case of the Lanczos algorithm, there is extensive knowledge on the nature and structure of its numerical error when applied with finite-precision arithmetics, starting from the seminal analysis presented in the PhD thesis by \citet{Paige_PhD}, and there exist efficient methods that can be used to cure these errors to a very good extent, even without the need to increase the machine's precision, as reviewed by \citet{Parlett}. 

In light of the above discussion, in this section we will discuss qualitatively the numerical error in the Lanczos algorithm, and will present efficient methods to counter it. Let us start by noting that there are two sources of numerical error that can build up across the successive Lanczos steps, which have a qualitatively different nature:
\begin{itemize}
    \item An immediate source of numerical error that one can identify is the loss of orthogonality between the successively constructed Krylov basis elements, which is a well known effect in finite-precision implementations \cite{Paige_PhD}. Since each new Krylov basis element is constructed by the Lanczos algorithm in terms of the two previous ones, this error propagates and amplifies, eventually resulting in unreliable Lanczos coefficients. 
    
    \item Another unwanted phenomenon that can take place in a finite-precision implementation of the Lanczos algorithm, identified by \citet{Eckseler:2025cgy}, is the development of spurious projections of the successive Krylov basis elements over the orthogonal complement of the (analytically correct) Krylov space. In other words, the Krylov elements numerically constructed may leak out of Krylov space, besides developing a spurious overlap over the previous Krylov elements as described in the previous bullet point.
\end{itemize}

Achieving an efficient numerical implementation of the Lanczos algorithm amounts to finding an algorithm that minimizes the numerical error described above and at the same time does not compromise the memory requirements at time complexity. As mentioned above, in the numerical diagonalization literature there are several analyses and proposals on how to get around the loss of orthogonality described in the first point of the list above. The first option to cure this problem is the so-called \textit{Full Orthogonalization} (FO) algorithm \cite{Golub_FO,Wilkinson_1965}, which consists on explicitly orthogonalizing each newly constructed Krylov elements in the Lanczos algorithm against the previously constructed ones. This algorithm increases the time complexity of the original Lanczos recursion (besides requiring the storing of the Krylov basis in the RAM). Due to this, a subsequent modification dubbed the \textit{Selective Orthogonalization} (SO) algorithm was developed by \citet{Parlett_Scott_SO}, who proposed to use the results by \citet{Paige_PhD}, which allow to identify the predominant direction in which orthogonality spurious projections develop, and selectively subtract them when their norm crosses a customizable threshold. This method involves diagonalization of tridiagonal matrices at each step of the Lanczos algorithm, and as such it becomes computationally rather expensive time-wise. For this reason, the latter method was eventually superseded by the \textit{Partial Re-orthogonalization} (PRO) algorithm \cite{Simon1984TheLA}, where the specific direction in which the spurious projections develop is not explicitly computed, but uses a recursion that keeps track of the numerical overlap between the Krylov elements, which runs in parallel to the Lanczos iteration, in such a way that a full orthogonalization step is invoked only if the overlap recursion crosses a customizable threshold. This algorithm ensures orthogonality of the numerically output Krylov basis up to a customizable precision. As mentioned above, a comprehensive review of the Lanczos algorithm in the context of the numerical eigenvalue problem, together with its convergence properties, numerical error, and re-orthogonalization algorithms, can be found in the book by \citet{Parlett}.

The loss of orthogonality of the Lanczos algorithm has also been tackled by other, more heterodox algorithms with a good degree of success in numerical diagonalization. An example is the \textit{Implicitly Restarted Lanczos Algorithm}, cf. \citet{CALVETTI_RestartedLan}, which uses an iterative scheme to update the starting vector of the Lanczos algorithm after some number of Lanczos iterations in order to minimize numerical error. This method is efficient and successful for tasks such as the fast determination of low-lying, or isolated eigenvalues of a certain matrix, but it is not directly applicable in the context of Krylov complexity, where the starting vector needs to remain fixed through the full algorithm in order to obtain the full sequence of Lanczos coefficients associated to it. Additionally, it is worth to mention that in the context of matrix \textit{tridiagonalization} there is a very efficient method due to \citet{Householder_1958}, which concatenates some specific similarity transformations, nowadays dubbed \textit{Householder reflections} (cf. \citet{KrishnapurNotesRMT,Menon_Notes_RMT,Dumitriu_Matrix_Models_beta_ensembles}) that is able to bring any hermitian matrix to tridiagonal form. This method was shown by \citet{sau1979connection} to be equivalent to performing the Lanczos algorithm with the starting vector $(1,\dots,0)^T$ (in coordinates over the computational basis). Hence, in order to apply Householder reflections to compute the Lanczos coefficients associated to an arbitrary starting vector, it is needed to first perform a unitary rotation that has the effect of changing coordinates to a new basis in which the desired vector takes the canonical form described above. The numerical construction such an unitary boils down to a Gram-Schmidt process, which hinders the efficiency of this method, although the performance of this specific combination of the Gram-Schmidt protocol and Householder reflections has not been systematically studied in the literature. 
 
Let us now move on to a discussion on the second type of numerical error described at the beginning of this subsection. While re-orthogonalization algorithms may succeed in generating a Krylov basis which is orthonormal to a customizable precision, they are blind to whether this basis actually spans the correct Krylov space or an accidentally bigger, or slightly rotated, space, since the re-orthogonalization methods described above do not necessarily remove the spurious projections of the Krylov basis elements over the orthogonal complement of the analytically correct Krylov space. The authors \citet{Eckseler:2025cgy} propose to circumvent this problem by restricting the Hamiltonian or Liouvillian to the predicted Krylov space, by explicit computation of what we referred to as eigenspace representatives earlier in this text, cf. equation \eqref{eq:eigenspace_representatives} in the case of the operator formalism. However, the numerical computation of these eigenspace representatives and, in particular, distinguishing whether they are zero (in which case the corresponding eigenspace is not contained in the Krylov space) or not, is rather sensitive to the working precision, which may need to be increased to values higher than double floating point precision for the study of large system sizes. It is worth to mention that this ``Krylov space leakage'' effect described by \citet{Eckseler:2025cgy} has only been studied recently because it is mostly of relevance in situations where the expected Krylov space is not equal to the full Hilbert space, which in the context of the computation of Krylov complexity is a feature that may often be encountered depending on the symmetries of the theory and how the state or operator probes such a structure, as discussed in section \ref{subsect:KrylovSpaceOps}. Historically, in the context of numerical diagonalization where the Lanczos algorithm is used to compute the spectrum of a certain matrix, the approach is, instead, to seed the algorithm with a sufficiently generic vector completely uncorrelated to the matrix of interest, such as an element of the computational basis, in which case the Krylov dimension is generally expected to be maximal. In these situations, in which the Krylov space is not confined to a certain invariant subspace of the full Hilbert space, the loss or orthogonality is essentially the only source of numerical error.

In summary, in applications of the Lanczos algorithm to physical situations of interest for the computation of Krylov complexity, the optimal implementation of the algorithm involves a combination of high precision arithmetics and re-orthogonalization algorithms. In order to check how acceptable the obtained Lanczos sequence is, the two following tests are useful \cite{I}:
\begin{itemize}
    \item The length of the obtained sequence of Lanczos coefficients should match the expected Krylov space dimension, computed as described in \ref{subsect:KrylovSpaceOps}. We remind the Reader that, as explained in sections \ref{sec:Lanczos_operator} and \ref{sect:State_formalism_framework}, if the Krylov dimension $K$ is finite, then the Lanczos algorithm halts at the $K$-th step by reaching a zero $b$-coefficient, i.e. $b_K=\Vert \mathcal{A}_K\rVert=0$ (resp. $b_K=\lVert A_K\rVert=0$) in the operator (resp. state) formalism. 
    \item The eigenvalues of the tridiagonal matrix constructed out of the Lanczos coefficients, cf. expressions \eqref{eq:Liouvillian_tridiag} and \eqref{eq:H_tridiag}, should coincide with the eigenvalues associated to the eigenspaces contained in Krylov space, as also discussed in section \ref{subsect:KrylovSpaceOps}. 
\end{itemize}
Note that, while in the exact diagonalization community the Lanczos coefficients are intermediate tools used to reach the final goal, namely the spectrum of a certain matrix of interest, in the context of Krylov complexity the roles are reversed: We assume knowledge of the spectrum and seek the accurate evaluation of the Lanczos coefficients, which have a central physical role because they directly determine the dynamics in Krylov space.

Let us now provide, for reference, the explicit formulation of the two numerical algorithms that have been used in the simulations that constitute the state of the art in regards to the computation of Krylov complexity in many-body systems at finite but large system sizes \cite{I}: The FO and PRO algorithms.

\subsubsection{Full Orthogonalization (FO)}

Let us begin describing the Full Orthogonalization (FO) procedure. This algorithm is easy to implement, as it is a combination of the Lanczos iteration with the Gram-Schmidt (GS) algorithm: It performs a direct re-orthogonalization of every new Krylov element with respect to all the previous ones. It is nevertheless time-consuming because of the need to perform explicit re-orthogonalization at every Lanczos step, and also memory-wise expensive because it needs to keep the Krylov basis stored in the RAM, as it is used in every Lanczos step for the re-orthogonalization subtractions. For comparison, in the pure Lanczos recursion, the $n$-th Krylov element only depends on the $n-1$-th and $n-2$-th elements. Hence, if one is only interested in the Lanczos coefficients, only two (iteratively updated) Krylov elements would need to be stored in the RAM.

\begin{roundedboxw}[The Full Orthogonalization (FO) algorithm]
Let us give the precise formulation of the FO algorithm \cite{Golub_FO,Wilkinson_1965,Parlett}. We shall write it in the operator formalism, but including diagonal $a_n$ coefficients (which in principle can be required in the operator formalism for certain choices of inner product and initial operator). This way, the translation into the states formalism is straightforward and only a matter of notational change. \\

\noindent{\textbf{Input:} $\mathcal{O}, \mathcal{L}$}

\begin{itemize}
    \item Define $|\mathcal{O}_0) = \frac{1}{\|\mathcal{O}\|}|\mathcal{O})$. \\
    Compute $a_0 = \big(\mathcal{O}_0\big|\mathcal{L}\big|\mathcal{O}_0\big)$.
   
    \item For $n\geq 1$:\\
    Compute $|\mathcal{A}_n) = \mathcal{L}|\mathcal{O}_{n-1})$.\\
    GS step: Reassign $|\mathcal{A}_n)\longmapsto |\mathcal{A}_n) - \sum_{m=0}^{n-1} \big(\mathcal{O}_m\big|\mathcal{A}_n\big)|\mathcal{O}_m)$.\\
    Repeat the GS step. \\
    If $\|\mathcal{A}_n\|=0$ stop. Otherwise:\\
    Compute $b_n = \|\mathcal{A}_n\|$ and $|\mathcal{O}_n)=\frac{1}{b_n}|\mathcal{A}_n)$  \\
     Compute $a_n = \big(\mathcal{O}_n\big|\mathcal{L}\big|\mathcal{O}_n\big)$.
\end{itemize}
\textbf{Output:} \textit{Lanczos coefficients} $\{b_n\}_{n=1}^{K-1} $ and $\left\{a_n\right\}_{n=0}^{K-1}$, and orthonormal \textit{Krylov basis} $\{|\mathcal{O}_n)\}_{n=0}^{K-1} $.\\

Just like the pure Lanczos algorithm, this process ends at $n=K$, when $\|\mathcal{A}_K\|=0$ (within working precision). Note that at infinite precision the GS step would reduce to the Lanczos two-term subtraction, all other terms being zero. The repetition of the GS step is optional but it is often a good practice to improve numerical convergence at finite precision.
\end{roundedboxw}

\subsubsection{Partial Re-Orthogonalization (PRO)}
The FO algorithm described in the previous section is arguably an overly burdensome re-orthogonalization method: Certainly the GS steps should not be needed straight from the very first Lanczos step, as Krylov basis orthogonality is only gradually lost. Likewise, if at a Lanczos step $n$ the Krylov basis has been duly orthogonalized by FO subtractions, then one would expect that it should be possible to perform a few pure Lanczos steps thereafter while retaining orthogonality to some extent.  
This intuition is formalized and systematized by the Partial Re-Orthogonalization (PRO) procedure \cite{Simon1984TheLA,Parlett} which performs a re-orthogonalization step with respect to all previous Krylov elements \textit{only} when the overlap between the new Krylov element and the previously constructed ones exceeds a certain threshold value that can be customized. The crucial feature that makes this algorithm efficient is that such overlaps do not need to be computed, and instead their value and its buildup across the Lanczos steps is \textit{estimated} by noting that they follow a specific recursion that can be derived from the Lanczos algorithm itself. Depending on the above-mentioned customizable error threshold, PRO may reduce significantly the number of re-orthogonalization steps, which makes it more efficient than the FO algorithm. It nevertheless still requires the storage of the full Krylov basis, to be used when a GS subtraction needs to be performed, but depending on how frequently such subtractions need to be implemented (to be judged on a case-by-case basis), it might be possible to store the basis in the disk memory instead of the RAM. 

Let us now give an abridged derivation of the error recursion, based on the presentation given by \citet{I}. Just like in the presentation of the FO algorithm, we will use the notation of the operator formalism including $a_n$ coefficients for the sake of generality. For the rest of this discussion, symbolic objects such as Krylov basis elements $|\mathcal{O}_n)$ and Lanczos coefficients $a_n$, $b_n$ (or even the Krylov dimension $K$) shall be understood as the \textit{numerically computed} objects, rather than their analytical counterparts. With this, we define the Krylov basis overlaps as:
\begin{equation}
    \label{eq:Krylov_Basis_Overlaps_Wkn}
    W_{mn} := \big(\mathcal{O}_m\big|\mathcal{O}_n\big)~.
\end{equation}
Analytically (i.e. at infinite precision), the overlap matrix should be equal to the identity, that is $W_{mn}=\delta_{mn}$, but at finite precision this is no longer the case, and the matrix $W:=(W_{mn})_{m,n=0}^{K-1}$ contains the information of the spurious overlaps that reflect the loss of orthogonality of the basis. In order to estimate the value of these spurious overlaps, let us start by noting that the numerically computed Krylov basis elements satisfy a modify Lanczos recursion that contains an extra term accounting for spurious components generated by numerical error, namely:
\begin{equation}
    \label{eq:Lanczos_Recursion_With_error}
    b_n|\mathcal{O}_n) =(\mathcal{L}-a_{n-1})|\mathcal{O}_{n-1})-b_{n-1}|\mathcal{O}_{n-2})+|\xi_{n-1})~,
\end{equation}
where the extra vector $|\xi_{n-1})$ accounts for the built-up numerical error, and therefore vanishes at infinite precision. Conveniently acting with $(\mathcal{O}_m|$ on \eqref{eq:Lanczos_Recursion_With_error}, one reaches a recursion for $W_{mn}$ that involves matrix elements of $\mathcal{L}$. Subtracting two versions of the same recursion generated by a convenient relabeling of the indices, it is possible to cancel out (up to numerical errors that can be absorbed into the $|\xi_n)$ vectors) the factors of the form $(\mathcal{O}_m|\mathcal{L}|\mathcal O_n)$, yielding a recursion that purely involves elements of the $W$ matrix and error terms, namely (see \citet{I} for details, and \citet{Sanchez-Garrido:2024pcy} for a more general presentation):
\begin{equation}
    \label{eq:Recursion_Wkn_summary}
    W_{mn} = \frac{1}{b_n}\Bigg(\widetilde{W}_{mn}+\mathcal{E}_{m,n-1}\Bigg)~,
\end{equation}
where we have defined:
\begin{eqnarray}
   & \widetilde{W}_{mn}:= b_{m+1}W^{*}_{m+1,n-1} + a_m W^*_{m,n-1} \label{eq:Wmn_Tilde_line1}\\
   &+b_m W^*_{m-1,n-1}-a_{n-1}W_{m,n-1}-b_{n-1}W_{m,n-2}~ \label{eq:Wmn_Tilde_line2}\\
   & \mathcal{E}_{m,n-1}:= \big(\mathcal{O}_m\big|\xi_{n-1}\big)-\big(\mathcal{O}_{n-1}\big|\xi_m\big)~.\label{eq:VarEpsilon_m_nminus1}
\end{eqnarray}
The above expressions define a recursion for the numerical error $W_{mn}$, to be run simultaneously as the Lanczos recursion, which can then be used to estimate how numerical error propagates given an initialization that assumes orthonormality to machine precision. A few comments are in order:
\begin{itemize}
    \item At a given Lanczos step $n$, we are interested in estimating the overlaps $W_{mn}$ for $m=0,\dots,n$, as these are the inner products between the newly constructed Krylov basis element $|\mathcal{O}_n)$ and the previous ones and with itself, i.e. $|\mathcal{O}_m)$ with $m\leq n$. 
    \item Expression \eqref{eq:Recursion_Wkn_summary} is a two-term recursion. We thus need to initialize the two starting overlap lists $\left\{W_{mn}\right\}_{m=0}^n$ to: $W_{00}=1$ and $W_{01}=\varepsilon_M$, $W_{11}=1$, i.e. the first two Krylov elements are orthonormal within machine precision, denoted $\varepsilon_M$. Note that in order for this initialization to be justified, the step $n=1$ of the PRO algorithm will be required to explicitly orthogonalize $|\mathcal{O}_0)$ and $|\mathcal{O}_1)$, which is anyways already implemented by the subtraction term involving the $a_0$ coefficient, as we shall see further below.
    \item The recursion \eqref{eq:Recursion_Wkn_summary} contains at each $n$-step a new error term $\mathcal{E}_{m,n-1}$ defined in \eqref{eq:VarEpsilon_m_nminus1}, which is \textit{a priori} unknown. An acceptable strategy is to estimate it as being proportional to machine precision and to the norm of the Liouvillian\footnote{One may take the norm of the Liouvillian super-operator naturally induced by the norm in operator space $\widehat{\mathcal{H}}$ over which it acts.}. In general, it can be a complex number, and in order to give the most conservative estimate one may choose its argument to be equal to that of $W_{mn}$, this maximally amplifying error in \eqref{eq:Recursion_Wkn_summary}. More specifically, the estimate we are describing here amounts to posing:
    \begin{equation}
        \label{eq:VarEpsilon_estimate_norm}
        \mathcal{E}_{m,n-1}\approx \frac{\widetilde{W}_{mn}}{\big|\widetilde{W}_{mn}\big|}\cdot 2\varepsilon_M\lVert\mathcal{L}\rVert~,
    \end{equation}
    where the $2$ comes from the fact that there are two terms in \eqref{eq:VarEpsilon_m_nminus1}.
    \item In physical applications of interest, such as the ones studied by \citet{I,II,III}, it might be the case that the system's Hamiltonian is normalized in a way such that its spectral width does not scale with system size, and therefore so does its norm, implying the same feature for the norm of the Liouvillian. In these situations, $\lVert\mathcal{L}\rVert$ in \eqref{eq:VarEpsilon_estimate_norm} will just be some order-one number, and it is acceptable to consider that the error term is in absolute value merely controlled by the machine precision $|\mathcal{E}_{m,n-1}|\approx \varepsilon_M$. However, in a practical situation, the numerical implementation of \eqref{eq:Recursion_Wkn_summary}, will already generate errors of order $\varepsilon_M$ themselves simply due to finite-precision arithmetics. One can therefore safely ignore the error term $\mathcal{E}_{m,n-1}$, as even if it is not explicitly implemented in the recursion, the computer will produce spurious errors of the same order that effectively account for this term. This was the approach taken by \citet{I,II,III}.
\end{itemize}

With this, the logic of the PRO algorithm is the following: While running the usual Lanczos iteration, the duly initialized $W_{mn}$ recursion runs in parallel using the progressively computed Lanczos coefficients. At each Lanczos step $n$ the overlaps $W_{mn}$ with $m=0,\dots,n$ are sought. However, inspection of \eqref{eq:Recursion_Wkn_summary} shows that only $m=0,\dots,n-2$ can be determined in terms of overlap lists $\{W_{m,n^\prime}\}_{m=0}^{n^\prime}$ with $n^\prime < n$, computed in previous Lanczos steps. One is then forced to explicitly re-orthogonalize $|\mathcal{O}_n)$ against $|\mathcal{O}_{n-1})$ and to normalize it in every Lanczos step, a task that is anyway achieved by the computation of the $a_n$ and $b_n$ coefficients, respectively, but which would have to be implemented ad-hoc even in cases in which the $a_n$ coefficients are expected to be zero. This allows to set $W_{n-1,n}=\varepsilon_M$ and $W_{n,n}=1$. Once the Lanczos step is concluded, including its simultaneous $W$-iteration, the overlap list $\{W_{mn}\}_{m=0}^n$ is inspected and, if any overlap exceeds in absolute value a customizable numerical threshold $\varepsilon_T$, then a Gram Schmidt orthogonalization (i.e. an FO subtraction) is performed for $|\mathcal{O}_n)$, and the $\{W_{mn}\}_{m=0}^n$ is re-assigned to $W_{mn}=\varepsilon_M$ if $m<n$ and $W_{nn}=1$. As a final refinement, we note that, since the Lanczos recursion is a two-term recursion, in order to enter the $n+1$ step, where $|\mathcal{O}_{n+1})$ will be computed in terms of $|\mathcal{O}_n)$ and $|\mathcal{O}_{n-1})$ with error-free basis elements (within tolerance), then the subtraction described above for $|\mathcal{O}_n)$ also needs to be performed for $|\mathcal{O}_{n-1})$.

The precise formulation of the algorithm is the following:
\begin{mdbox}
\mdboxtitle{The Partial Re-Orthogonalization (PRO) algorithm}
\begin{itemize}
    \item Step $n=0$:
    \begin{itemize}
        \item Set $|\mathcal{O}_0)=\frac{1}{\lVert \mathcal{O}\rVert}|\mathcal{O})$.
        \item Set $W_{00}=1$.
        \item Compute $a_0 = \big( \mathcal{O}_0\big|\mathcal{L}\big|\mathcal{O}_0 \big)$.
    \end{itemize}
    \item Step $n-1$:
    \begin{itemize}
        \item Compute $|\mathcal{A}_1) = (\mathcal{L}-a_0)|\mathcal{O}_0)$.
        \item Compute $ b_1=\lVert \mathcal{A}_1\rVert$. \textbf{If} $b_1 <\varepsilon_T$ \textbf{stop}. \textbf{Otherwise} continue.
        \item Compute $|\mathcal{O}_1) = \frac{1}{b_1}|\mathcal{A}_1)$.
        \item Assign $W_{01}=\varepsilon_M$ and $W_{11}=1$.
        \item Compute $a_1=(\mathcal{O}_1|\mathcal{L}|\mathcal{O}_1)$.
    \end{itemize}
    \item \textbf{For} $n\geq 2$:
    \begin{itemize}
        \item Compute $|\mathcal{O}_n)=(\mathcal{L}-a_{n-1})|\mathcal{O}_{n-1})-b_{n-1}|\mathcal{O}_{n-2})$.
        \item Compute $b_n=\lVert\mathcal{O}_n\rVert$. \textbf{If} $b_n<\varepsilon_T$ \textbf{stop}. \textbf{Otherwise} continue.
        \item Assign $W_{n-1,n}=\varepsilon_M$ and $W_{n,n}=1$.
        \item \textbf{For} $m=0,\dots,n-2$ \textbf{do}:
        \begin{itemize}
            \item Compute $\widetilde{W}_{mn}$ as in \eqref{eq:Wmn_Tilde_line1}-\eqref{eq:Wmn_Tilde_line2}.
            \item Compute $\mathcal{E}_{m,n-1}$ as in \eqref{eq:VarEpsilon_estimate_norm}, or set to zero.
            \item Compute $W_{mn}$ as in \eqref{eq:Recursion_Wkn_summary}.
        \end{itemize}
        \item \textbf{If} $\max_{~0\leq m \leq n-2}\{|W_{mn}|\}\geq\varepsilon_T$ \textbf{do}:
        \begin{itemize}
            \item GS: $|\mathcal{A}_{n-1})\longmapsto |\mathcal{A}_{n-1}) - \sum_{m=0}^{n-2} \big(\mathcal{O}_m\big|\mathcal{A}_{n-1}\big)|\mathcal{O}_m)$
            \item Repeat previous GS.
            \item Recompute $b_{n-1}=\lVert\mathcal{A}_{n-1}\rVert$.
            \item \textbf{If} $b_{n-1}<\varepsilon_T$, \textbf{then} clear the variable $b_n$ and \textbf{stop}. \textbf{Otherwise} re-compute $|\mathcal{O}_{n-1})=\frac{1}{\Vert b_{n-1} \rVert}|\mathcal{A}_{n-1})$ and re-compute $a_{n-1}=(\mathcal{O}_{n-1}|\mathcal{L}|\mathcal{O}_{n-1})$.
            \item GS: $|\mathcal{A}_n)\longmapsto |\mathcal{A}_n) - \sum_{m=0}^{n-1} \big(\mathcal{O}_m\big|\mathcal{A}_n\big)|\mathcal{O}_m)$.
            \item Repeat the previous subtraction.
            \item Re-compute $b_n=\lVert\mathcal{A}_n\rVert$.
            \item \textbf{If} $b_n<\varepsilon_T$ \textbf{stop}. \textbf{Otherwise} compute $|\mathcal{O}_n)=\frac{1}{b_n}|\mathcal{O}_n)$ and compute $a_n=(\mathcal{O}_n|\mathcal{L}|\mathcal{O}_n)$.
            \item Re-assign $W_{m,n-1}=\varepsilon_M$ for $m=0,\dots,n-2$.
            \item Re-assign $W_{mn}=\varepsilon_M$ for $m=0,\dots,n-1$.
        \end{itemize}
        \item \textbf{Otherwise}:
        \begin{itemize}
            \item Compute $|\mathcal{O}_n) = \frac{1}{b_n}|\mathcal{A}_n)$.
            \item Compute $a_n=(\mathcal{O}_n|\mathcal{L}|\mathcal{O}_n)$.
        \end{itemize}
    \end{itemize}
\end{itemize}
\end{mdbox}

The value of $n$ at which the algorithm is halted gives the Krylov dimension $K$. Note that the error tolerance shall be taken to be bigger than the machine precision, $\varepsilon_T>\varepsilon_M$, so that numerical error is allowed to build up in a controlled manner. For instance, one may choose $\varepsilon_T = \sqrt{\varepsilon_M}$ (for $\varepsilon_M<1$), although the optimal value of $\varepsilon_T$ needs to be decided on a case-by-case basis. We have also included a repetition of the GS subtraction each time it is invoked in order to improve numerical convergence. For completeness, the computation of the error terms $\mathcal{E}_{m,n-1}$ has been included in the algorithm, but we remind that it may be dropped (i.e. set to zero) if the finite-precision arithmetic is expected to generate errors of the same order anyway. As a useful technical remark, we mention that, since the $W$-recursion is a two-term recursion, only two iteratively updated overlap lists $W_n:=\{W_{mn}\}_{m=0}^n$ need to be stored in the RAM at each Lanczos step. Namely, the $n$-th step only requires $W_{n-1}$ and $W_{n-2}$ in order to construct $W_n$.

The PRO algorithm described in this section is significantly more efficient than the FO algorithm, as it reduces the number of re-orthogonalization to the minimum required to preserve orthogonality of the Krylov basis to the desired accuracy $\varepsilon_T$. Both PRO and FO were implemented by \citet{I} for the computation of the Lanczos coefficients of a many-body system, the complex Sachdev-Ye-Kitaev, for system sizes such that the obtained Krylov dimensions were up to $K\sim 60~000$. For these system sizes, the re-orthogonalization algorithms succeeded in curing the numerical instability at double floating point precision, not requiring high-precision manipulations, and the authors noted that PRO reduced the number of re-orthogonalization subtractions by a factor of ten as compared to FO. These algorithms were also applied by \citet{II,III} to large spin chains, with similar conclusions regarding their efficiency and performance. For the physical analysis of these results, see section \ref{Sec:Krylov_Pheno}. 

Let us close this section by noting, for completeness, that the PRO algorithm reduces to FO in the limit $\varepsilon_T=\varepsilon_M$, as in that case the GS subtractions are performed at every single Lanczos step\footnote{Strictly speaking, in this limit PRO reduces to FO with the slight modification that in every Lanczos step $n$ \textit{both} $|\mathcal{A}_n)$ and $|\mathcal{A}_{n-1})$ are re-orthogonalized, while in FO only $|\mathcal{A}_n)$ is re-orthogonalized at the $n$-th step.}.

\subsection{Extensions of the Lanczos algorithm} \label{subsect:Extensions}

There are by now several extensions to the Krylov methods for operators and states described above, which allow to generalize the notion of Krylov complexity to contexts that were not previously accessible with the original formulation of the Lanczos algorithm. These include the multiseed method \cite{Craps:2024suj}, Krylov methods for time-dependent Hamiltonians \cite{Takahashi:2024hex}, Krylov methods for open systems \cite{Bhattacharya:2022gbz, Bhattacharya:2023zqt}; Unitary circuit dynamics \cite{Suchsland:2023cmb}, and Floquet systems \cite{Yates:2021asz, Nizami:2024ltk}.
These extensions demonstrate the versatility of Krylov methods to various frameworks and purposes. In this section we shall give brief summaries of these directions of research and point the reader to the relevant references.

\subsubsection{Multiseed Krylov complexity}
This method, introduced by 
\citet{Craps:2024suj}, adapts the notion of K-complexity to a collection of initial seeds rather than a single initial operator.
This allows to characterize the complexity growth in a given theory as an average over the Krylov complexities of a collection of seeds belonging to a suitably defined class, such as single-site operators in a $k$-local spin chain, all measured with respect to an \textit{extended Krylov basis} adapted to the time evolution of all the seeds considered. We view this as a systematization of the notion of typicality in the system. 
At the technical level, the algorithm that builds this extended Krylov basis is the so-called \textit{block Lanczos algorithm} \cite{GOLUB1977361,Ruhe_Block_Lanczos}. 
The starting point is a collection of $m$ seed operators, $\Omega_0:=\{|\mathcal{O}_{0,n})\}_{n=0}^{m-1}$, taken to be mutually orthogonal. Iteratively acting with the Liouvillian over this set and orthogonalizing layer by layer, the result is an \textit{extended} Krylov basis $\{\Omega_0, \Omega_1,\dots,\Omega_{M-1}\}$ where $\Omega_J=\{|\mathcal{O}_{J,0}),|\mathcal{O}_{J,1}), \dots , |\mathcal{O}_{J,p_{J}-1})\}$ consists of $p_J$ operators where $p_J$ is a non-increasing function of $J$.
In coordinates over this basis, it can be shown that the Liouvillian takes a \textit{block-tridiagonal} form. 
The authors \citet{Craps:2024suj} propose to define Krylov complexity with respect to this extended basis as the expectation value of the $J$-level.
With this, the K-complexity of each seed operator $|\mathcal{O}_{0,n})$, \textit{measured with respect to the extended basis}, is defined as:
\begin{eqnarray}
\label{eq:multiseed_KC_one_instance}
    C_K^{(n)}(t):=\sum_{J=0}^{M-1}\sum_{k=0}^{p_J-1} J|\phi^{(n)}_{J,k}(t)|^2
\end{eqnarray}
where $\phi^{(n)}_{J,k}(t):=(\mathcal{O}_{0,n}(t)|\mathcal{O}_{J,k})$. Finally, the \textit{multiseed Krylov complexity} is defined as the average over the $m$ seed operators:
\begin{eqnarray} \label{eq:KC_mult}
    C_\text{mult}(t) :=\frac{1}{m}\sum_{n=0}^{m-1}C_K^{(n)}(t).
\end{eqnarray}
This method is useful in assessing the generality of results for K-complexity without the issue of possible specific features of a single seed operator. In section \ref{subsect:Krylov_localization} we will discuss the application of these results to the analysis of the complexity saturation value in integrable and chaotic quantum systems.

It is worth to note that, by construction, the single-seed Krylov basis of each seed operator $|\mathcal{O}_{0,n})$ is contained in the extended multiseed Krylov basis. However, the single-seed Krylov complexity of $|\mathcal{O}_{0,n})$ need not be quantitatively the same as $C_K^{(n)}(t)$ defined in \eqref{eq:multiseed_KC_one_instance}. See \citet{Craps:2024suj} for a quantitative analysis of the differences between both prescriptions. From the perspective of quantum complexity, the authors claim that having a multiseed basis simultaneously adapted to the time evolution of a collection of simple operators is morally closer to notions of complexity \textit{à la} Nielsen. 

\subsubsection{Time-dependent Hamiltonians}
The first successful and elegant treatment of Krylov methods for time-dependent Hamiltonians was achieved by \citet{Takahashi:2024hex}. The Krylov methods we have reviewed so far in this review are applicable to time evolution generated by constant Hamiltonians. However, the most general form of time evolution in quantum mechanics is governed by a Schrödinger equation with a time-dependent Hamiltonian $H(t)$, namely: 
\begin{eqnarray}
    \label{eq:Schr_eq_time_dep}
    i\partial_t|\psi(t)\rangle = H(t)|\psi(t)\rangle~.
\end{eqnarray}
Given some initial condition $|\psi(t=0)\rangle = |\psi\rangle$, the solution of \eqref{eq:Schr_eq_time_dep}, $|\psi(t)\rangle = U(t)|\psi\rangle$ is implemented by some unitary operator that consists on a time-ordered exponential of $H(t)$. This object cannot be straightforwardly Taylor-expanded in terms of powers of some constant operator $H$, which makes it unsuitable for a naive application of the Lanczos algorithm. There exists, however, a formalism that allows (at some costs) to rewrite the solution of \eqref{eq:Schr_eq_time_dep} in terms of the exponential of a ``constant'' operator, in some sense: The \textit{double-time formalism} \cite{Howland:1974,Peskin:1993,Peskin:1994}, often referred to as double-time, or $(t,t^\prime)$ formalism. Following the notation used by \citet{Takahashi:2024hex}, we shall denote the two time variables as $(s,t)$. The double-time formalism starts by choosing an initial one-parameter family of states, $|\phi(t)\rangle$, where the argument $t$ simply labels the states in the family, i.e. translations in $t$ are not generated by a time evolution, and instead $|\phi(t)\rangle$ is a family that one can just \textit{choose}\footnote{See e.g. \citet{Peskin:1993} for examples of physically motivated choices of the one-parameter family $|\phi(t)\rangle$, which are often motivated by scattering setups or by the adiabatic approximation.}. The only constraint $|\phi(t)\rangle$ needs to satisfy is $|\phi(0)\rangle = |\psi\rangle$. One now considers time evolution with respect to a so-called \textit{quasi-time} variable $s$, defined as:
\begin{equation}
    \label{eq:Two-time_evol}
    |\widetilde{\psi}(s,t)\rangle = e^{-is(H(t)-i\partial_t)}|\phi(t)\rangle~.
\end{equation}
Note that $t$ above may be understood as a direction in an ``extended'' Hilbert space \cite{Peskin:1993}, whose quasi-time $s$ evolution is generated by the $s$-independent Hamiltonian $H(t)-\partial_t$. It is possible to prove that
\begin{equation}
    \label{eq:Double_time_sol_s_equal_t}
    |\psi(t)\rangle = |\widetilde{\psi}(t,t)\rangle 
\end{equation}
is a solution of \eqref{eq:Schr_eq_time_dep} with the initial condition $|\psi(0)\rangle = |\psi\rangle$, therefore successfully describing evolution generated by a time-dependent Hamiltonian. With this hindsight, \citet{Takahashi:2024hex} propose to implement the Lanczos algorithm for states described in section \ref{sect:State_formalism_framework}, for the $s$ time evolution, i.e. the initial state, or seed of the algorithm, is $|K_0(t)\rangle=|\phi(t)\rangle$, and the $s$-independent time-evolution generator is $H(t)-\partial_t$. The corresponding Lanczos recursion reads as follows:
\begin{eqnarray}
    \label{eq:t_dependent_Lanczos_recursion}
   b_{n+1}(t) \big|K_{n+1}(t)\big\rangle&=&(\hat{H}(t)-i\partial_t)\big|{K_n(t)}\big\rangle \nonumber \\
   &&-a_n(t)\big|{K_n(t)}\big\rangle-b_{n}(t)\big|{K_{n-1}(t)}\big\rangle~,\nonumber\\
\end{eqnarray}
where the Lanczos coefficients are defined as usual.
This yields a $t$-dependent Krylov basis $|\left\{|K_n(t)\rangle\right\}$ which is orthonormal at all values of $t$. We recall that at this stage $t$ is a parameter not related to $s$-evolution. It can then be shown, using \eqref{eq:Double_time_sol_s_equal_t}, that the state $|\psi(t)\rangle$ that solves the Schrödinger equation \eqref{eq:Schr_eq_time_dep} can be expressed, at arbitrary time $t$, as a linear combination of the basis elements $|K_n(t)\rangle$ (note that after having set $s=t$ in order to obtain $|\psi(t)\rangle$ we can again refer to $t$ as the \textit{actual} time), with coefficients $\psi_n(t)=\langle K_n(t)|\psi(t)\rangle$. The authors \citet{Takahashi:2024hex} thus propose to use these wave functions to define Krylov complexity as $C_K(t)=\sum_n ~n~ |\psi_n(t)|^2$. We note that, in analogy to \eqref{eq:Diff_rec_states}, the wave functions $\psi_n(t)$ satisfy the differential recursion
\begin{eqnarray}
    \label{eq:rec_t_dep}
    i\dot{\psi}_n(t)= a_n(t) \psi_n(t)+b_n(t)\psi_{n-1}(t)+b_{n+1}\psi_{n+1}(t)~,\nonumber\\
\end{eqnarray}
mapping to a hopping problem with time-dependent hopping amplitudes $b_n(t)$ and potential terms $a_n(t)$. All this formalism can be understood as a sort of adiabatic approach to Krylov space dynamics\footnote{See \citet{Takahashi:2023nkt} for related work on shortcuts to adiabaticity using Krylov space methods.}. For instance, in the concrete systems studied by \citet{Takahashi:2024hex}, the seed family $|\phi(t)\rangle$ is taken to be $t$-independent and, in particular, equal to an instantaneous eigenvalue of $H(t)$. A systematic analysis of the properties of the Krylov complexity and their dependence on the seed family $|\phi(t)\rangle$ has not been carried out yet in the literature.

Let us close this discussion by pointing out that the formalism described is, as presented, only applicable to time evolution in the Schrödinger picture, as Heisenberg evolution of operators would require of the computation of the time-dependent Hamiltonian in the Heisenberg picture, $H_H(t)=U(t)^\dagger H(t)U(t)$, and the Heisenberg-picture Liouvillian $\mathcal{L}_H(t)\equiv [H_H(t),\cdot] $ out of it. This is why the Krylov states formalism has been used throughout the discussion, but note that this double-time formalism can also be applied to density matrix operators, since they do evolve in the Schrödinger picture \cite{Takahashi:2024hex}.

\subsubsection{Open systems and non-Hermitian Hamiltonians} Open systems are systems which interact with their environment. The time evolution of such systems takes into account the loses into, and interaction with, the environment. Time evolution in these systems is often non-unitary, generated by a non-hermitian Hamiltonian, see for example \cite{Ashida:2020dkc, Naghiloo_nonHermitian2019} and references therein.
Under certain assumptions (see e.g.~\citet{Breuer2007Theory}), the evolution of density matrices in open systems can be seen as generated by a \textit{Lindblad} super operator $\mathcal{L}$, which, unlike the Liouvillian, in general is not necessarily a hermitian super-operator with respect to the operator-space inner product. This non-unitarity requires a modification of the Lanczos algorithms as originally formulated. \citet{Bhattacharya:2022gbz} suggested using the Arnoldi algorithm to incorporate the non-hermiticity of the Lindbladian.  However, a more suitable treatment was introduced by the same authors \cite{Bhattacharya:2023zqt}, via the \textit{bi-Lanczos} algorithm. This algorithm, already introduced by \citet{Lanczos:1950zz} as a general version of his algorithm applicable to non-symmetric matrices, produces two sets of bi-orthogonal Krylov basis elements, $\{|p_n)\}$ and $\{(q_n|\}$ such that $( q_m|p_n)=\delta_{mn}$, while separately each set need not be internally orthonormal. The initial operator $|\mathcal{O})$ seeds both sets, such that $|\mathcal{O})=|q_0 )= |p_0)$, and the bases $\{|p_n)\}$ and $\{(q_n|\}$ are constructed out of the Lanczos algorithm applied to the right and left action of the Lindbladian, respectively. With these bases one can construct the matrix $(q_n|\mathcal{L}|p_n)$, which takes a tridiagonal form with Lanczos coefficients $\{a_n\}$ on the diagonal and $\{b_n\}, \{c_n\}$ on the upper and lower diagonals, respectively. 
The time evolution of the operator $|\mathcal{O}(t))$ admits the two simultaneous decompositions
\begin{equation}
    \label{eq:Open_Op_decomps}
    |\mathcal{O}(t)) = \sum_{n=0}^{K-1}\phi_n(t)|p_n)~,\quad (\mathcal{O}(t)|=\sum_{n=0}^{K-1}(q_n|\psi(t)^*~,
\end{equation}
where it is assumed that the right and left bases have the same length, cf. \citet{Bhattacharya:2023zqt}, who used \eqref{eq:Open_Op_decomps} as a motivation to define Krylov complexity in this open-system setup as $C_K(t)=\frac{1}{\mathcal{N}(t)}\sum_{n=0}^{K-1}~n~\psi(t)^*\phi(t)$, where we have explicitly added a time-dependent normalization factor $\mathcal{N}(t)$ that accounts for the fact that the probability $P_n(t):= \psi(t)^*\phi(t) $ would need to be renormalized at each value of time due to the non-unitarity of time-evolution. 

Krylov methods have been applied to the description of dynamics of open systems by numerous authors, such as
\citet{Liu:2022god, Bhattacharjee:2022lzy, Bhattacharyya:2023zda, Bhattacharjee:2023uwx, Bhattacharya:2024uxx, Srivastav:2024apk,Bhattacharya:2023yec, Carolan:2024wov, Nandy:2024mml, Ganguli:2024uiq, Zhou:2025ozx, Medina-Guerra:2025rwa, Baggioli:2025ohh, Medina-Guerra:2025wxg,Li:2025gil,Mohan:2023btr}.
For a more detailed exposition, the Reader may consult Sections 11-13 of the review by \citet{Nandy:2024htc} and references therein.

\subsubsection{Discrete unitary dynamics}

There exist a number of contexts of physical interest where time evolution is not continuous and generated by a Hamiltonian, or Liouvillian, and instead it is discrete and implemented by some unitary operator. Examples of this are quantum circuit dynamics, and periodically-driven (Floquet) systems. In order to quantify the growth of a Krylov-like complexity, the Lanczos algorithm has been extended to these situations by e.g. \citet{Yates:2021asz}, \citet{Suchsland:2023cmb}, \citet{Nizami:2023dkf,Nizami:2024ltk}, \citet{Yeh:2023fek} \citet{ Sahu:2024urf} and \citet{Scialchi:2024zvq}, uplifting the Lanczos algorithm to the Arnoldi iteration, which we briefly review here.

The generator of evolution is a unitary superoperator $\mathcal{U}$ which acts on an operator $|O_0)$ as $\mathcal{U}|O_0)=|U^\dagger OU)$. The `time' evolution is generated by discrete action of the unitary on the operator, $|O_t)=\mathcal{U}^t|O_0)$ where $t\in \mathbb{N}_0$. Thus the Krylov subspace of an initial operator $|O_0)$ is spanned by $\{|O_0),|O_1),\dots\}$ and the Krylov basis is obtained via a Gram-Schmidt procedure over this set. In the resulting orthonormal Krylov basis, $\{|\mathcal{O}_0), |\mathcal{O}_1),\dots\}$ the evolved operator can be expanded as $|O_t)=\sum_{n=0}^t \beta_{t,n}|\mathcal{O}_n)$ and the
unitary superoperator has an upper Hessenberg form: $\mathcal{U}_{mn}=0$ for $m>n+1$, $\mathcal{U}_{mn}=b_m$ for $m=n+1$ and $\mathcal{U}_{mn}=a_m c_n/c_m$ if $m<n+1$. K-complexity is defined as $C_K(t) = \sum_{n=0}^t n |\beta_{t,n}|^2$.

\section{K-complexity as probe of chaos}
\label{Sec:Krylov_Pheno}

This Section will give an overview of the phenomenology of Krylov complexity that is found as an emergent picture out of extensive studies of systems of interest for many-body quantum chaos and holography, focusing on the question on whether K-complexity is a sensible probe of quantum chaos. In particular, we will discuss the time evolution of Krylov complexity in generic strongly-coupled many-body systems of finite size, highlighting the relationship between the profile of the Lanczos coefficients $b_n$ as a function of $n$ and the resulting dynamics of the wave function $\phi_n(t)$ and time evolution of K-complexity, and identifying the relevant time scales. This will eventually unveil a typical behavior of Krylov complexity as a function of time that, as we shall discuss, is entirely compatible with holographic expectations, which motivated the works where explicit instances of bulk/boundary correspondence have been proved in low-dimensional holography as we shall review in Section \ref{sect:Holography}. To contrast, the section will also discuss the behavior of Krylov complexity in strongly-interacting integrable systems.

All the works reviewed in this Section made use of the techniques collected in Section \ref{sec.CloserKrylov}, which will allow us to directly present the relevant results and directly dive into their discussion and interpretation. The emergent phenomenological picture anticipated in the previous paragraph will be complemented by studies of K-complexity in the states formalism, as well as instances of (conformal) quantum field theories.

This section is structured as follows: Subsection \ref{subsect:Operator_KC_chaotic} will give an overview of the generic behavior of the K-complexity of typical operators in quantum chaotic systems, while subsection \ref{subsect:Krylov_localization} will review the corresponding phenomenology in strongly-interacting integrable systems, elaborating on the phenomenon of Krylov localization; subsection \ref{Sec:TFD_state} will review the results on the state Krylov complexity (also referred to as spread complexity) of the thermofield double state and its relation to the spectral form factor; subsection \ref{subsect:SpeedLimits_ComplexityAlgebra} reviews some works on quantum speed limits in K-complexity and their relation to a complexity algebra and generalized coherent states; finally, subsection \ref{subsect:KC_QFT} reviews analyses of Krylov complexity in instances of (conformal) quantum field theory.

\subsection{Operator Krylov complexity in chaotic systems}\label{subsect:Operator_KC_chaotic}
We will begin with the generic picture of operator K-complexity in many-body chaotic systems. The reason we start with \textit{operators} rather than \textit{states} is that the results for operators are richer and perhaps more physically illuminating, especially at early time scales where K-complexity is related to the OTOC, size-complexity and scrambling.
For state complexity in chaotic systems, see Section \ref{Sec:TFD_state} where we discuss the Lanczos coefficients and K-complexity for the TFD state. 

The most important physical quantity in this discussion will be the number of degrees-of-freedom, $S$, which can be identified with the entropy of the system.   The Hilbert space dimension of a generic many-body quantum system with $S$ degrees-of-freedom is given by $D\sim e^S$ (think of a qubit or spin system) and thus, the entropy of the system understood as the logarithm of the number of microstates is given by $S$. For operators, the relevant Hilbert space dimension is $D^2 \sim e^{2S}$, cf. subsection \ref{subsect:KrylovSpaceOps}.

In generic chaotic $k$-local many-body systems there need not be degeneracies in the energy spectrum \cite{HaakeBook} and additionally, after modding out discrete symmetries, we expect and any local operator to be dense in the energy basis (i.e. its matrix elements in the energy basis will all typically be non-zero). This implies that the Krylov space dimension will saturate the bound \eqref{Krylov_dimension_bound}.  We thus expect that $K= D^2-D+1 \ \sim e^{2S}$. 

During the operator's exploration of its exponentially large Krylov space, the following time scales have been found to leave an imprint in its Krylov complexity  \cite{Parker:2018yvk,Barbon:2019wsy,I}, which we summarize below for the case of a typical\footnote{In quantum chaotic systems, the notion of \textit{typicality} is linked to the eigenstate thermalization hypothesis \cite{Srednicki:1994mfb,PhysRevA.43.2046}, expected to be satisfied by local operators in $k$-local systems.} operator in a chaotic system:

\begin{enumerate}
    \item \textit{Early times} $t\lesssim\frac{1}{\lambda_K}\log S$, where $\lambda_K$ is a parameter with dimensions of energy. These time scales describe times up to the \textit{scrambling time}, defined as $t_{s}:=\frac{1}{\lambda_K}\log S$, which may be understood as the time it takes an operator to grow to maximal size $S$, cf. subsection \ref{subsect:KC_and_size}. Within this time regime, as we shall explain further below, K-complexity of a typical operator in a chaotic system grows exponentially as $C_K(t)\sim e^{\lambda_K t}$. In the thermodynamic limit $S\to+\infty$ the scrambling time gets pushed to infinity, and hence K-complexity grows exponentially forever.
    \item \textit{Intermediate times} $\log S< \lambda_K t < e^{2S}$. During these time scales the operator is fully grown in the sense of size, but it continues to explore orthogonal elements of the Krylov basis. The typical behavior of K-complexity in this case is a linear growth as a function of time, $C_K(t)\propto t$.
    \item \textit{Late times} $\lambda_K t>e^{2S}$. At time scales of order of the \textit{Heisenberg time}, $t_H:=\frac{1}{\lambda_K} e^{2S}$, the operator has already fully explored the Krylov basis and is generically fully spread over it. Consequently K-complexity saturates at some constant value proportional to the Krylov space dimension, $C_K\sim K\sim e^{2S}$. The operator is however still changing, since the coefficients $\phi_n(t)$ in front of each Krylov element in \eqref{Op_Krylov_basis} can be fluctuating. This results in complexity oscillations around the late-time saturation value, which have thus-far not received much attention.
\end{enumerate} 

These time scales are precisely the time scales of black hole physics discussed in subsection \ref{subsect:Time_And_Space_Scales}, in particular the regimes they delimitate and the K-complexity profile within each regime agree with the typical profile of bulk lengths, or volumes, computed in black hole backgrounds. This is a very strong motivation for considering Krylov complexity as a well-defined notion of holographic complexity.

The behavior of K-complexity within each time regime is closely related to the corresponding $n$-dependent regimes of the Lanczos coefficients $b_n$, since at the end of the day they are the hopping amplitudes governing the Krylov wave function dynamics as discussed in subsection \ref{subsect:Krylov_Chain_Hopping}. We summarize the profile of the Lanczos sequence for typical operators in chaotic systems we summarize below:
\begin{enumerate}
    \item $n<S$. In this regime, \cite{Parker:2018yvk} proposed that in generic, chaotic systems, $b_n \sim \alpha n$. This is a reflection of operator growth. In the thermodynamic limit, $S\to \infty$, the Lanczos sequence is infinite and grows asymptotically linearly at large $n$. As we will review below, linear growth of the Lanczos coefficients implies the early exponential growth of Krylov complexity up to the scrambling time $t_s\sim \lambda^{-1} \log S$.
    \item $S<n \leq K$. This regime is present in the Lanczos sequence only when $S$ is finite. In such situations, the energy spectrum has a finite width due to the finite Hilbert space dimensionality. A spectral function \eqref{eq:Spectral_function_def} with compact support is known to imply an asymptotic plateau in the Lanczos sequence \cite{ViswanathMuller}, $b_n\to b$ at large $n$.
    \citet{Barbon:2019wsy} argue that the Lanczos coefficients of operators satisfying the eigenstate thermalization hypothesis (ETH) enter the plateau phase at $n\sim S$ after the initial linear growth, implying a post-scrambling linear behavior of Krylov complexity. This article was also the first to propose the holographic suitability of K-complexity. However, the aforementioned plateau regime is necessarily an asymptotic statement that cannot hold in finite-size systems, where the Lanczos coefficients should terminate at $n=K\sim e^{2S}$. \citet{I} computed for the first time the full Lanczos sequence of a typical operator in a chaotic many-body system and found that after $n\sim S$ the Lanczos coefficients enter a \textit{descent} regime, where they approach zero with an average slope $-e^{-2S}$ (which, in particular, is non-perturbative in a $\frac{1}{S}$-expansion). This in turn implies a gradual deceleration of the K-complexity growth as it approaches saturation.
    \item Finiteness of the Lanczos sequence and complexity saturation: As already discussed, we expect the Krylov space dimension of a typical operator in a chaotic system to be finite, $K\sim e^{2S}$. Given that Krylov complexity grows linearly after scrambling, one may estimate the time at which the Krylov wave packet reaches the edge of Krylov space by the Heisenberg time $t_H\sim \lambda^{-1} e^{2S}$. By that time, Krylov complexity has no further scope of growth and the wave packet typically spreads and eventually becomes a uniform distribution in Krylov space, whose K-complexity may thus be roughly estimated to be $C_K\sim K/2\sim e^{2S}/2$, which we may refer to as the typical K-complexity saturation value.
\end{enumerate} 

The phenomenology we have just described is summarized in Table \ref{tab:KC_Pheno}. Let us now provide a more detailed elaboration on the Krylov complexity regimes in terms of the Lanczos sequence profile and the Krylov wave function dynamics.

\begin{table*}[]
\centering
\begin{tabular}{c|c|c|c|c}
   $n$  & $b_n$ & wave function & K-complexity & time scales \\
    \hline 
    $n < S$ & $b_n \sim \alpha n$  & \adjustbox{valign=c}{\includegraphics[width=1.2in]{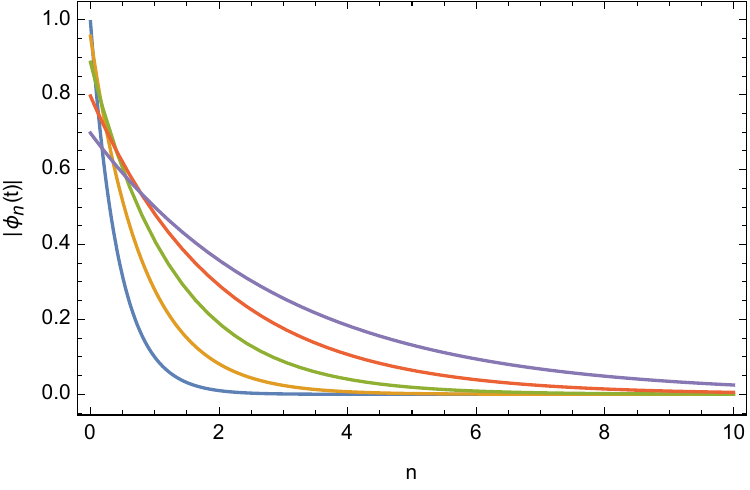}} & \adjustbox{valign=c}{\includegraphics[width=1.2in]{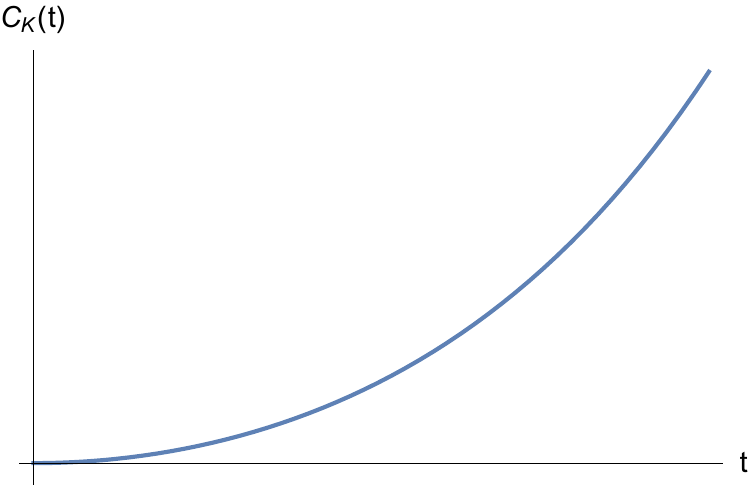}} & $t \lesssim \log S$\\
    \hline 
    $n > S$ & $b_n \sim \Lambda S$ & \adjustbox{valign=c}{\includegraphics[width=1.2in]{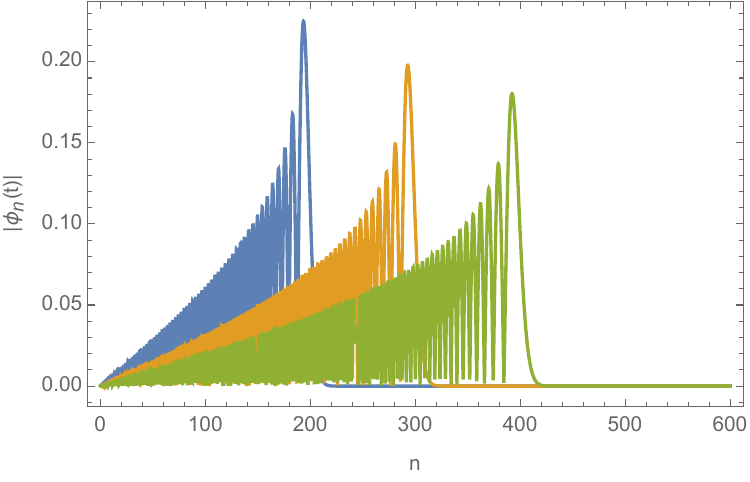}} & \adjustbox{valign=c}{\includegraphics[width=1.2in]{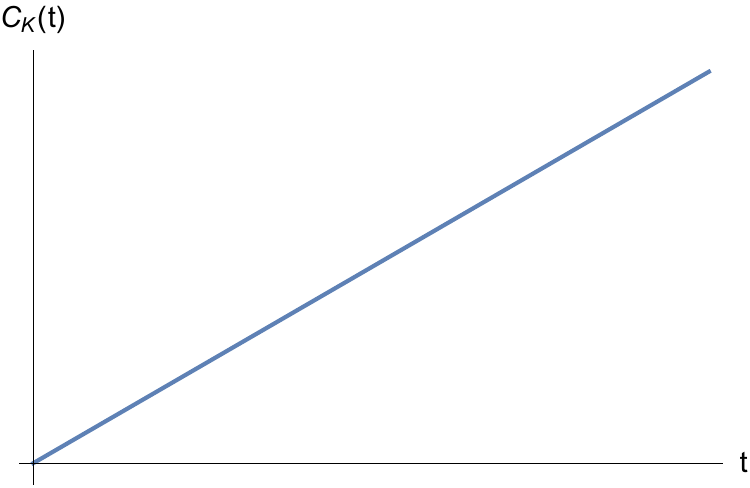}} & $\log S < t < e^{2S}$ \\
    \hline 
    $S \ll n \lesssim e^{2S}$ & ``descent" & \adjustbox{valign=c}{\includegraphics[width=1.2in]{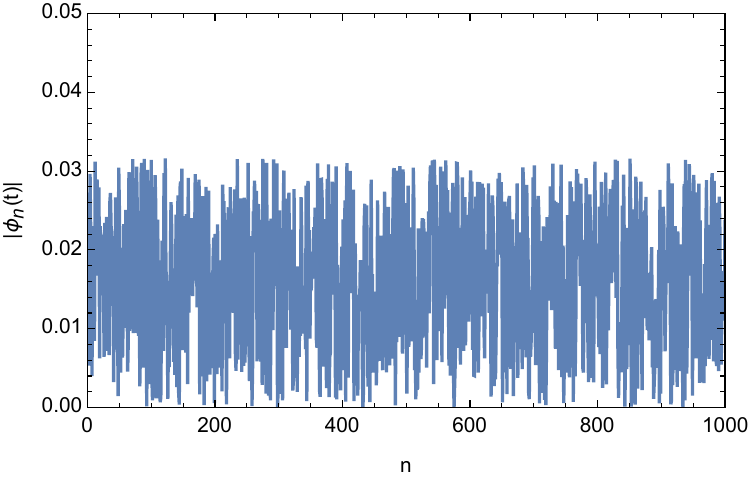}} & \adjustbox{valign=c}{\includegraphics[width=1.2in]{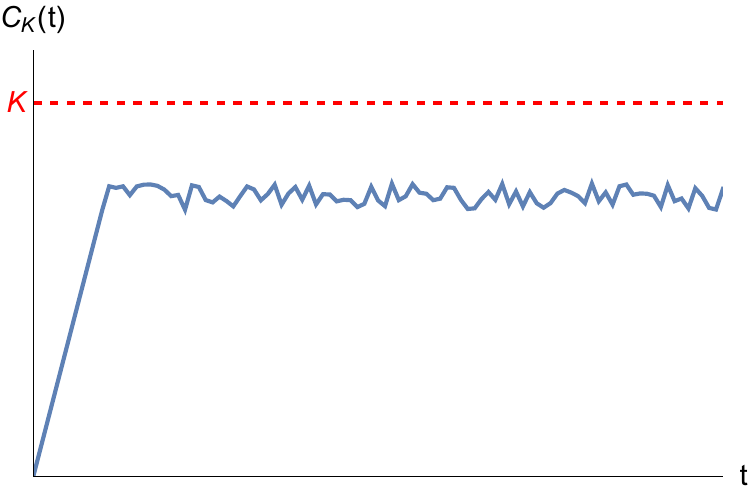}} & $t\gtrsim e^{2S}$
\end{tabular}  
    \caption{K-complexity phenomenology for local operator in generic many-body chaotic system with $S$ degrees of freedom.}
    \label{tab:KC_Pheno}
\end{table*}

\subsubsection{Early times: linear growth of Lanczos coefficients and exponential growth of K-complexity}
\citet{Parker:2018yvk} conjectured that in generic, \textit{non-integrable}, many-body systems, where $H$ is infinite and $\mathcal{O}$ is a local operator, the Lanczos coefficients exhibit asymptotic linear growth\footnote{In 1-dimensional models there are logarithmic corrections to this linear growth \cite{Parker:2018yvk}.}:
\begin{eqnarray}
\label{linear_LCs}
    b_n = \alpha n +O(1)
\end{eqnarray}
where $O(1)$ denotes term that behave asymptotically as a constant\footnote{In this discussion, we set such a constant to $1$ for the sake of simplicity.} when $n\to\infty$. $\alpha$ is an energy scale which determines the decay of the spectral function, namely $\Phi(\omega)=e^{-\frac{\pi |\omega|}{2\alpha}}$. In particular, \citet{Parker:2018yvk} proved that at infinite temperature the asymptotic growth of the Lanczos coefficients at large $n$ is at most linear, using known bounds on the high frequency tails of the spectral function, and proposed that such a bound is saturated by typical operators in maximally chaotic systems, providing both analytical and numerical evidence in favor of this so-called \textit{universal operator growth hypothesis}. To understand the resulting K-complexity, the wave function equation, \eqref{DifRecEq}, with $b_n$ given by \eqref{linear_LCs}, is solved with initial condition $\phi_{n}(t=0)=\delta_{n0}$ and the wave function is found to be given by $\phi_n(t) = i^n~ \mathrm{sech} (\alpha t)\, \tanh^n (\alpha t)$. This wave function spreads very quickly over the Krylov basis, see the first row of Table \ref{tab:KC_Pheno}.
The resulting time-dependent behaviour of K-complexity \eqref{KC_def} is that of exponential growth 
\begin{eqnarray}
    C_K(t) = \sum_{n=0}^\infty n\, |\phi_n(t)|^2 = \sinh^2(\alpha t) \sim e^{2\alpha t} ~.
\end{eqnarray}
Given that \cite{Parker:2018yvk} show that linear growth \eqref{linear_LCs} is the maximal growth rate of the Lanczos coefficients, K-complexity's maximal growth rate is exponential in time.  

As was already mentioned in the Introduction, the slope, $\alpha$, of the asymptotically linear growth of Lanczos coefficients \eqref{linear_LCs} is closely related to the quantum Lyapunov exponent, $\lambda$, as diagnosed from out-of-time order correlations functions \eqref{eq.OTOC4pt}:
As we have said, \citet{Parker:2018yvk} proved that the Lanczos coefficients $b_n$ of a system at infinite temperature and in the thermodynamic limit grow at most asymptotically linearly in $n$ with some slope $\alpha$, implying a Krylov complexity profile $C_K(t)$ which grows exponentially in time with exponent $2\alpha$, which in turn implies (see section \ref{subsect:theorems}) the bound on the OTOC Lyapunov exponent in  \eqref{eq:bound_Lyapunov_KrylovExp_intro}, which we repeat here for convience:
\begin{equation}\label{eq:Altman_Bound}
    \lambda \le 2\alpha\,.
\end{equation}
Note that this bound improves on the MSS bound $\lambda\leq \frac{2\pi}{\beta}$ \cite{Maldacena:2015waa}, which becomes trivial at infinite temperature. At generic finite temperatures, however, the relation \eqref{eq:Altman_Bound} is not proven but conjectured, due to the ambiguity in the choice of thermal operator inner product required to define both the OTOC and the Krylov complexity. \citet{Parker:2018yvk} present numerical evidence that this bound is tight and saturated at finite temperature in chaotic many-body systems such as the SYK model, suggesting the hierarchy $\lambda\leq 2\alpha(\beta)\leq 2\pi T$ at finite temperature, which has been the subject of study of several works since then \cite{Avdoshkin:2019trj,Gu:2021xaj,Dymarsky:2021bjq,Avdoshkin:2022xuw,Chapman:2024pdw}.

Important objections to the universal operator growth hypothesis extrapolated to finite temperatures have been raised, as we shall review in subsection \ref{subsect:KC_and_size}, in the realm of quantum field theory. In particular, \citet{Dymarsky:2021bjq} pointed out that in certain regularization schemes the Krylov complexity exponent $\lambda_K=2\alpha$ is entirely fixed by symmetry and by the choice of regulator, not allowing to distinguish chaotic from integrable dynamics. In such situations finite-size corrections away from the thermodynamic limit become relevant to distinguish quantum-chaotic behavior to integrable dynamics, as we shall also discuss in section \ref{subsect:KC_QFT}.
Relatedly, \citet{Chapman:2024pdw} approached the potential relation between the linear growth of the Lanczos coefficients and chaotic dynamics
by tracking both Lyapunov and Krylov exponents ($\lambda$ and $\lambda_K$, respectively) in chaotic-to- integrable flows of deformed SYK models. They reported that unlike the Lyapunov exponent itself, the Krylov exponent behaves monotonically as a function of the deformation parameter, and claimed that this means that $\lambda_K$ fails to diagnose periods of sub-maximal chaos, which are apparent in the Lyapunov exponent $\lambda$. The authors conjecture the monotonicity of the Krylov exponent to be a more general property also for quantum systems beyond (deformed) SYK. Another counterexample of the universal operator growth hypothesis was provided by \citet{Bhattacharjee:2022vlt}, who studied Krylov complexity in the Lipkin-Meshkov-Glick (LMG) model, a system which is known to be integrable (in particular, solvable via the Bethe Ansatz) but which nevertheless features an unstable saddle in its classical phase space: The authors found that Krylov complexity grows exponentially at early times despite the integrable nature of the model. Related analyses have been carried out by \citet{Bento:2023bjn, Huh:2023jxt}.

There exist abundant studies of the proposed relation between (maximal) chaos and the early-time growth of Krylov complexity; some examples are the ones by \citet{Kim:2021okd,Chen:2024imd,PG:2025ixk,He:2022ryk,Hashimoto:2023swv,Li:2024tej,Camargo:2023eev,Iizuka:2023pov, Iizuka:2023fba,Vardian:2024fsp,Bhattacharyya:2023dhp,Bhattacharya:2023xjx, Alishahiha:2024rwm, Murugan:2024ory, Toga:2025nyi,Tang:2023ocr, Loc:2024oen, Gorsky:2024xnl,Fan:2022xaa,Qi:2023jad,Gamayun:2025hvu,Fan:2022mdw,Du:2022ocp}. In any case, in light of the various caveats and special cases that provide counter-examples to the universal operator growth hypothesis, the analysis of post-scrambling time scales appears to be necessary in order to investigate the relationship between Krylov complexity dynamics and quantum chaos.

\subsubsection{Intermediate times: constant Lanczos coefficients and linear growth of K-complexity}

For finite systems, \citet{Barbon:2019wsy} showed that the linear growth of the Lanczos coefficients necessarily stops once the operator has grown to size of $O(S)$, and the Lanczos coefficients plateau.  Their argument goes as follows: in a system with $S$ degrees-of-freedom, and assuming the operator satisfies the ETH, the main contribution to the moments for $n\gg S$, comes from the largest energy difference of the system, of order $\omega_{\text{max}}\sim \Lambda S$ where $\Lambda$ is some intensive energy scale of the system.  
Thus a rough estimate results in $\mu_{2n}\sim (\Lambda S)^{2n}$. The relationship between the Lanczos coefficients and moments, Eq.~\eqref{eq:moments_Lanczos_Dyck},
reveals that the Lanczos coefficients for $n>S$ become effectively constant
\begin{eqnarray}
    b_n \approx \Lambda S, \quad n>S ~.
\end{eqnarray}

Assuming constant Lanczos coefficients, $b_n=b$, equation \eqref{DifRecEq} is solved by \cite{Barbon:2019wsy}
\begin{eqnarray}
    \phi_n(t) = i^n\, \frac{n+1}{b\,t} J_{n+1}(2bt)
\end{eqnarray}
where $J_n(x)$ is the Bessel function of the first kind. This function has a peaked front at $n \approx 2bt$ which moves linearly with velocity $2b$ with a long tail behind it, see the second row of Table \ref{tab:KC_Pheno}. It turns out that most of the weight of the function is located at its peak, but the tail has an effect of making the weighted average smaller and K-complexity is seen to grow linearly,
\begin{eqnarray}
    C_K(t) \approx \frac{16}{3\pi} b \, t \propto \Lambda S\, t~.
\end{eqnarray}
We are expecting K-complexity to grow linearly from times of order $t\sim \lambda_K^{-1} \log S$. This linear growth describes the operator exploring further and further into Krylov space and is expected to continue until times of order $e^{2S}$. Interestingly, circuit complexity also predicts an exponentially long period of linear growth. In the context of circuit complexity, the length of the circuit needed to describe the operator after scrambling time, grows linearly, see heuristic arguments in \cite{Brown:2017jil} showing that no more `cancellations' of gates occur after scrambling time. A more detailed mathematical treatment was given in \cite{Haferkamp:2021uxo}. We nevertheless remind the Reader that these results in circuit complexity require, unlike Krylov complexity, the introduction of a tolerance parameter in order to take finite values, even in Hilbert spaces of finite dimension. 
 
\subsubsection{Long times: the `descent' of the Lanczos coefficients and saturation of K-complexity}
Numerical analysis of many-body, strongly coupled finite systems, shows that the constant behaviour of the Lanczos coefficients described above is in fact part of a slow descent towards zero \cite{I,II,III}. This causes the velocity peak of the wave function to slow down. Simultaneously, the Krylov wave packet gradually develops a tail behind the wavefront that makes it disperse. 
In chaotic systems, by the time the wave function reaches the end of the Krylov basis it is completely dispersed. Given that the length of the Krylov chain is $O(e^{2S})$, the time it takes the wavefront to reach the end of the Krylov chain, assuming it is moving forwards with roughly constant velocity of $O(S)$, is of the order of the Heisenberg time, $t_H$, which scales as $O\left( e^{2S}\right)$.
In order to estimate the K-complexity late-time saturation value, attained at the aforementioned Heisenberg time scale, we may assume that the Krylov wave function is completely uniform in Krylov space, namely:
\begin{eqnarray}
    \phi_n = \frac{1}{\sqrt{K}} i^n  r_n~,
\end{eqnarray}
for some real number $r_n\sim O(1)$ that does not scale parametrically with the Krylov dimension $K$, and where we have pulled out an oscillating factor $i^n$ because it can be shown that $i^n\phi_n\in\mathbb{R}$. 
This implies the following saturation value:
\begin{eqnarray}
     C_K (t\gtrsim t_H)  \sim  \sum_{n=0}^{K-1} n\, \frac{r_n^2}{K}  \sim \frac{K-1}{2} \sim \frac{K}{2} ~,
\end{eqnarray}
whose scaling with system size is $C_K\sim e^{2S}$ for a maximal Krylov space dimension $K\sim D^2\sim e^{2S}$, cf. section \ref{subsect:KrylovSpaceOps}.
As we discuss in Section \ref{subsect:Krylov_localization}, this estimate for the saturation value of K-complexity can change quantitatively in many-body, strongly interacting integrable systems where the wave function propagation is hindered by a localization, which can be traced to stronger fluctuations in the Lanczos coefficients. 

Finally, in this regime where the wave function is fully spread over the Krylov basis, K-entropy can be estimated as
\begin{eqnarray} \label{eq:KS_sat}
    S_K \sim -\sum_{n=0}^{K-1} \frac{r_n^2}{K} \log\frac{r_n^2}{K} \sim \log(K) \sim O(S),
\end{eqnarray}
where we used the fact that $K\sim e^{2S}$.

As a practical example for the algorithms, bounds, and definitions presented in this review we will focus on numerical results for the time-evolution of Krylov complexity in the complex Sachdev-Ye-Kitaev (SYK) model (\citet{Sachdev:1992fk}; \citet{kitaev2015kitp}; \citet{Sachdev:2015efa}, \citet{Gu:2019jub}). The SYK model is a strongly interacting, chaotic \cite{Garcia-Garcia:2016mno, Cotler:2016fpe}, many-body system which saturates the MSS bound chaos bound \cite{Maldacena:2015waa}, constituting a benchmark of quantum chaos with a very tractable two-dimensional gravitational dual \cite{kitaev2015kitp,Maldacena:2016hyu}. This model satisfies the generic phenomenology described above, as we shall review next.

\subsubsection*{K-complexity in complex SYK\texorpdfstring{$_4$}{TEXT}}
In this Section, we follow the evolution of K-complexity for a local operator in the complex Sachdev-Ye-Kitaev (cSYK) model of $L$ complex fermions described by fermionic ladder operators $\{c_i\}_{i=1}^L$ which satisfy  $\{c_i,c_j^\dagger\}=\delta_{ij}$ and $\{c_i,c_j\}=0=\{c_i^\dagger,c_j^\dagger\}$. The Hamiltonian describes all-to-all interactions between every four fermions in the system:
 \begin{equation}
 \label{H_cSYK4}
    H_{\textrm{SYK}} = \sum_{i,j,k,l} J_{ij,kl} \, c_i^\dagger \, c_j^\dagger \, c_k \, c_l
\end{equation}       
where the coupling constants $J_{ij;kl}$ are constrained by hermiticity of the Hamiltonian ($J_{ij;kl}=J_{kl;ij}^*$) and the anti-commutation properties of the fermionic operators ($J_{ij;kl}=-J_{ji;kl}=-J_{ij;lk}$). 
The coupling strengths that are not related by these transformations are independent and identically distributed complex random variables described by a Gaussian distribution with zero mean,  $\overline{J_{ij,kl}} = 0$, and variance given by $\overline{|J_{ij,kl}|^2}=\frac{3! J^2}{L^3}$; and the sum in \eqref{H_cSYK4} ranges through the maximal set of index tuples $(i,j,k,l)$ that are independent with respect to the permutations that are symmetries of $J_{ij;kl}$.

The model described by the Hamiltonian \eqref{H_cSYK4} is not local in the sense that the latter is built out of all-to-all interactions. Nevertheless, since such interactions are four-site, operators whose size (as defined in subsection \ref{subsect:KC_and_size}) is intensive in system size shall still undergo a gradual growth. In other words, in SYK the notion of local operators is traded by that of intensive operators. An example of such operators, which we shall consider in the present discussion, is the hopping operator:
\begin{equation}
\label{Op_SYK4}
    \mathcal{O} = c_{L-1}^\dagger c_L +c_L^\dagger c_{L-1}~,
\end{equation}
where, again, the sites for which $\mathcal{O}$ is defined are not relevant due to the all-to-all nature of the interactions.
This operator was shown by \citet{Sonner:2017hxc} to satisfy the eigenstate thermalization-hypothesis (ETH), thus providing an instance of a typical operator in cSYK$_4$.

The Hamiltonian \eqref{H_cSYK4} commutes with the number operator, $\hat{N} = \sum_{i=1}^L c_i^\dagger c_i$, and thus preserves occupation number. Since also the hopping operator \eqref{Op_SYK4} commutes with the number operator, it is possible to restrict both $H$ and $\mathcal{O}$ to a symmetry sector with a fixed occupation number, $0\leq N\leq L$. The total Hilbert space dimension is $2^L$, while the Hilbert space dimension of an $N$-fermion sector is $D=\binom{L}{N}$. We shall consider the largest sector, for which $N=\left \lceil{L/2}\right \rceil$. The dimension of this Hilbert space sector is smaller than the total Hilbert space dimension but it still scales exponentially with system size $L$. 

\begin{figure}
    \centering
    \includegraphics[width=1\linewidth]{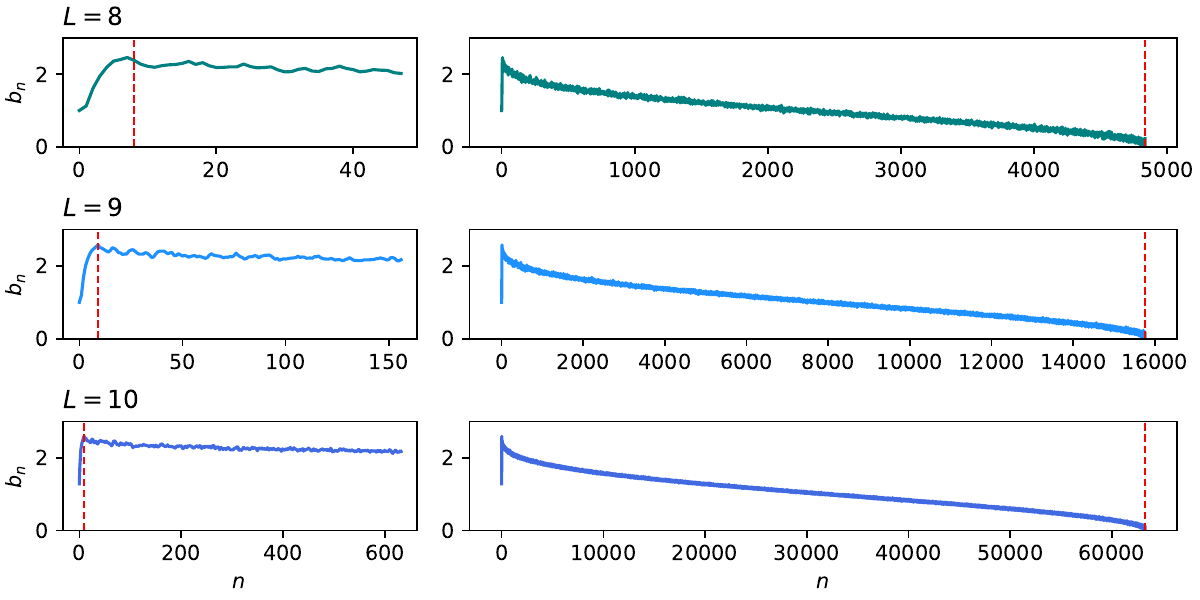}
    \caption{Lanczos coefficients of a two-site hopping operator in complex SYK$_4$ for $L=8,9,10$ fermions in the half occupation sector. Energy units are set by the coupling strength $J$. On the \textbf{left} we show the initial part of the Lanczos sequence which exhibits a (linear) growth up to $n\sim L$ (dashed red vertical line) and then a plateau at a value of the order of the spectral width (which does not scale with system size due to the normalization of the Hamiltonian). On the \textbf{right} we show the full Lanczos sequence where a persistent descent to zero is seen to operate after $n\sim S$. The dashed vertical line represents the value of the predicted Krylov dimension, $K=D^2-D+1$, which is seen to be in agreement with the value at which the Lanczos sequence terminates. The Partial Re-Orthogonalization (PRO) algorithm (cf. section \ref{subsect:numerical_implementations}) was implemented to obtain these results, which were reported by \citet{I}.}
    \label{fig:SYK4_Lanczos}
\end{figure}

Within the symmetry sector described above, the Krylov space dimension associated to the hopping operator saturates the upper bound \eqref{Krylov_dimension_bound}. This happens because (\textit{i}) the eigenvalues of the Hamiltonian \eqref{H_cSYK4} are non-degenerate within the symmetry sector\footnote{In fact, the spectrum of cSYK$_4$ is chaotic, featuring level repulsion well described by random matrix universality \cite{Sonner:2017hxc,Altland:2020ccq}.} and do not have any rational relations among themselves, such that the spectrum of the Liouvillian is also non-degenerate (apart from the $D$-fold degeneracy of the zero frequency); and (\textit{ii}) the local operator \eqref{Op_SYK4} satisfies the ETH, and thus is dense in the energy basis\footnote{We remind the Reader that in this context we use the term \textit{dense} as opposed to \textit{sparse} when referring to the matrix representation of an operator in terms of its coordinates over some basis.}, cf. section \ref{subsect:KrylovSpaceOps}. \citet{I} used the Lanczos re-orthogonalization algorithms described in section \ref{subsect:numerical_implementations} in order to compute the Lanczos sequences of the hopping operator \eqref{Op_SYK4} in the aforementioned half-occupation sector of cSYK$_4$ for sizes $L=8,9,10$. As predicted, in all cases the Krylov space dimension was found to be maximal and in the case of the largest system size analyzed it was given by $K=63253$. 
The results are shown in Figure \ref{fig:SYK4_Lanczos}.
Given that numerics were performed in half the occupation sector at all system sizes, the number of degrees of freedom can still be taken to be proportional to system size, i.e. $S\sim L$, and with this the features discussed throughout the current subsection \ref{subsect:Operator_KC_chaotic} are seen to apply in this paradigmatic model of a maximally chaotic many-body system, namely:
A (linear\footnote{Numerically, early linear growth of the Lanczos coefficients is difficult to distinguish from other behaviors because it is expected to apply only up to $n\sim S$ while the Krylov dimension is $K\sim e^{2S}$. In their numerical analysis, \citet{I} focused on the features of the Lanczos coefficients at large values of the Krylov index $n$, in order to probe non-perturbative effects due to finite system size.}) growth of the Lanczos coefficients up to $n=S$ that transitions to an apparent plateau (see left part of Figure \ref{fig:SYK4_Lanczos}) corrected by a persistent non-perturbative descent down to zero with an average negative slope of order $O(K^{-1})\sim O(e^{-2S})$. The Lanczos sequence $b_n$ is seen to terminate at $n\sim K \sim e^{2S}$.

\begin{figure*}
    \centering
    \includegraphics[width=0.32\linewidth]{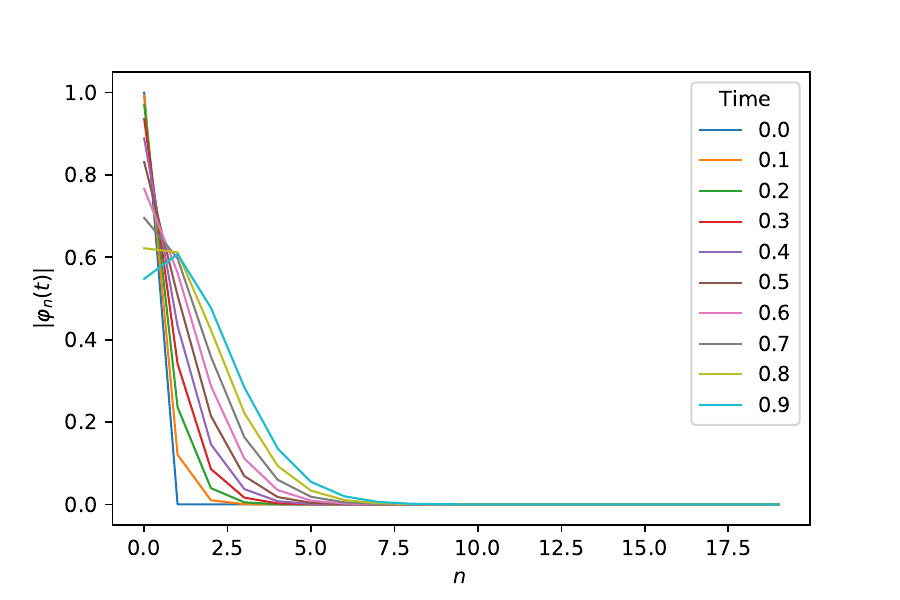}
    \includegraphics[width=0.32\linewidth]{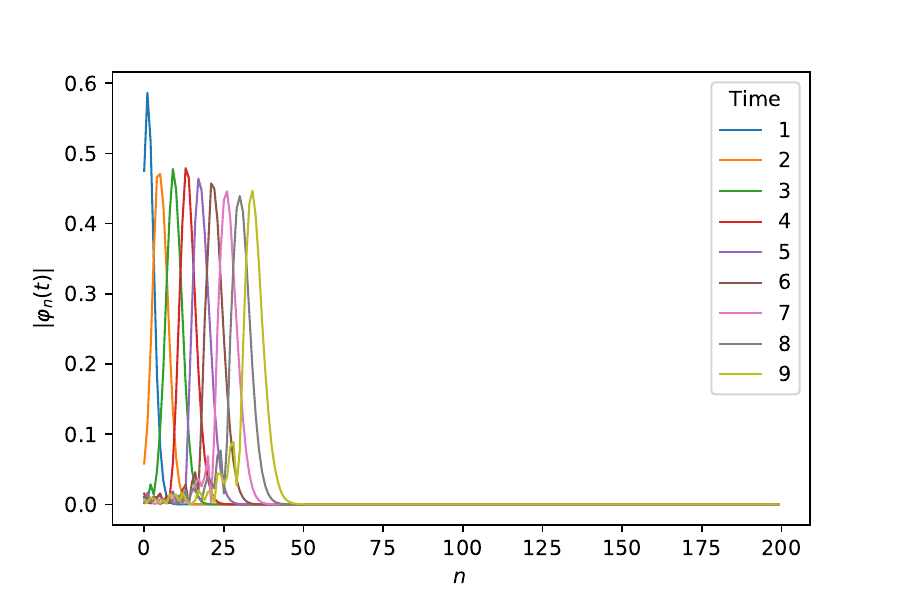}
    \includegraphics[width=0.32\linewidth]{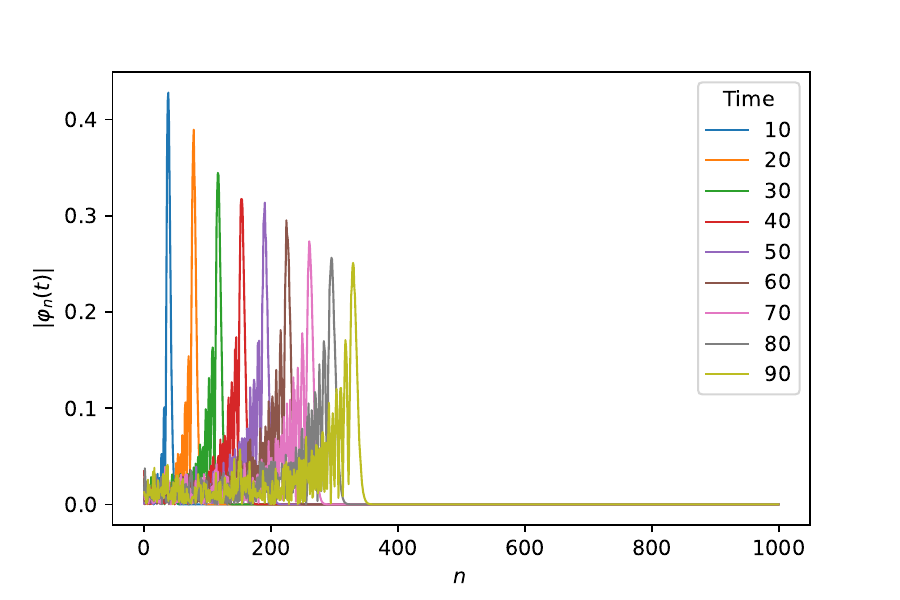}
    \includegraphics[width=0.32\linewidth]{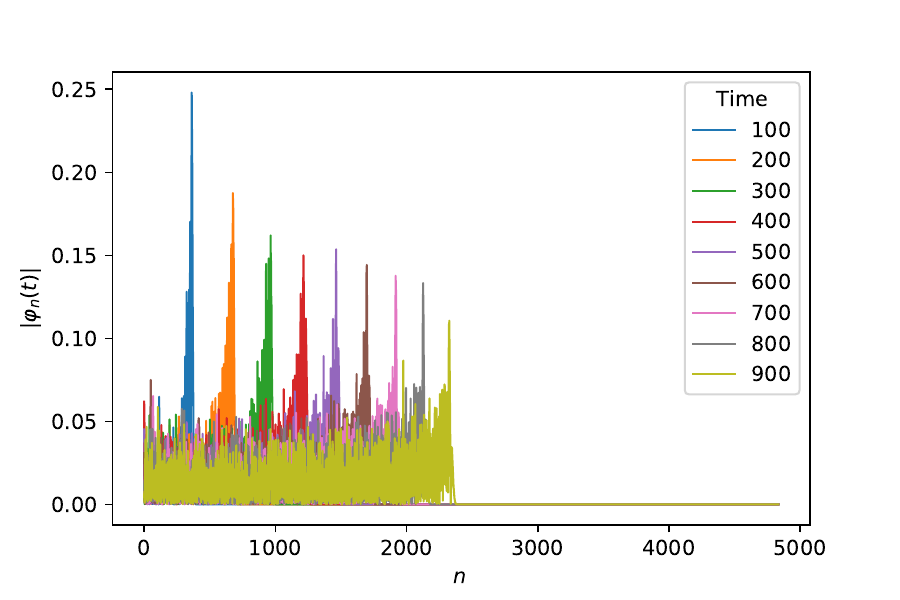}
    \includegraphics[width=0.32\linewidth]{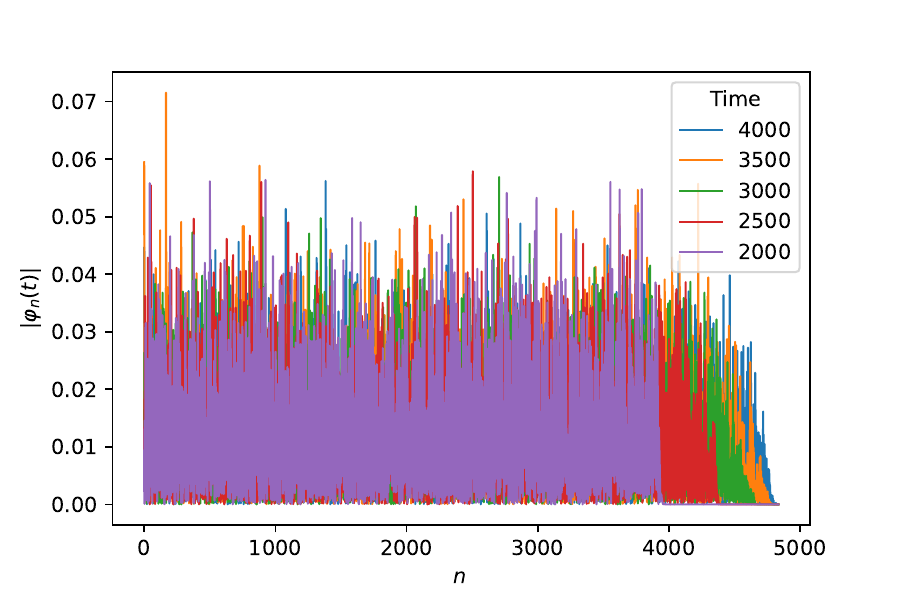}
    \includegraphics[width=0.32\linewidth]{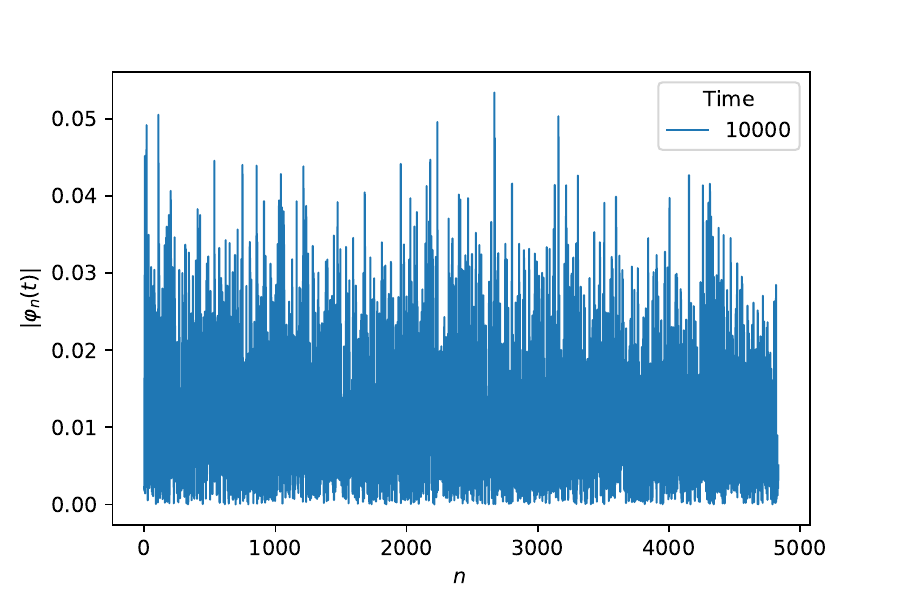}
    \caption{Absolute value of the Krylov space wave function, $\varphi_n(t)=i^{-n}\phi_n(t)\in\mathbb{R}$, of a hopping operator in complex SYK$_4$ with $L=8$ fermions at half occupation, evaluated at various times.  Initially, fast quasi-ballistic spreading of the wave function is observed; it then becomes a propagating wave with a peaked wave-front. As the wave propagates forwards along the Krylov basis, it becomes more and more dispersed. Eventually, the wave function is fully dispersed over the Krylov basis. Energy units are set by the coupling strength $J$, which was itself set to $1$ for the numerics.}
    \label{fig:Wave_function_SYK4_L8}
\end{figure*}

\begin{figure*}
     \begin{subfigure}[t]{0.45\textwidth}
         \centering
         \includegraphics[width=\textwidth]{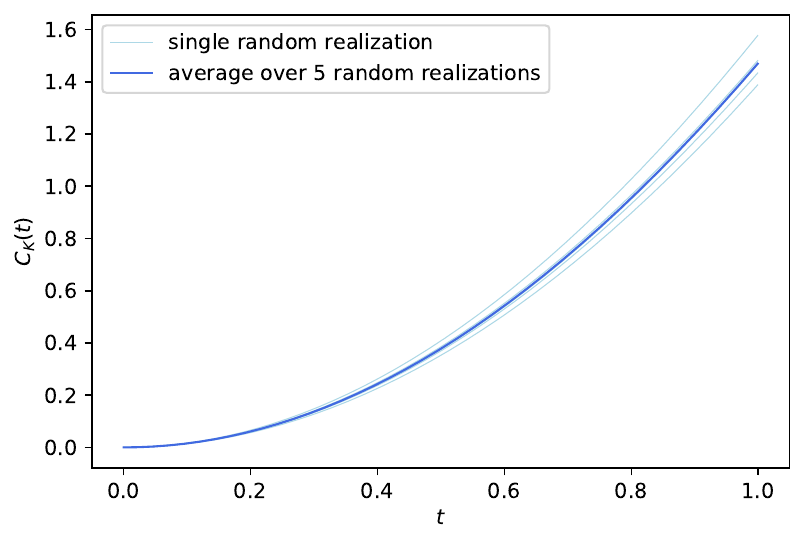}
         \caption{At very early times $t<\log S$, a non-linear growth of K-compleixty is observed. This is compatible with the initial linear growth of the Lanczos coefficients.}
         \label{fig:KC_cSYK_L10_N5_5R_early}
     \end{subfigure}
     \hfill
     \begin{subfigure}[t]{0.45\textwidth}
         \centering
         \includegraphics[width=\textwidth]{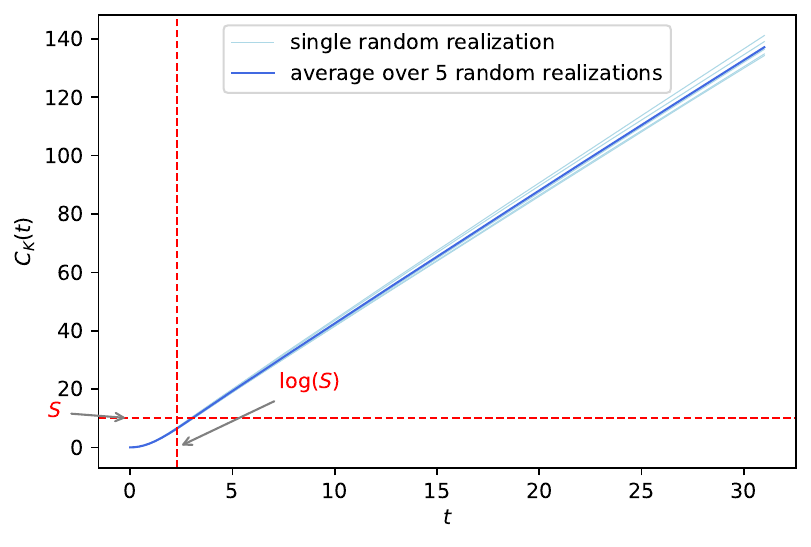}
         \caption{At early times $t>\log S$, the non-linear growth of K-complexity becomes a linear growth. This is compatible with the apparent plateauing of the Lanczos coefficients.}
         \label{fig:KC_cSYK_L10_N5_5R_medium_K_notation}
     \end{subfigure}
     \vfill
     \begin{subfigure}[t]{0.45\textwidth}
         \centering
         \includegraphics[width=\textwidth]{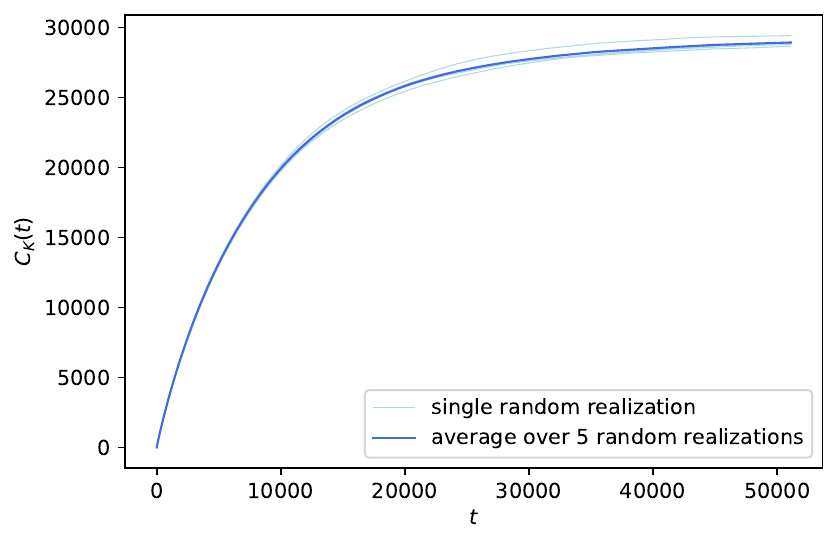}
         \caption{At late times approaching $t\sim e^{2S}$, the descent of the Lanczos coefficients causes K-complexity to slow down. The fluctuations in the Lanczos coefficients cause the wave function to become dispersed and K-complexity begins to show signs of saturation.}
         \label{fig:KC_cSYK_L10_N5_5R_late}
     \end{subfigure}
      \hfill
     \begin{subfigure}[t]{0.45\textwidth}
         \centering
         \includegraphics[width=\textwidth]{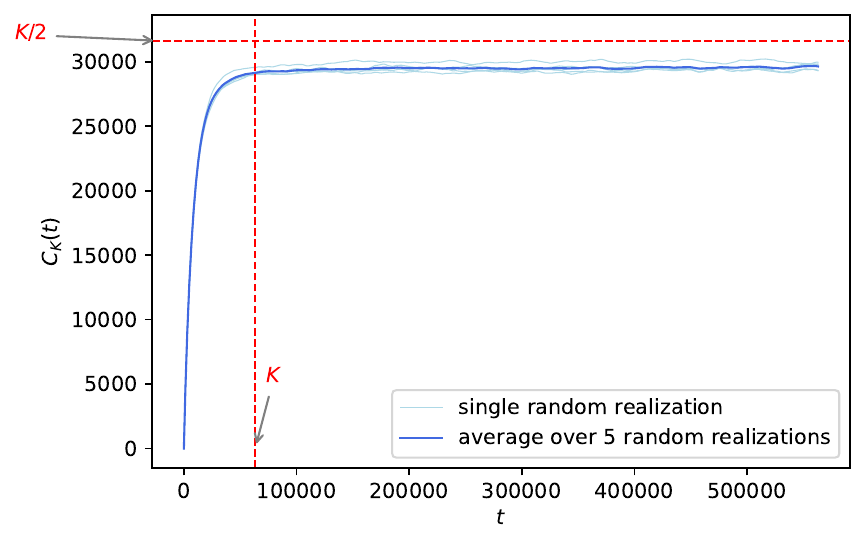}
         \caption{At time scales of order $t=K\sim e^{2S}$, K-complexity is fully saturated at a value close to $K/2$.}
         \label{fig:KC_cSYK_L10_N5_5R_vlate_K_notation}
     \end{subfigure}
        \caption{K-complexity of a hopping operator in complex SYK$_4$ at various time scales for $L=10$ fermions at half occupation, computed out of the Lanczos coefficients in Figure \ref{fig:SYK4_Lanczos}. 
        We identify $S=L$ as the number of degrees-of-freedom. Five random realizations and an average over them are shown. Energy units are set by the coupling strength $J\equiv 1$.
        These results were reported by \citet{I}.}
        \label{fig:KC_SYK4_L10}
\end{figure*}

\begin{figure}
    \centering
    \includegraphics[width=0.9\linewidth]{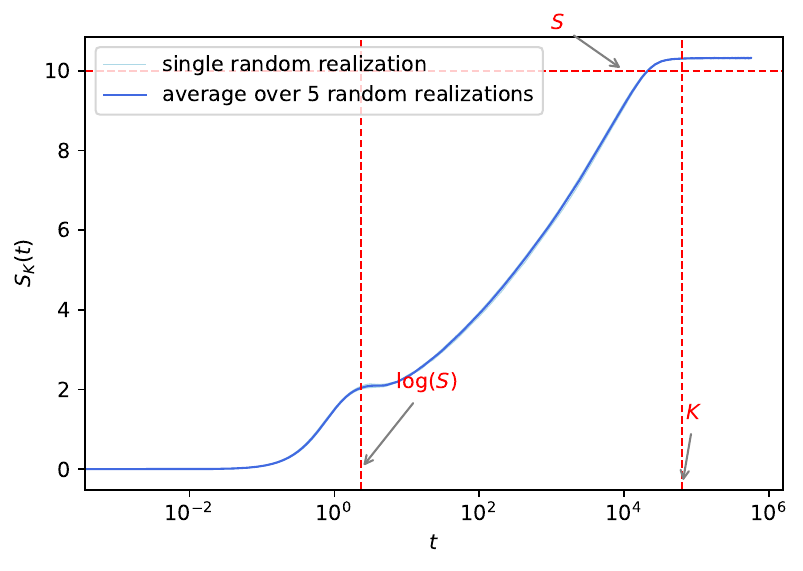}
    \caption{K-entropy of a hopping operator in complex SYK$_4$ as a function of time (in units set by the coupling strength $J\equiv 1$) with $L=10$ fermions at half occupation, computed out of the Lanczos coefficients in figure \ref{fig:SYK4_Lanczos}. After early initial growth, at $t\sim \log(S)$ it transitions to a logarithmic growth phase which continues up to saturation at $t\sim e^{2S}$ at a value of order $O(S)$ as predicted by \eqref{eq:KS_sat}, cf. \citet{I} for details.}
    \label{fig:KS_SYK4_L10}
\end{figure}

For reference, in Figure \ref{fig:Wave_function_SYK4_L8} we show the evolution of the (absolute value) of the wave function, $|\phi_n(t)|$, for cSYK$_4$ with $L=8$ fermions at half occupation. Numerical results for K-complexity and K-entropy at various time scales obtained by \citet{I} for cSYK$_4$ with $L=10$ fermions (also at half filling) are shown in Figures \ref{fig:KC_SYK4_L10} and \ref{fig:KS_SYK4_L10}, respectively. In particular, Krylov complexity is seen to grow linearly after scrambling, up to saturation at $t_H\sim e^{2S}$ at a value close to half of the Krylov space dimension $K/2\sim e^{2S}$, consistently with the fact that the Krylov wave function is essentially uniform at exponentially late times (cf. figure \ref{fig:Wave_function_SYK4_L8}).

\subsection{Integrable systems. Krylov localization}\label{subsect:Krylov_localization}

Integrable quantum systems have an extensive number of conserved charges \cite{1931ZPhy...71..205B,Samaj_bajnok_2013}, and generally do not spread information among degrees of freedom as efficiently as chaotic systems do. Simple operators in these systems fail to satisfy the eigenstate thermalization hypothesis\footnote{In fact, to the extent that \textit{typical operators} in chaotic system can be defined precisely thanks to the universal applicability of the ETH, in integrable systems there is no ``canonical'' definition of typical operators, and in our discussion we shall use the working definition of typical operators as \textit{simple} operators in terms of the elementary building blocks of the Hamiltonian, e.g. local operators in a system governed by a $k$-local Hamiltonian.}, cf. \cite{Rigol_XXZ}. In free, quadratic systems, commutators of the Hamiltonian with an initially local operator will not make it grow in size, and therefore one can expect the Krylov space dimension to be linear in system size, $K\sim S$, much smaller than its upper bound \eqref{Krylov_dimension_bound}, which in turn implies a bound on Krylov complexity that also scales as $\mathit{O}(S)$ \cite{I}.
However, strongly-interacting integrable systems need not have exact degeneracies in their spectrum (the most general statement is that their energy eigenvalues are uncorrelated) and, as studied by \citet{Rigol_XXZ}, simple operators can still be dense in the energy basis, yet not respecting the ETH. These facts imply that the Krylov dimension of a simple operator in a strongly-interacting integrable system may still be exponentially large in system size, \textit{a priori} not preventing K-complexity from reaching values of order $\mathit{e^{2S}}$ at late times just like chaotic systems. However, it is the specific features of the \textit{dynamics} in Krylov space what makes them differ from chaotic systems and feature quantitatively smaller late-time K-complexity saturation values, as we are going to review here.

\begin{figure}
     \centering
     \includegraphics[width=0.9\linewidth]{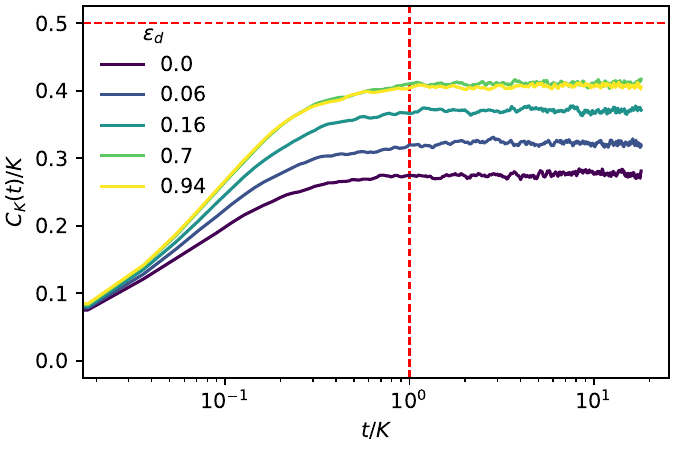}
     \caption{Krylov complexity for the operator $\mathcal{O}=\sigma^z_5+\sigma_7^z$ evolving under the Hamiltonian $H_{\textrm{xxz}}+\varepsilon_d\sigma_6^z$ of $N=11$ spins in a particular symmetry sector with an exponentially large Hilbert space dimension $D$, of order $D\sim e^N/N$. The Krylov space dimension is $K=D^2-D+1=55461$. The saturation value of K-complexity increases with the growth of the integrability-breaking parameter, $\varepsilon_d$. Note that the $H_{XXZ}$ part of $H$ is defined through a dimensionful parameter $J$ which sets the energy units (taken to be $J\equiv1$ in the numerics), and some dimensionless anisotropy parameter that is not relevant to this discussion. Reported by \citet{III}.}
     \label{fig:KC_Hinterpol}
 \end{figure}

As reviewed in section \ref{subsect:Krylov_Chain_Hopping}, dynamics in Krylov space are described by a one-dimensional hopping model with constant (zero) potential and hopping amplitudes given by the Lanczos coefficients. This directly maps Krylov space dynamics to an Anderson localization model, where it is known that disorder is responsible for (exponential) localization of the wave function. In particular, the original model by \citet{Anderson_AbsDiff} involves disorder on the diagonal (which can be interpreted as on-site potentials) and no disorder on the off-diagonal hopping amplitudes, the description of an operator's wave function spreading on the Krylov chain is mapped to an Anderson-like problem with disorder in the hopping amplitudes. Localization due to off-diagonal disorder was studied in works such as \citet{Fleishman_1977, PhysRevB.24.5698}. Based on this, \citet{II} proposed that the statistical properties of the Lanczos coefficients of typical operators in strongly-interacting integrable systems are such that they are effectively more disordered than their chaotic counterparts, yielding a localization effect \textit{on the Krylov chain} that hinders the Krylov wave function propagation and results in a smaller late-time K-complexity saturation value. Simultaneously, \citet{Trigueros:2021rwj} reported a similar localization effect in Krylov space observed for Hamiltonians with disordered couplings. The difference between the two works is that \citet{II} studied integrable systems \textit{without} disorder in their defining Hamiltonian, and proposed that the amount of disorder in the obtained Lanczos sequences is related to their integrable nature. In particular, they introduced the variance $\sigma^2=\textrm{Var}(x_i)$ of $x_i=\ln|b_{2i-1}/b_{2i}|$ as an indicator of the amount of disorder in the Lanczos coefficients. This choice was inspired by the work by \citet{Fleishman_1977}, where this quantity was observed to control the localization length of the center-of-the-band eigenstate, via $l_{ \rm loc}\propto \sqrt{K}/\sigma$, where we recall that the Krylov dimension $K$ is equal to the length of the so-called Krylov chain where the localization phenomenon operates. In fact, it is well known \cite{Kramer_1993,Thouless_1972,Fleishman_1977,IZRAILEV2012125,PhysRevB.72.174207} in the Anderson model that, provided that the disordered hopping amplitudes and/or potential energy terms are sufficiently uncorrelated, \textit{any} amount of disorder is just enough to make eigenstates localize exponentially, and the disorder strength, no matter how big or how small it is, controls the value of the localization length. From this point of view, \citet{II,III} noted that the Krylov localization effect operates, to some extent, both in chaotic and in integrable systems, as in both cases the dynamics in Krylov space are described by a one-dimensional hopping model. In particular, \citet{III} studied a local operator in a spin chain described by an XXZ Hamiltonian \cite{Heisenberg:1928mqa} supplemented with an integrability-breaking defect that allowed to interpolate between integrable and chaotic regimes \cite{Santos_2004,PhysRevE.84.016206,PhysRevB.80.125118,PhysRevB.98.235128}: They observed that approaching the integrable regime implied an increase in the disorder of the Lanczos coefficients $\sigma$ and a consistent reduction of the late-time saturation value of Krylov complexity, cf. Figure \ref{fig:KC_Hinterpol}. Specifically, \citet{III} observed that the long-time average of the Krylov wave function as defined in \eqref{eq:Wavefn_long_time} is more uniform in the chaotic regime of the model as compared to the integrable regime, yielding a K-complexity saturation value \eqref{KC_late-time_average} closer to $\sim K/2$ the stronger the integrability defect is. In the integrable regime the Krylov wave packet features an exponentially localized profile at late times that pushes down quantitatively the complexity saturation value.

Similarly, \citet{Menzler:2024ifs} studied $\sigma^2$ for models with an ergodicity-breaking parameter, for various operators, and found a universal increase in $\sigma^2$ with the ergodicity-breaking parameter.
It is worth to stress that the behavior described so far is not expected to apply to all operators in the theory, but only to the ones that can be considered as representatives of a \textit{typical}, or \textit{simple} operator in the model.
For example, the `flat' operator $\mathcal{O}=\sum_{a,b}\ket{E_a}\bra{E_b}$ will always result in $\overline{C_K}\approx K/2$ \cite{III} without depending on the details of the Lanczos coefficients, but this is a very fine-tuned operator and no local operator in a spin chain is expected to have such a profile in the energy basis \cite{Rigol_XXZ}. Other studies have also shown that the initial operator/state may affect the late-time saturation value of K-complexity, see e.g. \citet{Espanol:2022cqr}, which consider operators that probe global properties of the spin chain. These issues may be alleviated by the multiseed formalism \cite{Craps:2024suj} and its associated multiseed K-complexity, \eqref{eq:KC_mult} which averages over a set of initial operators comprised of all simple operators. This formalism consistently finds lower late-time complexity saturation values in integrable systems as compared to chaotic systems.

These findings have highlighted the relevance and usefulness of finite-size effects visible at (exponentially) late time scales of K-complexity, which are sensitive to small energy differences and spectral correlations as discussed in section \ref{subsubsect:Lanczos_Spectral_Probes}, in order to probe chaotic dynamics. This is in resonance with studies of late-time physics of quantum chaos performed for other observables, such as the spectral form factor or the operator two-point function \cite{Cotler:2016fpe,Garcia-Garcia:2016mno,Altland:2020ccq,Altland:2021rqn}. There exist numerous studies in the K-complexity literature along these lines, such as e.g. \cite{Baggioli:2024wbz,Alishahiha:2024vbf,Huh:2024lcm,Scialchi:2023bmw,Menzler:2024ifs,Sahu:2024kho,Kannan:2025xpt,Ganguli:2024myj}, both in the operator and in the states formalism, which confirm the general picture where K-complexity of simple operators in integrable systems saturates at late times at values that are smaller than chaotic counterparts in the same universality class and with comparable system size. We will elaborate further on the latter formalism in the next subsection.

\subsection{State formalism and the TFD} 
\label{Sec:TFD_state}

The thermofield double (TFD) state is a particularly interesting state in holographic setups: it is the boundary state dual to an eternal black hole in AdS \cite{Maldacena:2001kr}. The Krylov state complexity, aka spread complexity \cite{Balasubramanian:2022tpr}, of the suitably-defined time evolution of this state is directly related to the spectral form factor (SFF) and is sensitive to spectral statistics at exponentially late times. This study was initiated by \citet{Balasubramanian:2022tpr}, who analyzed the K-complexity of the TFD state at inverse-temperature $\beta$, defined as
\begin{eqnarray}\label{eq:TFD_def}
    \ket{\psi_\beta} := \frac{1}{\sqrt{Z_\beta}}\sum_{n=1}^D e^{-\frac{\beta E_n}{2} }\ket{E_n,E_n}
\end{eqnarray}
where $E_n$ (resp. $|E_n\rangle$) are the eigen-energies (resp. eigenstates) of the system's Hamiltonian $H$ and $Z_\beta:=\sum_n e^{-\beta E_n}$ is the associated partition function. The state \eqref{eq:TFD_def} belongs to the doubled Hilbert space $\mathcal{H}\otimes\mathcal{H}$ that describes two identical copies of the same theory\footnote{Strictly speaking, the two theories should be the related by an anti-unitary operator \cite{Maldacena:2001kr}, such as CPT. Making an abuse of notation we shall omit this operator and indicate the direction of time evolution in each theory by explicitly writing the corresponding sign in the exponent of the time-evolution operator.}, where $\dim(\mathcal{H})=D$. The state $|E_n,E_m\rangle$ denotes $|E_n\rangle\otimes|E_m\rangle$, and we shall denote operators inserted in the left or right copy of the Hilbert space $\mathcal{H}$ by using the subscripts L or R, respectively; for instance, $H_L:=H\otimes \mathbb{1}$, and similarly for $H_R$. We shall consider the time-evolution of this state generated by a single-sided Hamiltonian, e.g. $H_L$ \cite{Balasubramanian:2022tpr}, since any two-sided Hamiltonian $aH_L+bH_R$ with $a+b=1$ will generate the same time evolution of the unperturbed state \eqref{eq:TFD_def} as $H_L$ due to its diagonal structure\footnote{In particular, note that the state $|\psi_\beta\rangle$ is invariant under evolution generated by $H_R-H_L$.}. In particular, $\ket{\psi_{\beta}(t)}:=e^{-iH_Lt}\ket{\psi_\beta}=\ket{\psi_{\beta+2it}}$. The associated survival amplitude, $\braket{\psi_{\beta+2it}|\psi_{\beta}}=Z_{\beta-it}/Z_\beta$, where $Z_{\beta-it}$ is the analytically-continued partition function, is related to the finite-temperature SFF given by $\mathrm{SFF}_\beta(t):=|Z_{\beta-it}/Z_\beta|^2$ which can be interpreted as the survival probability $|\braket{\psi_{\beta+2it}|\psi_{\beta}}|^2$ \cite{delCampo:2017bzr}. The resulting Krylov space dimension is bound by the dimension of a single Hilbert space, i.e. $K\leq D$, since the complete time evolution of the TFD is contained within the span of $\{\ket{E_n,E_n}\}_{n=1}^D$, cf. section \ref{sect:State_formalism_framework}. For systems with no exact degeneracies in their spectrum, such as chaotic systems, this upper bound is saturated.

\begin{figure*}
    \centering
    \includegraphics[width=1\linewidth]{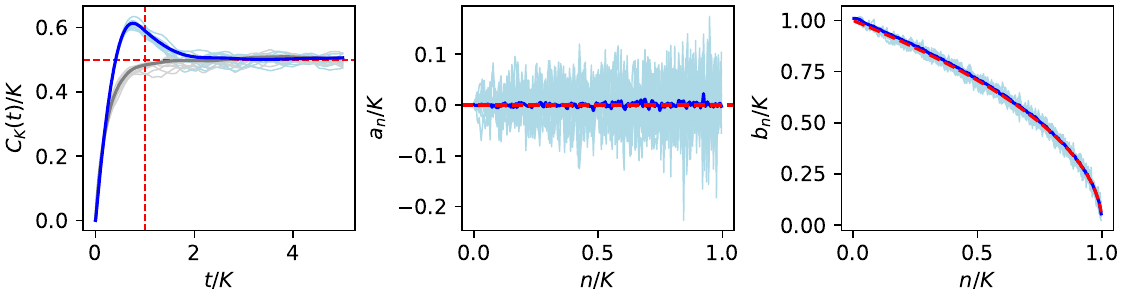}
    \caption{Krylov complexity (left) and Lanczos coefficients (middle and right) for the infinite-temperature TFD state in the GUE. The blue lines show an average over 100 realizations for the TFD state evolving under a Hamiltonian taken from the GUE of dimension $D=250$; light blue lines show results for 10 individual realizations. The Krylov space dimension is $K=D$. Krylov complexity (left), saturates at value $K/2$ at times of $O(K)$ and exhibits a \textit{peak} and \textit{slope} before saturation. For comparison, we also plotted K-complexity for random matrices without eigenvalue correlations (light grey lines for 10 individual realizations; dark grey line for an average over 50 realizations), these indeed do not feature the peak and slope. The $a_n$ Lanczos coefficients (middle) fluctuate around zero with an average value close to zero; and the $b_n$ Lanczos coefficients (right) exhibit a \textit{descent} with average profile given by $\sqrt{1-n/K}$ as predicted in \eqref{eq:bn_RMT_asympt} up to the choice of normalization (red dashed line). Each Hamiltonian was normalized by the value of $b_1$ for that realization.}
    \label{fig:Krylov_TFD_GUE}
\end{figure*}

\citet{Balasubramanian:2022tpr} analyzed the dynamics of Krylov state complexity analytically in various symmetry-dominated systems, in the spirit of previous analyses in the operator formalism reviewed in section \ref{subsubsect:coherent_states}; we refer the Reader to their paper for details. Furthermore, as a benchmark of typical quantum-chaotic behavior at late times, they studied numerically the general behavior of the Lanczos coefficients and the resulting K-complexity of the TFD
for the case in which the Hamiltonian $H$ discussed above is a $D\times D$ random matrix.
For reference, let us describe the results they obtained when the random matrix is drawn from the Gaussian Unitary Ensemble (GUE): Both\footnote{ We remind the Reader that that, in the states formalism, the Lanczos algorithm yields non necessarily vanishing diagonal Lanczos coefficients, $a_n$, cf. section \ref{sect:State_formalism_framework}.} the $a_n$ Lanczos coefficients and the $b_n$ ones exhibit a linear growth up to $n\sim O(\beta)\ll D$. The $a_n$ coefficients then plateau to $a_n\approx 0$ up to $n=D-1$, and the $b_n$ coefficients exhibit a gradual descent down to zero at $n=D-1$ with average slope of $O(1/D)$, akin to the one described for operators in section \ref{subsect:Operator_KC_chaotic}. These regimes imply an initial quadratic growth of K-complexity up to times of $O(D^0)$, followed by a linear growth and finally a transition to a late-time complexity plateau at time scales of $O(D)$, with a value of $O(D)$.
The K-complexity time scales and regimes for the TFD reported by \citet{Balasubramanian:2022tpr} differ from the operator case reviewed in subsection \ref{subsect:Operator_KC_chaotic} differ at early times, where the initial scrambling growth is replaced, in the operator case, by an initial quadratic growth up to a system-size-independent time scale which is potentially sensitive to the temperature \cite{Balasubramanian:2022tpr}; but linear growth up to $t\sim K$ (note that $K=D$ in this case) and subsequent complexity saturation are also a feature of state complexity.
Nevertheless, \citet{Balasubramanian:2022tpr} reported an additional feature which they conjectured to be related to the spectral rigidity of random matrix theory: At the transition between the linear complexity growth and the saturation regime, they found a \textit{peak} at times of $O(D)$ followed by a \textit{slope} down to the plateau. As noted by the authors, this feature is absent from K-complexity for uncorrelated spectra, see Figure \ref{fig:Krylov_TFD_GUE} (left) where K-complexity was computed for the infinite-temperature TFD state evolving under Hamiltonians with eigenvalues drawn randomly and independently from a normal distribution.
For completeness, we refer the reader to the article by \citet{Balasubramanian:2022tpr} for numerical analyses on the other Gaussian universality classes, namely the Gaussian Orthogonal and Symplectic Ensembles (GOE and GSE, respectively), where the sharpness or smoothness of the complexity peak is found to be dependent on the universality class, morally similarly to the dip of the spectral form factor \cite{Cotler:2016fpe,HaakeBook} albeit at a very different time scale (namely, Heisenberg time vs Thouless time). 

Let us now review the treatment of the K-complexity of the TFD by \citet{Erdmenger:2023wjg}. While the previous treatment used the finite-temperature TFD as the seed state for the Lanczos recursion, resulting in temperature-dependent Lanczos coefficients and Krylov basis elements, this treatment delegates the temperature-dependence to the imaginary component of the time evolution parameter, fixing the initial condition to be the infinite-temperature TFD (i.e. the maximally entangled state), which we may denote here as $|\Omega\rangle:=|\psi_0\rangle$, thanks to the relation $|\psi_\beta\rangle=e^{-\beta H/2}|\Omega\rangle$. With this approach, only a single set of Lanczos coefficients and Krylov basis elements, namely the ones output by the Lanczos algorithm applied to the Hamiltonian $H_L$ and the seed state $|\Omega\rangle$, can be used for all temperatures. In particular, the TFD at finite temperature $1/\beta$ evolved to Lorentzian time $t$ can be written as $|\psi(\tau)\rangle = e^{-\tau H_L}|\Omega\rangle$, with $\tau=\beta/2+it$. Denoting the temperature-independent Krylov basis elements as $|\psi_n\rangle = P_n(H_L)|\Omega\rangle$, where $P_n(E)$ are the orthogonal polynomials defined in equation \eqref{eq:Orthog_polyns_def}, albeit in the states formalism, we can write a non-normalized version of the Krylov wave function \eqref{eq:states_Krylov_wavefn} as:
\begin{equation}
    \label{eq:TFD_wavefn_Erdmenger_non-normalized}
    \widetilde{\psi}_n(t;\beta)=\braket{\psi_n|e^{-\tau H}|\Omega}=(1/D)\mathrm{Tr}[P_n(H_L)e^{-\tau H_L}]~,
\end{equation}
where the tilde denotes that the wave function is not normalized.
At fixed $\beta$, the norm of the state $|\psi(\beta/2+it)\rangle$ is constant in $t$ and equal to $\sqrt{\frac{Z(\beta)}{D}}$, as one can verify e.g. using \eqref{eq:TFD_wavefn_Erdmenger_non-normalized} and the orthogonality property of the Krylov polynomials written in \eqref{eq:Polynimials_orthog}. Renormalizing the Krylov wave function as $\psi_n(t,\beta) = \sqrt{\frac{D}{Z(\beta)}}\widetilde{\psi}(t,\beta)$, one may now compute thermal K-complexity as \cite{Erdmenger:2023wjg}:
\begin{eqnarray}
    C_K(t;\beta) = \frac{\sum_{n=0}^{K-1}n\Big|\mathrm{Tr}[P_n(H_L)e^{-(\beta/2+it) H_L}] \Big|^2}{D\, \mathrm{Tr}(e^{-\beta H_L})},
\end{eqnarray}
where the dependence on the spectrum of the Hamiltonian is manifest. This definition of thermal K-complexity, which uses the maximally entangled state $|\Omega\rangle$ as the seed regardless of $\beta$, has the feature of taking a $\beta$-dependent value at $t=0$. We may return to this point in section \ref{subsec.BulkComplexity}.

To summarize, the time-dependent profile of K-complexity for the TFD exhibits a chaos-sensitive peak before saturation, as discovered by \citet{Balasubramanian:2022tpr}. This was observed for RMT ensembles as well as chaotic systems such as the SYK model; see Figure \ref{fig:Krylov_TFD_GUE} for a summary of results in the GUE. 

\subsubsection*{Lanczos coefficients in RMT}
Let us now provide some further details on the Lanczos coefficients in random matrix theory (RMT), given its importance in holography and as a benchmark of quantum chaos.
Random matrices play a crucial role in understanding and defining the notion of quantum chaos, see for example \cite{Brody:1981cx, HaakeBook,stoeckmann_1999,MehtaBook,dyson1962statistical}. In particular, properties such as spectral rigidity in quantum chaotic systems can be modeled by random matrix ensembles where correlations between eigenvalues are responsible for an effect of level repulsion between them \cite{wigner1957statistical}. For closed quantum systems, the Hamiltonian's spectral statistical fluctuations are conjectured to fall into one of the universality classes of unitary RMT \cite{BGSpaper}. In contrast, the spectrum of integrable systems is uncorrelated and its level spacing statistics are typically Poissonian \cite{BerryTabor}. 

\citet{Edelman_RMT} found a tridiagonal representation for the Gaussian random matrix ensembles, namely the GOE, GUE and GSE, which we review next. Given a random $D\times D$ Hamiltonian, we recall that the joint distribution of eigenvalues of these RMT ensembles is given by
\begin{eqnarray}\label{eq:DOS_RMT}
    \rho(\{\lambda_i\}_{i=1}^D)\propto \prod_{i\neq j} |\lambda_i-\lambda_j|^\nu e^{-\frac{1}{2}\sum_i \lambda_i}~,
\end{eqnarray}
where the Dyson index $\nu=1,2$ and $4$ corresponds to the GOE, GUE and GSE, respectively. We use equation \eqref{eq:DOS_RMT} to clarify the normalization convention for the random matrix ensembles that is used in this discussion.

Using Householder reflections \citet{Edelman_RMT} were able to transform dense random matrices taken from the above-mentioned ensembles to the form
\begin{eqnarray}  \label{eq:tridiag_RMT}
\frac{1}{\sqrt{2}}
    \begin{pmatrix}
        \mathcal{N}(0,2) & \chi_{(D-1)\nu}\\
        \chi_{(D-1)\nu} & \mathcal{N}(0,2) & \chi_{(D-2)\nu} \\
        & \ddots & \ddots & \ddots \\
         & & \chi_{2\nu} & \mathcal{N}(0,2) & \chi_\nu\\
        & & &\chi_{\nu} & \mathcal{N}(0,2)
    \end{pmatrix} \nonumber \\
\end{eqnarray}
where $\mathcal{N}(\mu,\sigma^2)$ denotes the normal distribution with mean $\mu$ and variance $\sigma^2$, and $\chi_{k}$ is the chi distribution. Note that the pdf of $\frac{1}{\sqrt{2}}\chi_k$ is given by $\frac{2}{\Gamma(k/2)}x^{k-1}e^{-x^2}$. As we discussed in section \ref{subsect:numerical_implementations}, the Housholder method is equivalent to performing the Lanczos algorithm with initial vector $v\overset{*}{=}(1,0,0,\dots,0)$.
The ensemble-averaged  Lanczos coefficients that this construction yields are, \cite{Balasubramanian:2022dnj, Erdmenger:2023wjg}: 
\begin{eqnarray}
    \braket{a_n} &=& 0 \label{eq:RMT_Average-an}\\
    \braket{b_n} &=& \frac{\Gamma\left[\frac{\nu}{2}(D-n)+\frac{1}{2}\right]}{\Gamma\big[\frac{\nu}{2}(D-n)\big]}, \quad n=1,\dots, D-1 . \label{eq:RMT_averaged_bn}
\end{eqnarray}
In the large $D,n$ limit, with fixed $n/D$, the last expression reduces to:
\begin{eqnarray} \label{eq:bn_RMT_asympt}
    \braket{b_n} \approx \sqrt{\frac{\nu D}{2}}\sqrt{1-\frac{2n}{\nu D}}.
\end{eqnarray}
These GUE averaged Lanczos coefficients were already written by \citet{Kar:2021nbm}, cf. \citet{Menon_Notes_RMT}.
In GOE (resp. GUE, GSE), expressions \eqref{eq:RMT_Average-an}-\eqref{eq:RMT_averaged_bn} may be extrapolated to seed states given by orthogonal (resp. unitary, symplectic) rotations of the initial state $v\overset{*}{=}(1,0,\dots,0)^T$, because of the invariance of the ensemble under such rotations. With this hindsight, equations \eqref{eq:RMT_Average-an}-\eqref{eq:RMT_averaged_bn} may be used as a probe expression of the general behavior of the Lanczos coefficients in chaotic systems of the corresponding universality class. There is, however, a caveat: The seed vector $v$ whose Lanczos coefficients have been computed is uncorrelated with the eigenvectors of the random matrix, which is not the case in our previous discussions, where the Lanczos was seeded by the TFD state. Nevertheless, Figure \ref{fig:Krylov_TFD_GUE} shows good agreement between numerical results for the TFD state and the Ansätze \eqref{eq:RMT_Average-an}-\eqref{eq:RMT_averaged_bn} in GUE. 

The results gathered in this section have led to numerous studies on the chaotic properties of systems through the Lanczos coefficients of the thermofield double state \cite{Balasubramanian:2023kwd}. For instance, \citet{Bhattacharjee:2024yxj} analyzed the Lanczos coefficients in the \textit{Rosenzweig-Porter} random matrix model. This model interpolates between ergodic-fractal-localized phases. They provide an Ansatz for the Lanczos coefficients that interpolates between these phases. In the localized phase (where the eigenvalues are essentially uncorrelated) the $b_n$ Lanczos coefficients follow the inverse of a shifted binomial function \cite{Zuker:1999ik}. Related applications of the states formalism of Krylov complexity to the thermofield double state in the context of quantum-chaotic or integrable systems devoting attention to, among other issues, the relation to the SFF and the role of the state complexity \textit{peak}, include \cite{Jha:2024nbl,Bhattacharya:2024hto,Chakrabarti:2025hsb,Ganguli:2024myj,Tan:2024kqb,Camargo:2024rrj,Muck:2024fpb,Basu:2024tgg,Basu:2025mmm,Caputa:2022eye,Caputa:2022yju,Afrasiar:2022efk,Beetar:2023mfn,Baggioli:2024wbz}.

\subsection{Speed limits and complexity algebra}\label{subsect:SpeedLimits_ComplexityAlgebra}
\citet{Hornedal:2022pkc} derived a dispersion bound on operator\footnote{More specifically, thy assumed vanishing diagonal Lanczos coefficients, i.e. $a_n=0$.} Krylov complexity. Based on the Robertson uncertainty relation, they showed that the growth of Krylov complexity is bounded by a constant times the dispersion of the complexity operator $\hat{n}$:
\begin{eqnarray} \label{eq:Disp_bound}
    |\dot{C}_K(t)| \leq 2b_1 \Delta \hat{n}(t)
\end{eqnarray}
where $\Delta\hat{n}:=\sqrt{\braket{\hat{n}^2}-\braket{\hat{n}}^2}$, $\hat{n}$ being shorthand notation for the K-complexity operator \eqref{eq:KC_nhat} and schematically $\braket{\cdot}\equiv \big(\mathcal{O}(t)|\cdot|\mathcal{O}(t)\big)$. 
They showed that the dispersion bound \eqref{eq:Disp_bound} is saturated if and only if the complexity algebra defined and discussed below, is closed.

\citet{Caputa:2021sib} pointed out that the Krylov problem simplifies if one can show that a certain algebra is satisfied. Following the discussion in section \ref{subsubsect:coherent_states}, we introduce the generalized ladder operators, $\mathcal{L}_+|\mathcal{O}_n):= b_{n+1}|\mathcal{O}_{n+1})$ and $\mathcal{L}_-|\mathcal{O}_n):= b_{n}|\mathcal{O}_{n-1})$, with which the Liouvillian takes the form $\mathcal{L} =L_+ + L_-$, along with two other operators 
$\mathcal{B} := L_+ - L_-$ and $\tilde{\mathcal{K}} :=[\mathcal{L},\mathcal{B}]$. The so-called \textit{simplicity hypothesis} is the assumption that the (super-)operators defined above satisfy the following \textit{complexity algebra}:
\begin{equation}
\label{eq:complexity_algebra}
    [\mathcal{L},\mathcal{B}]=\tilde{\mathcal{K}},\quad [\tilde{\mathcal{K}},\mathcal{L}] = \alpha \mathcal{B}, \quad [\tilde{\mathcal{K}}, \mathcal{B}] = \alpha \mathcal{L}~,
\end{equation}
for some constant\footnote{Note that the $\alpha$ in this discussion is not the same as that in \eqref{linear_LCs}.} $\alpha$. As discussed in section \ref{subsubsect:coherent_states}, whether \eqref{eq:complexity_algebra} is satisfied or not depends on the Lanczos coefficients. In fact, \citet{Hornedal:2022pkc} proved that \eqref{eq:complexity_algebra} is satisfied \textit{if and only if} $\tilde{\mathcal{K}}=\alpha \hat{n}+\gamma$ \textit{and}
\begin{eqnarray}\label{eq:Lanczos_Algebra}
    b_n = \sqrt{\frac{1}{4}\alpha\, n(n-1)+\frac{1}{2}\gamma\, n}
\end{eqnarray}
for some $\alpha\in\mathbb{R}$ and $\gamma>0$. Furthermore, as we announced earlier, \citet{Hornedal:2022pkc} proved that the dispersion bound \eqref{eq:Disp_bound} is saturated \textit{if and only if} \eqref{eq:complexity_algebra}, which, as we have mentioned, is equivalent to \eqref{eq:Lanczos_Algebra}. Interestingly, this optimal Lanczos sequence had already appeared in the article by \citet{Parker:2018yvk} as an analytically tractable example\footnote{In fact, the analytical tractability of this Lanczos sequence is related to the fact that it furnishes a suitable representation of the Lie algebra $su(1,1)$, cf. \citet{Hodges:2007,Sukumar:2007,Hetyei_2009,viennot1983combinatorial}.} of Lanczos coefficients that grow asymptotically linearly in $n$ and yield exponential K-complexity growth in time.  

By construction, the initial operator is the lowest-weight state of the complexity algebra, $\mathcal{L}_-|\mathcal{O}_0)=0$ and, as reviewed in section \ref{subsubsect:coherent_states}, it consequently evolves in time as a generalized coherent state according to the generalized displacement operator
\begin{eqnarray}
    D(\xi) = e^{\xi \mathcal{L}_+ - \bar{\xi}\mathcal{L}_-}
\end{eqnarray}
with $\xi=it$. Combining the results by \citet{Caputa:2021sib} and \citet{Hornedal:2022pkc}, we can now classify the different K-complexity dynamics resulting from this complexity algebra:
\begin{itemize}
    \item $\alpha<0$ corresponds to a finite Krylov space dimension $K\geq 1$. \citet{Hornedal:2022pkc} showed that $\alpha<0$ necessarily requires $\alpha=-2\gamma/(K-1)$. In this case, K-complexity oscillates as $C_K(t) = (K-1) \sin^2(\omega t)$, with $\omega=\sqrt{\frac{\gamma}{2(K-1)}}$, and the Lanczos coefficients \eqref{eq:Lanczos_Algebra} give an irreducible representation of the $su(2)$ algebra with total angular momentum\footnote{Note that the identification $j=(K-1)/2$ is consistent with the equality of the Krylov space dimension and the dimension of an irreducible representation of $su(2)$, i.e. $2j+1=K$.} $j=\frac{K-1}{2}\in\frac{1}{2}\mathbb{N}_0$, for which \citet{Caputa:2021sib} find that the curvature associated to the Fubini-Study metric on the complex projective manifold through which the generalized coherent state propagates (cf. section \ref{subsubsect:coherent_states}) is positive and given by $R=\frac{4}{j}=\frac{8}{K-1}>0$.
    \item $\alpha=0$ corresponds to an infinite Krylov space, and the Lanczos coefficients $b_n = \sqrt{\frac{\gamma n}{2}}$ furnish the Heisenberg-Weyl algebra (whose irreducible representation is unique), where K-complexity grows quadratically as $C_K(t)=\gamma t^2/2$. In this case, the information metric is flat, i.e. $R=0$.
    \item Finally, $\alpha>0$ corresponds to an infinite Krylov space whose Lanczos coefficients \eqref{eq:Lanczos_Algebra} describe an irreducible representation of the $sl(2,\mathbb{C})$ algebra with dimension $h=\gamma$, for which K-complexity grows as $C_K(t)=(2\gamma/\alpha)\sinh^2(\sqrt{\alpha}t/2)$. In particular, note that these Lanczos coefficients grow asymptotically linearly with $n$, saturating the bound by \citet{Parker:2018yvk} and yielding a complexity profile that grows asymptotically exponentially in time. The Fubini-Study metric in this case is hyperbolic, with negative curvature $R=-\frac{4}{h}=-\frac{4}{\gamma}<0$.
\end{itemize}

In addition to this discussion, \citet{Hornedal:2023xpa} studied the complexity super-operator $\hat{n}$ in the the \textit{super-Heisenberg} picture, in which the initial super-state $|\mathcal{O})$ is kept fixed, while the complexity super-operator evolves as
\begin{eqnarray}
    \hat{n}(t)=e^{-i\mathcal{L}t}\hat{n}(0) e^{i\mathcal{L}t}.
\end{eqnarray}
\citet{Hornedal:2023xpa} showed that in the case of closed complexity algebra, where the dispersion bound \eqref{eq:Disp_bound} is saturated, the evolution of $\hat{n}(t)$ is contained in a 3-dimensional space $\hat{n}(t)\in \textrm{span}\{\mathbb{1}, \hat{n}, \mathcal{B}\}$ which may be understood as the \textit{super-Krylov space} for the K-complexity super-operator itself. This manifestly simplifies the K-complexity dynamics for the cases of closed complexity algebras \eqref{eq:complexity_algebra}, and leads further to the super-operator $\hat{n}$ saturating certain (super-)operator quantum speed limits. For details, we refer the Reader to their paper.

The works reviewed in this subsection provide a deep justification for the interest in complexity algebras and symmetry-driven systems, not only because they are benchmarks of extremal dynamics in the senses discussed, but because they may provide accurate early-time descriptions of physical systems of interest, such as the SYK model \cite{Parker:2018yvk,Caputa:2021sib,III}. For related lines of research, see e.g. \citet{Haque:2022ncl, Adhikari:2023evu, Caputa:2024xkp, Chowdhury:2024qaj, Fu:2024fdm, Zhai:2024tkz, Das:2024tnw,Gill:2024acg,Caputa:2022eye,Caputa:2022yju,Bhattacharjee:2022qjw, Nandy:2023brt, Hu:2025zvv}.

\subsection{Krylov complexity in field theories}\label{subsect:KC_QFT}
The results and phenomenology that have been reviewed so far in this Section characterize Krylov complexity as a very rich probe of quantum chaos which displays, in the case of chaotic systems, the expected profile of a holographic complexity, featuring the relevant time scales of black hole physics. However, these results have not been exempt of debate: They have mostly been derived in instances of quantum-mechanical many-body systems, and the study of higher-dimensional instances compulsorily needs to pass by the computation of Krylov complexity in the context of quantum field theory (QFT). This challenging topic, which has been explored by authors like \citet{Dymarsky:2021bjq}, \citet{Avdoshkin:2022xuw}, \citet{Camargo:2022rnt} and \citet{Kundu:2023hbk}, is the subject of this subsection.

The first problem to address in this context is, as already discussed in section \ref{subsubsect:Toda}, the fact that in quantum field theory there is no obvious way to apply the Lanczos algorithm due to the more intricate nature of its Hilbert space and Hamiltonian. The aforementioned works circumvent this issue by avoiding an explicit construction of the Krylov basis elements and directly evaluating the Lanczos coefficients out of the moments of the operator two-point function, either via numerical implementation of the recursion method or by using the Toda method. 

Another subtlety that one encounters in the study of operator Krylov complexity in quantum field theory is the choice of the operator inner product: Generally, position-space correlation functions in QFT suffer from contact singularities which need to be regulated away in order to have an auto-correlation function $C(t)$ that is smooth at $t=0$, thus admitting a Taylor expansion at that point with well-defined moments $\mu_{2n}$. Often in field theory contexts the preferred regulator is the Wightman correlator, namely:
\begin{equation}
    \label{eq:Wightman_correlator}
    C_\beta^{W}(t):=\langle\mathcal{O}(t-i\beta/2)\mathcal{O}(0)\rangle_\beta~,
\end{equation}
where we suppress the dependence of the operator on the spatial coordinates\footnote{In the context of conformal field theory, one may equivalently consider timelike separated operator insertions.}. This two-point function can be obtained as the overlap $(\mathcal{O}|\mathcal{O}(t))_\beta^W$ computed through the Wightman inner product, given by \eqref{eq:InProd_beta} with $g(\lambda)=\delta(\lambda-\beta/2)$. \citet{Dymarsky:2021bjq} were the first to consider this regulated inner product in the context of the computation of Krylov complexity in conformal field theory (CFT). In particular, they noted that the Wightman thermal correlator of a scalar primary field in two-dimensional conformal field theory is fixed by symmetry to be
\begin{equation}
    \label{eq:2d_CFT_two_pt}
    C_\beta^W(t) = \frac{1}{\cosh^{2\Delta}(\pi t/\beta)}~,
\end{equation}
where $\Delta$ is the primary dimension. The main observation pointed out by \citet{Dymarsky:2021bjq} is the fact that the two-point function \eqref{eq:2d_CFT_two_pt} has poles at $t=\pm i\beta/2$ which are entirely due to the Wightmann regularization prescription applied to a thermal correlator which was itself entirely determined by symmetry. No assumptions on the chaotic nature of the theory, or on the specific value of $\Delta$, are needed to derive it. And yet, \textit{if one assumes a smooth Lanczos sequence} $b_n$, such a singularity in the two-point function already determines the asymptotics of the Lanczos coefficients out of the asymptotics of the two-point function moments $\mu_{2n}$, along the lines of the analyses given by \citet{ViswanathMuller,Parker:2018yvk,Avdoshkin:2019trj}. This asymptotic behavior is found by \citet{Dymarsky:2021bjq} to be a linear growth $b_n\sim \alpha n$, with $\alpha =\pi / \beta$, in turn implying a Krylov complexity exponent given by $\lambda_K=2\alpha = 2\pi/\beta$, thus coinciding with the MSS bound. The smoothness assumption of the Lanczos coefficients was justified by \citet{Dymarsky:2021bjq} via explicit computation of the $b_n$ sequence using the Toda method, not only for scalar primaries in two-dimensional conformal field theories but also for free scalars and fermions in $d$ dimensions, as well as holographic CFTs\footnote{In this case, the two-point function can be computed solving the wave equation in the bulk, or using the geodesic approximation if applicable.}. The authors presented this result as a counter-example to the universal operator growth hypothesis at finite temperature (where, as we discussed earlier in Section \ref{subsect:Operator_KC_chaotic}, it is a conjecture): All the instances of Lanczos coefficients computed either correspond to manifestly free theories, or have been computed with no assumption about the chaotic or integrable nature of the theory, and in all cases a maximal linear growth of the $b_n$ sequence is obtained, which is entirely determined by symmetry and/or the regularization prescription adopted in the definition of the inner product.

At the time of writing this review, a number of replies to the controversy raised by \citet{Dymarsky:2021bjq} have already appeared in the literature. As discussed in subsections \ref{subsect:Operator_KC_chaotic} and \ref{subsect:Krylov_localization}, one might hope that the eventual integrable nature of the field theories discussed above shows up at post-scrambling time scales due to finite-size effects that may be accessible if one considers a field theory framework such that the relevant Hilbert space is effectively finite. A study in this direction was performed by \citet{Avdoshkin:2022xuw} and \citet{Camargo:2022rnt}, where the following protocols were considered:
\begin{enumerate}
    \item Implementation of a UV cutoff e.g. by considering field theories on a lattice.
    \item Discretization of the Hilbert space, achieved e.g. by placing the field theory on a compact manifold. Technically speaking, at least for the free theories considered by \citet{Avdoshkin:2022xuw}, this effective discretization of the Hilbert space implies a spectral gap. A similar gap in the spectral function is identified by \citet{Camargo:2022rnt} considering massive free theories in non-compact space.
\end{enumerate}

An effectively finite Hilbert space would be achieved in a framework where both 1 and 2 above apply simultaneously. Despite not considering such a situation, separate analyses of both protocols by \citet{Avdoshkin:2022xuw} and \citet{Camargo:2022rnt} are in line with the general relation between the features of the Lanczos coefficients and the spectral function $\Phi(\omega)$ \eqref{eq:Spectral_function_def} detailed by \citet{ViswanathMuller} and summarized in appendix A of the paper by \citet{Kar:2021nbm}, namely:
\begin{itemize}
    \item A bounded spectrum induces a saturation of the Lanczos coefficients, $b_n\to b$ as $n\to\infty$.
    \item A spectral gap induces a staggering of the Lanczos coefficients as described in subsection \ref{subsect:Cont_Approx}, namely $b_{2n}$ and $b_{2n+1}$ follow distinct subsequences.
\end{itemize}

In particular, considering free theories on compact manifolds at finite temperature, \citet{Avdoshkin:2022xuw} observe a staggering of the Lanczos coefficients that differs from the symmetry-dominated linear growth described by \citet{Dymarsky:2021bjq} when analyzing \eqref{eq:2d_CFT_two_pt}. As we now understand in light of the discussions in subsection \ref{subsect:Krylov_localization}, this staggering induces a localization effect in Krylov space which makes K-complexity in these theories freeze at late times, even in the absence of a UV cutoff that would make the Hilbert space finite. In conclusion, compactification of the manifold eventually allows to distinguish integrable (free) theories from chaotic ones, as they feature a Krylov localization effect and consequently K-complexity fails to grow maximally (i.e. exponentially). We stress that this compactification allows for the conformal two point function to not be entirely dominated by symmetry.

In their work, \citet{Avdoshkin:2022xuw} also propose to compute ``holographic Lanczos sequences'' as obtained from the moments of CFT two-point functions computed for heavy primary operators using the geodesic approximation via the method of summing over images, cf. for instance \citet{Keski-Vakkuri:1998gmz}, \citet{Maldacena:2001kr}, or \citet{Alday:2020eua}. Let us focus on their results for three bulk dimensions: These thermal correlation functions are computed by summing geodesic lengths for which the bulk is taken to be thermal AdS$_3$ or the BTZ black hole depending on whether the temperature is below or above the Hawking-Page (HP) transition (cf. section \ref{sec.HoloComplexity}), respectively. Without going into the details of the computation, we shall just say that below the HP transition the Lanczos coefficients are found to feature staggering subsequences, inducing a Krylov localization effect in the K-complexity dynamics, while above the HP transition the Lanczos coefficients are smooth and grow asymptotically linearly with slope $\alpha=\pi/\beta$, implying an exponential growth of K-complexity with exponent $\lambda_K=2\pi/\beta$. In their paper, \citet{Avdoshkin:2022xuw} interpret the fact that operator Krylov complexity behaves differently (in particular, not growing exponentially) in different holographic setups as an argument against considering Krylov complexity as a holographic complexity. Contrarily, we understand their result as a very strong indication that Krylov complexity \textit{is} holographic, since it only features early exponential growth when the bulk dual contains precisely a black hole. 

In a separate work, \citet{Kundu:2023hbk} proposed an alternative approach to the computation of Lanczos coefficients and Krylov complexity in conformal field theory that is not entirely dominated by symmetry even without compactifying the manifold on which the theory is defined, and which succeeds at distinguishing chaotic from integrable dynamics and can be used to probe the holographic behavior of Krylov complexity. The main ingredient of their computation is the use of a different operator inner product as compared to the Wightman inner product used to define \eqref{eq:Wightman_correlator}: Instead, \citet{Kundu:2023hbk} proposed to define the inner product of two operators as their expectation value in a primary operator. Using the state-operator correspondence, this in turn implies that the auto-correlation function $C(t)$ of the operator becomes a vacuum four-point correlator, which is no longer determined by symmetry in the same way the vacuum two-point function is. In holographic two-dimensional conformal field theories with large central charge $c$, $C(t)$ can be computed using the Virasoro identity block \cite{Hartman:2013mia,Fitzpatrick:2014vua}, from which the Lanczos coefficients may be computed using the Toda method. Denoting the scaling dimension of the primary used to define the inner product as $\Delta$, \citet{Kundu:2023hbk} find that above the black hole threshold, i.e. $\Delta>\frac{c}{12}$, K-complexity of a light primary operator grows exponentially, while for $\Delta<\frac{c}{12}$ it is oscillatory (and, in particular, bounded). \citet{Kundu:2023hbk} compared this behavior to similar computations, with their prescribed inner product, in free massless scalar theory and the two-dimensional Ising CFT, both of which are instances of integrable theories (the former is free, while the latter is strongly interacting) which do not feature a black-hole threshold in their spectrum and for which the four-point function can be analytically computed. In both cases the authors obtain oscillatory Krylov complexities.

The behavior of Krylov complexity in quantum field theory is a promising area of study which is not yet extensively explored, although there exist several works \cite{Caputa:2021ori,Caputa:2022zsr, Malvimat:2024vhr, Chattopadhyay:2024pdj, Caputa:2025dep} that have made use of the techniques described in this subsection, or more generic manipulations of generalized coherent states (cf. section \ref{subsubsect:coherent_states}) to analyze aspects of chaotic complexity dynamics, mainly in the realm of conformal field theory.

Let us close the section by pointing out that an analysis of the Krylov complexity saturation value in quantum field theory would require, as already mentioned, the introduction of an effectively finite Hilbert space by combining the compactification of the manifold with, for instance, the restriction to a micro-canonical window of the spectrum, using the micro-canonical inner product \eqref{eq:MicroCanonical_Inner_Kar}, as suggested by \citet{Kar:2021nbm}. In two-dimensional CFT on the torus, this would result in an effectively finite Hilbert space of dimension $D(E)\sim e^{S(E)}$, where $S(E)$ can be approximated by the Cardy scaling due to modular invariance \cite{Cardy:1986ie}. Such a combined analysis is still to be performed.

\section{Krylov complexity as a probe of the gravitational bulk}\label{sect:Holography}

The quantum Hilbert space of gravitational theories is generally unknown. There is however an example in 2d gravity where such a Hilbert space can be constructed explicitly, namely the 1+1 dimensional Jackiw-Teitelboim (JT) gravity \cite{Harlow:2018tqv} at disk level. Furthermore, there exists a holographic dictionary mapping states in the Hilbert space of two-dimensional gravity and states in the so-called \textit{double-scaled SYK model} (DSSYK), cf. \citet{Lin:2022rbf}. This explicit bulk-boundary map has also given rise to the first exact bulk-boundary correspondence for any notion of complexity, realized by Krylov complexity \cite{IV}. This is the topic of the current Section. It may be interesting to note that AdS$_2$ gravity, especially JT gravity, also often arises when reducing higher-dimensional gravitational theories, see e.g. \cite{Nayak:2018qej}, and explored in a complexity context e.g. in \cite{Gautason:2020tmk}.

This Section is structured as follows: Subsection \ref{subsec.KC_in_DSSYK} will review the computation of both state and operator Krylov complexity in double-scaled SYK, providing an explicit construction of the Krylov basis elements; subsequently, subsection \ref{subsec.BulkComplexity} will review the bulk dual of the DSSYK model in terms of two dimensional gravity and will provide the explicit bulk dual of Krylov complexity in terms of bulk length by applying the bulk-boundary map between Hilbert spaces in various regimes of the theory, including the low-energy regime in which the bulk dual is the theory of JT gravity; finally, subsection \ref{sec.BulkSaturation} will give an outlook on the properties and signatures of wormhole length saturation and the possibility of probing them via Krylov complexity.
\begin{figure*}[ht]
\justifying
\begin{roundedbox}[From chords to states]
\begin{center}
\includegraphics[width=0.8\textwidth]{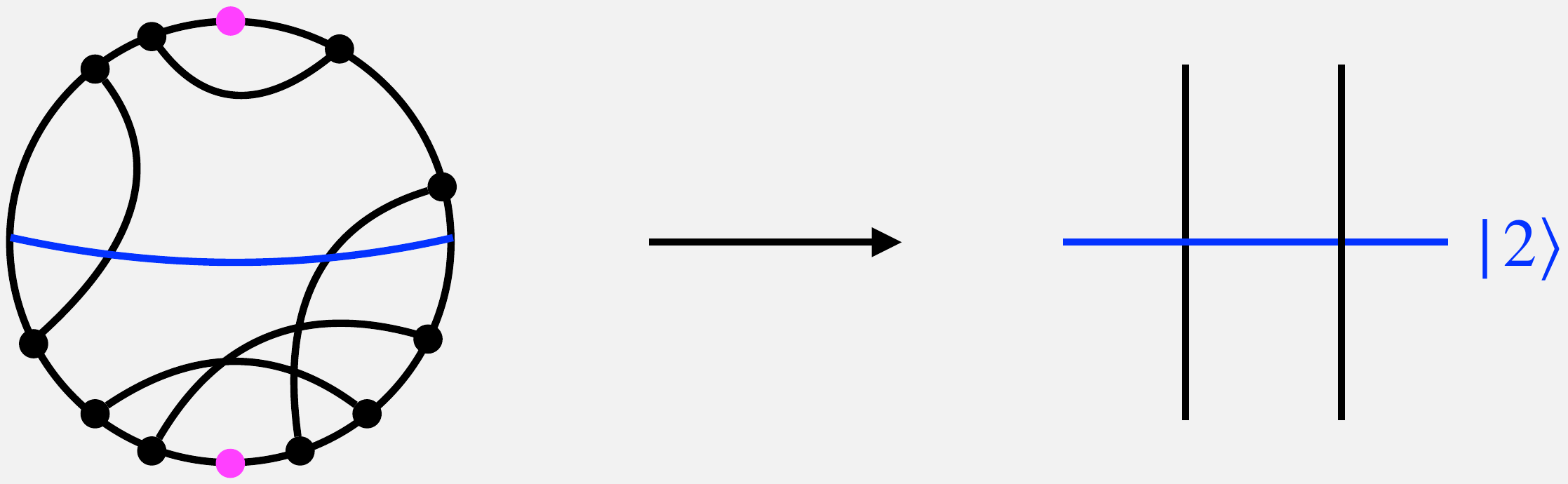}\end{center}
    \vskip1em
    A major technical advantage of working in the DSSYK model is that moments of the Hamiltonian and more general correlation functions can be evaluated using so-called chord diagrams (as reviewed in \citet{berkooz2024cordial}).
   The figure above shows an example of a chord diagram representing a specific Wick contraction contributing to $\big\langle\text{Tr}[H_{\rm SYK}^{10}]\big\rangle$. The circle represents the trace, where black dots are Hamiltonian insertions, and the black lines (chords) are Wick contractions. A slicing of the chord diagram can be defined by choosing two points on the circumference (purple dots) defining the (Euclidean) infinite past and future which split the diagram into two distinct left and right sides, such that constant time slices (of which the blue line is an example) have anchoring points on opposite sides. States defined on a time slice therefore admit a \textit{two-sided interpretation} \cite{Lin:2022rbf}, and are labeled by the number of \textit{open chords} that intersect the slice. As shown on the right of the figure, the state defined on the specific slice considered is the two-chord state. In particular, open chords crossing a time slice are defined such that they do not intersect each other in the past of the slice. As mentioned in the main text, this construction can be extrapolated to the case of chord diagrams involving chords corresponding to operator insertions. For further details on the slicing of chord diagrams and the rules for evaluating them, we refer the Reader to \citet{Lin:2022rbf}.   
\end{roundedbox}
\end{figure*}
\subsection{Krylov complexity of states and operators in DSSYK} \label{subsec.KC_in_DSSYK}

In this discussion, we shall consider the real version of the complex SYK model treated in section \ref{subsect:Operator_KC_chaotic}.
The real SYK model is a many-body system of $N$ Majorana fermions, satisfying $\{\psi_{i},\psi_j\}=2\delta_{ij}$, with $p$-body interactions:
\begin{eqnarray} \label{eq:H_SYK}
    H_{\text{SYK}}=i^{p/2}\!\!\!\! \sum_{1\leq i_1<i_2<\dots<i_p\leq i_N}\!\!\!  J_{i_1 i_2 \dots i_p}\psi_{i_1} \psi_{i_2} \dots \psi_{i_p} ~,
\end{eqnarray}
where the coupling coefficients, $J_{i_1 i_2 \dots i_p}$, are disordered, drawn from a Gaussian distribution with zero mean and a particular variance (see below). This model was studied extensively in the literature, in particular its large $N$ limit \cite{Sachdev_1993, Sachdev:2015efa, kitaev2015kitp, Garcia-Garcia:2016mno}. 

In the so-called double-scaling limit \cite{Cotler:2016fpe,Erdos_2014}, we take:
\begin{eqnarray} \label{eq:DSSYK_lim}
    N, p \to \infty, \quad \text{and} \quad \lambda = \frac{2p^2}{N}  \text{fixed},
\end{eqnarray}
where $\lambda$ plays the role of a 't Hooft coupling for the double-scaled SYK (DSSYK) model. This limit provides an analytical leverage to perform several non-trivial computations of various quantities. We will only review the objects relevant to our discussion and refer the interested Reader to \cite{Berkooz:2018qkz, Berkooz:2018jqr,  berkooz2024cordial}. In the context of DSSYK the coupling strengths are taken from a distribution that satisfies
\begin{eqnarray} \label{eq:DSSYK_stats}
    \braket{J_{i_1 i_2 \dots i_p}} = 0, \quad \braket{J^2_{i_1 i_2 \dots i_p}} = \frac{1}{\lambda} \binom{N}{p}^{-1}J^2
\end{eqnarray}
where $J$ provides the energy scale and the $\braket{\cdot}$ denotes disorder average. With \eqref{eq:DSSYK_lim} and \eqref{eq:DSSYK_stats} the disorder-averaged moments of the Hamiltonian  \eqref{eq:H_SYK}, $M_{2k}:= \braket{\mathrm{Tr}(H^{2k})}$, can be computed by summing the so-called \textit{chord diagrams}:
\begin{eqnarray} \label{eq:DSSYK_moments}
    M_{2k} = \frac{J^{2k}}{\lambda^k}\!\!\!\!\! \sum_{\substack{\text{chord diagrams}\\{\text{with $k$ Hamiltonian chords}}}} \!\!\!\!\!\!\!q^{\text{number of intersections}} \nonumber \\
\end{eqnarray}
where $q:= e^{-\lambda}$. Each chord diagram consists of a circle, representing the trace, with $2k$ points on its circumference and $k$ chords connecting them which represent a specific Wick contraction. The chord diagrammatic helps to define an \textit{effective Hilbert space} of DSSYK \cite{Berkooz:2018jqr,Lin:2022rbf}.
We shall take the two-sided Hilbert space construction spelled out by \citet{Lin:2022rbf}, where states are defined through a convenient slicing of the chord diagram and labeled by the number of \textit{open} chords that cross the slice on which a given state is defined, open chords being those whose starting and end point are respectively in the (Euclidean) past and future of the slice, respectively; the slicing can be chosen such that open chords only intersect in the future of the slice. 
This effective Hilbert space admits the interpretation of a \textit{two-sided} (yet not factorizable in the absence of matter) space \cite{Lin:2022rbf}, and it is spanned by the so-called chord basis, denoted by $\{\ket{n}\}_{n=0}^\infty$, which are states of $n$ open chords. In terms of the chord basis the effective DSSYK Hamiltonian can be written as
\begin{eqnarray}\label{eq:H_eff_DSSYK}
    H = \frac{J}{\sqrt{\lambda}} (a+a^\dagger)
\end{eqnarray}
where $a^\dagger$ and $a$ create and annihilate open chords. They act on the chord basis in the following way:
\begin{eqnarray}\label{a_ad_DSSYK}
    a^\dagger\ket{n}=\ket{n+1}, \quad a\ket{n} = [n]_q \ket{n-1}
\end{eqnarray}
where $[n]_q:= (1-q^n)/(1-q)$ is the $q$-number, cf. \cite{Arik:1976}. The inner product in the chord basis, which accounts for the chord intersections \cite{Lin:2022rbf}, is defined by
\begin{eqnarray}\label{eq:InProd_DSSYK}
    \braket{m|n} = [n]_q! \delta_{mn}, \quad \braket{0|0}=1,
\end{eqnarray}
where $[n]_q!$ is the $q$-factorial, cf. \cite{Arik:1976}. The ladder operators $a$ and $a^\dagger$ are known to satisfy a quantum deformation of the $sl(2,\mathbb{R})$ algebra \cite{Lin:2023trc}.

In the double-scaling (DS) limit, the Hamiltonian \eqref{eq:H_SYK} is described effectively by \eqref{eq:H_eff_DSSYK} in the following sense:
\begin{equation}
    \label{eq:DSSYK_effective_Ham}
   \bigg\langle \text{Tr} [f(H_{\rm SYK})]\bigg\rangle \overset{DS}{\longrightarrow} \langle 0| f(H)|0\rangle~,
\end{equation}
for any function $f$ and
where the large angular brackets on the left-hand side denote disorder average. We also see that in this limit the infinite-temperature thermofield-double state, or maximally entangled state, cf. equation \eqref{eq:TFD_def}, is identified with the zero-chord state, $\ket{0}$. Importantly, the evolution $e^{-iH_{\textrm{SYK}}t}\ket{\psi_{\beta=0}}$ is mapped to $|\psi(t)\rangle = e^{-iHt}\ket{0}$ in the effective averaged theory. This effective description of SYK yields an analytical understanding of the Krylov basis and Krylov complexity of the TFD state in SYK, as we review next.

Performing the Lanczos algorithm starting with the zero chord state, $\ket{\psi_0}=\ket{0}$, as the initial Krylov element, evolving under the effective Hamiltonian \eqref{eq:H_eff_DSSYK} with the inner product \eqref{eq:InProd_DSSYK}, produces the Kryov basis elements and Lanczos coefficients for this problem \cite{IV}:
\begin{eqnarray} \label{eq:Krylov_DSSYK}
    \ket{\psi_n} = \frac{\ket{n}}{\sqrt{\braket{n|n}}}, \quad b_n =\frac{J}{\sqrt{\lambda}}\sqrt{\frac{1-q^n}{1-q}},
\end{eqnarray}
and $a_n=0$ due to the Gaussian nature of the model.
In other words, the Krylov basis elements are equal to normalized chord number eigenstates.
The relationship between chord-number states and Krylov elements had been qualitatively suggested by \citet{Lin:2022rbf}.
The Krylov complexity operator for the seed state $|0\rangle$ is therefore equal to the chord number operator, i.e. $ \widehat{C}_K=\hat{n}$ where $\hat{n}\ket{\psi_n}=n\ket{\psi_n}$. Defining additionally the unit displacement operator on the Krylov chain as $D$ such that $D\ket{\psi_n}=\ket{\psi_{n-1}}$ and $D^\dagger\ket{\psi_n}=\ket{\psi_{n+1}}$, the effective Hamiltonian \eqref{eq:H_eff_DSSYK} becomes:
\begin{eqnarray} \label{eq:H_DSSYK_Krylov}
    H = \frac{J}{\sqrt{\lambda(1-q)}}\left(D \sqrt{1-q^{\hat{n}}}+\sqrt{1-q^{\hat{n}}}D^\dagger\right)~,
\end{eqnarray}
which is by construction the same Hamiltonian derived by \citet{Lin:2022rbf} out of chord number manipulations.
With this all, K-complexity of the infinite-temperature TFD in DSSYK is equal to total chord number \textit{as an operator} and for arbitrary $\lambda$ \cite{IV}, $\widehat{C_K}=\hat{n}$ thanks to the identification between their respective (normalized) eigenstates \eqref{eq:Krylov_DSSYK} and associated eigenvalues $n$, which in particular implies that their expectation values, through which K-complexity is defined, also agree, i.e. $C_K(t) = \braket{\psi(t)|\hat{n}|\psi(t)}=n(t)$. Concatenating this statement with the bulk-boundary map relating chord eigenstates to bulk length eigenstates implies directly the correspondence between Krylov complexity and bulk length at the quantum level, as we will review in \ref{subsec.BulkComplexity}.

Furthermore, \citet{IV} noted that the parameter $\lambda$ plays the role in this model of a small $\hbar$-like parameter in the precise sense described in section \ref{subsect:Cont_Approx}, allowing a continuous approximation of Krylov space with which K-complexity can be computed as the classical trajectory of a point particle in Krylov space with velocity profile dictated by the continuous limit of the Lanczos coefficients. For reference, the expression of the state K-complexity in this semiclassical limit of DSSYK is:
\begin{eqnarray}\label{eq:DSSYK_KC_matterless}
   \lim_{\lambda\to 0} \lambda C_K(t) = 2\log\left[\cosh\left( J\,t\right) \right].
\end{eqnarray}

The DSSYK chord diagram formalism lends itself to provide an effective description of operator evolution. Operators in DSSYK are taken to be of the following generic form:
\begin{eqnarray} \label{eq:Operator_DSSYK}
    \mathcal{O} = i^{{\tilde{p}}/2}\!\!\!\! \sum_{1\leq i_1<i_2<\dots<i_{\tilde{p}}\leq i_N}\!\!\!  O_{i_1 i_2 \dots i_{\tilde{p}}}\psi_{i_1} \psi_{i_2} \dots \psi_{i_{\tilde{p}}} ~, 
\end{eqnarray}
where $O_{i_1 i_2 \dots i_{\tilde{p}}}$ are independent random real coefficients drawn from a normal distribution with zero mean and variance given by $\braket{O_{i_1 i_2 \dots i_{\tilde{p}}}^2} =  \binom{N}{\tilde{p}}^{-1}$. 
Within the double-scaling limit \eqref{eq:DSSYK_lim}, the size of the operator $\widetilde{p}$ is taken to infinity together with $p$ and $N$, keeping $\tilde{\lambda} = 2 \tilde{p}p/N$ fixed. It is additionally useful useful to introduce the parameter $\Delta:=\frac{\widetilde{p}}{p}$, which is fixed in this limit and verifies $\widetilde{\lambda} = \Delta \lambda$. This parameter plays the role of the scaling dimension of the operator $\mathcal{O}$ in the quantum algebra of DSSYK \cite{Lin:2023trc}.

The moments of the two-point autocorrelation function at infinite temperature, defined as $C(t)=(\mathcal{O}|\mathcal{O}(t))=\mathrm{Tr}(\mathcal{O}e^{iHt}\mathcal{O}e^{-iHt})$, are given in the disorder-averaged theory by\footnote{Given that the algebra of observables of DSSYK is a type II$_1$ von Neumann algebra \cite{Xu:2024hoc}, the trace is normalized as $\text{Tr}[\mathbb{1}]=1$.}:
\begin{eqnarray} \label{eq:Op_moments_DSSYK}
    \mu_{2n} = \sum_{k=0}^{2n} (-1)^k \binom{2n}{k} \braket{\mathrm{Tr}(\mathcal{O} H^{2n-k} \mathcal{O} H^k)}~,
\end{eqnarray}
and objects of the form $ \braket{\mathrm{Tr}(\mathcal{O} H^{k_1} \mathcal{O} H^{k_2})}$ are also computed in DSSYK by summing over marked chord diagrams \cite{Berkooz:2018jqr, Berkooz:2018qkz, berkooz2024cordial}:
\begin{eqnarray}
    \braket{\mathrm{Tr}(\mathcal{O} H^{k_1} \mathcal{O} H^{k_2})} =\!\!\!\!\!\!\!\!\!\!\!\sum_{\substack{\text{chord diagrams with}\\{\text{$(k_1+k_2)/2$ $H$-chords}}\\{\text{and one $\mathcal{O}$-chord}}}} \!\!\!\!\!\!\!\!\!\!\!q^{k_{HH}} \tilde{q}^{k_{H\mathcal{O}}}
\end{eqnarray}
here, $\tilde{q}:= e^{-\tilde{\lambda}}=e^{-\Delta\lambda}=q^\Delta$~, and $k_{HH}$ counts the number of intersections between Hamiltonian ($H$) chords while $k_{H\mathcal{O}}$ counts the number of intersections between Hamiltonian chords and the single operator ($\cal O$) chord. Such marked chord diagrams give rise to a one-particle chord basis, $\{\ket{n_L,n_R}\}_{n_L,n_R=0}^\infty$, by using a slicing such that the state on a given slice is labeled by the number of open $H$ chords to the left and right of the $\cal O$ chord is $n_L$ and $n_R$, respectively\footnote{In this specific slicing, Euclidean infinite future and past are selected by the operator insertions \cite{Lin:2022rbf}.}. These states have a non-trivial inner-product, described recursively by \cite{Lin:2023trc}:
\begin{align} 
    \langle n_L', n_R' | n_L,n_R\rangle 
    =& [n_L]_q \langle n_L'-1, n_R' | n_L-1,n_R\rangle \nonumber\\&+q^{\Delta+n_L}[n_R]_q \langle n_L'-1, n_R' | n_L,n_R-1\rangle \label{eq:Op_InProd_DSSYK_line1} \\
    =& [n_R]_q \langle n_L', n_R'-1 | n_L,n_R-1\rangle \nonumber\\& +q^{\Delta+n_R}[n_L]_q \langle n_L', n_R'-1 | n_L-1,n_R\rangle. \label{eq:Op_InProd_DSSYK_line2}
\end{align}
The time-evolution of an operator of the form \eqref{eq:Operator_DSSYK} under the SYK Hamiltonian, is effectively described in DSSYK by a set of left and right Hamiltonians:
\begin{align}
    H_L &= \frac{J}{\sqrt{\lambda}}(a_L+a_L^\dagger)\\
    H_R &= \frac{J}{\sqrt{\lambda}}(a_R+a_R^\dagger),
\end{align}
where the operator $a_{L/R}^\dagger$ creates a left/right chord and $a_{L/R}$ annihilates it, in the following way:
\begin{align}
    a_L^\dagger\ket{n_L,n_R} &= \ket{n_L+1,n_R}\\ a_R^\dagger\ket{n_L,n_R} &= \ket{n_L,n_R+1}\\
    a_L\ket{n_L,n_R} &=[n_L]_q\ket{n_L-1,n_R}+q^{n_L+\Delta}[n_R]_q\ket{n_L,n_R-1}\\
    a_R\ket{n_L,n_R} &=[n_R]_q\ket{n_L,n_R-1}+q^{n_R+\Delta}[n_L]_q\ket{n_L-1,n_R},
\end{align}
Using the chord algebra discussed by \citet{Lin:2023trc} it can be shown that $[H_L,H_R]=0$. The evolution of an operator of the type \eqref{eq:Operator_DSSYK} under the SYK Hamiltonian \eqref{eq:H_SYK}, inserted on the maximally entangled state, $e^{iH_{\mathrm{SYK}}t}\mathcal{O}e^{-iH_{\mathrm{SYK}}t}\ket{\Omega}$ is effectively described in DSSYK as the evolution $e^{-it(H_R-H_L)t}\ket{0,0}$: It can be shown that the ensemble-averaged moments \eqref{eq:Op_moments_DSSYK} are equal to $\braket{0,0|(H_R-H_L)^{2n}|0,0}$. For more details on this, we refer the reader to \citet{Ambrosini:2024sre}. 

In order to have an analog statement to the equivalence between the Krylov basis elements and chord number eigenstates previously derived for states in the zero-particle (i.e. matterless) sector of the chord algebra, one now needs to solve the Lanczos algorithm in the one-particle sector that describes operator dynamics \cite{Lin:2023trc,Xu:2024gfm}.
The Krylov problem for an operator of the form \eqref{eq:Operator_DSSYK} evolving in the Schrödinger picture under the SYK Hamiltonian is translated in DSSYK into the Krylov problem of the state $\ket{\psi_0}=\ket{0,0}$ evolving under the Hamiltonian $H_R-H_L$ with inner product \eqref{eq:Op_InProd_DSSYK_line1}-\eqref{eq:Op_InProd_DSSYK_line2}. Carrying out the Lanczos algorithm reveals that the Krylov basis elements take the form:
\begin{align} \label{eq:Op_KE_DSSYK}
    \ket{\psi_n} = \frac{1}{\prod_{i=1}^{n}\bar{b}_i}\ket{\chi_n}+ \sum_{m=1}^{\lfloor n/2\rfloor}\ket{\xi_{n-2m}}
\end{align}
where for simplicity we denote $\bar{b}_n:= b_n/(J/\sqrt{\lambda})$, and $b_n$ are the Lanczos coefficients; $\ket{\chi_n}$ is the following state of total chord number $n_L+n_R=n$:
\begin{eqnarray} \label{eq:binom_state}
    \ket{\chi_n} := \sum_{k=0}^n (-1)^k \binom{n}{k}\ket{k,n-k},
\end{eqnarray}
and $\ket{\xi_{n-2m}}$ are states of total chord number smaller than $n$ which we shall not specify. For $n=0,1,2,3$ the second term in \eqref{eq:Op_KE_DSSYK} does not appear, and the Lanczos coefficients are given by $\bar{b}^2_n=\braket{\chi_{n-1}|\chi_{n-1}}/(\bar{b}_1^2 \dots \bar{b}^2_{n-1})$. For $n>3$, and general $\lambda$, the \textit{tails} $\ket{\xi_{n-2m}}$ that capture the projection of the Krylov elements $|\psi_n\rangle$ over sectors of total chord number smaller than $n$ are in general not vanishing and can be computed numerically, cf. \citet{Ambrosini:2024sre} for details. Therefore, in particular, at arbitrary finite $\lambda$ the Krylov complexity eigenstates are \textit{not} total chord number eigenstates.

In the semiclassical limit $\lambda\to 0$ (keeping $\widetilde{q}$ fixed), a significant simplification occurs as shown by \citet{Ambrosini:2024sre}. In this limit\footnote{Formally speaking, the semiclassical limit is defined as the limit in which $\lambda\to0$ and $n\to\infty$ with $\lambda n$ fixed, turning the discrete Krylov chain into a continuous semi-infinite interval, cf. \citet{IV,Ambrosini:2024sre}.} it can be inductively proved that the Krylov elements take the schematic form
\begin{eqnarray}
    \ket{\psi_n}=\frac{1}{\prod_{i=1}^{n}\bar{b}_i}\ket{\chi_n}+\text{terms subleading in $\lambda$}~,
\end{eqnarray}
where the statement that the tail terms are subleading refers to their contribution to the norm of the Krylov element. Thus, in this limit the Krylov basis elements \textit{are} total chord number eigenstates, and operator K-complexity can be identified with total chord number at the level of operators restricted to the relevant Krylov space.
The Lanczos coefficients are given by
\begin{eqnarray}
    b_n \sim_{\lambda\sim 0} \frac{2J}{\sqrt{\lambda}}\sqrt{\frac{1-q^{n/2}}{1-q}\left(1-\tilde{q} q^{n/2}\right)}~,
\end{eqnarray}
where the sign ``$\sim_{\lambda\sim 0}$'' denotes asymptotic equivalence in the limit $\lambda\to 0$.
The operator K-complexity can be computed using the continuous Krylov space approximation controlled by the small parameter $\lambda$:
\begin{eqnarray}\label{eq:DSSYK_KC_minusEvol}
   \lim_{\lambda\to 0} \lambda C_K(t) = 2\log\left[ 1+ (1-\tilde{q}) \sinh^2(J\,t)\right].
\end{eqnarray}
Importantly, for $\tilde{q}\to1$, which represents a small operator, there is an initial period of exponential growth that results in a \textit{time delay} effect, after which K-complexity transitions to linear growth. This is the imprint of operator scrambling. See \citet{Ambrosini:2024sre} for a more detailed canonical analysis of this effect in terms of an unstable potential in the classical Krylov chain.

Another operator-related result for K-complexity in DSSYK, is the Krylov problem for the operator inserted on the maximally entangled state, $\mathcal{O}\ket{\Omega}$ with time evolution generated by the two-sided Hamiltonian $H_L+H_R$. In the effective averaged theory of DSSYK, the latter evolution gets mapped to $|\psi(t)\rangle = e^{-it(H_R+H_L)}|0,0\rangle$. Just like in the previous analysis, Krylov basis elements, denoted $|\psi_n^{+}\rangle$, are found to be eigenstates of total chord number \textit{only} in the semiclassical $\lambda\to$ limit with $\widetilde{q}$ fixed:
\begin{eqnarray}
    \ket{\psi_n^{+}}&=&\frac{1}{\prod_{i=1}^{n}\bar{b}_i^{+}}\ket{\chi_n^{+}}+\text{terms subleading in $\lambda$}, \nonumber\\
    \\
     b_n^{+} &\sim_{\lambda_\sim 0}& \frac{2J}{\sqrt{\lambda}}\sqrt{\frac{1-q^{n/2}}{1-q}\left(1+\tilde{q} q^{n/2}\right)},
\end{eqnarray}
where $b_n^{+}$ are the Lanczos coefficients, $\bar{b}_n^{+}= b_n^{+}/(J/\sqrt{\lambda})$ and $\ket{\chi_n^{+}}:=\sum_{k=0}^n\binom{n}{k}\ket{k,n-k}$. K-complexity in this limit can in turn be computed via the continuum Krylov space approximation:
\begin{eqnarray}\label{eq:DSSYK_KC_plusEvol}
   \lim_{\lambda\to 0} \lambda C_K(t) = 2\log\left[ 1+ (1+\tilde{q}) \sinh^2(J\,t)\right]~,
\end{eqnarray}
and does not exhibit any scrambling period, as may be expected from the complexity growth of a state. In particular, when $\widetilde{q}\to1$ it reduces to the matterless result \eqref{eq:DSSYK_KC_matterless}.

We emphasize that the fact that the operator Krylov elements are eigenstates of total chord number in the semiclassical limit implies that operator K-complexity, namely equations \eqref{eq:DSSYK_KC_minusEvol}-\eqref{eq:DSSYK_KC_plusEvol}, is itself equal to total chord number in the same regime for both the time evolution protocols considered in the one-particle sector.

\subsection{A microscopic bulk-boundary correspondence for complexity}\label{subsec.BulkComplexity}

In the previous subsection we have reviewed the following two identifications:
\begin{itemize}
    \item K-complexity of the maximally entangled state equals chord number, as an operator statement, in the zero particle sector of the chord algebra and for arbitrary $\lambda$.
    \item K-complexity of the maximally entangled state perturbed by an operator insertion evolving under either $H_L-H_R$ and $H_L-H_R$, and total chord number in the one-particle sector, \textit{only} in the semiclassical limit $\lambda\to 0$. 
\end{itemize}
Combining these identifications with the bulk-boundary Hilbert space correspondence \cite{Berkooz:2018jqr,Lin:2022rbf} that maps chord number to bulk length in two-dimensional gravity will result in the proof of correspondence between Krylov complexity and bulk length. We will review this next.

We begin by noting that, in the matterless sector, the DSSYK Hamiltonian \eqref{eq:H_DSSYK_Krylov}, written in terms of the chord number operator, which we have shown to be equal to the position operator on the Krylov chain, admits a suggestive rewriting, already noted by \citet{Lin:2022rbf}, by defining a (dimensionless) length operator $\hat{l}:=\lambda n$ and conveniently writing the displacement operator $D^\dagger$ in terms of its canonically conjugate momentum $k$. The result is: 
\begin{eqnarray} \label{eq:H_DSSYK_k_l}
    H = \frac{J}{\sqrt{\lambda(1-q)}}\left(e^{i\lambda \hat{k}} \sqrt{1-e^{-\hat{l}}}+\sqrt{1-e^{-\hat{l}}}e^{-i\lambda \hat{k}}\right)~.\nonumber\\
\end{eqnarray}
The can be understood as the bulk Hamiltonian \cite{Lin:2022rbf},
where by construction $\hat{l}$ can be simultaneously interpreted as the bulk length operator (in AdS units) and as $\lambda \widehat{C_K}$. This results in the identification
\begin{equation}
    \label{eq:KC_length_corresp}
    \widehat{l} = \lambda \widehat{C}_K
\end{equation}
in the one-particle sector. In particular, note that the above is an operator identification that was reached thanks to the correspondence between Krylov, chord and eventually length eigenstates. Additionally, the fact that the K-complexity eigenvalue of the zero-chord state is equal to zero \textit{fixes} the regularization ambiguity in the bulk length operator such that its eigenvalue is also zero for the same state. We shall see below that this specifically selects the bulk length behind the black hole horizon\footnote{See appendix B of \citet{IV} for some further details about the relation between K-complexity and the length regularization scheme in the specific JT limit of the Hamiltonian \eqref{eq:H_DSSYK_k_l}.\label{footnote_reg}}.

Towards the end of the section we will say some words on the bulk gravitational theory described by the Hamiltonian \eqref{eq:H_DSSYK_k_l}. For now, let us begin by giving an explicit presentation of its low-energy regime, relevant to JT gravity.

\subsubsection*{Triple-scaled SYK and JT gravity}

JT gravity on the disk is the low-energy description of the Hamiltonian \eqref{eq:H_DSSYK_k_l}. As noted by \citet{Lin:2022rbf}, it is possible to zoom in to this energy regime by the means of the so-called \textit{triple-scaling} limit of SYK \cite{Lin:2022rbf}, which is a double scaling of DSSYK itself:
\begin{eqnarray}
    \lambda\to0, \quad l\to \infty,\quad \text{and}\quad  \frac{e^{-l}}{(2\lambda)^2}=: e^{-\tilde{l}}\quad\text{fixed},
\end{eqnarray}
where the renormalized length $\tilde{l} = l - 2 \log[1/(2\lambda)]$ was defined\footnote{See footnote \ref{footnote_reg}.}.

In this limit the Hamiltonian \eqref{eq:H_DSSYK_k_l} takes the form
\begin{eqnarray} \label{eq:H_TS}
    H= E_0+2 \lambda J \left(\frac{\tilde{k}^2}{2}+2e^{-\tilde{l}} \right)+O(\lambda^2),
\end{eqnarray}
where $E_0 = -\frac{2J}{\lambda}+O(\lambda^0)$ is the ground state energy\footnote{In going from \eqref{eq:H_DSSYK_k_l} to \eqref{eq:H_TS} the replacement $H\to -H$ has been done to ensure that the triple-scaled Hamiltonian is bounded from below \cite{Lin:2022rbf}.  This may also be seen as an expansion around the momentum modes that minimize energy.}. This is the low energy regime of DSSYK at $\lambda\to 0$, where the spectral density is proportional to $\sinh(2\pi \sqrt{E})$ \cite{Berkooz:2018qkz, Bagrets:2016cdf}. As we shall review below, both this Hamiltonian and the associated density of states are equal to their counterparts in JT gravity.

The identification between bulk length eigenstates and Krylov basis elements remains correct after the triple scaling limit \cite{I}. One may qualitatively understand this given that at the end of the day the triple-scaling limit is no more than a zooming-in on the low-energy regime of DSSYK. More technically, the eigenstates of renormalized length, $\ket{\tilde{l}}=\Big|l-2\log(1/2\lambda)\Big\rangle$, are identified with the position eigenstates on the correspondingly shifted Krylov basis, since we had $\widehat{l}=\lambda \widehat{C}_K$. They are the Krylov basis elements of the $\beta=0$ thermofield-double state\textit{ of the low-energy Hamiltonian}, which itself is a low-temperature TFD state of the full Hamiltonian, see \citet{IV} for details, including an explicit derivation of its temperature, which is of order $\lambda$. 
As mentioned above, at small $\lambda$, $l$ (and thus $\tilde{l}$) becomes continuous and additionally the wave function dynamics becomes classical. This allows for a classical computation of K-complexity by solving the equations of motion for $\tilde{l}(t)$ as determined by the Hamiltonian \eqref{eq:H_TS}, which is equivalent to the continuum approximation described in section \ref{subsect:Cont_Approx} \cite{IV}. The relevant initial conditions, given that this is a K-complexity computation, are $\dot{\tilde{l}}(0)=0$ enforcing symmetry of K-complexity\footnote{This has been shown to be a property of K-complexity of unitary evolution, cf. \citet{Sanchez-Garrido:2024pcy}.} around $t=0$ and $\tilde{l}(0)=\tilde{l}_0$; strictly speaking K-complexity starts at zero by definition, but we allow for some shift freedom $\tilde{l}_0$ in the definition of the triple-scaled Krylov chain. The result is:
\begin{eqnarray} \label{eq:KC_TS}
    \lambda \tilde{C}_K(t)\! =\! \tilde{l}(t) = 2 \log \left[\cosh\left(\sqrt{\lambda J E}\,t\right)\right]- \log\Big( \frac{E}{4\lambda J}\Big)\nonumber\\
\end{eqnarray}
where $E=4\lambda J e^{-\tilde{l}_0}$ is the energy of the solution.

By construction, the above results match the JT gravity prediction because they have been computed out of the triple-scaled Hamiltonian generating dynamics on the Krylov chain, whose position operator is equal to the length operator. Nevertheless, for completeness, we now turn to provide an independent description in the context of JT gravity \cite{Jackiw:1984je, Teitelboim:1983ux}, see \citet{Mertens:2022irh} for an instructive review. This is a theory of two-dimensional gravity with a dilaton scalar field $\Phi$, which plays the role of a Lagrange multiplier fixing the metric $g_{\mu\nu}$ to be locally AdS$_2$:
\begin{align}\label{eq.JTAction}
    S_{JT} =& \int_\mathcal{M} d^2 x \sqrt{-g}\Big[\Phi_0 R+\Phi(R+2) \Big]\nonumber\\
    &+2 \int_{\partial\mathcal{M}}dx\sqrt{\gamma}\Big[\Phi_0 K +\Phi(K-1) \Big]~,
\end{align} 
where $\gamma$ is the induced metric on the boundary and $K$ is the extrinsic curvature. The following boundary conditions fix the induced metric and the value of the dilaton on the boundary, while $\epsilon$ acts as boundary regulator:
\begin{equation}
    ds^2\Big|_{\partial \mathcal{M}} = -\frac{dt_b^2}{\epsilon^2}\, , \quad \Phi\Big|_{\partial \mathcal{M}} = \frac{\phi_b}{\epsilon}.
\end{equation}
As mentioned, variation with respect to the dilaton sets the curvature to $R=-2$ which in two-dimensions fully determines the metric to be AdS$_2$, while varying the metric yields an equation of motion that determines the profile of the dilaton over spacetime. Although there is no singularity, an observer-dependent horizon can be defined as seen by an accelerated observer on each boundary\footnote{As we shall discuss towards the end of this review, these coordinate patches can be obtained in schemes of dimensional reduction from higher-dimensional black holes. Nevertheless, from the exclusive point of view of two-dimensional gravity, there are no black hole solutions and one instead has an accelerated observer.}. Such observers will define left- and right- boundary times and each will see only a wedge of spacetime up to their respective horizons; on the horizon they will measure the value of the dilaton to be $\Phi_h$. The length of a spacelike geodesic connecting the left boundary with the right boundary at the same boundary time $t_L=t_R=t_b$ can be computed using global coordinates which cover the entire spacetime. This length is infinite because the boundaries are at infinity, but with the help of the $\epsilon$ regulator it can be normalized to a finite value:
\begin{eqnarray} \label{eq:renorm_length_JT}
    \tilde{l}(t_b) = 2 \log \left[\cosh\left(\sqrt\frac{E}{2\phi_b}\,t_b \right)\right]-2\log \left(\frac{\phi_b\,E}{2}\right),
\end{eqnarray}
where $E=2\Phi_h^2/\phi_b$ is the energy on the boundary\footnote{This energy is the sum of the values of the ADM Hamiltonians on the two boundaries.}. We refer to this length as the \textit{wormhole length} because it connects two asymptotic boundaries which are each inaccessible to the accelerated observer on the other boundary. 

The classical phase space of JT gravity  can be expressed in terms of the renormalized length \eqref{eq:renorm_length_JT} and its conjugate momentum, $P$. With this, \citet{Harlow:2018tqv} constructed the Hamiltonian of JT gravity and found that it is given by a Liouville Hamiltonian:
\begin{eqnarray}\label{eq:H_JT}
    H = \frac{1}{\phi_b} \left( \frac{P^2}{2} +2 e^{-\tilde{l}} \right),
\end{eqnarray}
which is indeed the triple-scaled Hamiltonian of SYK, cf. \eqref{eq:H_TS}. At the quantum level, the Hilbert space of JT gravity can be described by states $\ket{\tilde{l}}$ of renormalized wormhole length, evolving under the Hamiltonian \eqref{eq:H_JT}
\cite{Harlow:2018tqv}. 

\subsubsection*{K-complexity of infinite-temperature TFD state in triple-scaled SYK = wormhole length in JT gravity}
The title of this section summarizes its contents. We remind the Reader of the details and further elaborate on the consequences and matching of parameters.

\citet{Lin:2022rbf} presented a bulk-to-boundary map between the chord states in triple-scaled SYK and the length eigenstates in JT gravity. As described above, \citet{IV} showed that the (normalized) chord states in triple-scaled SYK are the Krylov basis elements of the infinite-temperature TFD state of the JT Hamiltonian. These two relations map the Krylov basis elements in this problem with length eigenstates in JT gravity. As expected by construction, Krylov complexity in triple-scaled SYK \eqref{eq:KC_TS} and wormhole length in JT gravity \eqref{eq:renorm_length_JT} are seen to match exactly provided the suitable identification between the parameters of both bulk and boundary theories \cite{IV}. See the table below for a summary of this matching.

\begin{roundedboxw}[Boundary-bulk dictionary for K-complexity]
\begin{center}
\includegraphics[width=0.5\linewidth]{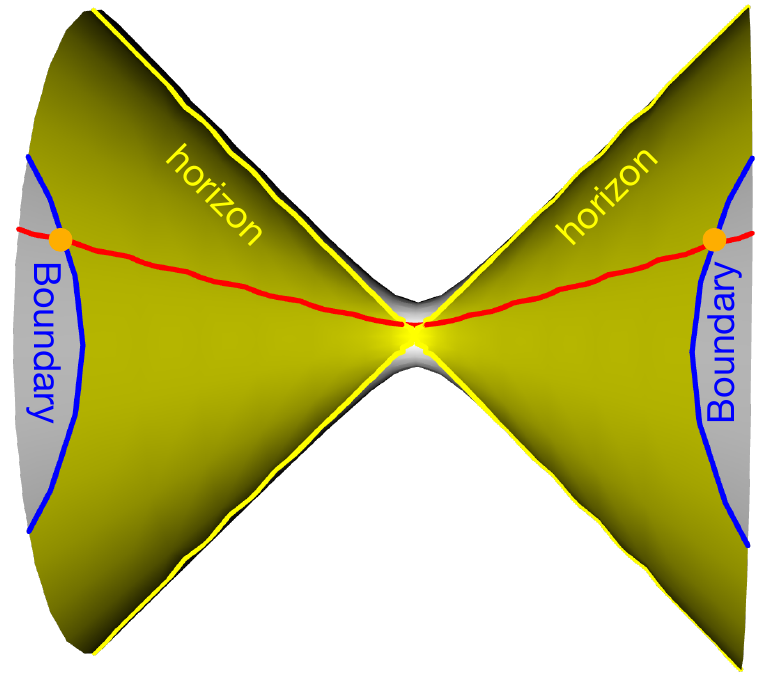}
\end{center}
The wormhole (red line) in JT gravity is a space-like geodesic extending from one of AdS$_2$'s regularized boundaries to the other (blue lines). It is a wormhole in the sense that it connects two causally disconnected space-time regions with respect to an observer accelerating on the boundary; such an observer sees a horizon (yellow lines).
    \begin{center}
\begin{tabular}{|p{3.5cm}|p{3.5cm}|}
\hline
    \textbf{Boundary} & \textbf{Bulk}  \\ \hline
    triple-scaled SYK & JT gravity  \\ \hline
    Hamiltonian Eq.~ \eqref{eq:H_TS} & 
    Hamiltonian Eq.~\eqref{eq:H_JT}\\ \hline
    Krylov basis are $|\tilde{l}\rangle$ states & Hilbert space consists of states with well defined wormhole length $|\tilde{l}\rangle$ \\ \hline
    K-complexity in the semiclassical limit: Eq.~\eqref{eq:KC_TS} & Wormhole length in JT gravity: Eq.~\eqref{eq:renorm_length_JT}\\
    \hline
    initial condition $\tilde{l}_0$ & $-2\log \Phi_h$\\
    \hline
    Energy $4\lambda J e^{-\tilde{l}_0}$ & ADM energy $2\Phi_h^2/\phi_b$\\
    \hline
\end{tabular}
\end{center}
\end{roundedboxw}

\subsubsection*{Operator K-complexity and JT gravity}

The analyses described earlier about operator K-complexity established that the Krylov bases associated to the evolution of the infinite-temperature thermofield double state perturbed by an operator insertion and evolving under either $H_R-H_L$ or $H_R-H_L$ are eigenbases of total chord number only in the semiclassical limit $\lambda\to 0$, cf. \citet{Ambrosini:2024sre}. Further taking the triple-scaling limit in order to go the the JT regime, the latter authors used the Lanczos coefficients to construct the Krylov space Hamiltonian by expressing, in terms of the position operator on the Krylov chain and its conjugate momentum, the continuous limit of the tridiagonal Hamiltonian \eqref{eq:H_tridiag} in operator form, additionally requiring that $\Delta$ in $\widetilde{q}=e^{-\lambda \Delta}$ is kept fixed as $\lambda \to 0$ in order to avoid strong backreaction due to the operator insertion. We refer the Reader to their article for details; here we may flash the result obtained for the specific choice of Hamiltonian $H=H_R+H_L$:
\begin{equation}
    \label{eq:triple_scaled_Hplus}
    H=E_0+4\lambda J\left(\frac{\tilde{k}^2}{2}+\Delta e^{-\tilde{l}/2}+2e^{-\tilde{l}}\right)+\mathit{O}(\lambda^2)~.
\end{equation}
This Hamiltonian, derived for the Krylov space position variable $\widetilde{l}$, matches exactly the one derived by \citet{Lin:2022rbf} summing $H_L$ and $H_R$ expressed through chord number operators interpreted as bulk length operators and using a shockwave-like approximation that assumes no backreaction of the shockwave sourced by the operator, allowing for the identification $\tilde{l}_L=\tilde{l}_R=\tilde{l}/2$ that enables the cancellation of some terms and lands directly on \eqref{eq:triple_scaled_Hplus}, see footnote 10 in \citet{Lin:2022rbf} as well as \citet{Maldacena:2018lmt}.
The Hamiltonian \eqref{eq:triple_scaled_Hplus} features a Morse potential, which differs from the Liouville potential in \eqref{eq:H_TS} by the term corresponding to an energy insertion $E_{shock}\propto \Delta\lambda$ \cite{Ambrosini:2024sre}. 

The analogous construction of the triple-scaled Krylov Hamiltonian for the evolution under $H_R-H_L$ can be found in \cite{Ambrosini:2024sre}. Its interpretation in terms of the Hamiltonian describing a bulk shockwave setup is still not entirely settled, at the moment of writing this review.

\subsubsection*{Beyond JT gravity}
While the above bulk-boundary dictionary for K-complexity is an extremely encouraging development, it is so far confined to the setting of two-dimensional JT gravity. It is evidently of great importance to extend the validity of the complexity correspondence beyond this setting, both in terms of bulk gravity models, as well as -- in future work -- to higher dimensions. Interesting progress is achieved in
\citet{Heller:2024ldz}, who extend the relationship between K-complexity and bulk length to sine-dilaton gravity. Sine-dilaton gravity is a 2D theory of gravity with action\footnote{\citet{Heller:2024ldz} use a notational convention such that their factor $q$ is the square root of the factor $q$ used by \citet{IV}. For consistency, in this review we have uniformly adapted the notation to the convention used by the latter.}
\begin{equation}\label{eq.SineDilatonAction}
  S = -  \frac{1}{2} \int \mathrm{d}^2 x \, \sqrt{g} \left( \Phi R + \frac{2\sin\left( |\log q| \, \Phi \right)}{|\log q|} \right)\,,
\end{equation}
which gives a one-parameter family of deformations of JT gravity \cite{Blommaert:2023opb,Blommaert:2024ymv}, reproducing the linear-dilaton potential of JT for $q\rightarrow 1$. Note that we have not written down the boundary terms, that are needed to give a well-defined variational principle. The analysis by \citet{IV} reviewed earlier proved a correspondence between Krylov basis elements of the $\beta=0$ TFD and normalized chord eigenstates at arbitrary $\lambda$ and, in particular, away from the JT limit; however, the corresponding bulk theory was not identified and this was achieved by \citet{Heller:2024ldz} using sine-dilaton gravity. In particular, they exploited the fact that the Hamiltonian of \eqref{eq.SineDilatonAction} matches that of DSSYK at arbitrary $\lambda$. Furthermore, they extend the results to finite-temperature by fixing the Krylov basis to be the one adapted to the infinite-temperature TFD state, and incorporating the preparation of the finite-temperature one as imaginary time evolution, as described in section \ref{Sec:TFD_state}. To summarize their results, the relation between wormhole length $\hat{l}$ and K-complexity at arbitrary temperature $C_K(t)_\beta$ that they find is:
\begin{eqnarray}
    \braket{\hat{l}} = \lambda \sum_n \Big|\braket{\psi_n| \frac{e^{-iH(t-i\beta/2)}}{\sqrt{Z_\beta}}|\psi_0} \Big|^2= \lambda C_K(t)_\beta\nonumber\\
\end{eqnarray}
where $|\psi_0\rangle=|\Omega\rangle$ is the $\beta=0$ TFD state, the Krylov elements remain chord number eigenstates $|\psi_n\rangle=\frac{|n\rangle}{\sqrt{\braket{n|n}}}$ as in \eqref{eq:Krylov_DSSYK}, and $H$ is the sine-dilaton Hamiltonian obtained through canonical quantization of the theory's classical phase space \cite{Blommaert:2024ymv}: 
\begin{equation}
    \hat{H}_{\text{grav}} = \frac{1}{ |\log q| \sqrt{1 - q}} \left[ -\cos(\hat{p}) + \frac{1}{2} e^{i \hat{p}} e^{-l} \right]\,.
\end{equation}
which is the Hamiltonian
\eqref{eq:H_DSSYK_k_l}, after a similarity transformation accounting for the differing normalization of chord states in \cite{Heller:2024ldz} and \cite{IV}, cf. \cite{Berkooz:2018jqr,Lin:2022rbf}. As noted in the discussion around equation \eqref{eq:KC_length_corresp}, the matching between length and K-complexity holds as an operator statement, where the definition of K-complexity selects the length regularization scheme that makes it vanish on the $|0\rangle$ state. The generalization to sine-dilaton gravity is an important result in a number of aspects, not least because it significantly increases one's confidence in the geometric bulk interpretation of Krylov complexity as a length, including in the quantum regime.

The works reviewed so far are the starting point for the analysis of the bulk-boundary correspondence between Krylov complexity in SYK and length in two-dimensional gravity. An earlier work that compared length in JT to operator K-complexity in SYK was done by \citet{Jian:2020qpp}, while studies in DSSYK include \citet{Anegawa:2024yia,Aguilar-Gutierrez:2024nau,Aguilar-Gutierrez:2025mxf,Aguilar-Gutierrez:2025pqp}. The potential arbitrariness of the choice of bulk length regularization scheme, which in the preceding discussions was fixed by the requirement of positive semi-definiteness of the K-complexity operator, may also be circumvented by studying the time derivative of the latter in connection to bulk momentum as explored by e.g. \citet{Caputa:2024sux,Fan:2024iop,Aguilar-Gutierrez:2025kmw} in contexts not necessarily related to DSSYK.

Finally, it is worth to mention that probing the saturation of wormhole length out of K-complexity in SYK might require going away from the double-scaling limit in order to have a finite-size system, which is consequently no longer analytically tractable through chord techniques; nevertheless, works such as \cite{Balasubramanian:2024lqk,Nandy:2024zcd, Miyaji:2025ucp} have proposed to model this finite-Hilbert-space effect by trading the SYK Hamiltonian by a matrix model designed to approximate its density of states in a controlled manner \cite{Jafferis:2022wez,Jafferis:2022uhu}. We will elaborate further on these aspects in the next subsection.

\subsection{Outlook: non-perturbative bulk lengths and complexity dynamics}\label{sec.BulkSaturation}
As we have seen, the understanding of Krylov complexity in many-body systems includes epochs of time for which a non-perturbative understanding is an essential necessity (see Section \ref{sec.CloserKrylov}). It would be desirable to develop an equally non-perturbative understanding of the bulk representation we developed in Section \ref{sect:Holography}. We now review what is known about the bulk wormhole length non-perturbatively and discuss its connection to K-complexity.
\subsubsection{Operators and wormholes at late times}
For a certain class of 2D dilaton gravity theories, there now exists a fine-grained non-perturbative understanding of the gravity path integral \cite{Saad:2019lba}. For concreteness, we will focus on the simplest among them, the JT gravity theory, whose action has been introduced in Equation \eqref{eq.JTAction} above. In keeping with topological diversity, one can evalute the path integral as a sum over 2D surfaces satisfying asymptotically AdS boundary conditions, ordered by their topological counting paramter, the genus $g$. The action contains a term $\chi(M) S_0$, so that effectively a 2D ``universe" ${\cal M}_{g,n}$, of genus $g$ with $n$ boundaries receives a topological suppression factor of $e^{-S_0 \chi ({\cal M}_{g,n})}$. The quantity $\chi$ is the Euler number of the  manifold in question, given by 
\begin{eqnarray}
    \chi\left({\cal M}_{g,n} \right) = 2g - 2 + n\,.
\end{eqnarray}
Concretely in JT gravity, the gravitational path integral can be solved in terms of the topological expansion
\begin{equation}
    \left\langle {\cal Z}(\beta_1) \cdots {\cal Z}(\beta_n) \right\rangle_{\rm con} = \sum_{g=0} \frac{{\cal Z}_{g,n}(\beta_1\,,\cdots\,, \beta_n)}{e^{S_0 (2g -2 + n)}}\,.
\end{equation}
The left hand-side of this expression corresponds to the gravity calculation, where we specify $n$ asymptotically AdS$_2$ boundaries, whose lengths are given by the $\beta_i$. The right-hand side shows that this quantity can be evaluated by summing contributions genus by genus, where each ${\cal Z}_{g,n}$ is given by an integral over the moduli space of all genus-$g$ Riemann surfaces with $n$ geodesic boundaries. Without going into the details (see \cite{Saad:2019lba}), let us quote what is perhaps the most striking result, namely the computation of the spectral form factor, $F(t) = {\cal Z}_{2}(\beta + it, \beta - it)$, which we can alternatively define as the Fourier transform,
\begin{equation}\label{eq.SFF}
    F(t) = \int dE d\omega \left\langle\rho\left(  E+\tfrac{\omega}{2}\right) \rho\left(  E-\tfrac{\omega}{2}\right) \right \rangle_{\rm con}e^{-i\omega t-2\beta E}\,,
\end{equation}
of the connected two-point correlation function of energy levels. A quantity like the one in angled brackets is natural when evaluating the spectral densities with respect to an ensemble, such as a random-matrix theory. The fact that JT gravity also gives rise to a non-vanishing connected spectral correlation is more puzzling, but has been intensely discussed over the years, in the context of the so-called `factorization problem'. The behavior of this quantity in JT gravity has been elucidated \cite{Saad:2019lba}, \cite{Altland:2020ccq, Altland:2022xqx}, both perturbatively and non-perturbatively in the coupling $e^{S_0}$, giving rise to the universal behavior
\begin{equation}\label{eq.sineKernel}
    \left\langle\rho\left(  E+\tfrac{\omega}{2}\right) \rho\left(  E-\tfrac{\omega}{2}\right) \right \rangle_{\rm con} =  \rho(E) \delta(\omega) - \frac{ \sin^2 \left(\pi \rho(E) \omega  \right)}{\pi^2\omega^2}\,.
\end{equation}
often referred to as the `sine kernel', first discovered by Dyson, \cite{dyson1962statistical}. Note that we have focused on the connected contribution only, but it is easy to add the disconnected term, which takes the form `$+1$' in the chosen normalization (see for example \cite{Altland:2021rqn}). Here $\rho(E)$ is the spectral density at energy $E$, which in JT gravity takes the form
\begin{equation}
    \rho(E) \sim e^{S_0} \sinh \left(2\pi \sqrt{E}  \right)\,.
\end{equation}
Going back to \eqref{eq.SFF}, and substituting the sine kernel expression \eqref{eq.sineKernel}, the first term gives rise to a constant `plateau' value, while the Fourier transform of of the sinc-squared term gives a contribution that rises linearly in time $\propto t$, that gets cut off once the function reaches zero. Putting the two together gives the characteristic `ramp-plateau' behavior universally seen in quantum chaotic systems \cite{Cotler:2016fpe}, \cite{Altland:2020ccq}. Let us note that the plateau phase of the spectral form factor corresponds to doubly non-perturbative physics in the bulk, which in JT has been clarified in a formulation using Kodaira-Spencer theory in \cite{Altland:2022xqx} and via the so-called `tau-scaling limit' in \cite{Blommaert:2022lbh,Saad:2022kfe}.

\subsubsection{Non-perturbative bulk length = Krylov?}
\begin{figure}
    \centering
    \includegraphics[width=0.9\linewidth]{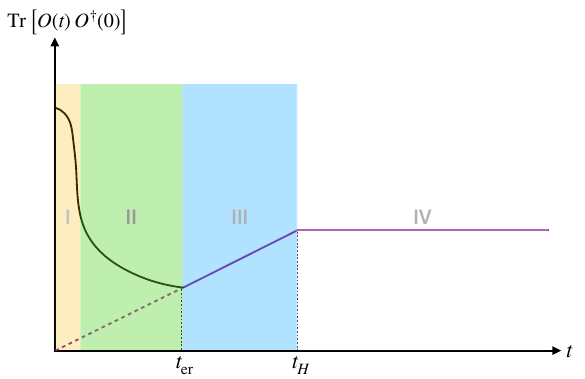}
    \caption{Non-perturbative behavior of an operator two-point function in a quantum chaotic system (figure taken from \cite{Altland:2021rqn}). We can see the ramp and plateau phase III and IV, which we focus on in this Section, preceeded by the decay of the disconnected and the connected two-point function at times prior to the Thouless time (sectors I and II).}
    \label{fig:OpTwoPtNonPert}
\end{figure}
Emboldened by these results, one may attempt to give a calculation of the non-perturbative physics of bulk Krylov complexity. An indication how this could proceed can be garnered by realizing that K-complexity is closely related to the non-perturbative operator two-point functions, via the moment expansion \eqref{Sec:LancCoef_Moments}, and such operator corelation functions are under control at late times in certain chaotic quantum systems and their holographic duals \cite{Saad:2019pqd}, \cite{Altland:2021rqn}. By using the (doubly) non-perturbative approach of the supersymmetric sigma model\footnote{The paper \cite{Saad:2019pqd} derives the late-time behavior of correlation functions from a combination of the ETH and RMT ansätze, and reproduces the late-time behavior from a gravity analysis in JT.},  these authors show that the late-time behavior of operator two-point functions can similarly be understood by convolving with a generalized form of the sine kernel, called the `operator sine-kernel', which takes the form
\onecolumngrid
\begin{align}
     \left\langle{\cal O}\left(  E+\tfrac{\omega}{2}\right) {\cal O}^\dagger
     \left(  E-\tfrac{\omega}{2}\right) \right \rangle_{\rm con} =  \left[\rho(E)\delta(\omega) + 1 - {\rm sinc}^2 \left( \pi \rho(E) \omega \right)  \right]
     \times \left({\rm tr}{\cal O} {\cal O}^\dagger - \frac{1}{D}{\rm tr}{\cal O} {\rm tr}{\cal O}^\dagger  \right)\,,
\end{align}
\twocolumngrid
where $D$ is the dimension of the Hilbert space. Let us take note that the expression is almost exactly the same as \eqref{eq.sineKernel} -- modulo the operator traces -- but that remarkably the `$+1$' has now snuck into the connected contribution.

While the correspondence between K-complexity in the boundary and bulk wormhole length is not yet established at the non-perturbative level, we can nevertheless get a good idea of how this should work, by constructing a non-perturbative bulk length $\langle  \ell (t) \rangle$ directly. We remind the Reader that in Section \ref{sect:Holography} we showed that bulk length corresponds to Krylov complexity in the DSSYK model, which roughly translates to the leading disk level in the present context. It is thus plausible, though not yet proven, that a suitably defined non-perturbative bulk length geometrises the non-perturbative behavior of Krylov complexity of Section \ref{sec.CloserKrylov} and \ref{Sec:Krylov_Pheno}. 

\begin{figure}
    \centering
    \includegraphics[width=0.9\linewidth]{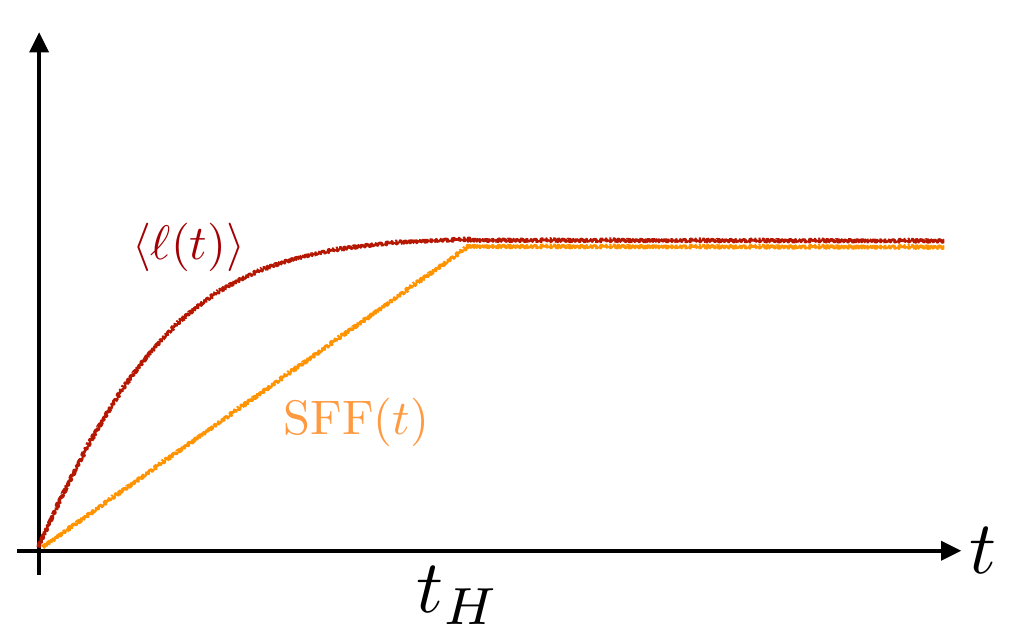}
    \caption{Non-perturbative bulk length, $\langle \ell(t)\rangle$ vs spectral form factor, SFF$(t)$. (figure adapted from \cite{Iliesiu:2021ari}). Note that both show a qualitatively similar behavior, plateauing at the Heisenberg time, but the detailed behavior of the ramp can differ (we show the ramp in the unitary symmetry class, and we have rescaled the plateau of both quantities so that they coincide).} 
    \label{fig:WHnonPert}
\end{figure}

The proposal of \cite{Iliesiu:2021ari}, which we review here, is to define $\langle \ell(t) \rangle$ as a sum over geodesic lengths stretching between to boundaries
\begin{equation}
 \left\langle \ell \right\rangle =  \lim_{\Delta \rightarrow 0} \left\langle \sum_\gamma \ell_\gamma e^{-\Delta \ell_\gamma}\right \rangle\,,
\end{equation}
where $\ell_\gamma$ is the length of any non self-intersecting geodesic stretching  between left and right boundaries and $\Delta$ serves as a regulator. The Reader can convince themselves that a disk level, this quantity becomes equivalent to the definition in Section \ref{subsec.BulkComplexity}. A key point is that one can now relate this geodesic sum to the single-boundary two-point function $G_\beta(t) =\left\langle {\cal O}_\Delta(\tfrac{\beta}{2} + it)  {\cal O}_\Delta^\dagger(0)\right \rangle $, via analytic continuation, and thus
\begin{eqnarray}\label{eq.nonpertBulkLength}
    \langle \ell(t)\rangle := -\lim_{\Delta\rightarrow 0} \frac{\partial}{\partial \Delta} \left\langle {\cal O}_\Delta(\tfrac{\beta}{2} + it)  {\cal O}_\Delta^\dagger(0)\right \rangle\,.
\end{eqnarray}
The resulting expectation value is evaluated in the bulk by summing over all single-boundary topologies as well as non-intersection geodesics anchored at boundary points $x_1$ and $x_2$ on each topology, as follows\footnote{This is true in the probe approximation of the operator ${\cal O}$, which we employ here.}
\onecolumngrid
\begin{equation}\label{eq.JTbulkLength}
    G_\beta(t) \sim \int dE_1 dE_2 \left\langle \rho(E_1) \rho(E_2)\right\rangle_{\rm JT}e^{-E_1(x_1-x_2)-E_2 (\beta - x_1 + x_2)} {\cal M}_{\Delta}(E_1, E_2)\,.
\end{equation}
\twocolumngrid
The expectation value in the angle brackets is just the two-point correlation of JT gravity we evaluated above, and the matrix element inside the integral reads
\begin{equation}
    {\cal M}_\Delta (E_1, E_2) = \frac{\Gamma(\Delta \pm is_1 \pm i s_2)}{2^{2\Delta+1} \Gamma(2\Delta)}\,,\quad s_i = \sqrt{2E_i}\,.
\end{equation}
The signs in the Gamma function are meant to indicate a product over four Gamma functions, one for each independent choice of the sign (see e.g. \cite{Jafferis:2022wez} for a more detailed explanation). Recalling the previous sub-section, the astute Reader may already be seeing in Equation \eqref{eq.JTbulkLength} the structure of the standard sine kernel -- in the guise of the JT level correlation, $\left\langle \rho(E_1) \rho(E_2)\right\rangle_{\rm JT}$, in conjunction with operator matrix elements. In fact, upon inserting the above expression into the expression for the non-perturbative bulk length \eqref{eq.nonpertBulkLength}, one finds exactly such an expression, namely
\begin{equation}
    \left \langle \ell(t)\right\rangle  \sim \int ds d\omega e^{-is\omega t - \beta s^2/2}\frac{\rho(s)^2}{\omega^2} \left[1 - \frac{\sin^2 \left( \pi \omega \right) }{\pi^2 \rho(s)^2\omega^2 } \right]\,,
\end{equation}
where we have absorbed the constant contribution together with the disconnected part into the value of the length at time zero. The authors of \cite{Iliesiu:2021ari} have performed the calculation in full, keeping track of all factors, and we display their result in Fig. \ref{fig:WHnonPert}

This non-perturbative saturation of the bulk length, and thus any associated complexity, is accompanied by quantum fluctuations. In fact, \cite{Iliesiu:2021ari} demonstrate that these fluctuations of the bulk length increase with time, rather than saturating, as is for example the case for the spectral form factor. In \cite{Stanford:2022fdt,Iliesiu:2024cnh} it is argued that these fluctuations can affect the experience of the infalling observer. In fact, the leading correcting to the bulk length, the disk partition function with a handle attached, implies that there are fluctuations that shorten the length of the wormhole, and thus reduce the associated complexity, by the emission of a baby universe. This process was previously analyzed in the context of late-time operator correlations functions at ETH \cite{Saad:2019pqd}. In the present case it is argued, that this shortening fluctuation at late times can be severe enough to make the black hole fluctuate into a state of `negative age', which would correspond to the infalling observer encountering a firewall \cite{Stanford:2022fdt}. It would be interesting to revisit this discussion in the context of a non-perturbative bulk dual of Krylov complexity. Progress in this direction would involve a bulk computation linking the results of \cite{Jafferis:2022uhu,Jafferis:2022wez}, \cite{Balasubramanian:2024lqk},\cite{Nandy:2024zcd}, to the wormhole shortening phenomenon described above as the bulk mechanism of the saturation.

\section{Discussion}\label{Sect:Discussion}
Before concluding this review, let us pause and reflect on what has been learned.

First and foremost, we hope to have convinced the Reader that the set of mathematical and numerical techniques associated to the names of Krylov and Lanczos give rise to a powerful new complexity measure, and moreover one that is useful in practice not only as a quantifier of quantum complexity but also a sensitive probe of chaotic dynamics. This relies on several features, which we developed in this review, but it is worth highlighting once again two of them: firstly, K-complexity is an extremely natural notion of measuring the complexity of time evolution of a quantum system, given that its very nature derives from quantum evolution and is defined via an optimal basis that iteratively only adds those directions in Hilbert space that are absolutely necessary to accurately capture time evolution of a given operator (in the Heisenberg picture), or initial state (in the Schrödinger picture). It is defined without the need of an a-priori arbitrary cut-off parameter, which would be the case for gate complexity, for example. The precise sense in which the Krylov base and its associated complexity are an ideal choice is captured in two optimality theorems which we presented in Section \ref{sec.CloserKrylov}.

Secondly, at least in the arena of low-dimensional gravity, K-complexity is a notion of complexity, defined from first principles in the boundary theory, which can be matched across the AdS/CFT dictionary. As we explained in Section \ref{subsec.BulkComplexity}, this complexity correspondence holds at the full quantum level, by matching the bulk length {\it operator} to the boundary operator whose expectation value gives the average position on the Krylov chain, that is Krylov complexity. It can then be shown  that in the semiclassical limit the holographic correspondence between gravity and QFT geometrizes complexity into classical bulk length, and achieves for complexity what the RT formula has achieved for entanglement entropy.

Both of these showcase results are well established for the case where the complexity is evaluated in many-body quantum systems and in the case of holographic duality, where the bulk is a lower-D quantum gravity theory. As we reviewed in Section \ref{subsect:KC_QFT} progress has also been achieved in evaluating K-complexity for QFTs in higher dimensions, but this area remains an active area of investigation with many open problems to attack. Similarly it would be highly desirable to extend the successes of matching K-complexity between bulk and boundary to higher-dimensional holographic setups, not least because because the setup of Section \ref{subsec.BulkComplexity} has not enough structure to distinguish between the various conjectures on holographic complexity.

Another major set of results that have been established over the years concern the dynamics of K-complexity, both for operators and for states, which reveal $C_K(t)$ as a sensitive probe of quantum chaos. This is true both at early times, where the exponential growth rate of complexity implied by a phase of linear growth of the Lanczos coefficients $b_n \sim \alpha n$ diagnoses Lyapunov chaos, as well as at intermediate to late times, where the linear behavior of $C_K(t)$ itself shows chaotic post-scrambling dynamics that eventually gives way to a saturation, whose `plateau' value is sensitive to whether the underlying system is integrable or chaotic. The late-time saturation value can be physically understood in terms of an effectively localising behavior of a pseudorandom off-diagonal Anderson-type hopping problem on the Krylov chain. Despite these promising successes, the use of Krylov complexity as a measure of chaos is still in its infancy and we expect a wealth of new developments in this direction in the coming years.  

The perspective of long-time dynamics of chaotic quantum systems examined under the lens of (K-)complexity also offers the opportunity to reflect further about the universal nature of the phenomenology that has been revealed and reviewed in this article. On a number of occasions we have discussed the early time behavior of complexity (up to scrambling time), with its relation to operator growth and scrambling dynamics, and pointed out that late-time behavior (post scrambling time) of complexity remained non-trivial despite the fact that probes of thermalization have reached their equilibrium value. The results reviewed in Section \ref{Sec:Krylov_Pheno} underpin the linear phase of post-scrambling complexity growth and later saturation with the physics of ETH and random-matrix theory, which are both manifestations of the chaotic structure of Hilbert space, and sensitive to non-perturbative ${\cal O}(e^{-S})$, and even doubly non-perturbative ${\cal O}(e^{-e^S})$ effects on the exact spectrum of the theory. This underlines the power of focusing on the physics of complexity of late-time physics in chaotic quantum systems. However, it is also worth mentioning that if one is willing to work with such exquisite accuracy, one will eventually also discern non-trivial effects in probes we previously discarded as insensitive: for example, while time-ordered correlation functions will indeed thermalize and decay to their equilibrium values after $t_\beta$, they will eventually start fluctuating chaotically (Section \ref{sec.HoloComplexity}), at characteristic values of ${\cal O}(e^{-S})$ and on average exhibit the famous `ramp-plateau' physics that reflects level correlations and operator matrix element statistics. Complexity remains a probe that is particularly sensitive to physics at such late times, but level correlations of quantum-chaotic systems can be seen as an overarching physical principle that helps us understand this behavior, as well as its associated far subtler manifestations in more ordinary probes, such as correlations functions and entanglement entropy.

Areas in which K-complexity and Krylov methods have produced new insights in recent years, but which we have not extensively covered in this review include open systems and non-unitary dynamics (which are reviewed in \cite{Nandy:2025ktk}), quantum simulation (see e.g. \cite{cortes2022quantum}), applications to many-body localization (see e.g. \cite{Balasubramanian:2024ghv}), black-hole thermalization (see e.g. \cite{Dodelson:2025rng}),  and a multitude of other fascinating and important contributions beyond the scope of this review. It is certainly an exciting time for Krylov complexity, and many important developments are appearing at an impressive pace. Thus, before ending this review we step back and attempt to give a panoramic overview of emergent directions in the field of Krylov and Lanzcos methods.

\subsection{Quo navigas, Krylov?}
 This review has given an overview of only a selection among the large amount of work and results that have already emerged in the nascent field based on Krylov and Lanczos methods and the associated theory of K-complexity. In addition this area is thriving and constantly expanding into new directions. Which directions would one expect the field to turn and expand into in the near, intermediate and long term? Such a `laundry list' of topics, as a part of a larger review, is of course condemned to become obsolete, especially in an area as active as K-complexity. We nevertheless attempt to round out the present article with a number of directions and areas that we would be delighted to see progress on in the coming years -- and perhaps even beyond. 

An area that is in need of further study is the theory, phenomenology and utility of K-complexity in QFT, where preciously few concrete results are available, especially on the highly interesting issue of long-term dynamics of K-complexity in chaotic and integrable field theories. Is K-complexity an interesting diagnostic of chaotic dynamics? Can one observe a similar Krylov-localization and under-saturation of late-time K-complexity as we have reviewed to be observed in many-body quantum dynamics? Developing K-complexity in QFT is a particularly interesting and important topic, given that a-priori its definition, e.g. in terms of moments of correlation functions or moments of the Hamiltonian itself in particular states, carries over to higher dimensions without apparent need for modification. On the other hand, quantum field theory is rife with subtle surprises, and even the notion of well-defined inner products on the space of operators will require a careful mathematical analysis before more significant progress can be achieved in higher-dimensional settings, to name but one potential obstacle one expects to have to clear along the path to higher dimensions.

In the same line of work, establishing the K-complexity-bulk duality in general dimensions would be highly desirable, elevating K-complexity to the first well-defined dual pair on both the gravity side and the dual field theory side of the AdS/CFT correspondence in all dimensions. Important groundwork has been laid in JT gravity, and DSSYK, as reviewed in Sections \ref{sec.CloserKrylov} and \ref{sect:Holography}.  In this vein, it would be enlightening to investigate the fate of the K-complexity duality in setups where a low-energy AdS$_2$ duality arises from dimensional reduction of a higher-dimensional gravity theory (see e.g. \cite{Nayak:2018qej}). Furthermore, as we have described in this article, the Krylov basis plays a distinguished role in the holographic encoding map in JT gravity, in fact (see Section \ref{sect:Holography}), the bulk Hilbert space can be efficiently constructed as being spanned by the chord-number basis which happens to coincide exactly with the boundary Krylov basis. It would very interesting to reveal and study the importance of the Krylov basis in higher-D bulk reconstruction, which is another area that we expect to see further progress going forward.

At their very origin Krylov and Lanczos methods (and later, K-complexity) have been developed in an effort to optimize the computational resources needed to perform certain numerical computations in linear algebra, and have consequently soon entered the realm of quantum dynamics. 
Given that notions of complexity as a barrier to bulk reconstruction has been an active topic of research in holographic duality in recent years, we may also expect to learn more concrete manifestations of the barrier to efficient computation implied by precise K-complexity-bulk geometry dualities, in particular at the non-perturbative level.

Regardless which research direction first leads to progress, or which unexpected surprises may turn up in the course of pursuing them, by now K-complexity has transcended its `humble origins' as a computational tool into a conceptually enlightening physical quantity of its own right, with an ever-increasing list of impressive applications and important implications across a large range of disciplines, charting the course from quantum complexity to quantum gravity. We look forward to much future progress, following the course outlined above, but above all heading in new surprising directions, that would have been, one suspects, entirely unforeseen by its originator, the naval engineer A. Krylov.

\section*{Acknowledgments}
We would like to thank our many colleagues who have shaped the material in this review both via discussions and with their own manifold contributions to the literature. We thank Sergio Aguilar-Gutiérrez, Marco Ambrosini, Vijay Balasubramanian, Jose Barbón, Rahel Baumgartner, Alexandre Belin, Martí Berenguer, Micha Berkooz, Elena Caceres, Hugo Camargo, Pawel Caputa, Shira Chapman, Chandramouli Chowdhury, Ben Craps, Nicolas De Ro, Gabriele Di Ubaldo, Anatoly Dymarsky, Johanna Erdmenger, Altay Etkin,  Oleg Evnin, Damián Galante, Felix Haehl, Michal Heller, Maria Knysh, Javier Magan, Javier Mas, Ioannis Matthaiakakis, Mark Mezei, Pratik Nandy, Jacopo Papalini, Gabriele Pascuzzi, Juan Pedraza, Pietro Pelliconi, Alfonso Ramallo, Andrew Rolph, Juan Santos-Suárez, Tim Schuhmann, Misha Smolkin, Michael Sonner, Benjamin Strittmatter, Leonard Susskind, Brian Swingle, Kazutaka Takahashi, Thomas Tappeiner and Benjamin Withers for numerous insightful conversation over the years.

This work has received support through the Swiss Quantum Initiative awarded by the State Secretariat for Economic Affairs, under the grant "HoloGraph". This research is supported in part by the Fonds National Suisse de la Recherche Scientifique (Schweizerischer Nationalfonds zur Förderung der wissenschaftlichen For\-schung) through the Project Grant 200021\_215300 and the NCCR51NF40-141869 The Mathematics of Physics (SwissMAP). JS thanks Harvard University for support through a Bershadsky Distinguished Visiting Fellow award. ASG is supported by the UKRI Frontier Research Guarantee [EP/X030334/1].

\newpage

\bibliography{ref}

\end{document}